\documentclass[11pt]{ruthesis}

\usepackage{notoccite}
\usepackage{graphicx}
\usepackage{rotating}
\usepackage{lscape,longtable}

\usepackage{latexsym}
\usepackage{amsmath}
\usepackage{topcapt}
\usepackage{amssymb}
\usepackage{multirow}
\usepackage{amsbsy}
\usepackage{rotating}
\usepackage[toc,page]{appendix}
\usepackage{lineno}
\usepackage{hyperref}
\usepackage{color}
\usepackage{url}
\hypersetup{linktocpage}
\hypersetup{
    colorlinks,
    citecolor=blue,
    filecolor=black,
    linkcolor=blue,
    urlcolor=blue
}

\DeclareMathAlphabet{\mathsfsl}{OT1}{cmss}{m}{sl}
\extrafloats{400}
\makeatletter
\renewcommand{\@chapapp}{}
\newenvironment{chapquote}[2][2em]
  {\setlength{\@tempdima}{#1}%
   \def\chapquote@author{#2}%
   \parshape 1 \@tempdima \dimexpr\textwidth-2\@tempdima\relax%
   \itshape}
  {\par\normalfont\hfill--\ \chapquote@author\hspace*{\@tempdima}\par\bigskip}
\makeatother

\newcommand{\gev}{$[\rm{GeV}/c]$} 
\newcommand{\gevc}{$[\rm{GeV}/c^{2}]$} 
\newcommand{\pt}{$p_{T}$} 
\newcommand{\raa}{$R_{\rm{AA}}$} 
\newcommand{\rpa}{$R_{\rm{pA}}$} 
\newcommand{\taa}{$T_{\rm{AA}}$} 
\newcommand{\akt}{anti-k$_{t}$}
\newcommand{\teta}{$\eta_{\rm{CM}}$}
\newcommand{\RpAPythia}{$R^{PYTHIA}_{pA}$}
\newcommand{\hd}{{\sc{HYDJET}}}
\newcommand{\pyhd}{{\sc{PYTHIA+HYDJET}}}
\newcommand{\py}{{\sc{PYTHIA}}}
\newcommand{\jw}{{\sc{JEWEL}}}
\newcommand{\jwpy}{{\sc{JEWEL+PYTHIA}}}

\newcommand{\NA}{---}
  
\begin{document}
\copyrightpage
\phd \title{Jetting Through The Primordial Universe}
\author{Raghav Kunnawalkam Elayavalli}
\program{Physics and Astronomy}
\director{Prof. Sevil Salur}
\approvals{4}
\submissionyear{2017}
\submissionmonth{October} 

\abstract{
Collisions of heavy ion nuclei at relativistic speeds (close to the speed of light), sometimes referred to as the ``little bang", can recreate conditions similar to the early universe. This high temperature and very dense form of matter, now known to consist of de-confined quarks and gluons is named the quark gluon plasma (QGP). An early signature of the QGP, both theorized and seen in experiments, was the aspect of ``jet quenching" and understanding that phenomenon will be the main focus of this thesis. The concept behind quenching is that a high energetic quark or gluon jet undergoes significant energy loss due to the overall structure modifications related to its fragmentation and radiation patterns as it traverses the medium. The term jet, parameterized by a fixed lateral size or the jet radius, represents the collimated spray of particles arising from an initial parton. In this thesis, Run1 experimental data from pp and heavy ion collisions at the CERN LHC is analyzed with the CMS detector. Analysis steps involved in the measurement of the inclusive jet cross section in pp, pPb and PbPb systems are outlined in detail. The pp jet cross section is compared with next to leading order theoretical calculations supplemented with non perturbative corrections for three different jet radii highlighting better comparisons for larger radii jets. Measurement of the jet yield followed by the nuclear modification factors in proton-lead at 5.02 TeV and lead-lead collisions at 2.76 TeV are presented. Since pp data at 5.02 TeV was not available in Run1, an extrapolation method is performed to derive a reference pp spectra. A new data driven technique is introduced to estimate and correct for the fake jet contribution in PbPb for low transverse momenta jets. The nuclear modification factors studied in this thesis show jet quenching to be attributed to final state effects, have a strong correlation to the event centrality, a weak inverse correlation to the jet transverse momenta and an apparent independence on the jet radii in the kinematic range studied.  These measurements are compared with leading theoretical model calculations and other experimental results at the LHC leading to unanimous agreement on the qualitative nature of jet quenching. This thesis also features novel updates to the Monte Carlo heavy ion event generator JEWEL (Jet Evolution With Energy Loss) including the boson-jet production channels and also background subtraction techniques to reduce the effect of the thermal background. Keeping track of these jet-medium recoils in JEWEL due to the background subtraction techniques significantly improves its descriptions of several jet structure and sub-structure measurements at the LHC. 
}

\beforepreface

\acknowledgements{
Teachers and mentors are amongst the most important catalyst any student needs to become a successful individual in society. I have been most fortunate to have had the chance to interact with some of the best teachers any young, restless and (often) troublemaking student could hope for. I shall always be indebted to Mrs. Vijayalakshmi Raman, Mr. Sadagopan Rajash, Prof T. R. Subramaniyam, Prof Raghavan, Prof Derin Sherman, Prof Abhay Deshpande, Prof Sevil Salur, Prof Ron Gilman and Dr. Korinna Zapp amongst many others for anchoring my wavering mind with an insatiable gift of logical curiosity, rigorous problem solving and the importance of a righteous character. Without their presence and guidance I truly believe I would not be the person I am today and for that I'm immensely grateful. 

I want to mention my special gratitude to my PhD advisor Prof. Sevil Salur. Not only did I learn from her how to think and approach research but more importantly, I learnt how to be an efficient communicator, an effective mentor and a respected colleague by shadowing her approaches. She mastered the art of letting me come up with solutions to my problems and nudge me on the right track when I was lost. Thank you Sevil for all the time you devoted towards my questions, ideas and last but definitely not the least, for providing the opportunity to choose my path.  

This thesis would not be possible but for the many CMS colleagues and LHC staff at CERN. Thank you so much for the upkeep of the detector and the ion and proton beams so essential for my studies. I would also like to thank the CMS Heavy Ion group for being the best CMS physics interest group to work with! We shared a lot of wonderful experiences together, albeit it was quite stressful during beam times as one would expect. I have learnt a lot working with you all and I hope I can carry it forward to my next step in life. Special thanks for Eric Applet, Kurt Jung, Marta Verweij, James Castle, Chris Ferrioali, Pawan Kumar Netrakanti, Marguerite Tonjes, Matthew Nguyen, Alex Barbieri and all of the CMS Heavy Ion high pT group.  

I'm thankful to the MCnet group and to Dr. Korinna Zapp for making my my brief stint in the CERN theory group a lot of fun. I really enjoyed the opportunity to take a side step from all the data analysis and spend long days working on the physics of MC generation and thinking about the big picture. 

In addition, I'm very thankful to my thesis committee members, Prof Sevil Salur, Prof Ron Gilman,  Prof Aram Mekjian, Prof Hauro Kojima and Dr. Korinna Zapp for your invaluable time in making my dissertation a whole lot readable and with the possibility to reach a wider audience.  

I'm a firm believer of John Donne's words ``No Man Is an Island" and thus I would like to thank my childhood friends, the now infamous 10B whatsapp group for keeping me grounded and providing a sense of home and nostalgia during all my years away from Chennai. To all my fellow PhD students at Rutgers, Ian Laflotte, Colin Rylands, John Wu, Kartheik Iyer, Ethan Cline, Caitlin Carpenter, John Bonini, David Walter, Dan Brennan and Jesus Rives (to name a few), I have thoroughly enjoyed our friendship and I wish you all the best of luck while you finish up and for your future stellar careers. I have to include a special mention to Ian for being a very good friend, philosopher and guide. I have nothing but fond memories of our shared time at Rutgers and thank you so much for accepting to officiate on the most important event of my life.   

I would like to take the opportunity here to thank my mom, dad and my brother for their infinite source of affection, resilience, humor, and strength that I can always rely on during dull times. Appa and Amma, I shall always be amazed by your farsighted thoughts and beliefs and if there is one thing I have learnt from you both, it is respect for oneself, our fellow creatures and towards nature. Sriva, words cant express how proud I am of where you are now. You have single handedly overcome adversity that crept up on you and I cannot wait to stand by and watch you be successful in whatever you choose to do.  

Finally, thank you so much my dear Amu! You have taught me that life is not siting in front of a computer and debugging ROOT macros, but rather, life is meant to be enjoyed with a special someone who I can laugh, argue, cry, and ultimately grow old with. If my mentors and colleagues have noticed a general brightness in my mood and increased productivity in my work these past couple of years, the credit should reach you. Thanks again and I cant wait for our day next year and in consequence, for our bright future!

}

\dedication{
	\begin{center}
		{\em To my dear brother Srivatsav} \\
		
		\begin{figure}[h] 
		   \centering
		   \includegraphics[width=0.6\textwidth]{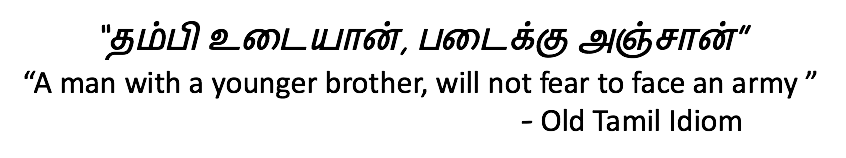} 
		\end{figure}
		
	\end{center}
}

\figurespage
\tablespage

\afterpreface

\chapter{Introduction and Motivation}
\label{ch_intro}
\begin{chapquote}{Lewis Carroll; Alice in Wonderland}
``Begin at the beginning,'' the King said, gravely, ``and go on till you come to an end; then stop.''
\end{chapquote}

\section{The big bang and the early universe}

\begin{figure}[h!]
	   \centering
	   \includegraphics[width=0.8\textwidth]{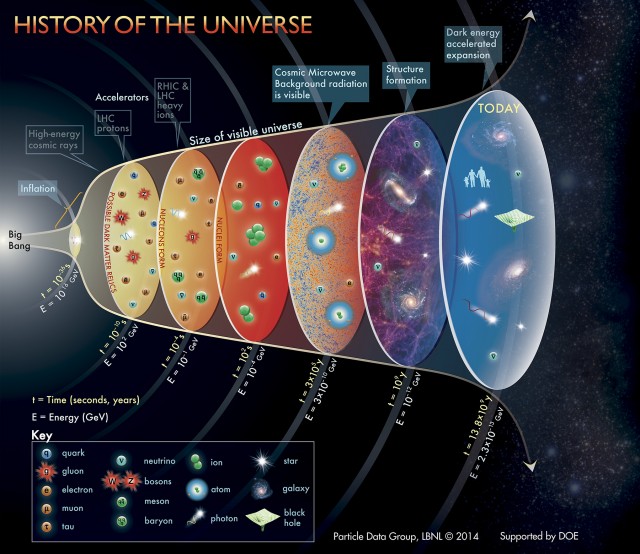}
	   \caption{Cartoon description of the universe timeline starting from the big bang till the present. The particle era is expected to have formed around a few microseconds. Figure courtesy Particle Data Group, LBNL.}
	   \label{fig:univHistory}
	\end{figure}

	The early universe has proven itself notoriously difficult and elusive to physicists who seek to understand its beginning and evolution. The current theory of universe formation begins with the big bang followed by a period of rapid inflation, corresponding to about a few pico seconds or an energy scale of $\approx 10^{16}$\gev~\cite{reviewEarlyUniverse} as shown in Fig~\ref{fig:univHistory}. As the universe begins to expand and cool, elementary particles begin to start forming from the high temperature vacuum leading to nucleons such as protons/neutrons. These nucleons then start interacting with other elementary particles (such as electrons) and start to form nuclei. At this epoch, the photons are no longer in thermal equilibrium with normal matter and the universe becomes photo-transparent leading to the famed cosmic microwave background (CMB)~\cite{Bucher:2015eia}. Our conventional (microwave, visible and radio) telescopes are limited in their ability to see beyond the CMB and study the early formation of the universe~\cite{highestredshiftpaper}. Recent results from LIGO~\cite{PhysRevLett.116.241103} with gravitational wave astronomy has the potential to see past the CMB and therefore generated a lot of excitement in the general community but the field is in its infancy. One way to recreate the conditions after the big bang, especially the particle physics epoch, is by colliding heavy element nuclei at relativistic speeds in particle colliders.  

\section{The little bang and relativistic heavy ion collisions}

	Relativistic heavy ion collisions at the CERN Large Hadron Collider (LHC) in Geneva, Switzerland and the Relativistic Heavy Ion Collider (RHIC) at Brookhaven National Laboratory (BNL), Long Island USA, provide the necessary conditions to recreate the particle epoch in the early universe. Studying how the collision remnants evolve using particle detectors such as trackers, calorimeters, muon chambers, time of flight counters etc. tantamount to understanding how fundamental particles navigate the primordial universe.   	

	\begin{figure}[h!]
	   \centering
	   \includegraphics[width=0.7\textwidth]{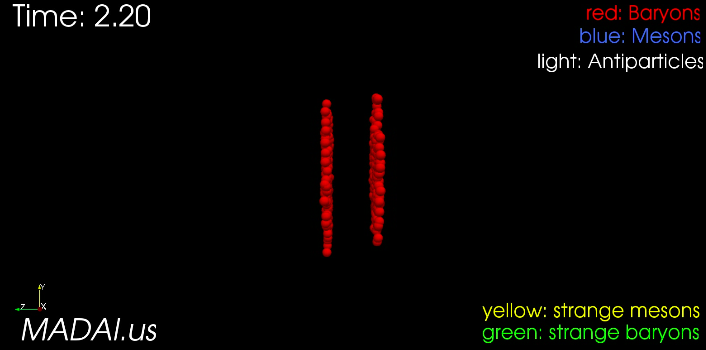}
	   \includegraphics[width=0.7\textwidth]{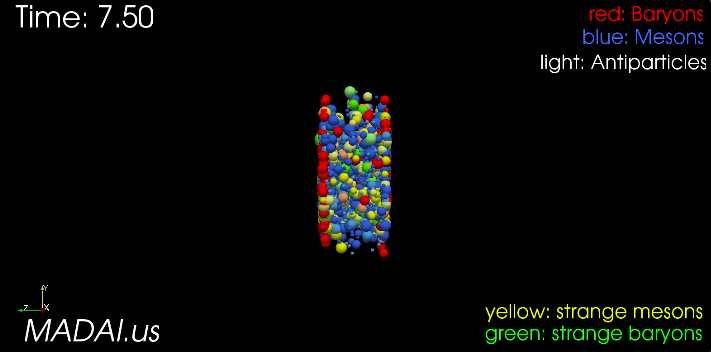}
	   \includegraphics[width=0.7\textwidth]{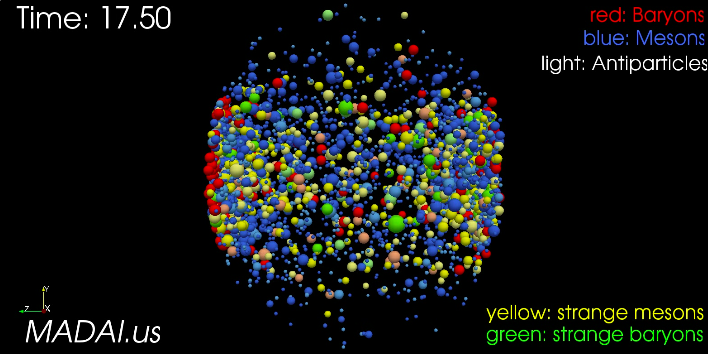}
	   \caption{Snapshots of a heavy ion collision shown in three different time stamps via the MADAI~\cite{doi:10.1117/12.912419} simulation package. The particles are color coded with red as baryons, blue as mesons and white, yellow and green as antiparticles, strange mesons and strange baryons respectively. Top panel shows the lorentz contracted nuclei just before the collision followed by the middle panel which is just after the collision and the bottom panel is after about 10-15 femtoseconds after the collision.}
	   \label{fig:MadaiHINCollision}
	\end{figure}

	A simulation of a single heavy ion collision in shown in Fig:~\ref{fig:MadaiHINCollision}, with the help of the MADAI framework~\cite{doi:10.1117/12.912419}, split up into three panels; before the collision (top), immediately after (middle) and 10-15 femtoseconds after the collision (bottom). Before the collision, we see lorentz contracted nuclei in the direction of the beam (along the z axis as shown in the bottom left of the panels). After the collision, we see the formation of all kinds of particles  and the whole system expanding reminiscent of a blast wave type pattern. It is by the reconstruction of these final state particles that we are able to extract the properties of the state of matter immediately after the collision. 
	
	\begin{figure}[h!]
	   \centering
	   \includegraphics[width=0.8\textwidth]{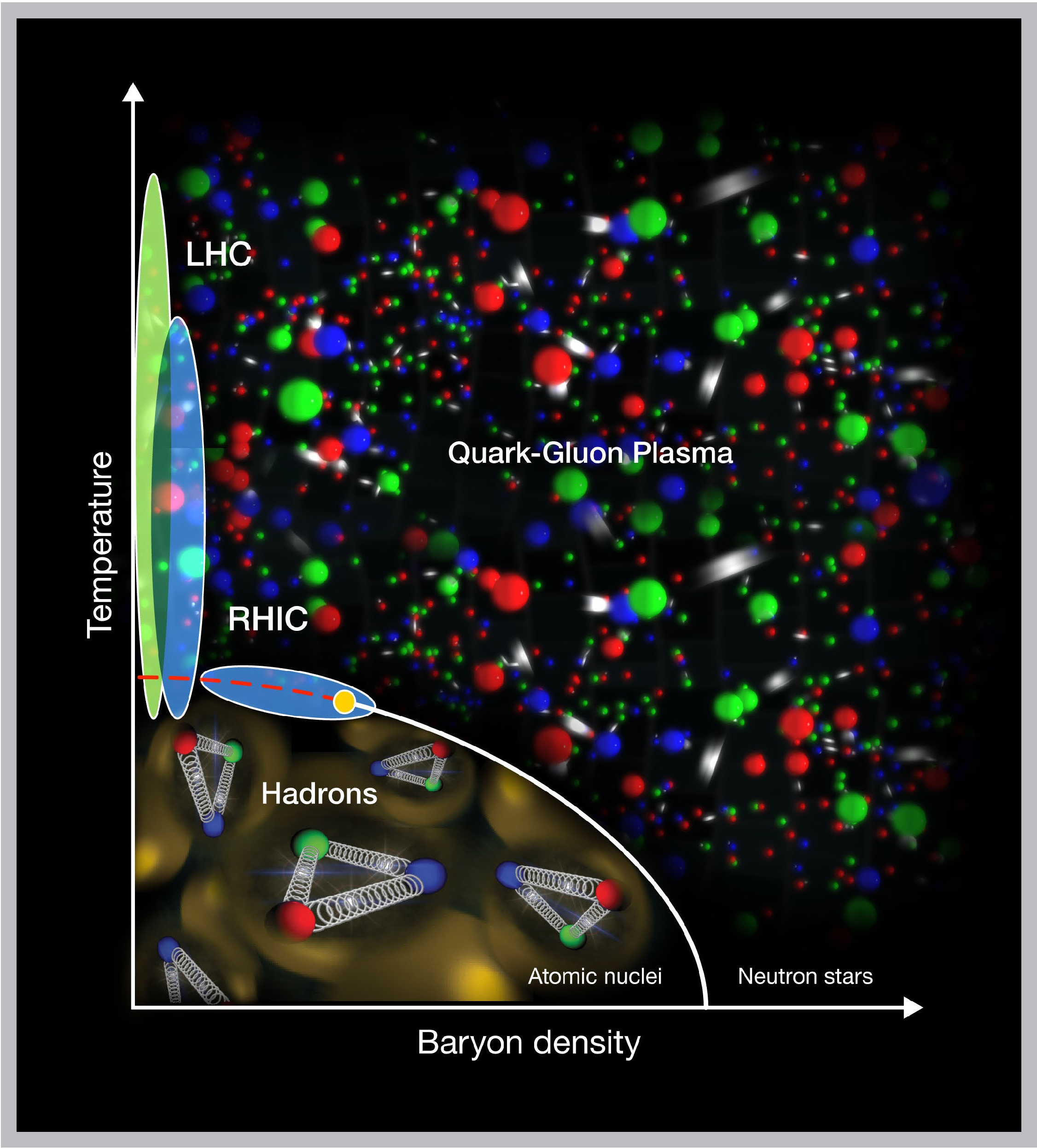}
	   \caption{Cartoon representation of the QCD phase diagram showing the baryon chemical potential on the x axis and the temperature on the y axis. The white line and yellow dot represents a first order and possible second order phase transition respectively. The region probed by LHC and RHIC are shown in the yellow and blue shaded regions. Figure courtesy BNL}
	   \label{fig:qcdPhaseDiagram}
	\end{figure}

	 In the collisions of heavy ions at relativistic speeds, the immediate aftermath can be composed of fundamental particles such as quarks or gluons, deconfined and moving around in whats now know as the quark gluon plasma (QGP)~\cite{Shuryak:2008eq}. The path taken by the deconfined quarks/gluons (henceforth referred to as partons) to color neutral hadrons constitutes a quantum chromo dynamics (QCD) phase transition, first proposed in the early 70s~\cite{CABIBBO197567} as shown in Fig:~\ref{fig:qcdPhaseDiagram}. In heavy ion collisions at the LHC, the baryon density is very low and and the temperature is high since the nucleons in the incoming beams at the LHC go through each other with minimal billiard-ball like contact. The area of the phase diagram being studied by the LHC is shown in the green shaded region in Fig:~\ref{fig:qcdPhaseDiagram}. On the other hand, RHIC has the capability to collide different particle species at a variety of beam energies (the lower the beam energy, the more probable it is to have direct collisions between nucleons in the beam) leading to a wider coverage in the phase diagram. A recent physics review of QCD phase diagram can be found here~\cite{0034-4885-74-1-014001}. The edges of the phase diagram correspond to different physical states of matter (as mentioned below) with the yellow dot in the middle representing a possible second order phase transition which is still the matter of current research and an open question probed via the Beam Energy Scan (BES) at RHIC (see ref.~\cite{1742-6596-455-1-012037} for a recent review of the physics from BES). 	
	 \begin{itemize}
	 	\item top left (high temperature and low baryon density): Early universe
		\item top right (high temperature and high baryon density): QGP 
		\item bottom right (low temperature and high baryon density): Neutron stars
		\item bottom left (low temperature and low baryon density): Hadron gas. 
	\end{itemize}

\section{Jet quenching and thesis overview}

	Parton energy loss was one of the predicted signatures of the QGP~\cite{Bjorken:1982tu,Gyulassy:1990ye,Wang:1992bz} with the amount of energy loss per distance travelled in the medium being a direct inference on the QGP properties. With the QGP being very short lived, it is not possible to study this particular signature in a laboratory, wherein one can shoot particles at it and study the energy loss. Rather, we use ultra-relativistic heavy ion collisions since both the QGP and the hard scattered partons are created around the same time with the parton traversing the medium in its path. By comparing such heavy ion collisions with proton-proton collisions, in similar parton kinematics, one can extract qualitative features of its energy loss or quenching. Experimentally, one cannot measure or detect individual partons, jets or collections of particles arising from a parton are used as first order proxies. Jets are ideal probes of the QGP since they are essentially involve both the hard scale (from the hard scattering) and the soft scale (from the QGP interactions). While there are other signatures of the QGP, jets are especially useful at the LHC due to their high production cross section and relative abundance leading to multiple interesting measurements to probe the effect of the QGP on the jet energy and structure. Jets are the main actors in this thesis and I will show measurements of the jet cross section from different systems at the LHC such as proton-proton, proton-lead and lead-lead aiding in supplementing our understanding of fundamental QCD and physics of the QGP. 
	
	There are excellent review articles and seminal papers referenced throughout the early chapters and the interested reader is encouraged to look up these references and further references within. The necessary theoretical framework/assumptions and background of QCD/jets is provided to the reader in Chapter 2. In addition, a quick introduction to Monte Carlo (MC) event generators and theoretical calculations is presented following a discussion of the signatures of the QGP in a heavy ion environment. The experimental point of view and technicalities of heavy ion events and in measuring a jet spectra are presented in chapters 3 and 4. We move to the physics of jet spectra measurements in pp and pPb collisions in chapter 5 and PbPb collisions in chapter 6. The heavy ion MC generator JEWEL is introduced in chapter 7 and we discuss the physics we can extract in comparing data to theoretical and MC models. We finally summarize the current state of the art in heavy ion jet measurements and ponder the future of the field in the coming years. The appendices contain useful discussion of particle physics kinematics, the technical details involving jet reconstruction algorithms, a list of abbreviations commonly used in the field and a short review of recent progress in jet structure measurements at the LHC.

\clearpage

\chapter{Theoretical Overview}
\label{ch_theorystatus}
\begin{chapquote}{Niels Bohr}
``We are all agreed that your theory is crazy. The question which divides us is whether it is crazy enough to have a chance of being correct. My own feeling is that it is not crazy enough."
\end{chapquote}

\section{Standard model of particle physics}

\begin{figure}[h] 
	   \centering
	   \includegraphics[width=0.75\textwidth]{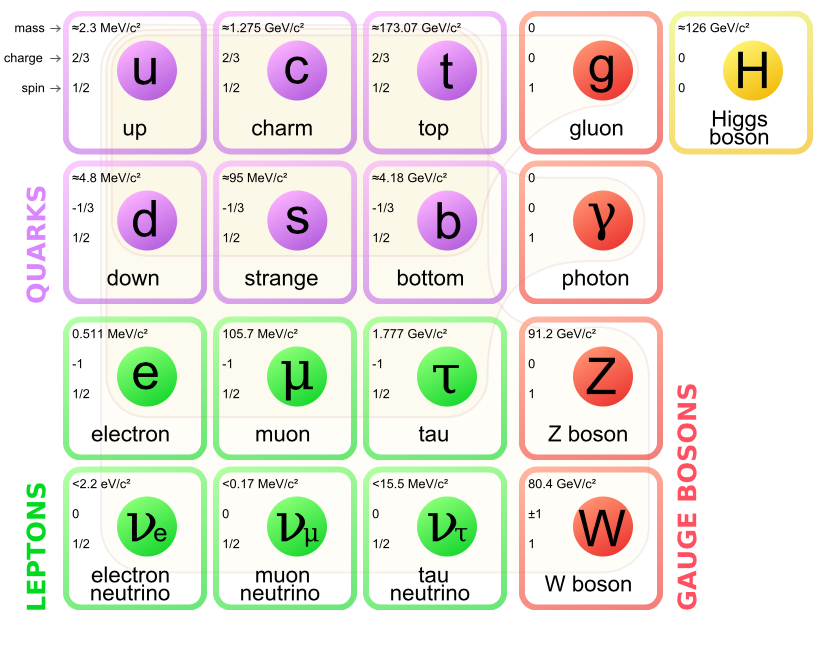} 
	   \caption{The fundamental particles in the standard model of particle physics, including the quarks (in purple), leptons (in green), electroweak bosons and the gluon in red along with the higgs particle in yellow. Figure courtesy wikipedia.}
	   \label{fig:standardModel}
	\end{figure}

	There are four fundamental types of interactions or forces that are present in nature; Gravity, Electromagnetism, Weak and the Strong force. Of the four, strong and electro-weak forces can be described by one theory, rather collection of theories called the Standard Model (SM) of particle physics. There are two kinds of fundamental particles in the SM, force mediators or gauge bosons and matter particles or fermions. The matter particles (and their corresponding anti-matter particles) shown in Fig:~\ref{fig:standardModel} in the pink and green boxes are the quarks and leptons. These matter particles are fermions and they appear in nature as couplets such as u, d and e, $\nu$ of which three subsequent generations have so far been experimentally confirmed. Amongst the bosons, shown in the red boxes, the photon mediates the electromagnetic force, and its heavier cousins, the Z and W$^{\pm}$ mediate the weak force. The gluons facilitate interactions between quarks and thus mediate the strong force.  The most recent addition to the table is the notable Higgs boson in the yellow box as the Higgs field being responsible for the individual particle masses. There are several textbooks and review papers to learn more about the standard model and its intricacies and the reader can find a few of those here~\cite{Herrero:1998eq, Halzen:1984mc, Olive:2016xmw, RevModPhys.71.S96}. 
	
	As mentioned above, the standard model only deals with the strong and electroweak forces with Gravity as the notable exception. Theories that include gravity are called Grand Unified Theories (GUT) (see~\cite{Raby:2006sk} for a recent review) and will not be discussed in this thesis. Our main focus in this chapter is studying the formation and behaviors of the QGP which involves the following; understanding the interaction of quarks and gluons, jet production cross sections, technicalities involved in MC event generation models and finally signatures of the QGP in heavy ion collisions.  

\section{Fundamentals of QCD and jets}

	The interaction of quarks and gluons makes up the theory of QCD and we will go over some of the fundamental properties of QCD from an experimentalist's point of view. Quarks were proposed in the early 60s by Murray Gell-Mann and George Zweig~\cite{Riordan:1992hr} as a way of understanding the fundamental particles that make up cosmic rays and their collision products with the atmosphere. Since quarks are fermions, they are spin one half and carry both electromagnetic (non integral) and color charge. Color charge is unique to quarks and gluons with each quark capable of carrying one of three possible options, which physicists in their infinite wisdom have called red, blue and green (with anti-quarks carrying either anti-red, anti-blue or anti-green charges). Gluons~\cite{PhysRevD.4.3418} carry color charge composed of quark-anti-quark configurations, leading to 8 varieties of gluons. The composite objects made up of quarks such as Baryons (three or five quarks) and Mesons (quark-antiquark) are by definition colorless (RBG = R$\bar{\rm{R}}$ = White). 
	
	\subsection{Asymptotic freedom and Confinement}
	
\begin{figure}[h] 
		   \centering
		   \includegraphics[width=0.7\textwidth]{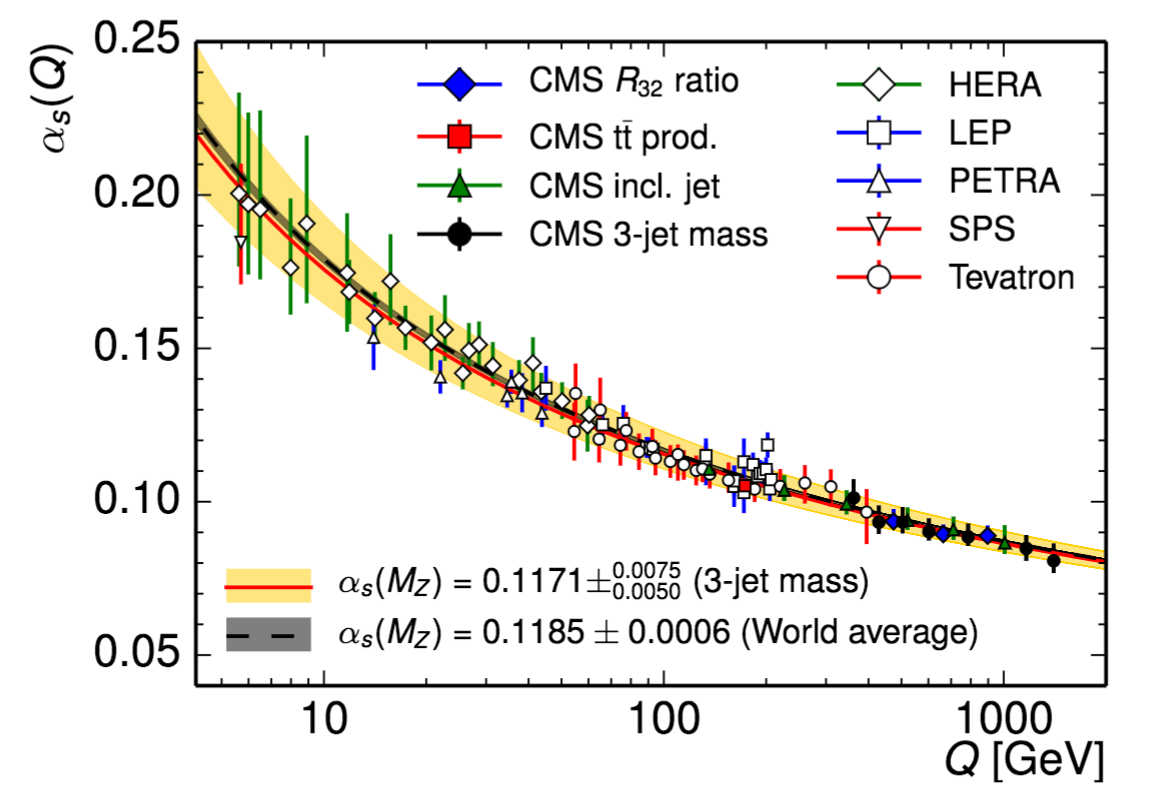} 
		   \caption{The characteristic running of the strong coupling constant at the Z mass scale as a function of momentum transfer extracted as fits to data from a variety of experiments. Figure taken from~\cite{CMS:2014mna}.}
		   \label{fig:runningCoupling}
		\end{figure}

		The strength of an interaction is primarily characterized by its coupling in the respective field. When you consider QED~\cite{Feynman:1986er}, the coupling constant is dependent on the beta function which is positive to several orders in perturbative theory. At low energies, we find that $\alpha_{ew} \approx 1/137$ and around the Z mass (90 GeV/c) $\alpha_{ew} \approx 1/127$. Since it is small, the overall sum across all orders in perturbative theory converges. In QCD however, the beta function non zero and thus the coupling constant is a function of the energy scale at which it is probed and is of order $\alpha_{s} \approx 0.12$ at 100 GeV/c and increasing to $ \alpha_{s} \approx 1$ as the energy scale decreases to 1 GeV/c or the proton mass scale. This dependence on the energy scale, referred to as asymptotic freedom~\cite{RevModPhys.77.837} in literature, has been extracted from several experimental measurements during the past decades is shown in Fig:~\ref{fig:runningCoupling}. The markers correspond to measurements and the shaded region are fits to the data showing the characteristic dependence of the $\alpha_s (M_Z)$ which is calculable in perturbation theory.   
	
		The running of the coupling constant also leads to infra-red slavery where quarks and gluons are bound to color singlet states~\cite{Wilczek:2000ih}. This is analogous to thinking about a quark-antiquark pair as two ends of a string and as one pulls one side of the string apart, the tension increases and at some point, the string breaks and multiple strings are formed. In simpler terms, it is not possible to isolate an individual quark or gluon and thus the QGP is a direct consequence of the asymptotic freedom. The formation of hadrons from quarks/gluons is called as hadronization. A theoretical and phenomenological review is available here~\cite{Marchesini:2006rg}. Thus, hadronization and infra-red slavery mean that the QGP is not directly observable and experiments utilize jets as internal hard probes to study the medium indirectly.  
		
	\subsection{Jets as evidence for quarks and gluons}
	
\begin{figure}[h] 
		   \centering
		   \includegraphics[width=0.9\textwidth]{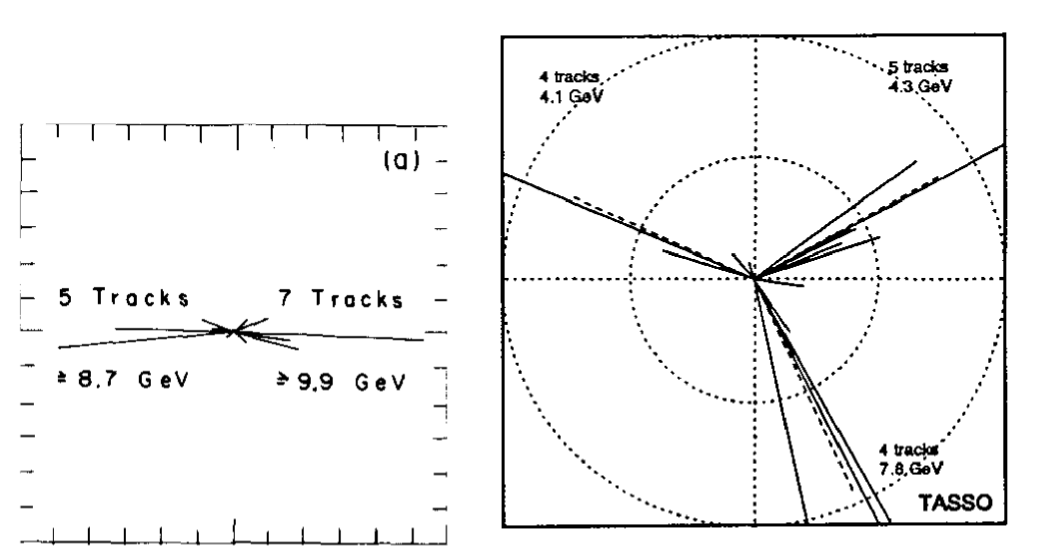} 
		   \caption{Characteristic events in $e^{+}e^{-}$ collisions showcasing quark-antiquark pair in the left and gluon bremstrallung at the TASSO experiment at DESY. Figures taken from the review~\cite{Ali:2010tw}.}
		   \label{fig:tassoquarksgluons}
		\end{figure}

		Quarks were first experimentally discovered in high energy electron-positron collisions in the PETRA (housed in DESY) accelerator ring at the TASSO experiment~\cite{Ali:2010tw} as shown in the left panel of Fig:~\ref{fig:tassoquarksgluons}. The presence of back to back collimated streams of particles confirmed the existence of the so-called quark ``jets". A more formal definition of a jet is given in the next chapter and in Appendix:~\ref{app_jetcomposition}. At the moment, we shall only consider a jet as a collimated spray of particles, limited by a given distance scale from a single parton. Three jet or Mercedes style events (right panel of Fig:~\ref{fig:tassoquarksgluons}) again at TASSO, were consistent with gluon bremmstrahlung from one of the quarks and thus verified the existence of gluons. Since these were $e^{+}e^{-}$ collisions, background contamination was negligible and the signals were very clean with almost all the tracks directly associated with the collision products. When we move over to pp and heavy ion collisions, we will soon see that the picture becomes quite complicated. 
		
		Jet studies at the LHC, due to the nature of the high energy beam, probe the high $Q^{2}$, or momentum transfer region where perturbative theory is mostly valid. This is an important point to note since we will be focusing on understanding the behavior of QCD at high density and temperature by comparing it to a baseline pp which we henceforth call vacuum. In order to understand our vacuum production of jets, we will in turn be comparing it with perturbative QCD (pQCD) calculations with the use of the factorization theorem. 

	\subsection{Theoretical calculations of inclusive hadron cross section}		
	
		To first order, a jet is a proxy for a hard parton and it has a characteristic structure arising from the QCD radiation pattern which can be calculated. Such calculations typically involves the factorized or perturbative part (hard process) and the non perturbative corrections (soft processes) including hadronization, multi-parton scattering and the underlying event. The QCD factorization theorem~\cite{Sterman:2014nua} relies on the ability to separate the calculable short scale quark gluon scattering from the long-distance hadronization process. Thus essentially, a factorized hadron cross section in hadron collisions is a direct generalization of the deep-inelastic scattering outlined as follows for the production of a hadron of type `H' 
		
		\begin{equation}
		\begin{split}
			\frac{d \sigma (A+B \rightarrow H(p))}{d^{3} p} & = \int dx_{a} dx_{b} f_{a/A} (x_{a}, \mu_{F}) f_{b/B} (x_{b}, \mu_{F}) \\
				& \times \int dz C \left(x_{a} p_{A}, x_{b} p_{B}, \frac{p}{z \mu_{F}} \right)_{ab \rightarrow c(p/z)} \\
				& \times D_{H/c} (z)
		\end{split}
		\end{equation}
	
		where $f_{a/A} (x_{a}, \mu_{F})$ is the parton distribution function, followed by the individual cross section of $a+b \rightarrow c$ and the fragmentation function $D_{H/c}(z)$ which is the probability distribution of hadrons with the fractional momentum z. In the above factorized formula, $x$ is the fraction of the parton momentum compared to that of the proton, referred to as Bjorken $x$, and $\mu_{F}$ is the energy scale of the factorization process. The factorization scale ($\mu_{F}$) is introduced to separate the long and short distance physics of the interaction and thus at best arbitrary. Systematic uncertainties of theoretical cross section often include variations on the factorization scale utilized in the calculation. 
		
		\subsubsection{Parton distribution functions}
		
\begin{figure}[h] 
			   \centering
			   \includegraphics[width=0.8\textwidth]{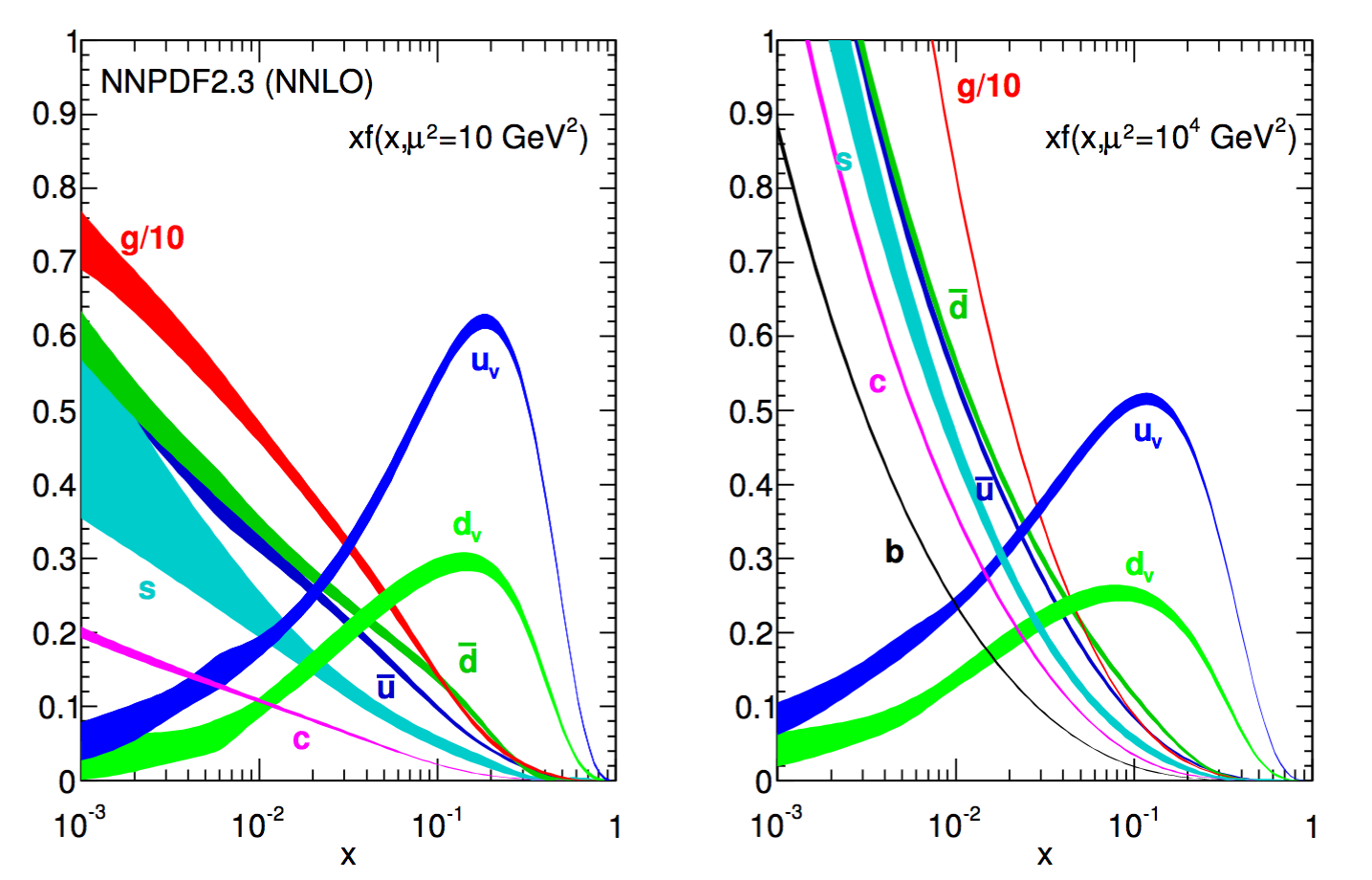} 
			   \caption{Proton PDFs for the quarks, and gluons from the NNPDF2.3 at the next to next to leading order for $Q^2 = 10$ \gevc (left) and $Q^2 = 10000$ \gevc (right). Figure taken from~\cite{Carrazza:2015dea}}
			   \label{fig:nnpdf23NNLO}
			\end{figure}

			The make up of a proton is characterized by the parton distribution functions (PDF)~\cite{Placakyte:2011az}, represented as a probability of finding a certain constituent at a given Bjorken $x$ as shown in Fig:~\ref{fig:nnpdf23NNLO} based on the NNPDF 2.3 PDF~\cite{Forte:2002fg, Ball:2012cx} for two particular $Q^{2}$ scenarios (left - low $Q^{2} \sim 10$ GeV$^{2}$, right - high $Q^{2} \sim 10^{4}$ GeV$^{2}$). The different curves are labeled to represent the contributions that make up a proton such as the valence quarks, up (blue) as $u_{v}$, down (green) as $d_{v}$ and the strange quarks (teal/turquoise) as $s$ and the gluon contribution in red, which is the largest and is divided by a factor of 10 in the panels. This tells us that the proton is predominantly "seen as composed " made up of gluons as we move to high $Q^{2}$ and low $x$ region (significant in our high \pt~jet studies). PDFs are universal and its scale dependence is calculable in QCD and thus only the x dependence is parameterized.  
		
			Experimentally, is not possible to measure the initial parton kinematics such as $x$ and $Q^{2}$ directly. Nevertheless by looking at a special class of events with two high \pt~hard scattered jets called a di-jet system we can extract strong correlations between the following; 
			\begin{itemize}
				\item average transverse momentum of the two jets and the $Q^{2}$ 
				\item di-jet pseudorapidity and the $x$. 
			\end{itemize}
			
			There are several options for choosing a PDF since they are essentially fits to data deep inelastic lepton+nucleus and hadron collisions utilizing different methods of fitting and fit functions. The NNPDF curves shown in Fig:~\ref{fig:nnpdf23NNLO} utilize a neural network to fit a variety of functions and more than 250 individual parameters on different datasets with a non linear activation function. 

		\subsubsection{Splitting functions}
		
			In QCD, splitting details how quarks or gluons can radiate additional quarks or gluons in vacuum. They are encompassed in the DGLAP evolution equations for parton densities, named after Dokshitzer, Gribov, Lipatov, Altarelli and Parisi~\cite{Martin:2008cn, ALTARELLI1977298}, which are essentially a matrix in flavor space where the splitting functions are given as follows (as a function of energy fraction in the splitting fraction) 
			\begin{equation}
			\begin{split}
				P_{q \rightarrow q g} (z) &=  C_{F} \frac{1+z^2}{(1-z)_+} \\
				P_{g \rightarrow g g} (z) &= 2 C_{A} \left[ \frac{z}{(1-z)_{+}} + \frac{1-z}{z} + z(1-z) \right] + \delta(1-z) \frac{11 C_{A} - 4n_{f} T_{R}}{6} \\
				P_{g \rightarrow q \bar{q}} (z) &= T_{R} [z^2+(1-z)^2]
			\end{split}
			\end{equation} 
			where the coefficients are the respective casimir color factors $C_{F} = 4/3, C_{A} = 3$ and  $T_{R} = 1/2$. It is important to note that the $P_{gg, qg}$ are symmetric with $z$ and $1-z$ excluding virtual emissions. The divergence for $P_{g \rightarrow g g} (z)$ and $P_{q \rightarrow q g}$ are for the conditions $z\rightarrow 1$ implying soft gluon emission and $z\rightarrow 0$ which has a direct consequence of the growth in the PDFs as we go to small $x$. 

		\subsubsection{Non-Perturbative corrections}
		
\begin{figure}[h] 
			   \centering
			   \includegraphics[width=0.4\textwidth]{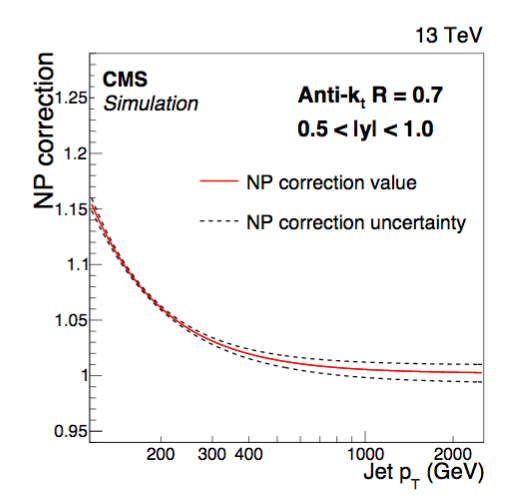} 
			   \includegraphics[width=0.4\textwidth]{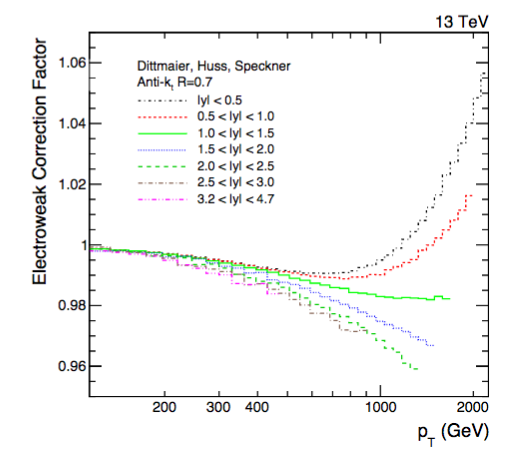} 
			   \caption{Non perturbative corrections for inclusive jet spectra with R = 0.7 and mid rapidity at 13 TeV~\cite{Khachatryan:2016wdh}. The left panel shows the NP correction based on MPI and HAD whereas the right panel shows the electroweak correction factor.}
			   \label{fig:npcorrectioncmsresult}
			\end{figure}

			 Experimentally Non-Perturbative (NP) effects are evaluated by using models for the underlying event~\cite{Gieseke:2007ad}, hadronization~\cite{PhysRevD.94.034016}, and multi-parton interactions~\cite{Sjostrand:2017cdm} in MC event generators. The underlying event primarily consists of proton remnants concentrated along the beam direction and multiple soft interactions between the nuclei~\cite{Khachatryan:2015jza}. The ratio between a nominal event generation using a well performing tune, including matrix element (either Leading Order or Next to Leading Order) and the parton shower (PS) and a sample with hadronization (HAD) and multi parton interactions (MPI) effects switched off is taken as correction. Note that NP corrections, so defined, are used to correct any available pQCD calculation at parton level to bring it to the jet level for direct comparison with experiment. The NP correction factors can be defined as:
			\begin{equation}
				C^{NP}_{LO} = \frac{d\sigma^{LO+PS+HAD+MPI}}{d\sigma^{LO+PS}}
			\end{equation}
			where in the superscript, the components of the simulations are listed and in the subscript the order of the matrix element is specified. These correction factors are usually estimated with different tunes, generators and PDF sets, and the envelope resulting from them is considered as theoretical uncertainty of the correction factors as shown in the left panel of Fig:~\ref{fig:npcorrectioncmsresult} estimated for CMS measurement of jet spectra with a given distance parameter or jet radius of R = 0.7, at 13 TeV~\cite{Khachatryan:2016wdh}. More details about the clustering algorithm and jet radius will be provided in the upcoming chapters. 

			As one goes to very high $Q^{2}$ processes, electroweak corrections to the jet spectra also become non-negligible. These are corrections due to virtual exchanges of massive guage bosons such as Z and W especially at central rapidity and high \pt~as shown in the right panel of Fig:~\ref{fig:npcorrectioncmsresult}. 

\begin{figure}[h] 
			   \centering
			   \includegraphics[width=0.6\textwidth]{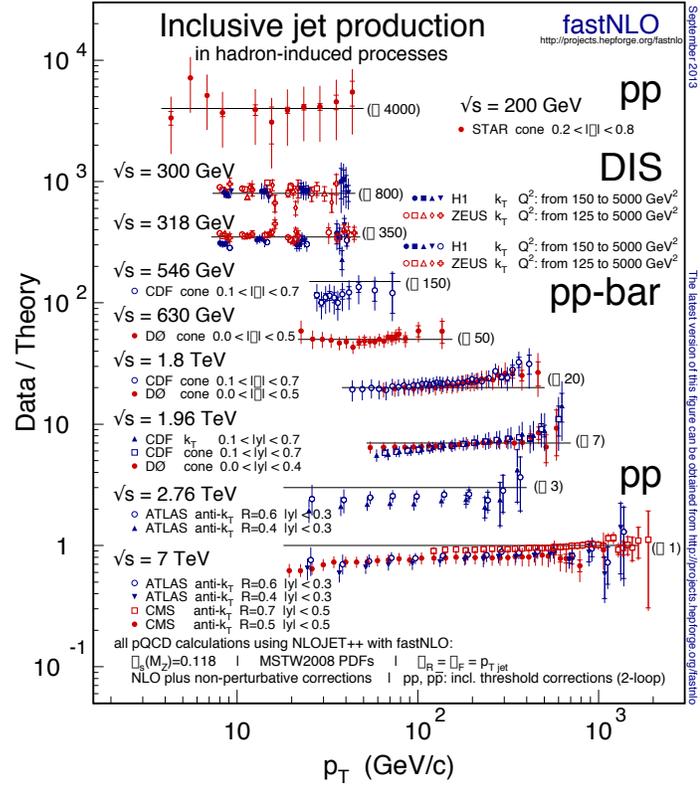} 
			   \caption{Inclusive jet production calculations in hadron induced hard processes from the fastNLO package. The plot shows the comparison of data from different experiments and center of mass energies as a function of the jet \pt. Figure courtesy of fastNLO~\cite{Wobisch:2011ij}.}
			   \label{fig:fastnloDataTheory}
			\end{figure}

			With the addition of these correction factors the factorized jet cross section calculation at NLO in the matrix element and parton showers are available in the fastNLO framework~\cite{Wobisch:2011ij} for a variety of collisions and center of mass energies. A comprehensive comparison of data with fastNLO predictions are shown in Fig:~\ref{fig:fastnloDataTheory}. This plot is quite complicated but the main takeaway message is that across a large scale of jet \pt, clustering algorithms, center of mass energy and rapidity regions, theoretical calculations can reproduce data at certain kinematic regions, but have notable exceptions for example \akt small jet radii and low jet \pt. There are several non trivial effects that arise in this particular region and as physicists, our sights now turn to study what causes this discrepancy and what is missing in our calculations which we will look at in detail in the upcoming chapters.   

\section{Standard particle physics MC generators}

	Ab-initio theoretical calculations for every observable in question are very complicated and take a lot of time and person power to materialize. Monte Carlo event generators employ the same fundamental rules of particle production and interactions in a probabilistic approach and asymptote to a general solution~\cite{Buckley:2011ms, srimanobhasMC}. The term Monte Carlo is honorary to the city in Europe famous for its gambling casinos and the usage in physics tantamount to repeated random sampling from a given probability distribution, for example, a proton PDF. 
	
	The state of the art simulation of a full proton-proton collision has to deal with several processes such as 
	\begin{itemize}
		\item Matrix Element for the hard scattering~\cite{Nason:2012pr}
		\item Initial State Radiation from the incoming protons 
		\item Parton showers and evolution towards jets provided by the DGLAP equations (see~\cite{Hoche:2014rga} for a  review)    
		\item Parton Fragmentation towards hadrons
		\item Hadronization - model dependent since the process is fundamentally non perturbative
		\item Underlying Event and MPI from the same proton-proton collision 
		\item Final State Radiation say from bremmstrahlung photons.
	\end{itemize}

	There are a lot of MC generators on the market and many of them have parameters fitted to a particular datasets leading to a variety of tunes. The most well known generator in particle physics is PYTHIA~\cite{Sjostrand:2006za, Sjostrand:2007gs}, regularly utilized by all the experiments at the LHC for comparisons and for removing detector resolution biases. Other commonly used generators are HERWIG~\cite{Corcella:2000bw, Bahr:2008pv}, SHERPA~\cite{Gleisberg:2008ta, Hoche:2014kca} which utilize an angular ordered parton shower stemming from the Seymor-Catani splitting functions~\cite{Catani:1996vz, Hasegawa:2014oya}. 


\section{Signatures of the QGP}

	Armed with a phenomenological introduction to QCD and pp collisions, we can begin to comprehend heavy ion collisions and any experimental signatures of the QGP. Since the QGP only exists in a very small time frame, it is not straight forward to say if a particular collision produced the deconfined state directly. It is also not possible to experimentally study the state of matter via say spectroscopy or some external probe. By analyzing the final state products, several possible signatures have emerged during the last decades that are discussed in Appendix:~\ref{app_qgpsignatures}. 
		
\begin{figure}[h]
	   \centering
	   \includegraphics[width=0.48\textwidth]{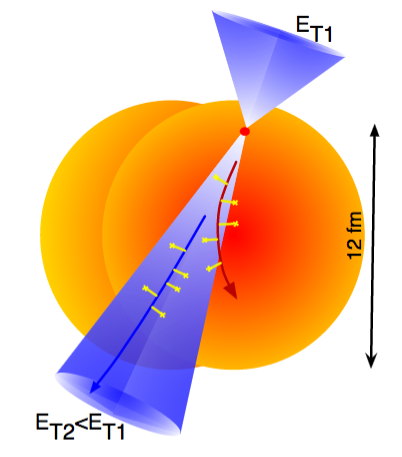}
	   \caption{Cartoon representation of a quenched event where the recoiling parton loses much of its energy traveling a longer distance through the medium. Figure taken from~\cite{CasalderreySolana:2011rq}.}
	   \label{fig:cartoonDijet}
	\end{figure}

	The main evidence for the presence of the QGP in heavy ion collisions is the concept of jet quenching. In relativistic heavy ion collisions, there is a significantly large cross section for hard scattering of two partons from the incoming beams. These hard scattered partons are produced back-to-back in the transverse plane and thus by reconstructing the jets arising from these partons, one expects them to be of roughly equal momentum due to momentum conservation. A naive expectation of medium induced modification is an asymmetric reduction in the jet energy depending on distance from the center of the collision. This is pictorially highlighted in Fig:~\ref{fig:cartoonDijet} where one jet with transverse energy $E_{T1}$ is seen to escape the interaction region fairly easily where as its recoil partner travels through much of the medium and loses additional energy, ending up with $E_{T2}<E_{T1}$~\cite{CasalderreySolana:2011rq}.  
	
\begin{figure}[h]
	   \centering
	   \includegraphics[width=0.8\textwidth]{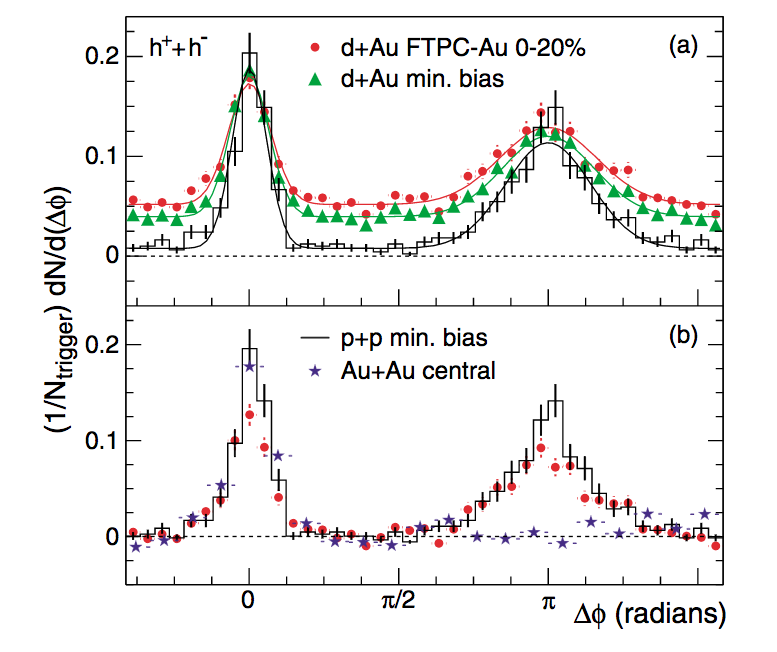}
	   \caption{Measurement of the angular correlation by the STAR collaboration~\cite{PhysRevLett.91.072304} between the leading charged hadrons and all other hadrons in the event for p-p minimum bias, Au-Au and d-Au collisions. The peak at zero corresponds to the near side particle yield and the stark disappearance of the back to back peak at $\pi$ in AuAu collisions is evidence for particles getting quenched in the medium.}
	   \label{fig:stardeltaphiHadronLoss}
	\end{figure}

	The STAR collaboration was able to directly measure this phenomenon in its landmark result as shown in Fig:~\ref{fig:stardeltaphiHadronLoss} where the azimuthal angular correlation ($\Delta \phi$) of all identified charged hadrons with the leading charged hadron in the event~\cite{PhysRevLett.91.072304}. The result shows this correlation for minimum bias pp (proton-proton), head-on AuAu (gold-gold) and dAu (deuteron-gold) collisions. The pp and dAu correlations both show the near side peak ($\Delta \phi \approx 0$) and the away side recoiling peak ($\Delta \phi \approx \pi$). The disappearance of the away side peak in AuAu collisions (shown in blue stars) confirms the presence of medium (or QGP) interaction where the recoiling jet appears to be quenched. 
	
\begin{figure}[h] 
	   \centering
	   \includegraphics[width=0.8\textwidth]{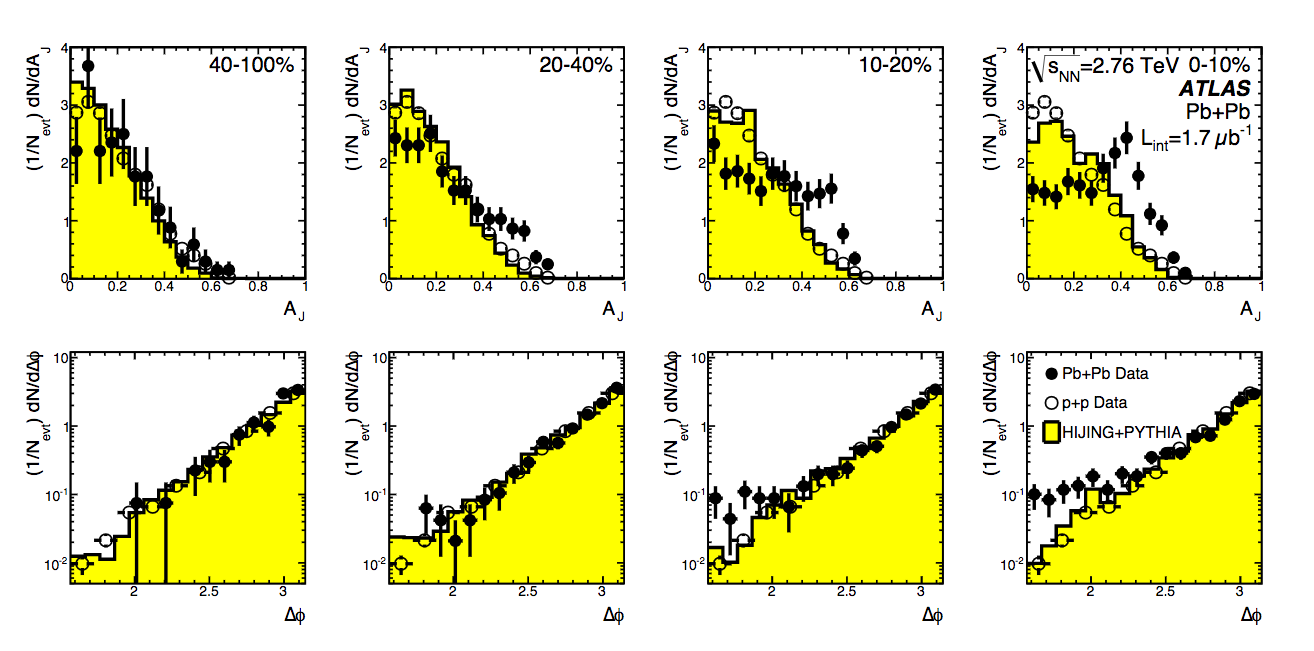} 
	   \caption{ATLAS dijet measurements of the $A_{J}$ (top panels) and $\Delta \phi$ (bottom panels) for peripheral to central PbPb colisions from left to right (see text for more details). Figure taken from~\cite{Aad:2010bu}}
	   \label{fig:atlasdijet}
	\end{figure}

	At the LHC, both ATLAS and CMS measured this particular phenomenon with fully reconstructed jets as opposed to charged tracks. The observable of interest was the dijet asymmetry ($A_{J}$) which is defined as the difference between the leading and the subleading jet \pt divided by the sum of the two. Both ATLAS and CMS published the asymmetry measurement as a function of centrality~\cite{Aad:2010bu} and the jet \pt~\cite{Chatrchyan:2012nia} respectively. The top panels of ~\ref{fig:atlasdijet} show the asymmetry distribution normalized to the number of events for head on collisions on the right and peripheral, glancing collisions on the left for PbPb data with the solid markers, pp data with the open markers and an MC sample without quenching in the yellow histograms. A clear deviation from pp collisions to PbPb is observed in the data with a significant fraction of dijet events being more asymmetric than symmetric. The bottom panels in Fig:~\ref{fig:atlasdijet} also shows the angular correlation between the two leading jets (ordered by their transverse momentum) showing very little deviation in the away side region but small modification in the near side. 
		
\begin{figure}[h] 
	   \centering
	   \includegraphics[width=0.8\textwidth]{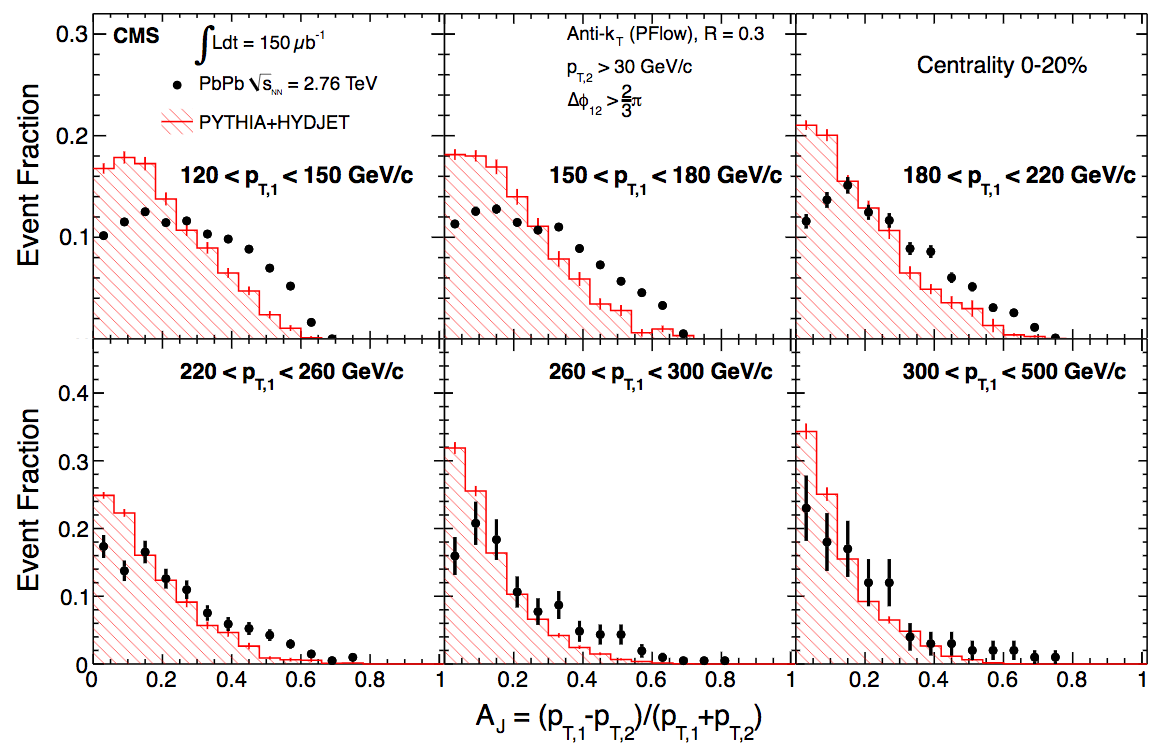} 
	   \caption{CMS measurement of the $A_{J}$ in central PbPb events with different panels representing leading jet \pt selections (see text for more details). Figure taken from ~\cite{Chatrchyan:2012nia}}
	   \label{fig:cmsdijet}
	\end{figure}

	The corresponding CMS measurement for the $A_{J}$ is shown in Fig:~\ref{fig:cmsdijet} for head on collisions where the different panels correspond to varying bins of leading jet momenta increasing from top left to bottom right.  The data is comparable with an embedded, unquenched MC at high \pt but the deviations are quite clear for the low leading jet \pt bins in the analysis. 
	
	While these results directly show evidence of jet quenching, both the jets in the $A_{J}$ measurement can be quenched and thus one cannot get an idea of how much energy loss is happening for a jet of a given \pt. In addition, the jets considered here are quite high in the \pt and that might constitute a selection bias in our events. Another conceptually simple way to measure an effect of the QGP is to compare a measurement in heavy ion collisions to proton proton collisions multiplied by the equivalent number of binary collisions ($n_{\rm{coll}}$). Such measurements are know as nuclear modification factors or \raa,
	\begin{equation}
		R_{\rm{AA}} = \frac{X^{\rm{AA}}}{n_{\rm{coll}} X^{\rm{pp}}}
	\end{equation}
	where the AA stands for Ion-Ion (including proton-Ion). 
	  The number of binary collisions is estimated with the use of the Glauber model that will be discussed in the coming sections. Thus, the \raa~provided early clear signals indicating the presence of the QGP or medium behavior when it is found to be less than or greater than 1 and this will be primary measurement of this thesis.

\section{Key features of heavy ion collisions}

	The complications in a heavy ion collision partly arises due to the theoretical difficulty in describing the following  
	\begin{itemize}
		\item Initial state 
		\item Event Centrality
		\item Hydrodynamic evolution
		\item Parton energy loss 
		\item Hadronization 
	\end{itemize} 	
	with the main question revolving around the symbiotic treatment of the soft and the hard scale in the collision. 

	There are two groups of theories based on the description of nuclear matter, right before or during the collision, described as the initial state~\cite{Venugopalan:2016vgm, Albacete:2013tpa, Lappi:2015jka}. Firstly, one can assume that its entirely composed of gluons with high coupling such as in the Color Glass Condensate (CGC) approach~\cite{Gelis:2010nm} and secondly, weak coupling gluons originating from a cascade of multiple splittings from the initial valence quarks as described in the EKRT~\cite{Eskola:1999fc, Niemi:2015qia, Eskola:2017imo} or EPOS~\cite{Pierog:2009zt, Pierog:2013ria} model. Both these approaches are able to reproduce the particle production dependent on the multiplicity under a variety of assumptions including a parameterized description of temperature and the entropy density in the initial state. 
	
	In earlier sections, we discussed the proton PDFs and in heavy ion collisions, one often talks about the use of nuclear PDFs~\cite{Honkanen:2013goa}. This approach is data driven where in one assumes the respective PDF is modified and the nuclear modification factor is often obtained by global fits to data from deep inelastic measurements~\cite{Qin:2015srf}. A well used example of such nPDFs are from the EPS09~\cite{Eskola:2009uj} and nCTEQ~\cite{hannunPDF, Kovarik:2015cma} collaborations with its fits for cold nuclear matter effects. These can describe jet data reasonably well at the LHC for pPb collisions as we shall seen in the upcoming chapters. One facet of a heavy ion event which we will cover next is the question of event centrality and its modeling.   
	
	In collisions of lorentz contracted nuclei, the impact parameter $b$ (distance between the center of the nuclei) varies from event to event. By looking at the simple geometry of the collision based on the event multiplicity or the energy deposits in the forward calorimeters we can identify events into several centrality or event activity classes. The most central collisions are identified with the smallest $b$ and peripheral collisions have the largest $b$. The Glauber model (or a variant thereof) is used to calculate the nuclear thickness function $T_{AA}$ (which we will use in our binary scaling of the pp reference). 
		
\begin{figure}[h]
	   \centering
	   \includegraphics[width=0.5\textwidth]{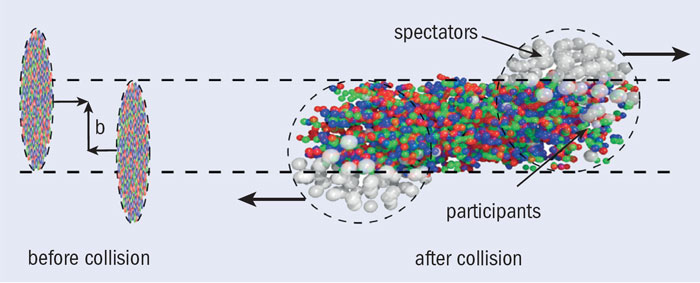} 
	   \caption{Graphical representation of the impact parameter of a heavy ion collision (left) and the corresponding participant and spectator nucleons (right). Figure taken from the CERN Courier article on April 16th 2013.}
	   \label{fig:Glauberdiagram}
	\end{figure}

	The Glauber model~\cite{Miller:2007ri} describes the interaction of two nuclei in terms of the interaction of its constituent nucleons as shown pictorially in Fig:~\ref{fig:Glauberdiagram}. The following assumptions are made to have a model of the interaction. 
	\begin{itemize}
		\item nucleus moves in a straight line
		\item classical scattering to find the number of interacting nucleons as a function of the given impact parameter
	\end{itemize}
			
	We start with the nuclear thickness function, $T_A (s)$ defined as the probability density of finding a nucleon at a given impact parameter
	\begin{equation}
		T_A(s) = \int \rho_A (s,z_A) dz_A .
	\end{equation}
	where $\rho_A$ is the nuclear density distribution of nucleus A. 
	Now we can talk about the thickness overlap function between two nucleus which is the effective overlap area for which a specific nucleon in A can interact with a given nucleon in B. This is defined as 
	
	\begin{equation}
		T_{AB}(b) = \int T_A(s) T_B(s-b)d^2s 
	\end{equation} 
		
\begin{figure}[h] 
	   \centering
	   \includegraphics[width=0.5\textwidth]{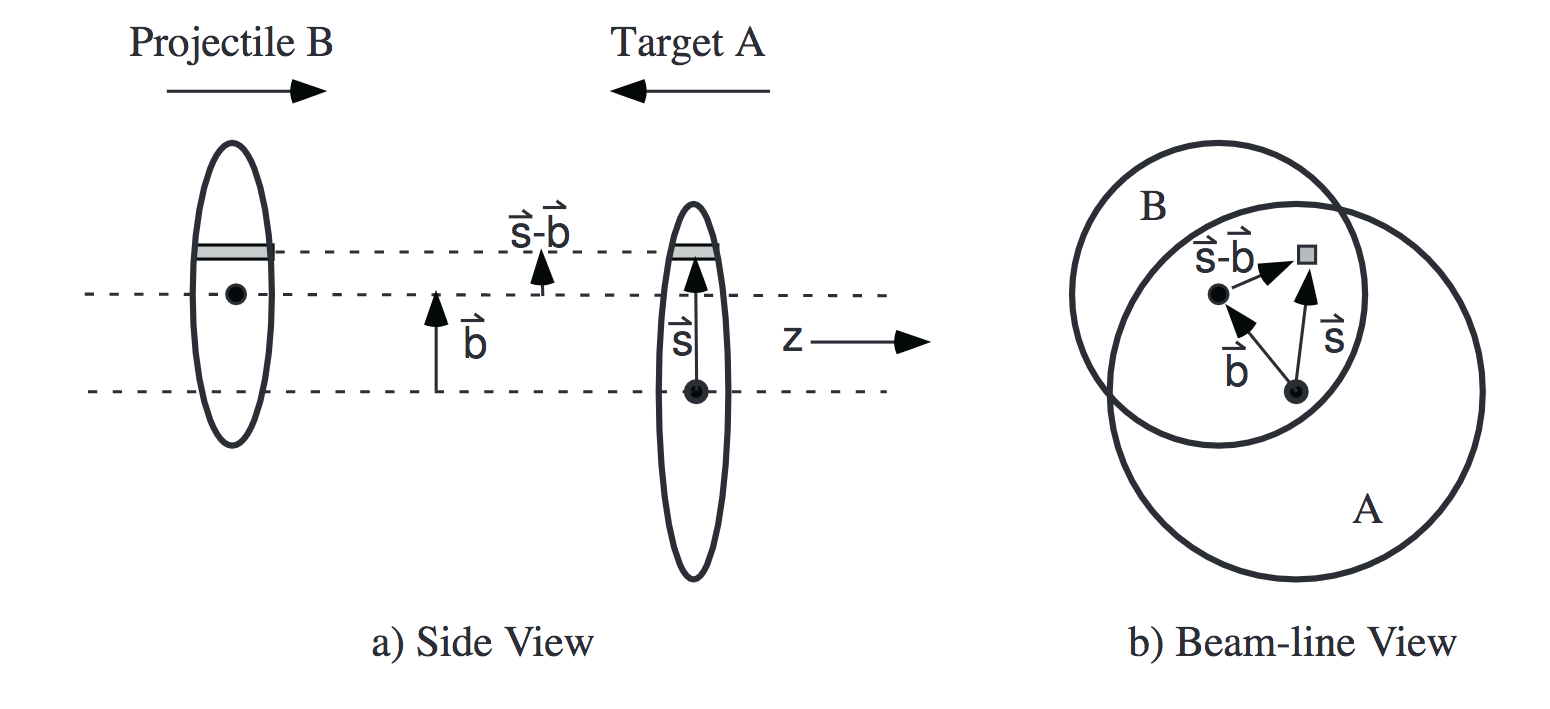} 
	   \caption{Nuclear collisions as viewed in a Glauber model calculation on the side view (a) and in the beam direction (b). Figure taken from ~\cite{Chaudhuri:2012yt}.}
	   \label{fig:GlauberModel}
	\end{figure}

	When integrated over the nucleons in A, one can calculate the expected number of hard collisions, $N_{coll}$ with the nucleons from B at the same two-dimensional position ($s-b$ in coordinate from the center of B, as shown in Fig:~\ref{fig:GlauberModel}) where $\sigma_{inel}^{NN}$ is the total inelastic nuclear nuclear scattering. 	An accurate description of the $N_{coll}$ variable is important to extract useful physics from the \raa. 
		
	\begin{equation}
		N_{coll} (b) = \sigma_{inel}^{NN} T_{AB}(b).
	\end{equation}

	Since a heavy ion collision is inherently a many-body QCD problem, it involves microscopic non-equilibrium physics (as we realized in the first chapter), and a full description or solution of its evolution is an area of active research. What we can do as physicists, is to describe a coarse grained collective motion of the system after thermal equilibrium with a fluid-like treatment. Thus it is intuitive that hydrodynamics is an integral part of the theoretical modeling~\cite{Qin:2015srf}. Almost all of the heavy ion MC models in market utilize some form of a hydrodynamics transport model for its particles and the interested reader can find an exhaustive review here~\cite{Jeon:2015dfa}. As was mentioned in the introductory chapter, the main focus of this thesis deals with jet quenching and so we will take a brief look at how individual parton energy loss is treated in both theory and MC generators.   
	
	\subsection{Jet Quenching}

\begin{figure}[h] 
		   \centering
		   \includegraphics[width=0.6\textwidth]{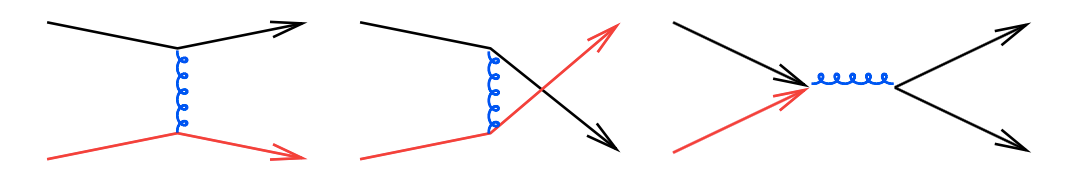} 
		   \includegraphics[width=0.6\textwidth]{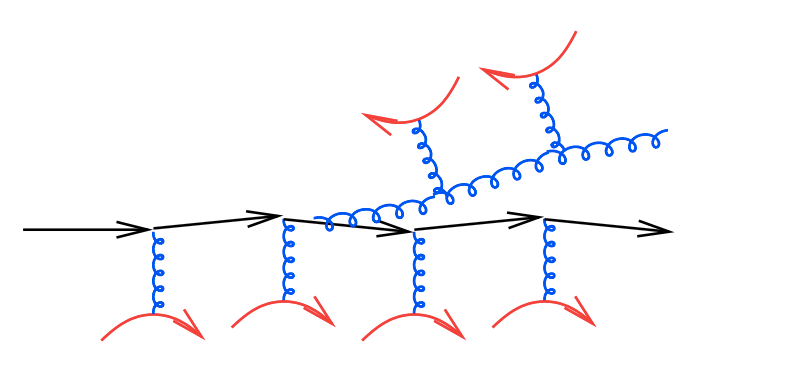} 
		   \caption{Leading order representation of elastic (top) and inelastic (bottom) energy loss. Black, red and blue represents the incoming, thermal medium and the interaction/radiation partons. Figure taken from Sangyong Jeon's QM2017 talk~\cite{JeonJetTheory}.}
		   \label{fig:diageloss}
		\end{figure}

		Since a jet is a composite object, understanding jet energy loss in the medium entails first studying single parton energy loss as was done in the early papers on QGP phenomenology~\cite{Bjorken:1982tu, Gyulassy:1990ye, Wang:1992bz, Wiedemann:2009sh, d'Enterria:2009am, CasalderreySolana:2007zz}. The primary idea revolved around how a single parton lost energy via a variety of processes such as radiative (inelastic) and collisional (elastic) as a function of its path length in the medium as shown in Fig:~\ref{fig:diageloss} at leading order. 
		
\begin{figure}[h] 
		   \centering
		   \includegraphics[width=0.8\textwidth]{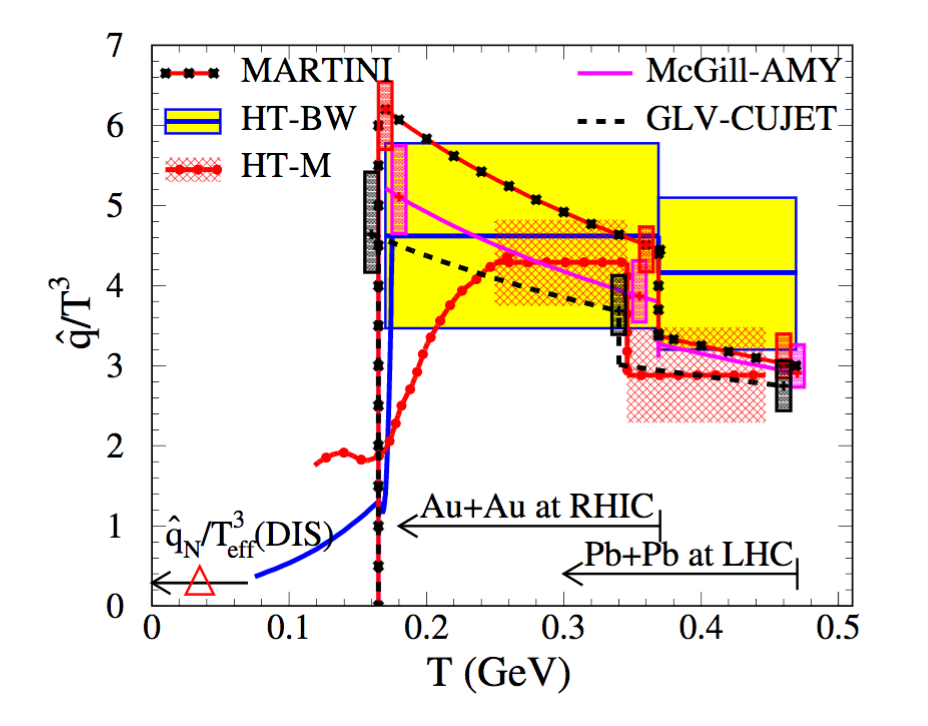} 
		   \caption{Temperature dependence of the jet transport parameter $\hat{q}/T^{3}$ scaled by the temperature for a 10 GeV quark extracted from different models comparing to single hadron suppression data as a function of the initial temperature. Figure taken from the JET collaboration~\cite{Burke:2013yra}.}
		   \label{fig:JETqhat}
		\end{figure}

		The direct calculation of these processes is not trivial since it involves a hot and dense medium. The medium is also finite size and is evolving and expanding in a non intuitive way. Such effects of the medium to first order can be included as corrections on top of the vacuum structure of jets since the soft interactions does not affect the hard scatter or the  production as soon on the bottom panels of Fig:~\ref{fig:atlasdijet}. Going back to our cross section for a single hadron from a hard scattering, we can add an additional term to the cross section as follows 
		
		\begin{equation}
		\begin{split}
			\frac{d \sigma (A+B \rightarrow H(p))}{d^{3} p} & = \int_{\rm{geometry}} \int dx_{a} dx_{b} f_{a/A} (x_{a}, \mu_{F}) f_{b/B} (x_{b}, \mu_{F}) \\
				& \times \int dz C \left(x_{a} p_{A}, x_{b} p_{B}, \frac{p}{z \mu_{F}} \right)_{ab \rightarrow c(p/z)} \\
				& \times P(c \rightarrow c\prime | T, Q) \\
				& \times D_{H/ c\prime} (z)
		\end{split}
		\end{equation}
		
		with medium modification $P(c \rightarrow c\prime)$ now dependent on the event geometry. This new term is related to the transport properties of the QGP, quantized via the $\hat{q}$ variable defined as the momentum transfer squared per elastic collision with the medium

		\begin{equation}
			\hat{q} = \frac{<(k^{g}_{T})^{2}>}{l_{coh}}
		\end{equation}
	
		where $l_{coh}$ is the coherent length and $k^{g}_{T}$ is the emitted gluon upon interaction with the QGP~\cite{JeonJetTheory}. The coherence length is effectively the threshold within which all scatterings count as a single scattering within the medium. When the coherence length is larger than the mean free path (the average maximal distance between scatterings), it implies a suppression of coherent radiation i.e. gluon bremsstrahlung. This is formally known as the Landau Pomeranchuck Migdal~\cite{Levin:1995mx} effect and has been very effective in MC implementation of parton energy loss~\cite{Zapp:2008af}. 		
		Phenomenologically, the $\hat{q}$ variable is very interesting and a lot of literature exists in that subject~\cite{CasalderreySolana:2007sw, Liu:2015vna}. Briefly, it not only plays an important role in describing medium induced radiative energy loss (as we saw above), it also quantifies the transverse momentum exchange between the propagating jet and the medium. 
		The JET collaboration utilized single hadron suppression data and extracted a parametrized temperature dependence of the temperature scaled jet transport parameter $\hat{q}/T^{3}$ at both LHC and RHIC energies~\cite{Burke:2013yra}. Their main result is shown in Fig:~\ref{fig:JETqhat} for a variety of jet energy loss models and with the summary as follows, for a quark with an energy of 10 \gev, it will lose around $1.2 \pm 0.3$ and $1.9 \pm 0.7$ of \gevc~per fermi of distance travelled in the plasma at RHIC and at the LHC with initial temperatures of 370 and 470 MeV respectively. 
		
		It is interesting to note that the value of $\hat{q}$ in LHC is larger compared to RHIC, essentially meaning that the two mediums are quantitatively different. While the overall effect is comparable, when say estimated by the nuclear modification factors, the corresponding energy loss is much larger at the LHC (and gets even larger when increased center of mass energy of the collision). 
		
		The prescription for energy loss for a given jet builds on the aforementioned hadron suppression, since a jet corresponds to its leading constituent in the limiting case of infinitely small lateral size of the jet. One intuitive picture is to take the vacuum branching of a jet and run the individual constituents through the energy loss prescription. The probabilistic nature of this branching causes several issues qualitatively since the medium interactions inherently depend on the nature of the splitting and the matter density in the path of the jet and its constituents. There are several p-QCD MC approaches that treat this prescription utilizing scattering centers and with either elastic or inelastic scattering with such centers. For example, the AMY~\cite{Arnold:2001ms}, ASW~\cite{Salgado:2003gb} model focusses on multiple soft scatterings and averages over the path length to define transport coefficients. Such a method excludes hard scatterings, whereas methods such as HT~\cite{Wang:2001ifa} or DGLV~\cite{Wicks:2005gt} take into account one scattering with interference terms calculated in multiple orders. The elastic scattering centers described in the BDMPS-Z formalism often includes the LPM suppression as discussed above for emitted gluons based on their formation times as given in ~\cite{Zapp:2008af}. See ~\cite{Qin:2015srf,Wiedemann:2009sh,Blaizot:2015jea} for detailed reviews on the theoretical aspect of jet quenching in heavy ion collisions. 
		
		With all these questions and complexities involved in the theoretical description of the QGP, any experimental measurement in heavy ions needs to keep in mind a few of fundamental points such as understanding the baseline behavior in pp (including at the NLO level) and relevance/connection to QGP transport properties or nuclear effects. 

\section{Jet measurements in the different systems}

\begin{figure}[h] 
	   \centering
	   \includegraphics[width=0.8\textwidth]{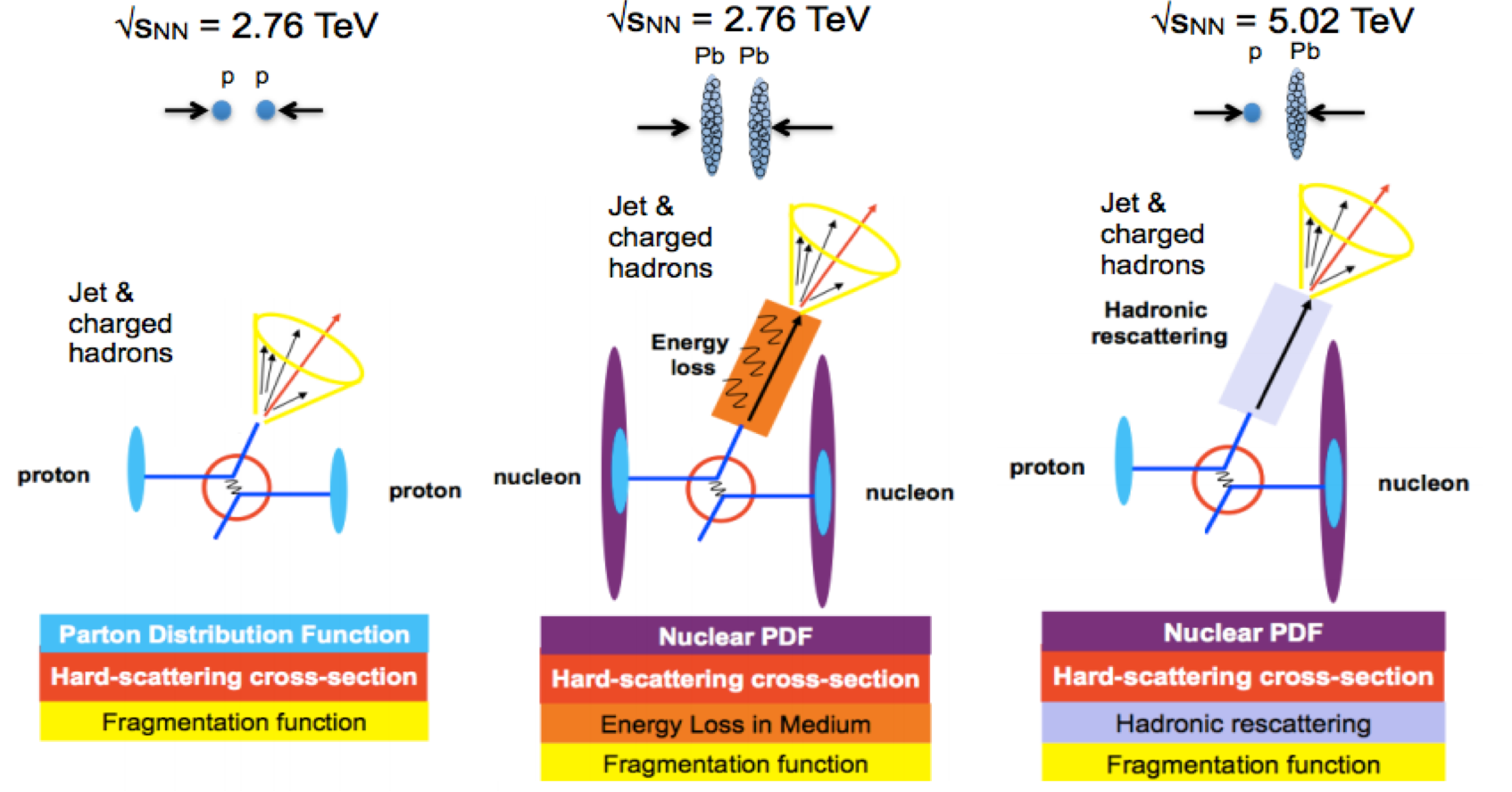} 
	   \caption{Cartoon representation of the collisions systems at the LHC Run1 and the different physics processes that can be accessed via jet studies in the respective systems. Figure taken from Eric Applet's talk at QM'14~\cite{ericjettalk}.}
	   \label{fig:cartoonDiffSystems}
	\end{figure}

	With what we learnt so far about QGP signatures, Fig:~\ref{fig:cartoonDiffSystems} shows a few of the standout measurements one can perform with the use of jets~\cite{ericjettalk}. 

	Jets in pp collisions are very useful to tune PDFs, cross check the hard scattered cross-section and probe the fragmentation functions, with the use of the charged hadrons. In pPb and PbPb collisions, we can study the nuclear modifications of the aforementioned measurements when comparing with pp, in addition to cold and hot nuclear matter effects respectively. We will go through some of these measurements in more detail in the upcoming chapters and see how we can extract physics and understanding about fundamental QCD and QGP behaviors by comparing data to theory.

\clearpage

\chapter{Experiment Details}
\label{ch_expDet}
\begin{chapquote}{Albert Einstein}
``A pretty experiment is in itself often more valuable than twenty formulae extracted from our minds." 
\end{chapquote}

\section{LHC and the CMS experiment}

	\begin{figure}[h]
	   \centering
	   \includegraphics[width=0.7\textwidth]{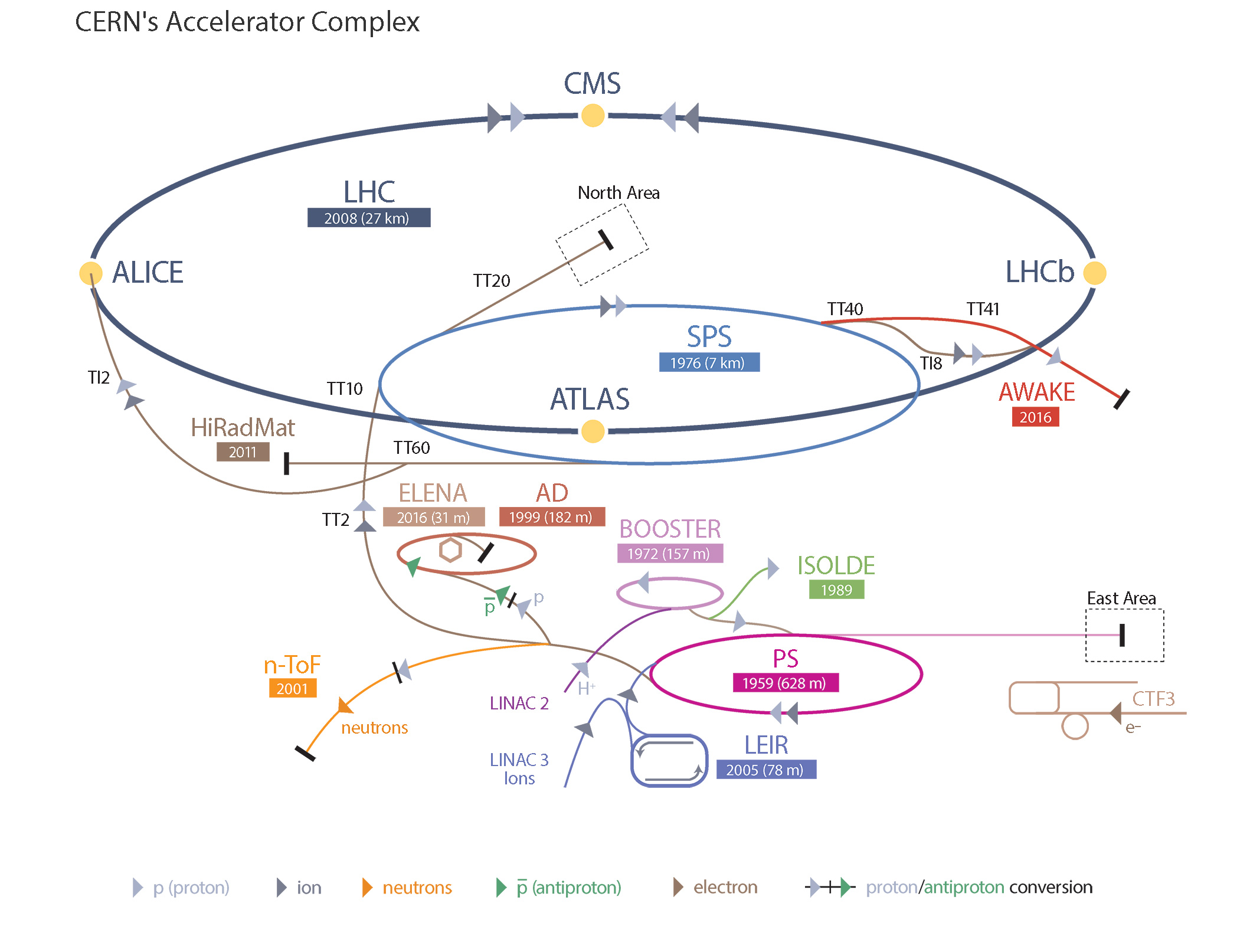} 
	   \caption{The CERN accelerator complex showing the production of the proton/ion beam at the LINAC/LEIR and its progressive acceleration towards the LHC. Figure courtesy of the Science and Technology Facilities Council, UK}
	   \label{fig:LHCAccelerators}
	\end{figure}
	
	The accelerator complex at CERN (\emph{Conseil Européen pour la Recherche Nucléaire} or European Organization for Nuclear Research), located just outside Geneva, Switzerland is uniquely capable of providing the most energetic particle collisions thus far in the world. The facility, as outlined in Fig:~\ref{fig:LHCAccelerators} has an impressive array of linear accelerators, ion rings (for heavy ion beams), synchrotrons, boosters, leading to the LHC ring with the ability to accelerate particles from gas cylinders to very nearly the speed of light. 
	
	In addition to proton beams, the LHC provides accelerated lead ions 208Pb for the possibility of colliding PbPb and pPb (and Pbp, please see Appendix~\ref{app_kinematics} for CMS convention) with the use of the same dipole magnets. The heavy ion collisions occur at a much smaller luminosity and the bunches contain less particles with a reduced frequency of the bunch crosses as shown in Table:~\ref{tab:ppvshinbeams}. This thesis deals primarily with data collected during the heavy ion running period of 2011 and 2013 where pp and PbPb were run at 2.76 TeV and pPb at 5 TeV respectively. One advantage in the pp reference runs is their low pileup\footnote{Number of additional primary vertices resulting from more than one hard scatter as the bunches cross each other} leading to a very clean data sample requiring only the very rudimentary event cleaning setups. For more details about the individual components and facilities at CERN and the LHC ring, please refer the to the technical design report~\cite{Bayatian:2006zz, PTDR2}. 
	
	\begin{table}[h]
	   \centering
	   \topcaption{Table highlighting the difference between LHC beam operational conditions for pp (13 TeV)~\cite{Sopczak:2017mvr} and PbPb (2.76)~\cite{Papaphilippou:2014qwa} collisions.} 
	   \begin{tabular}{@{} lcr @{}} 
	      \hline
	      Parameter    & pp (13 TeV in 2016) & PbPb (2.76 TeV in 2012) \\
	      \hline
	      beam energy & 6.5 TeV & 1.38 A TeV \\
	      $\#$ ions per bunch &  $1.6\times 10^{14}$   &  $1.6 \times 10^{8}$ \\
	      crossing frequency & 25 ns & 200 ns \\
	      luminosity & 40 f~b$^{-1}$ & 166 $\mu~\rm{b}^{-1}$\\
	      pileup per bunch crossing & $\approx 40$ &  $\ll$ 1\\
	      \hline
	   \end{tabular}
	   \label{tab:ppvshinbeams}
	\end{table}
		
	The Compact Muon Solenoid (CMS)~\cite{Chatrchyan:2008zzk} is one of the four main experiments on the LHC and is located in Annecy, France, farthest from the main CERN campus. The reason to have several experiments is one of fundamental necessity to all science i.e. reproducibility of results and verification of data. With proton-proton data at 8 TeV, experiments have discovered the Higgs particle (at both the CMS~\cite{Chatrchyan:2012xdj} and ATLAS collaborations~\cite{Aad:2012tfa}) along with an impressive array of results excluding the existence of beyond the standard model theories and particles, current status can be found in these reviews~\cite{Cakir:2015gya, Halkiadakis:2014qda}. The heavy ion running period is also an important time for CMS when operations shift from the large pp crew to the small but substantial heavy ion crew. We record, analyze and publish data leading to important results in the field of relativistic heavy ions, and quite recently, in pp collisions as well. 

	\begin{figure}[h]
	   \centering
	   \includegraphics[width=0.9\textwidth]{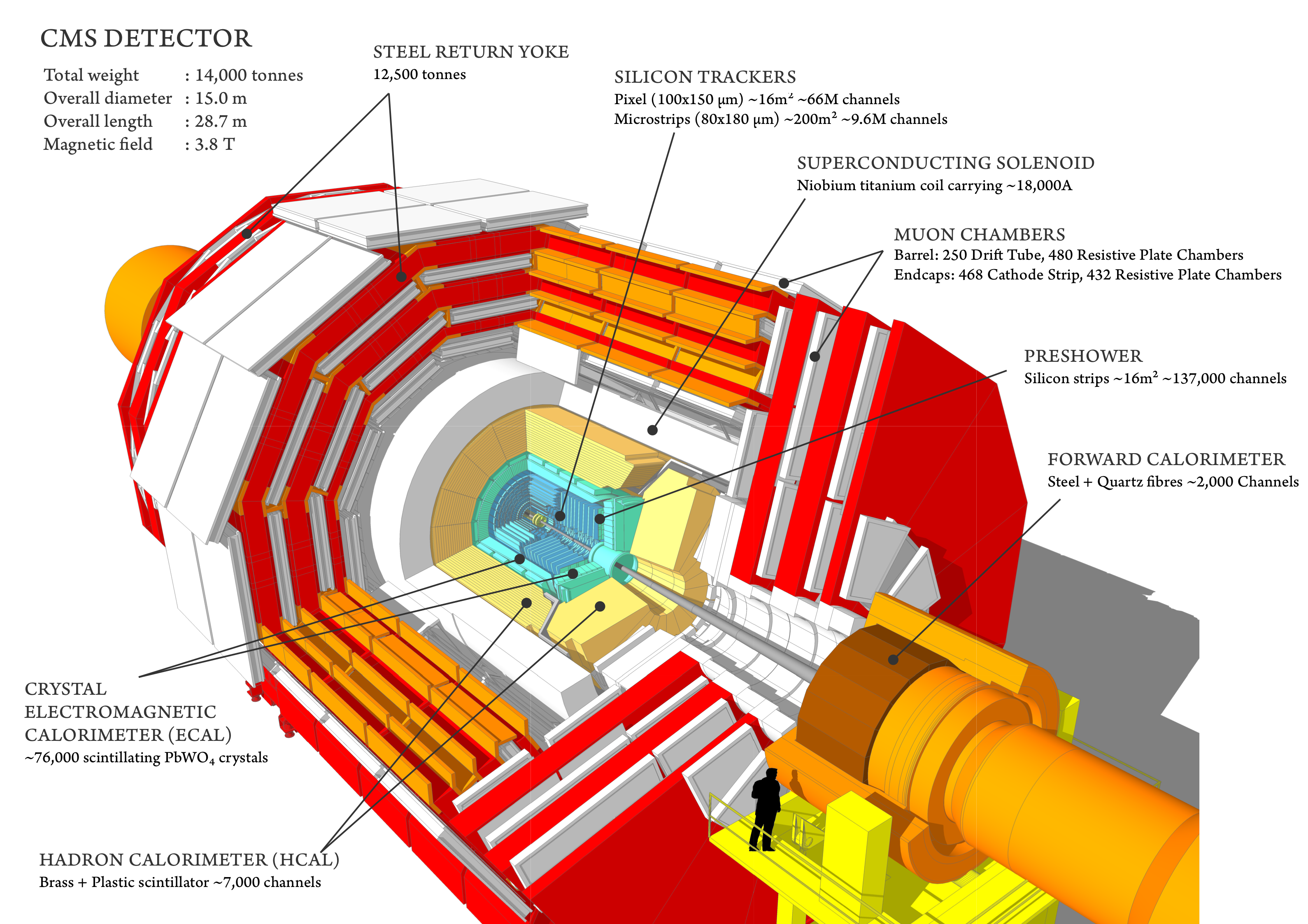} 
	   \caption[3D cutout of CMS]{3D cutout of the CMS experiment showcasing all the inner detectors from the silicon trackers (in blue),  electromagnetic and hadronic calorimeters (in cyan and yellow), the barrel solenoid (in white) to the muon chambers (in alternating red and white). Figure courtesy CMS collaboration.}
	   \label{fig:cmsDetectors}
	\end{figure}

	The CMS detector subsystems include the trackers, electromagnetic and hadronic calorimeters offering almost $4\pi$ hermetic coverage, magnets, the muon systems and the forward calorimeters as shown in the 3D rendition in Fig:~\ref{fig:cmsDetectors}. The central feature of the CMS apparatus is a superconducting solenoid providing a magnetic field of 3.8 T. Charged-particle trajectories are measured with the silicon tracker that allows a transverse impact parameter resolution of $\approx 15 \mu~m$ and a \pt~(transverse momenta) resolution of $\approx 1.5\%$ for particles with \pt~= 100 GeV/c. A PbWO$_{4}$ crystal electromagnetic calorimeter (ECAL) and a brass and scintillator hadron calorimeter (HCAL) surround the tracking volume. The forward regions are instrumented with an iron and quartz-fiber Hadronic Forward calorimeters (HF), which are very important for heavy ion events as we will see in the upcoming sections. A set of beam scintillator counters (BSC), used for triggering and beam halo rejection, is mounted on the inner side of the HF calorimeters. The very forward angles are covered at both ends by zero-degree calorimeters (ZDC). 
	
\begin{figure}[h]
	   \centering
	   \includegraphics[width=0.4\textwidth]{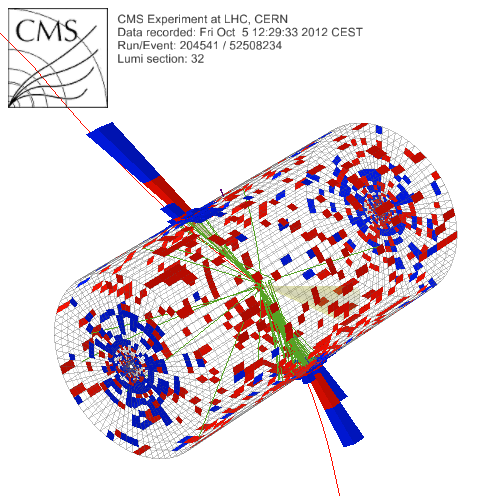} 
	   \includegraphics[width=0.4\textwidth]{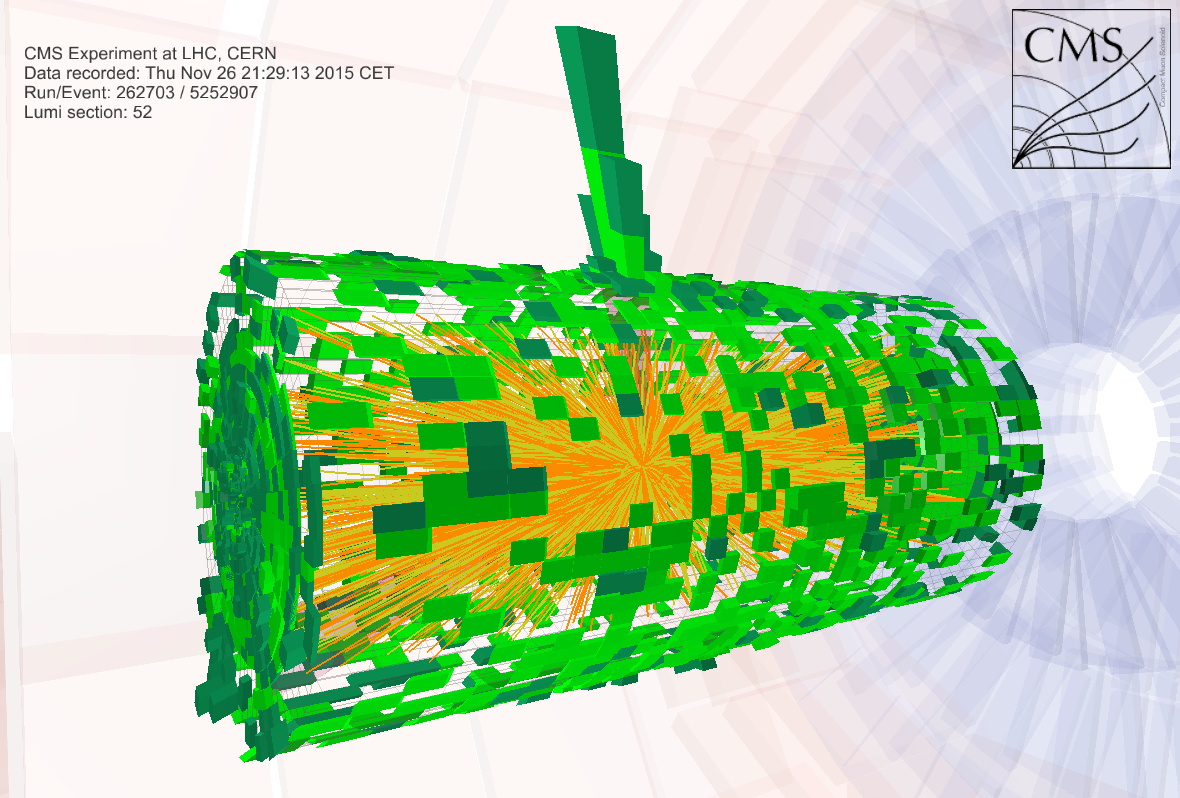} 
	   \caption{CMS event display of a di-jet system in proton proton collisions at 8 TeV recorded during Run1 at the LHC on the left and a high multiplicity PbPb collision from LHC Run2. Figure courtesy CMS Collaboration}
	   \label{fig:ppdijetevent}
	\end{figure}

	Example event displays for a dijet event in pp and PbPb collisions at CMS from Run1 and Run2 is shown in Fig:~\ref{fig:ppdijetevent}. As we discussed in the previous chapter, a dijet event is two back to back jets resulting from a hard quark/gluon from one proton scattering of another quark/gluon from the other proton, as shown on the left of Fig:\ref{fig:ppdijetevent}. Correspondingly, a quenched dijet system is shown on the right of Fig:~\ref{fig:ppdijetevent} in central PbPb collisions. Run1 at the LHC also had pPb collisions at 5.02 TeV in early 2013 but a pp reference run was not performed leading to an extrapolated reference which will be discussed in detail in the upcoming chapters. 
	
	 A separate suite of reconstructing software is installed in CMS to efficiently take data in the heavy ion environment. In comparison to pp events where the average track density in mid rapidity is around 16 tracks, heavy ion events average more than 1600 tracks for the most head on collisions with the same kinematic cuts, necessitating dedicated data operations. 

\section{Event selection and triggering}

	In order to clean our sample from background, beam gas, PKAM (Previously Known As Monster) and ultra peripheral (no significant hard scatter in the collision) events, the following cuts/filters are applied to select events off-line (at the analysis level). 
	\begin{itemize}
		\item BSC halo filter: events where any of the BSC halo bits fired were excluded. 
		\item Requirement of a reconstructed 2-track primary vertex was imposed. In peripheral events, all tracks above 75 MeV/c transverse momentum were used to reconstruct the vertex. In central events, the minimum \pt~requirement was increased, and the tracking region was narrowed down, to keep the maximum number of fitted tracks stable around 40 or 60, ensuring time efficient reconstruction. This requirement removes non-inelastic-collision events (e.g. beam gas, UPC, calorimeter) with large HF energy deposits but very few pixel hits. 
		\item A cut to remove PKAM events, which is a requirement of pixel cluster-length compatibility with the vertex.
		\item A requirement of an off-line HF coincidence, which requires at least 3 towers on each side of the interaction point in the HF with at least 3 GeV total deposited energy per tower.
	\end{itemize}
	
	Noise cleaning in the experiment also extends to the different reconstructed objects such as calorimeter clusters for example. Since you always have electronic noise, it is important to only select events with a calorimeter deposit due to particles from the collision (or depending on the run, cosmic rays). Shower shape, signal profile (in time and space) and, number of active channels are few of the most important quantities for a calorimeter deposit enabling distinction between signal and noise and these were specifically tuned and tested for heavy ion events for the analysis performed in this thesis. Detailed information about the calorimeter noise studies can be found here~\cite{1748-0221-5-03-T03014}.  

	\subsection{CMS trigger architecture}
		
\begin{figure}[h]
		   \centering
		   \includegraphics[width=0.8\textwidth]{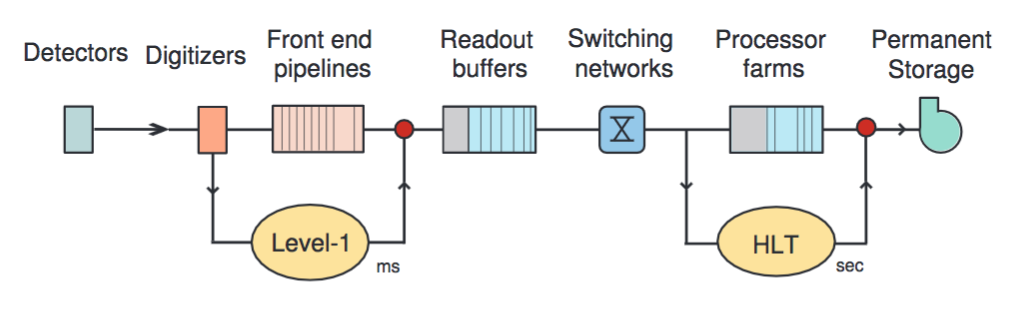} 
		   \caption{Flowchart of the CMS trigger systems, both L1 and HLT from detector digitization to permanent storage~\cite{Roland:2009wc}.}
		   \label{fig:cmsTrigger}
		\end{figure}

		Since the inelastic nucleon-nucleon scattering cross section is smaller than the elastic one, a lot of the events in heavy ion (and pp as well) collisions are not what we are interested in; high $Q^{2}$ momentum transfer. At the same time it is also not currently possible to write down every single collision to tape due to limits on data transfer. Thus one needs a workflow that makes a decision on a particular event as interesting or not based on a user defined criteria as soon as it occurs and store only those for analysis purposes. This selection process needs to happen before the next event occurs and at CMS, we utilize very quick hardware and software tools that have the physics details programmed in them. The CMS triggering architecture is designed on two levels, the Level 1 Trigger (L1) and High Level Trigger (HLT) as shown in the diagrammatic representation in Fig:~\ref{fig:cmsTrigger}. 
		
		There are several ways one can employ the use of L1 and HLTs in an event, such as triggering on high energy photons, muons, jets, track multiplicity and HF energy deposits etc. Due to the limit of writing data, several of the triggers are prescaled, which means that those events are only written a fraction of the time. A prescale of 2, means that once every two events, that particular event is written down to tape. As we go to lower \pt jet triggers, the prescale value increases due to the increased cross-section. More details on the implementation and usage can be found in this recent extensive review~\cite{Khachatryan:2016bia}. 
		
	\subsection{Collision centrality}

		The notion of event centrality was brought up and discussed in the last chapter with respect to the glauber model. In reality, it is essential to classify an event's centrality using an experimental observable since the impact parameter is not accessible directly. The centrality variable also has to be de-correlated from the main observable of interest, to avoid systematic biases in the measurement. For example, when the main observable is say the average track density in a certain detector region, the results would be biased if the events are split into different centrality classes based on the multiplicity in the event.  But there are other measurements such as the flow Fourier harmonics, which we looked at in the last chapter, where it is very interesting to study the effect of system size for events with the same overall multiplicity. 

		\begin{figure}[h]
		   \centering
		   \includegraphics[width=0.42\textwidth]{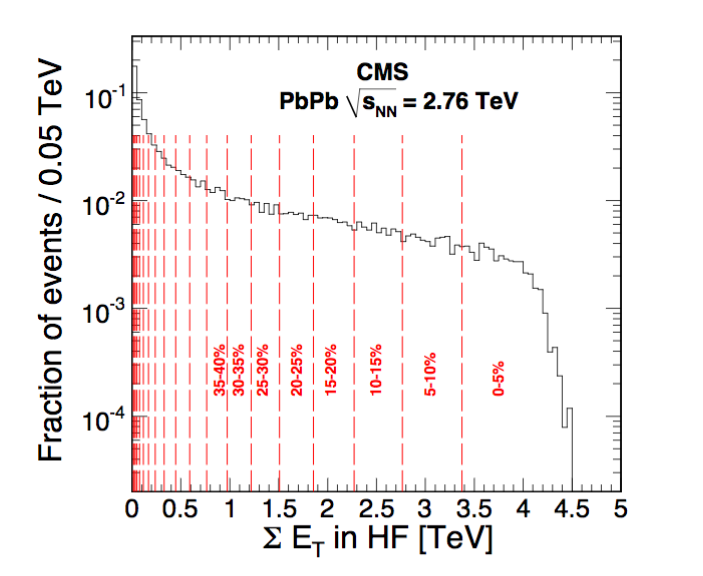} 
		   \includegraphics[width=0.4\textwidth]{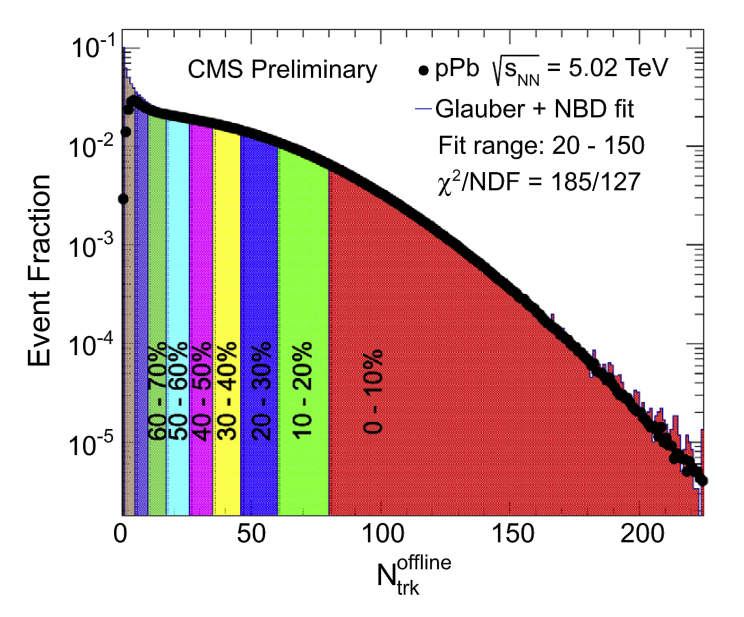} 
		   \caption{Event centrality determination in CMS for PbPb events~\cite{Chatrchyan:2011pb} using the energy deposit in the HF calorimeters (left) and for pPb events~\cite{Tuo2014258} using the total track multiplicity (right).}
		   \label{fig:cmscentralitypPbPbPb}
		\end{figure}
		
		In our published CMS papers, we have so far utilized two different estimations of an event centrality; total energy deposited in the HF calorimeters in PbPb~\cite{Chatrchyan:2011pb} and the offline track multiplicity in pPb~\cite{Tuo2014258} as shown in Fig:~\ref{fig:cmscentralitypPbPbPb}. In both these methods, the distributions are normalized to unity and bounds are chosen to represent a percentage fraction of the events. For example,  0-10\% represents events with the highest HF energy deposited (or largest number of tracks) and these are called as the most central collisions and similarly 80-100\% refers to peripheral collisions. Since the probability to have a central collision is smaller compared to peripheral collisions, there is a general need when studying hard probes to have maximal statistics possible.     
		
\section{Jets in the detector}
	
	Jets are the main actors in this thesis. The definition of a jet\footnote{Colloquially, one says that {\em a jet is a jet is a jet} and in practice, a jet is whatever the fastjet algorithm returns.} is primarily an agreement between experimentalists and theorists on their cluster algorithms as was done in the SNOWMASS accord~\cite{Huth:1990mi} in June 25th 1990 (which incidentally is only a couple of months after I was born indicating a rather curious connection). A jet represents a collimated spray of particles originating from a parton characterized by a lateral size limit and lower momentum cutoff. The lateral size limit is often called as the distance parameter or the jet radius and is described as a physical distance, for example in the $\eta-\phi$ plane. To get an idea of the jet radius, for a detector like CMS, a jet of R=4 would encompass the full barrel and cluster all final state particles in the event. The jet radii selection should be done carefully for a given analysis since it enforces an inherent selection on the fragmentation pattern and could also provide some handle on signal versus background in a very dense heavy ion environment. For searches of new physics in pp collisions, typically jets of a large size of R=0.7 or R=1.0 are used to fully contain the decay products of boosted objects. On the other hand, in heavy ion collisions, one seldom uses a radius larger than R=0.5 and we will quantify the effect in the upcoming chapters. 
	
	There many different ways one chooses to cluster these particles as outlined in App:~\ref{app_jetcomposition} but the main algorithm that is currently widely accepted and used is the \akt algorithm due to its infrared, collinear safety and rather pleasing overall shape. The algorithm clusters two objects if the distance between them $$ d_{i,j} = min(1/p_{t,i}^2 , 1/p_{t,j}^2)  \Delta R_{i,j}^2 / R^2$$ is less than the distance between the particle and the beam $$d_{i,B} = 1/p_{t,i}^2.$$ 
	
	
	Since CMS has both tracker and calorimeter systems, we can choose a variety of object collections for the jet clustering algorithms. Jets that are made from only tracks/calo towers are called track/Calo jets. There is also another option unique to CMS that uses all detector components which we shall go over in more detail in the coming sections. 				

	\subsection{Particle Flow}
		\begin{figure}[h]
		   \centering
		   \includegraphics[width=0.9\textwidth]{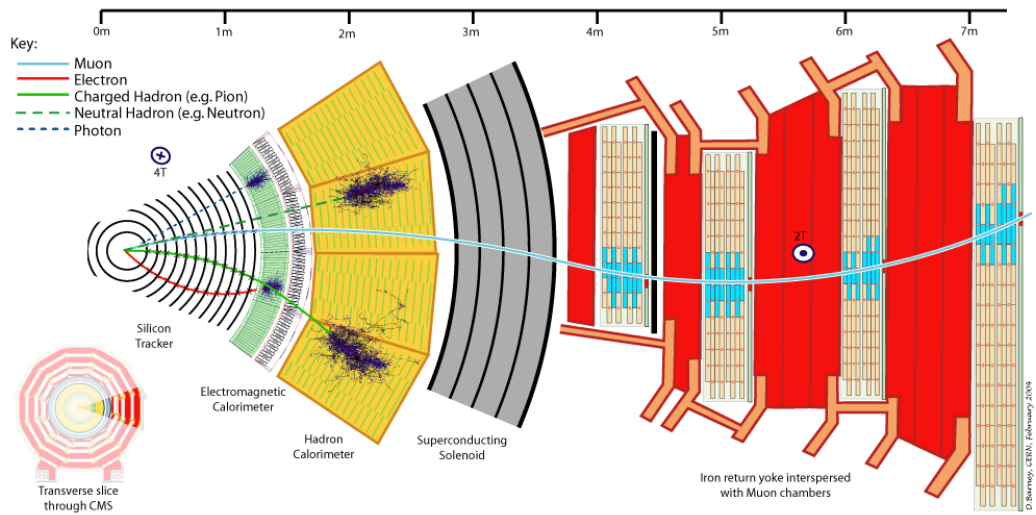} 
		   \caption{Cross sectional view of the CMS experiment and expected trajectories for different kinds of particles originating from the collision. Its is important the note the characteristic double bent path of the muon due to the solenoid magnetic field at the center of CMS. Figure taken from the public CMS collaboration website.}
		   \label{fig:cmsSideview}
		\end{figure}
		
		The CMS Particle Flow (PF) algorithm~\cite{Beaudette:2014cea} (sometimes referred to as Global Event Description) takes into account particle tracks, corresponding spatially matched calorimeter deposits (both ECAL and HCAL) and muon tracks and combines them into a single object representing as close to the original particle as possible. In its essence, it tries to follow the path of a given particle through the detector, collecting its energy deposits and momenta in every step of the way to globally identify the particle as one of the following types mentioned below and as shown in Fig:~\ref{fig:cmsSideview} 
		
		\begin{itemize}
			\item Photons ($\gamma$) : No tracks in the tracker but deposit in the ECAL 
			\item Electrons ($e^{\pm}$) : Tracks matched with a ECAL deposit
			\item Charged Hadrons ($h^{\pm}$) : Tracks matched with a HCAL deposit
			\item Neutral Hadrons ($h^{0}$) : No tracks in the tracker but energy deposited in HCAL 
			\item Muons ($\mu^{\pm}$) : Tracks in the inner tracker and the outer muon chambers (cleanest signal) and characteristic double bent signal due to the inner solenoid's magnetic field and the return iron yoke holding the muon systems.  
		\end{itemize} 
	
		
		In addition to the above mentioned categories of PF objects, the algorithm can also distinguish between hadronic style energy deposits in the HF calorimeter from electromagnetic style energy deposits based on the characteristic shower shape.  The interested reader can also refer Rishi Patel's PhD thesis~\cite{Patel:2014phdthesis} to learn how PF photons crucially helped in finding the Higgs boson.   
	
	
	
	\subsection{Jet Energy Corrections}
	
		\begin{figure}[h]
		   \centering
		   \includegraphics[width=0.9\textwidth]{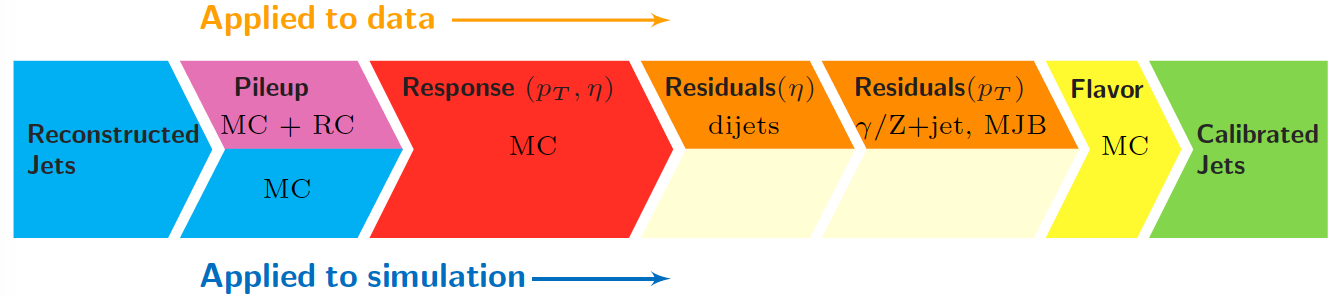} 
		   \caption{Schematic representation of the correction factors applied to the jet transverse momenta in data (top row) and MC (bottom row) and their meaning.}
		   \label{fig:JESincms}
		\end{figure}	
		
		Due to the finite and segmented detector response, it is imperative to apply corrections to the jets once they are clustered from reconstructed objects. The convention in CMS is outlined in Fig:~\ref{fig:JESincms} with several levels of jet energy corrections for data and MC~\cite{Chatrchyan:2011ds}. We briefly explain what they are and why they are needed. 
			
		\begin{figure}[h!]
		   \centering
		   \includegraphics[width=0.8\textwidth]{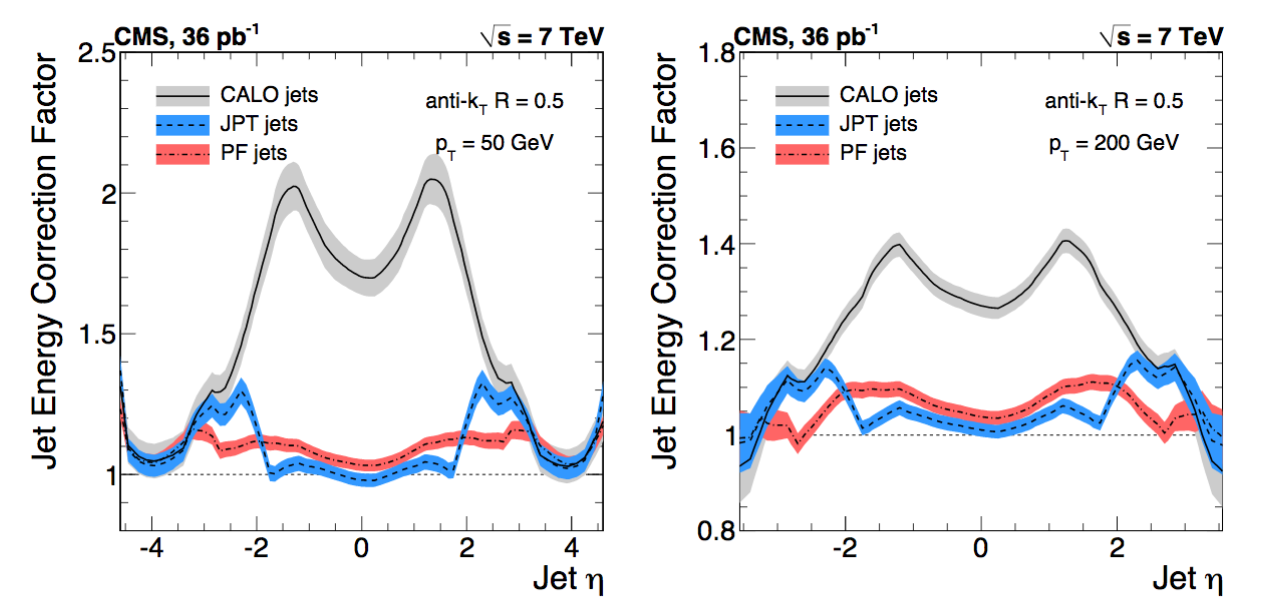} 
		   \includegraphics[width=0.4\textwidth]{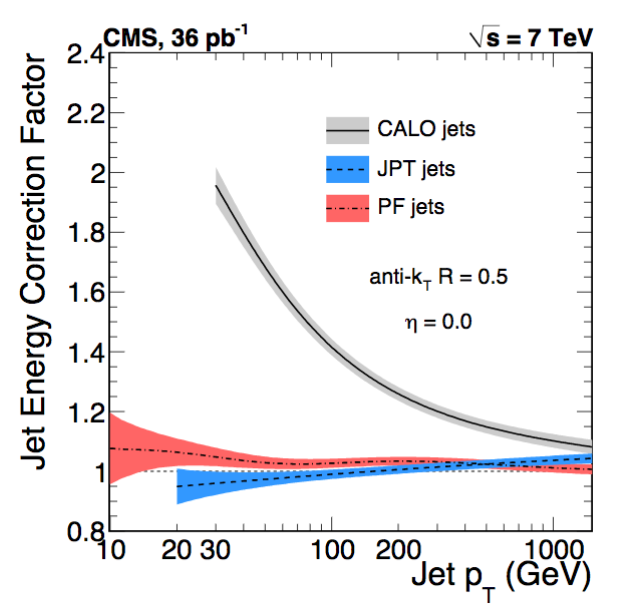} 
		   \includegraphics[width=0.39\textwidth]{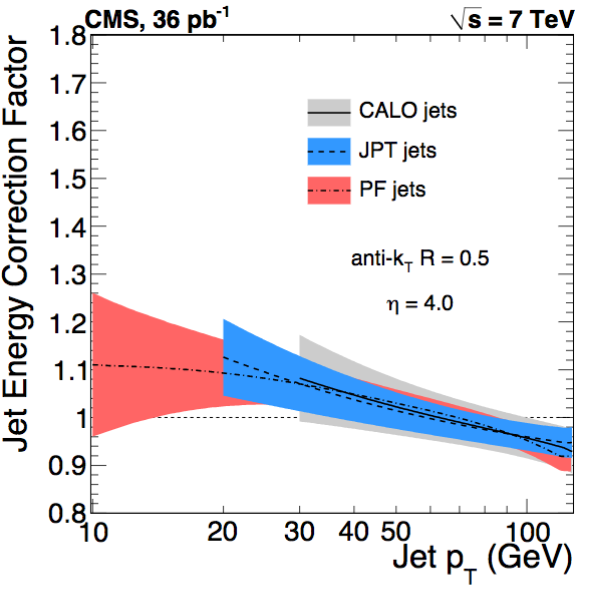} 
		   \caption{Jet energy correction factors~\cite{Chatrchyan:2011ds} for \akt PF (red), Calo (grey) and JPT (blue) jets shown as a function of jet pseudorapidity for low \pt~(top left panel), high \pt~(top right panel) and as a function of jet \pt~for mid pseudorapidity jets (bottom left panel) and forward jets (bottom right panel). The shaded regions represent the uncertainty in the correction factors.}
		   \label{fig:cmsJEC}
		\end{figure}	
				
		\begin{itemize}
			\item Pileup correction : This is the first correction applied to both data and MC jets based on the influence of pileup events. In the heavy ion samples we have very less pileup contribution and hence this correction is not applied in our case.  
			\item Response : Derived from MC samples and is the first major correction based on matching generator level (particle/gen) jets to reconstructed level (reco) jets. They are estimated as a function of the \pt~and $\eta$ and only correct the jet \pt.    
			\item Residuals : Secondary correction factors to be only applied on data by comparing MC to data. These residuals come in two categories as residuals as a function of $\eta$ and residuals as a function of \pt.
				\begin{itemize}
					\item Dijets : Utilizing the use of dijet even topologies where there is inherent momentum conservation in pp collisions. By identifying one of the dijets in a given detector $\eta$ region, we can measure the momentum asymmetry with the other dijet pair as they are found in different detector regions.     
					\item $\gamma/Z$+Jet : The $\gamma/Z$ in such events where a boson scatters of a jet can be used as a reference for the jet's \pt.  
				\end{itemize}
			\item Flavor : Purely MC based correction factor to account for the difference in the jet flavor (such as quarks vs gluons). 
		\end{itemize}
		
		The correction factors and their associated systematic uncertainty bounds for \akt PF jets with R=0.5 are shown in top panels of Fig:~\ref{fig:cmsJEC} as a function of the jet $\eta$ and the bottom panels as a function of the jet \pt. We see that the correction factor is the largest for CALO jets (in gray) and the least for PF jets (in red). The JPT jets (in blue) are essentially calo jets including the associated track information and have a better performance than just calo objects.     
		
		\begin{figure}[h]
		   \centering
		   \includegraphics[width=0.8\textwidth]{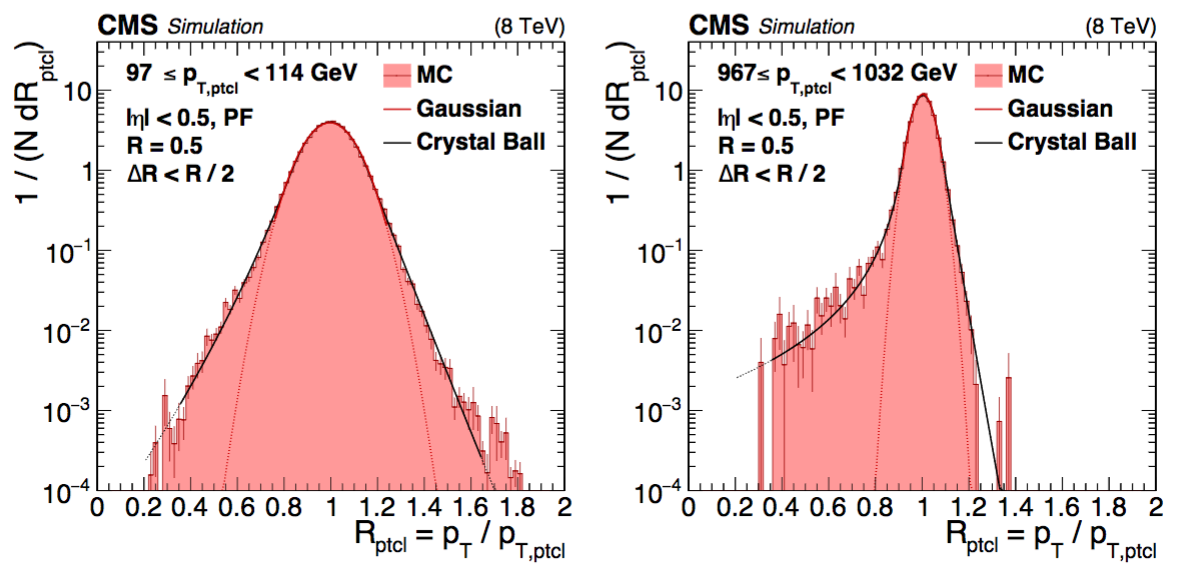} 
		   \caption{CMS Jet energy resolution for \akt PF R=0.5 jets for low (left) and high (right) \pt~ranges~\cite{Khachatryan:2016kdb}. The resolution from MC is fitted with a gaussian distribution and a crystal ball distributions to extract the tails.}
		   \label{fig:cmsJER}
		\end{figure}	
	
		While this is the standard case for pp collisions, for heavy ions (as we have begun to expect), things are not as straight forward. Since we have jet quenching, we cannot estimate residuals of any sort and the flavor dependent corrections are based on MC models which do not completely describe heavy ion collisions as saw earlier. An additional step for heavy ion jets involve background subtraction which we will look at in detail in the upcoming chapters. 		
	
	\subsection{Jet Energy Scale and Resolution}

		\begin{figure}[h]
		   \centering
		   \includegraphics[width=0.8\textwidth]{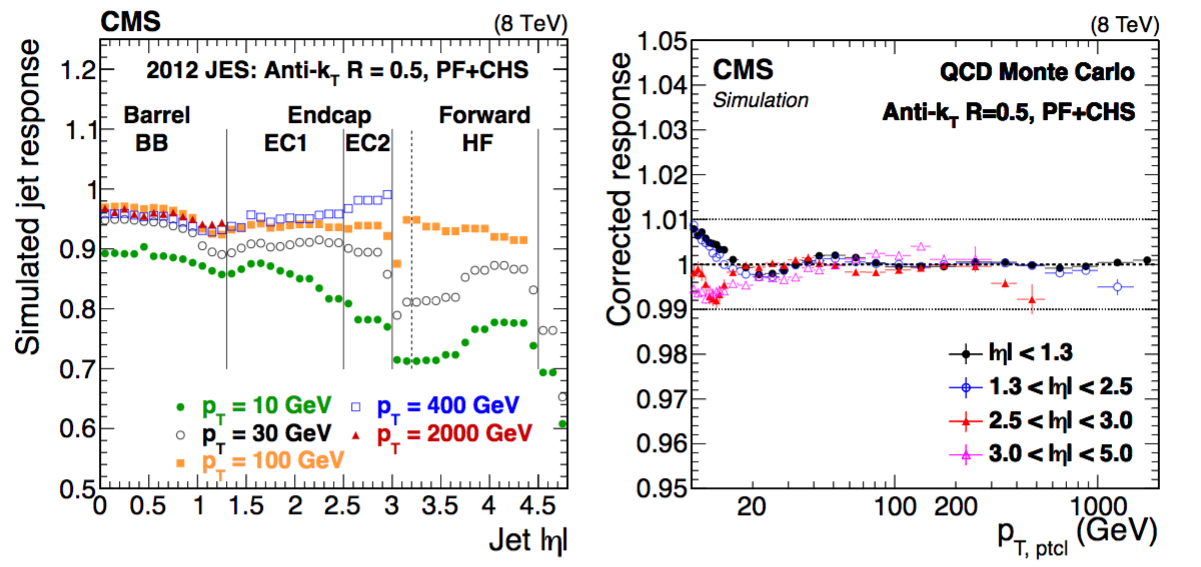} 
		   \caption{Simulated jet energy scale (left) as a function of the jet pseudorapidity and the excellent corrected response (right) within 1\%, as a function of the particle level jet \pt~across all pseudorapidity bins for \akt PF jets with R=0.5 and charged hadron subtraction to remove pileup contribution~\cite{Khachatryan:2016kdb}.}
		   \label{fig:cmsJESclosure}
		\end{figure}	

		Once these correction factors are applied to the reconstructed jets, the jet energy scale (JES) and jet energy resolution (JER) can be estimated by plotting the ratio of the corrected jet \pt~with the spatially matched particle level jet \pt~at the generator level~\cite{Khachatryan:2016kdb}. The mean (or $\mu$) of this distribution, as shown in Fig:~\ref{fig:cmsJER} is the overall jet energy scale or response and the width (or $\sigma/\mu$) is known as the resolution for a given jet \pt~window and $\eta$ selection. The distributions are usually fitted with a gaussian or a crystal ball function to extract these parameters.   

		The goal of the jet energy correction factors is to get the jet energy scale or response as close to 1 as possible at the same time reducing the resolution. The JES before and after the corrections is shown in Fig:~\ref{fig:cmsJESclosure}. The left panel shows the response from MC without any correction factors as a function of the jet $|\eta|$ for a variety of \pt~selections represented by different markers. The plot also shows the important detector regions covered such as the barrel, endcaps, detector gap in $3.0<|\eta|<3.2$ and the forward calorimeters. After applying JECs we can see the corrected response on the right panel of  Fig:~\ref{fig:cmsJESclosure} as a function of the particle jet \pt~and its within $1\%$ for all \pt~and $\eta$ ranges.

\clearpage

\chapter{Measuring the Jet Cross Section}
\label{ch_anaDetails}
\begin{chapquote}{Henri Poincare}
Experiment is the sole source of truth. It alone can teach us something new; it alone can give us certainty. 
\end{chapquote}

The main observable we are trying to measure is the double differential jet cross section defined as  
\begin{equation}
	\frac{d^{2}\sigma}{dp_{T} d\eta} = \frac{1}{L_{int}} \frac{dN^{jets}}{d p_{T} d \eta}.
\end{equation}   
where $L_{int}$ is in the integrated luminosity and its measured in bins of jet \pt~and $\eta$.  

In this chapter we shall go over the necessary analysis steps one has to undergo in order to measure the jet spectra in different systems such as pp, pPb and PbPb. We begin the chapter with event selection criteria and discuss the influence of pileup events on our sample followed by studies of the heavy ion background and the relevant background subtraction/noise suppression techniques for jets. 

\section{Event selection and pileup rejection}

\begin{figure}[htbp] 
		   \centering
		   \includegraphics[width=0.8\textwidth]{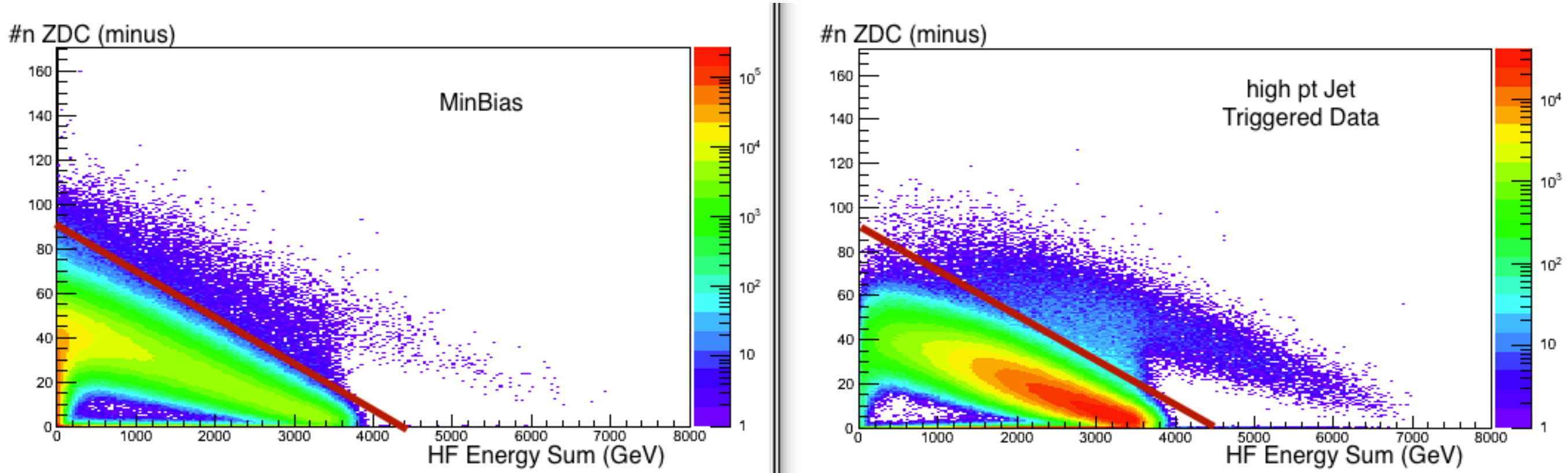} 
		   \caption{Number of neutrons in the ZDCminus vs HF energy Sum for minimum bias (left) and high \pt~Jet Triggered events (on the right) for PbPb collisions at $\sqrt{s_{NN}} = 2.76$ TeV. The solid red line indicates the cut above which events are identified as pileup events.}
		   \label{fig:ZDCm_vs_HF}
		\end{figure}

	In addition to the standard event selection criteria discussed in the last chapter, for all three kinds of events, pp, pPb and PbPb, we choose events where the primary vertex was situated in the barrel ($|v_{z}| < 15$cm). This selection ensured charged particles arising from the collision, would be measured by the tracker and thus offer better reconstruction performance. With regards to pileup, both the pp and pPb datasets were collected with negligible pileup and thus did not cause any problems. On the other hand, the PbPb dataset had a very small number of pileup events (affecting roughly $0.2\%$ of the dataset) and for this particular analysis, they were removed by applying a selection as follows. Firstly we can isolate them via correlations between deposits in the ZDC and the HF. Due to calibration issues of the ZDC-Plus during the entire run, we only rely on the ZDC-Minus information which is shown as the number of neutrons, assuming one neutron deposits 1.35 TeV of energy in Fig:~\ref{fig:ZDCm_vs_HF}. The ZDC Plus and Minus correspond to which side of the CMS detector the ZDC sits, with Plus along the direction of the counter-clockwise beam and Minus the opposite side. 

		Events that have two mid-central collisions instead of a single ultra-central event, i.e pileup events, are expected to have more spectator nucleons in the HF. Therefore we can isolate such events as having very large HF energy deposit but not corresponding ZDC events. Based on the band present in the minimum bias events shown on the left of Fig:~\ref{fig:ZDCm_vs_HF} we apply a straight line cut of shown in red and declare events above the line as pileup. 

		\begin{figure}
		  \begin{center}
		    \includegraphics[width=0.6\textwidth]{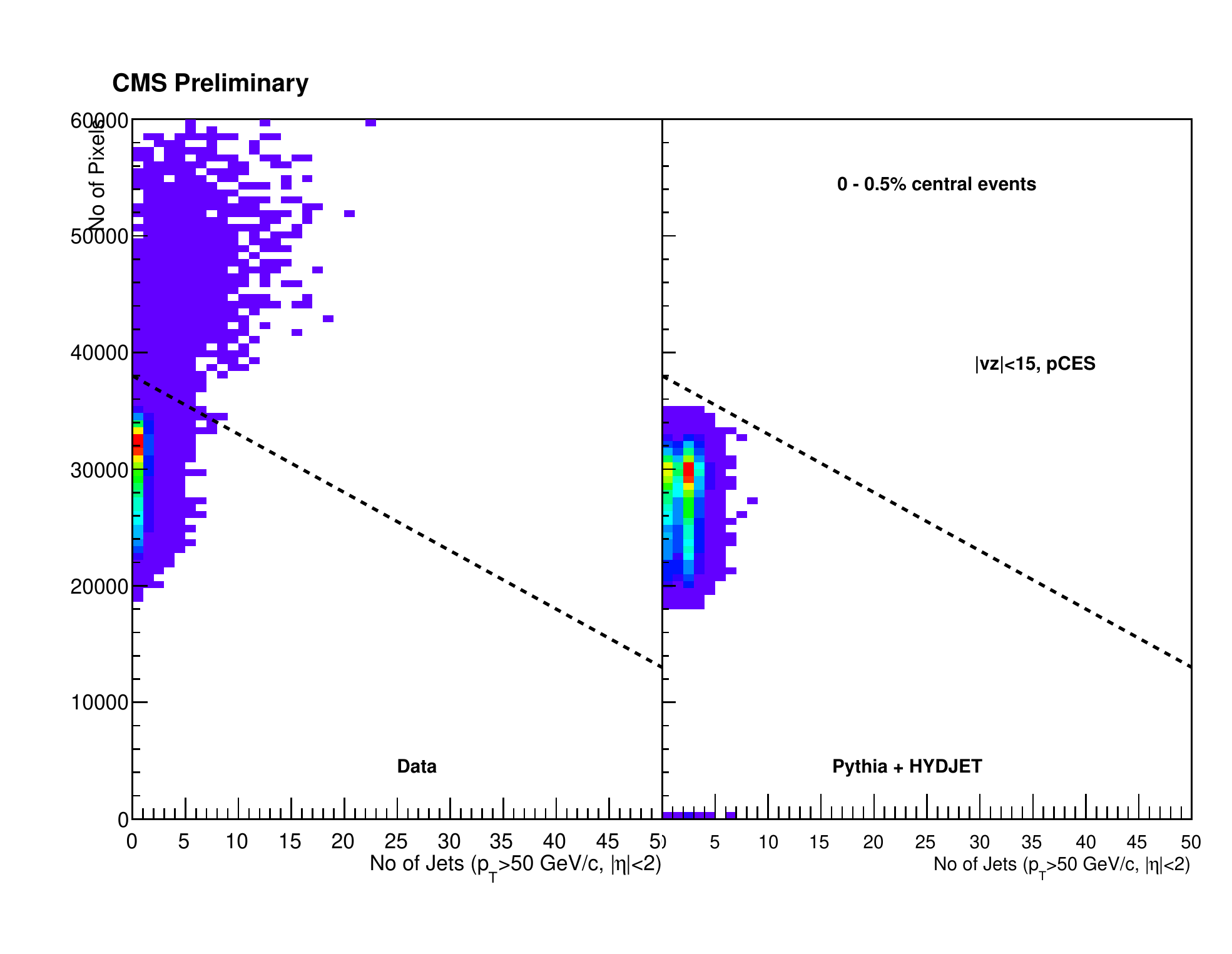}
		    \caption{Pixel hit counts as a function of the number of akPu3PF reconstructed jets with  \pt$ > 50$ GeV/c shown for the most central collisions ($0 - 0.05\%$) for data on the left and MC on the right.}
		    \label{fig:supernova_rejection}	
		  \end{center}
		\end{figure}

		 Secondly we can also tag events with large number of jets by comparing the number of pixel hits to the number of high \pt$>50GeV$ reconstructed jets as shown in Fig.~\ref{fig:supernova_rejection} for Data in the left panel and for MC simulations in the right panel for $0 - 0.05\%$ central collisions.  Area above the black dashed line represents the cut selection. As can be seen in Fig:~\ref{fig:supernova_rejection}, MC simulations and data differ for the large counts of pixel hits.  Events with a large pixel hit count (top of the black dashed line) in data are removed.  

\begin{figure}[h] 
		   \centering
		   \includegraphics[width=0.6\textwidth]{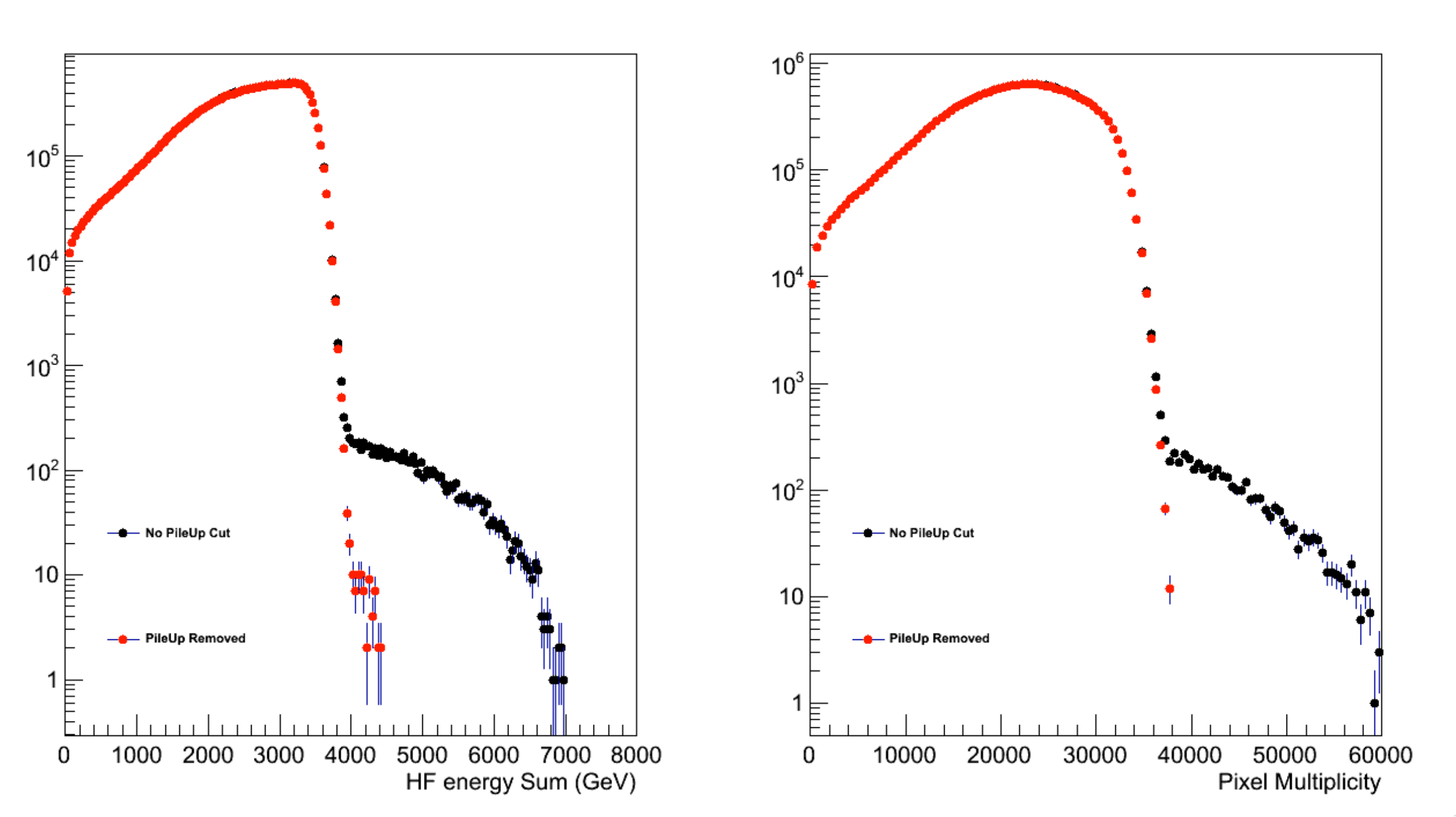} 
		   \caption{Comparison of HF energy sum (left) and pixel multiplicity (right) distributions before (black) and after (red) pile up rejections (pCES refers to events that pass standard event filters).}
		   \label{pileupCut_effect}
		\end{figure}

		We can see the effect of these two cuts on the HF Energy Sum (on the left) and the pixel multiplicity (on the right) of Fig:~\ref{pileupCut_effect}. The large tail of pileup events extending to the higher energies and multiplicities has indeed been removed by the application of both these cuts. 

\section{Underlying event and background subtraction in heavy ion collisions}

\begin{figure}[h!]
	   \centering
	   \includegraphics[width=0.9\textwidth]{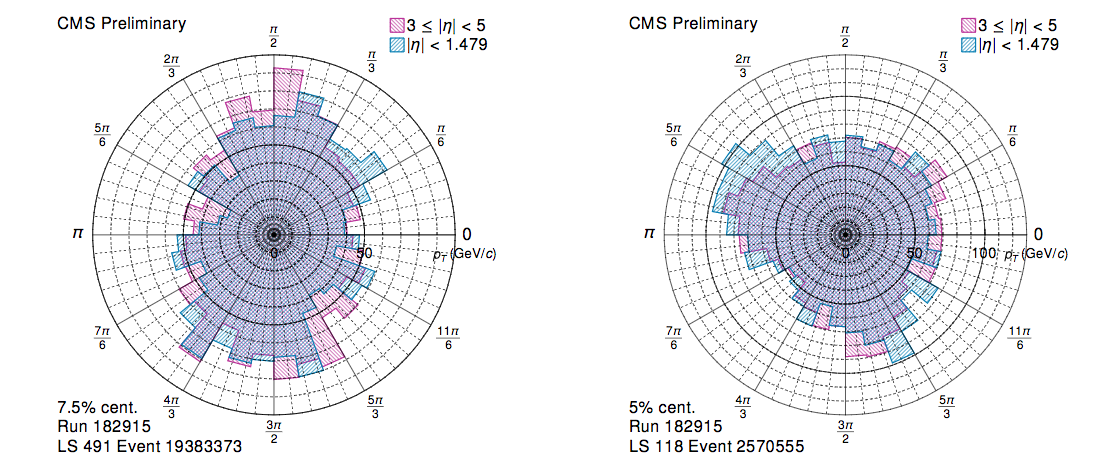} 
	   \caption{Transverse distributions of the flow modulations in PbPb collisions at CMS. Event with maximal elliptic (left) and triangular (right) flow properties. Figure taken from~\cite{CMS-DP-2013-018}.}
	   \label{fig:ueflowmodulation}
	\end{figure}

	In the context of heavy ion collisions, the underlying event is fluctuating on an event by event basis as one would expect, but also fluctuating inside a given event per centrality class since you can have a variety of flow modulations. As we see in Fig:~\ref{fig:ueflowmodulation}, we have two events that fall into the same central window, but end up having very different structures;  on the left we see an elliptical event but on the right we see a more triangular structure. Any estimate of the background needs to have a dependence, not only on the overall particle density but take into account the flow behavior as well.   
	
\begin{figure}[h!] 
	   \centering
	   \includegraphics[width=0.5\textwidth]{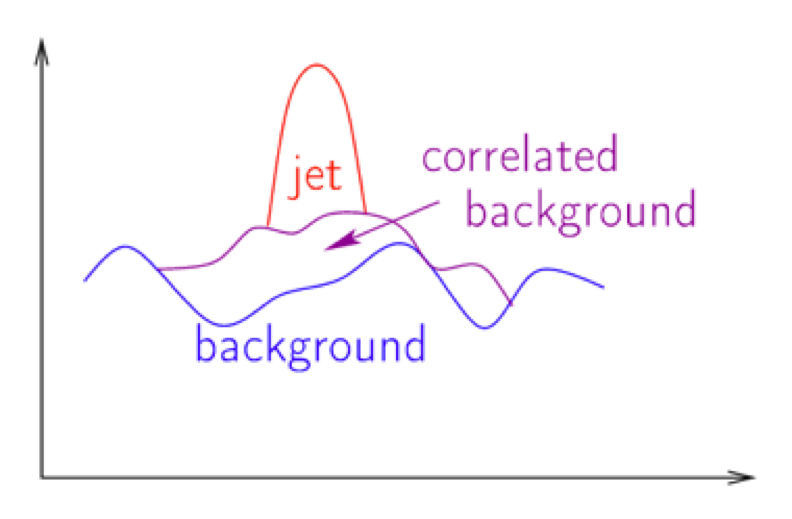} 
	   \caption{Cartoon drawing of the fluctuating background and the background correlated to the jet in a heavy ion event. Figure shown by Korinna Zapp at QM 2017~\cite{KZQM17}.}
	   \label{fig:cartoonjetbkg}
	\end{figure}

	Since we are particularly interested in studying the impact of jets and the medium, Fig:~\ref{fig:cartoonjetbkg} shows a cartoon-like, but meaningful insight into how a jet interacts with the background. In addition to sitting atop a fluctuating uncorrelated background, jets bring with them a correlated background. 
	
	\subsection{Embedded MC simulations}
	
		\begin{figure}[h!] 
		   \centering
		   \includegraphics[width=0.45\textwidth]{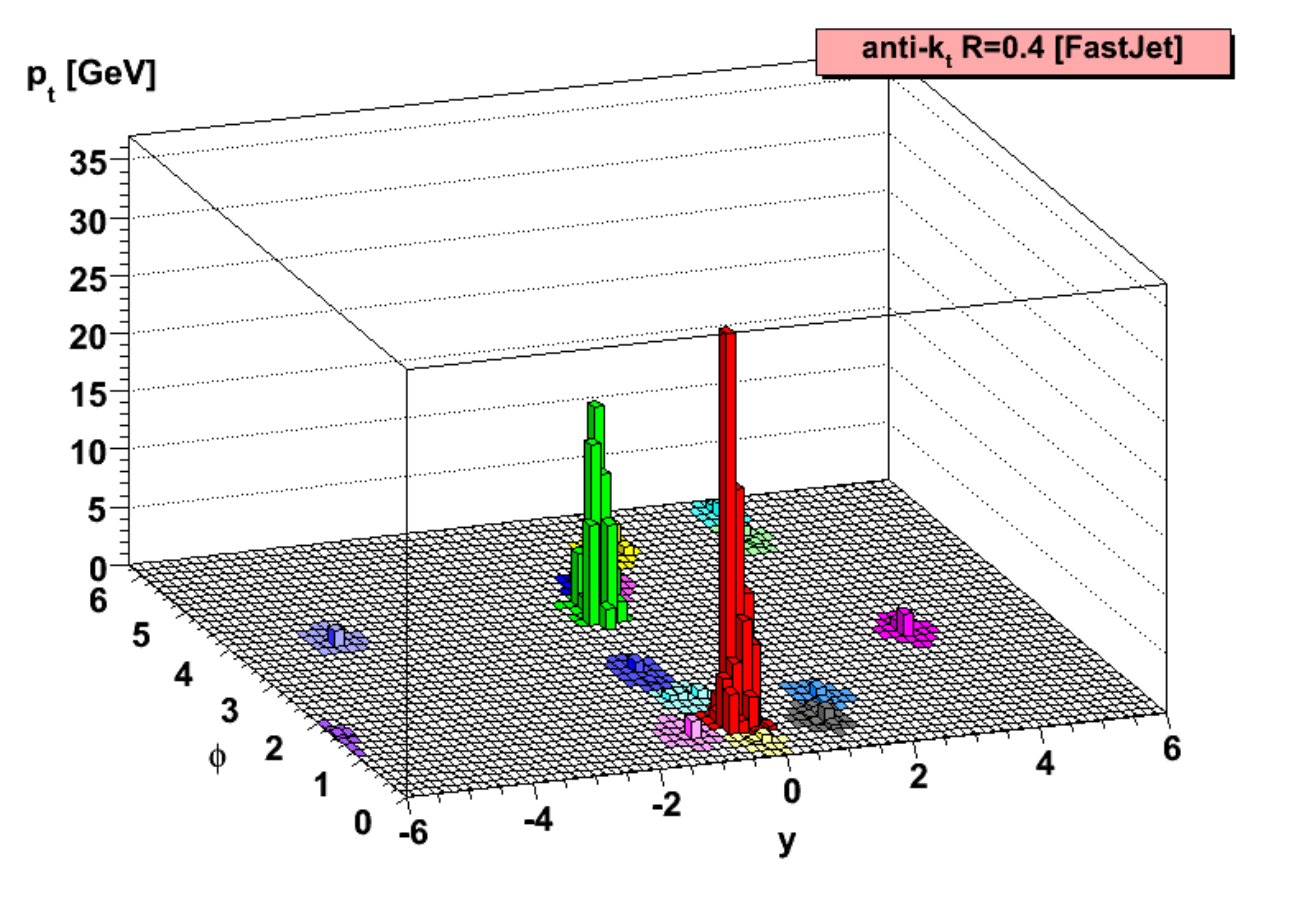} 
		   \includegraphics[width=0.45\textwidth]{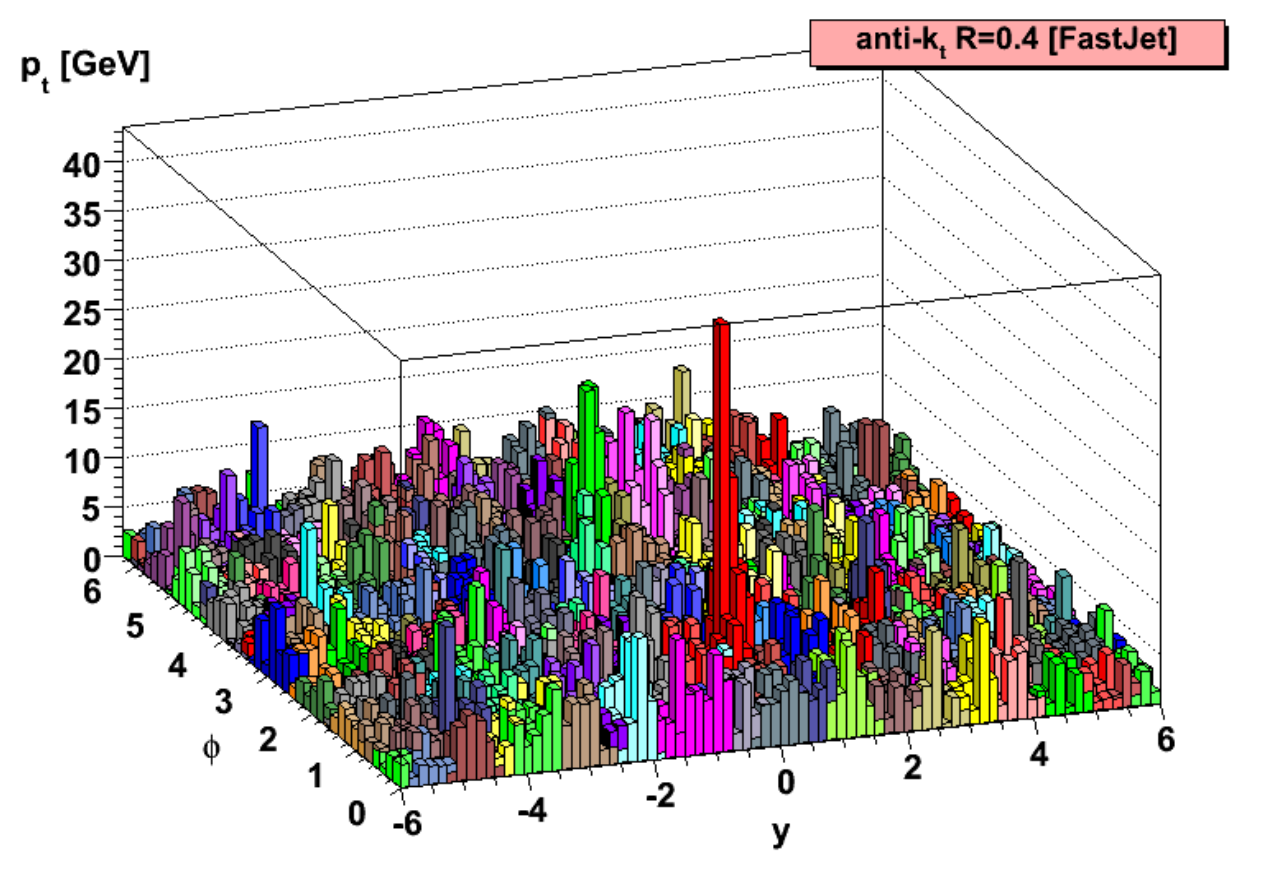} 
		   \caption{Top: Simulated pp \py~event ($y-\phi$ on the x-y axis and \pt~on the z axis) at 5.5 TeV. Bottom: Same event embedded in a \hd~background. Colors represent individual clustered jets with \akt and R=0.4. Figures taken from a talk by Gavin Salam~\cite{GSpyhdtalk2009}.}
		   \label{fig:pyhdembedding}
		\end{figure}
		
		The most popular underlying event MC models are \hd~(HYDrodynamics plus JETs) which treats heavy ion collisions as a superposition of the soft, hydro-type state and the hard state resulting from multi-parton fragmentation~\cite{Lokhtin:2008xi} and HIJING (Heavy Ion Jet INteraction Generator) which combines perturbative-QCD inspired models for multiple jet production with low pT multistring phenomenology~\cite{bib_hijing}. Both these primarily deal with the underlying event in complementary ways and in experiments at the LHC, we have utilized these for the study of jets by embedding \py~hard scattered events in them. A visual example of a single event embedding is shown in Fig:~\ref{fig:pyhdembedding}, the top panel a sample \py~dijet event and the bottom panel is the same event after embedding a \hd~background event. 
		
		The overall underlying density for a cone of R=0.4 in that event at 5.5 TeV, is around 100 GeV. Which means that, if we had a jet of say 150 GeV, it could be a 50 GeV jet sitting on top of the background and we need to compare it to a 50 GeV pp jet once we perform background subtraction. 

	\subsection{Background subtraction techniques}

		\begin{figure}[h!]
		   \centering
		   \includegraphics[width=0.9\textwidth]{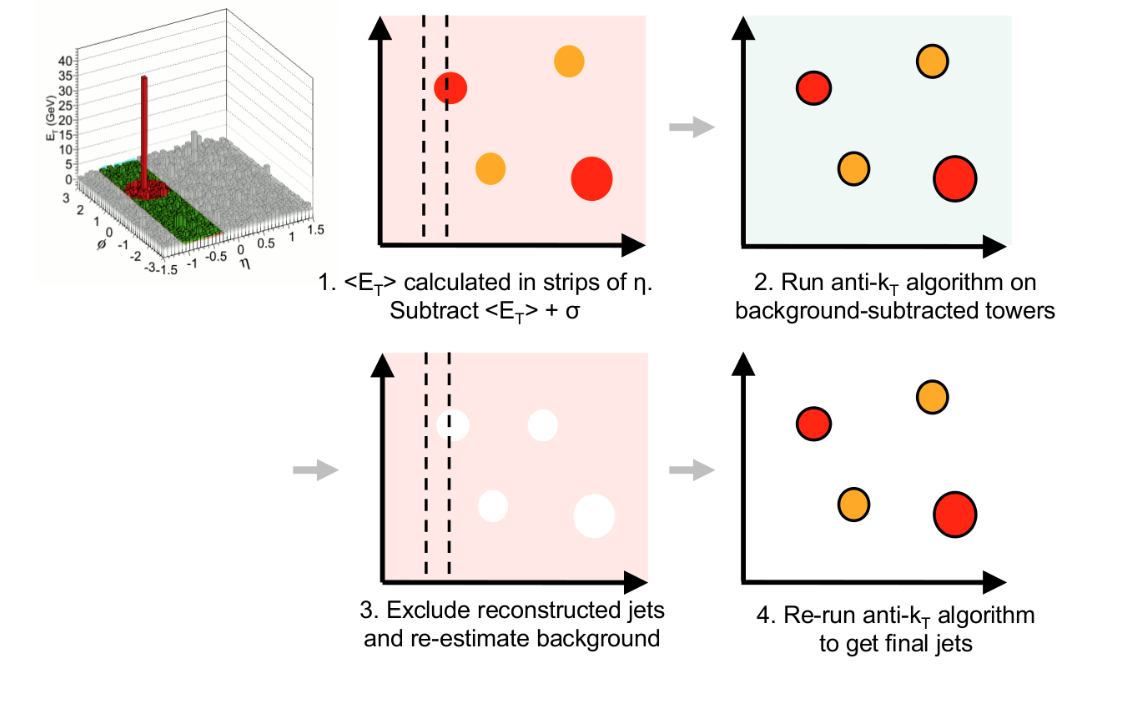} 
		   \caption{Diagrammatic representation of the PU algorithm. Details can be found in the text. Figure taken from a presentation by Matt Nguyen shown in QM 2012 ~\cite{mattpusub}.}
		   \label{fig:PUsubtraction}
		\end{figure}

		\begin{figure}[h!]
		   \centering
		   \includegraphics[width=0.7\textwidth]{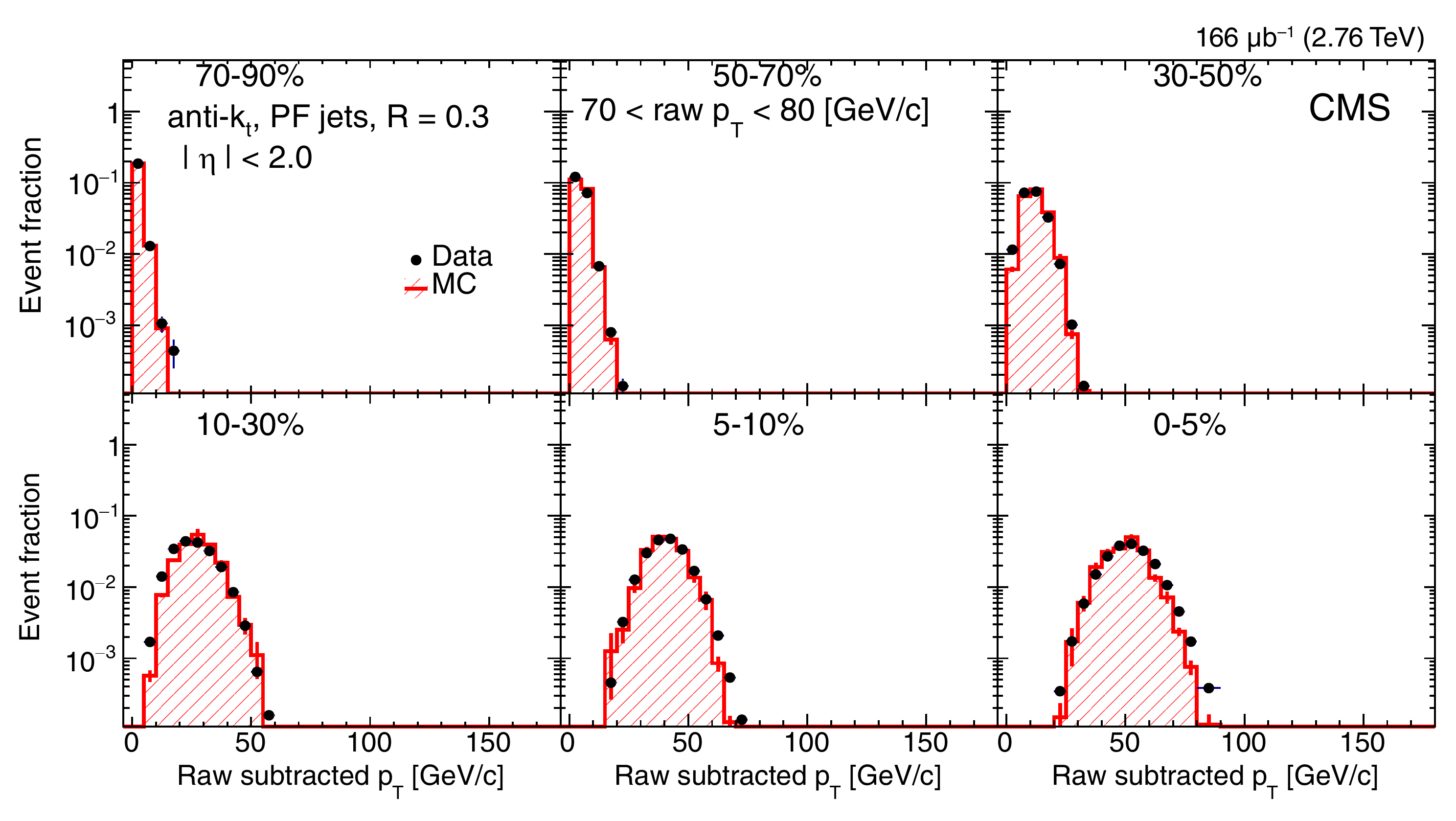} 
		   \includegraphics[width=0.7\textwidth]{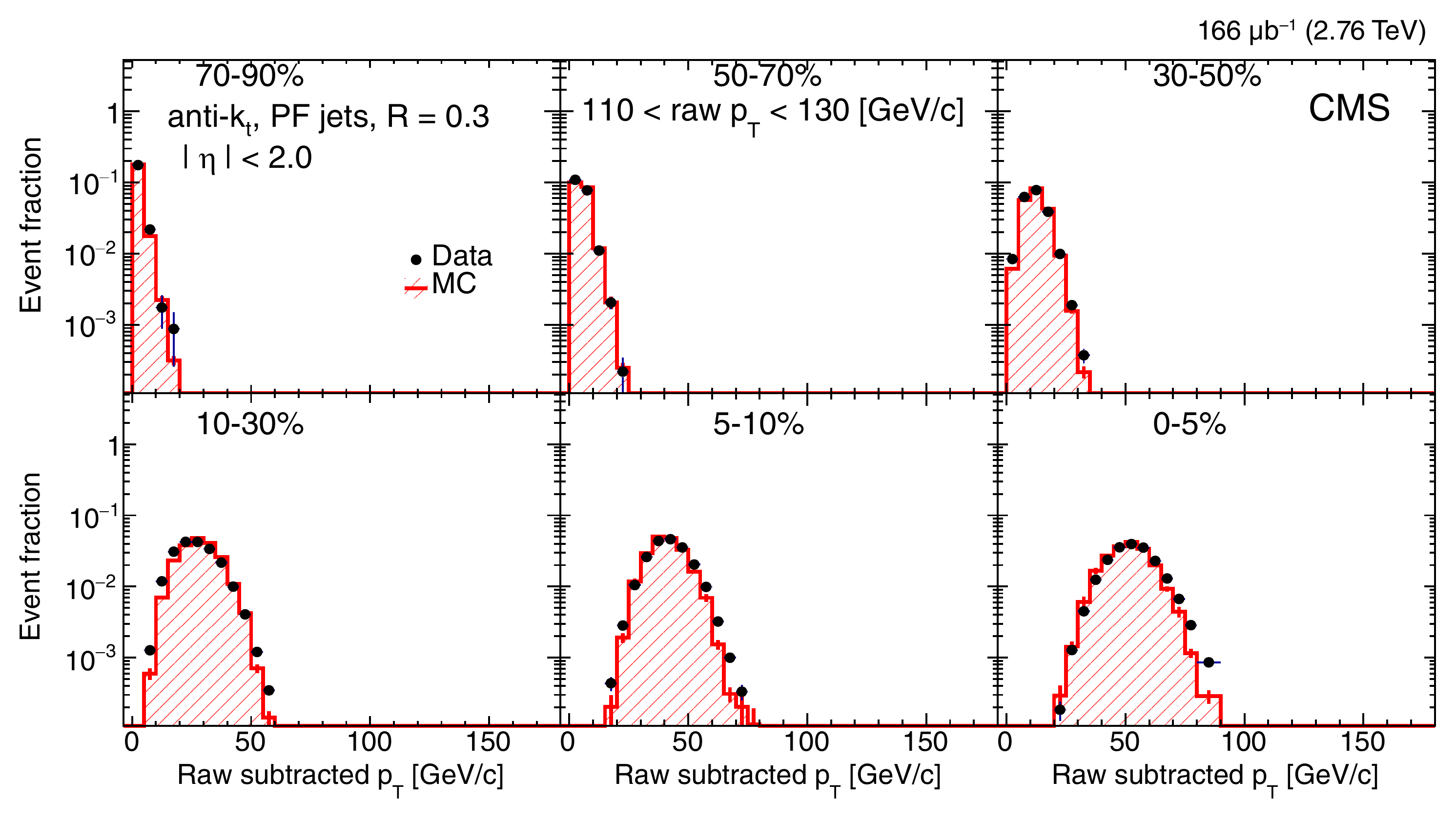} 
		   \caption{Raw subtracted \pt~for jets reconstructed with the \akt algorithm and a distance parameter of $R = 0.3$, in the ranges $70<  \text{jet \pt} < 80$\,\gev (top panels) and $110< \text{jet \pt} < 130$\,\gev (bottom panels). This quantity is found by taking the difference of the sum of PF candidates within the jet cone and raw jet \pt. Solid symbols show data, and the histogram is from \pyhd~generated events~\cite{Khachatryan:2016jfl}.}
		   \label{fig:rawsubjetpT}
		\end{figure}

		Fundamentally, the goal of any UE subtraction technique would be to remove the uncorrelated background leaving the jet structure intact. This is pretty hard in practice where there is no way to distinguish if one final state particle belongs to the initial hard scattered parton or just part of the UE. If we are just concerned about the jet's momenta, then a possible way would be employ what is known in the community as a $\rho$-subtraction meaning $p^{sub}_{T} = p_{T} - \rho \cdot A$ where $\rho$ is the UE density and $A$ is the area of the jet given by fastjet. This method is fine for first order estimations, but beyond that, there are several issues to consider; how $\rho$ is estimated, lack of account for fluctuations, insensitive to jet fragmentation etc. 

		\begin{figure}[h!]
		   \centering
		   \includegraphics[width=0.9\textwidth]{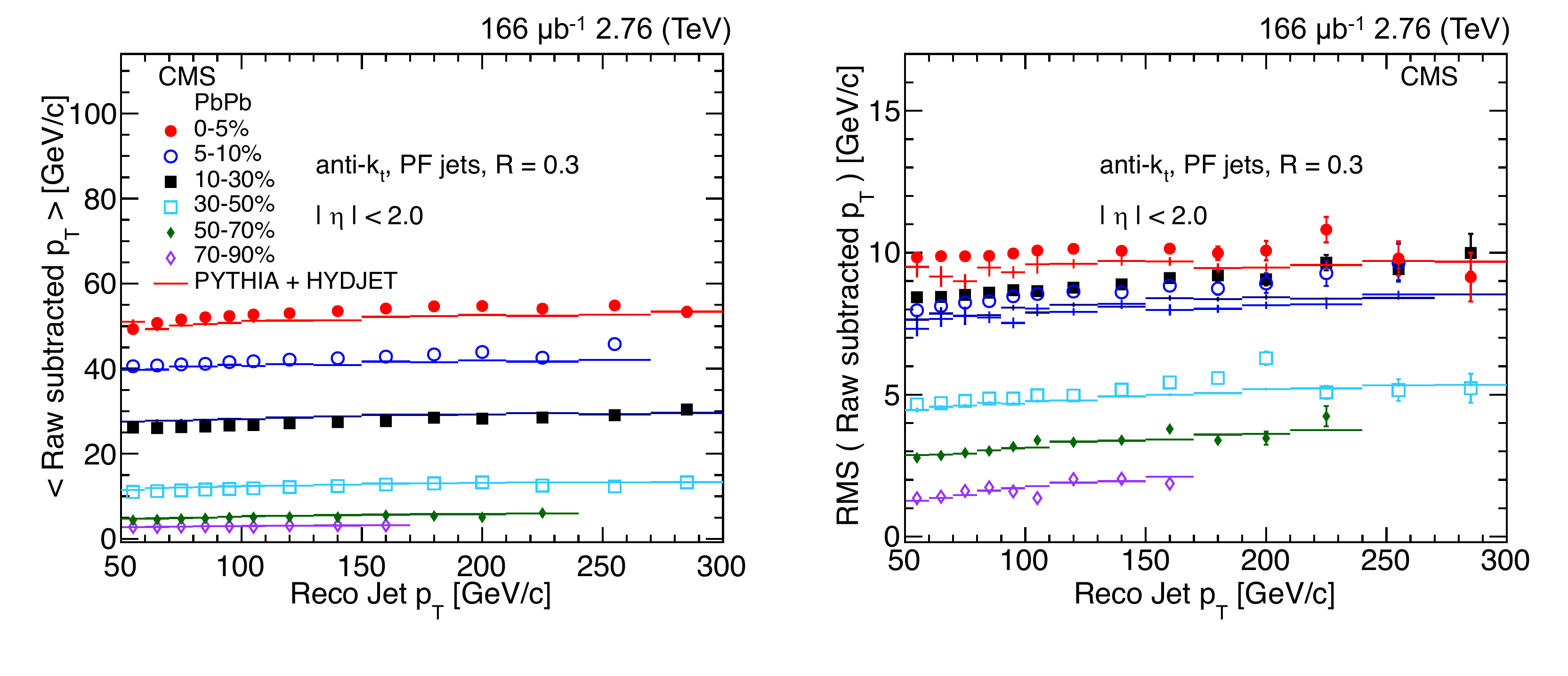} 
		   \caption{Average raw subtracted \pt~(Left) and its RMS (Right) for PF jets reconstructed with the \akt algorithm, with a distance parameter $R = 0.3$. Symbols represent data, and lines show \pyhd~simulated events~\cite{Khachatryan:2016jfl}.}
		   \label{fig:bkgmeanrms}
		\end{figure}
				
		For our analysis, we have utilized an algorithm designed to deal with backgrounds and their fluctuations called iterative pileup subtraction~\cite{Kodolova:2007hd}. The algorithm as its implemented in CMS can be seen in Fig:~\ref{fig:PUsubtraction} with 4 basic steps. It clusters the PF objects into pseudo-towers based on the HCAL geometry ($0.087 \times 0.087$) for an event and divides an event into strips of $\eta$. For each strip in eta, we take the mean and the standard deviation $\mu, \sigma$ from the distributions of the pseudotower energies and subtract from each tower the  $\mu + \sigma$. All towers that go negative are zeroed otherwise they would be unphysical during clustering. Then we cluster the event into jets and the jets, above a certain \pt~cutoff are removed from the event. After this step, we essentially recompute the mean and sigma for each strip and do the subtraction, following which the jets are put back into the event. This way, for a given jet radii, the effect of the subtraction to the correlated background is minimal.   

		The \hd~used in our simulations is tuned to match the distributions in data for particle/track density and the flow modulations. To study the background in PbPb events, data and \pyhd~simulations are compared. The correction to the jet \pt~obtained from this iterative subtraction technique (called ``raw subtracted \pt"), for a jet with distance parameter $R^\text{jet}$ is estimated by taking the difference between the sum of all the PF candidate \pt~in a $\Delta R < R^\text{jet}$ cone and the raw jet \pt. The $\Delta R$ is defined as the distance of the PF candidate from the reconstructed jet axis in the $\eta-\phi$ plane $$\Delta R=\sqrt{(\Delta\phi_\text{candidate, jet})^2+(\Delta \eta_\text{candidate, jet})^2}.$$

		The distributions of raw subtracted \pt~for $R = 0.3$ jets, from peripheral to central collisions are shown in Fig.~\ref{fig:rawsubjetpT} for two different reconstructed jet \pt~selections. Data are shown with filled circles and simulations with histograms. There is a good agreement between the two in all centralities and jet \pt~bins. A similar level of agreement is also seen for $R=0.2$ and $R=0.4$.
		
		The average raw subtracted \pt~and its root mean square (RMS) values are shown in Fig.~\ref{fig:bkgmeanrms} as a function of the reconstructed jet \pt, from central to the most peripheral collisions. Data are shown with markers and are compared with the \pyhd~generated events shown as histograms. The average raw subtracted \pt~decreases, from the most central to peripheral events, as expected, and distributions show reasonable agreement between data and \pyhd.

\section{Single jet triggers and the combined jet spectra}

\begin{figure}[h]
	   \centering
	   \includegraphics[width=0.8\textwidth]{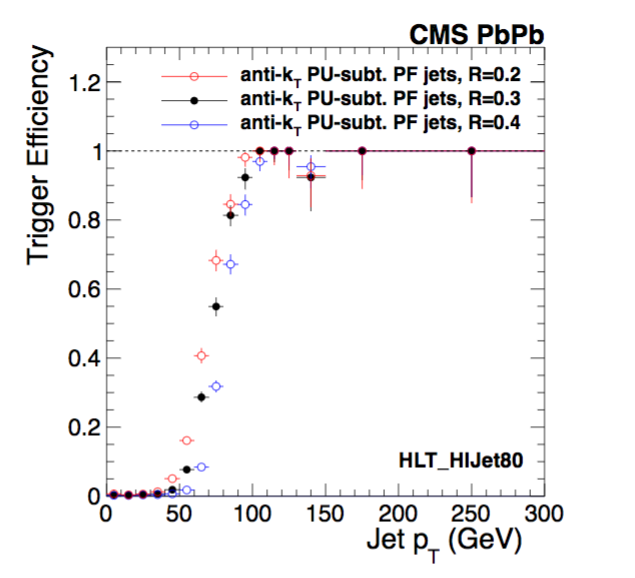} 
	   \caption{HLT$\_$Jet80 trigger efficiency curves for background subtracted \akt PF for different radii. Figure taken from the CMS trigger review~\cite{Khachatryan:2016bia}.}
	   \label{fig:cmsPbPbTriggerturnon}
	\end{figure}

	In the last chapter, we realized the need to collect data triggered on physics objects such as high \pt~jets since its very inefficient and sometimes impossible to collect all the data in a minimum bias way at CMS. For this thesis, HLT single jet triggers were primarily utilized in all three collision systems and the trigger efficiency for one such trigger (HLT$\_$Jet80) in PbPb events is shown in Fig:~\ref{fig:cmsPbPbTriggerturnon} as a function of the jet \pt~for different radii. The trigger HLT$\_$Jet80 corresponds to a high level trigger that selects an event if the corresponding reconstructed object, in this case a calo jet of R=0.5 clustered with the iterative cone algorithm, is greater than $80$ \gev. The jet spectra from such single jet triggered events are used to stitch together a minimum bias jet spectra with the use of trigger combination methods as described below. 

	\subsection{Primary method with three triggers}
	
		In order to extend the reach of the jet spectra, datasets from high-\pt\ single jet triggers with three different jet thresholds were combined together in both pp and PbPb collisions. The jet triggers that are used are:  HLT\_HIJet55, 65 and 80 for PbPb data and HLT\_PAJet40, 60 and 80 for pp data. 

		The entire sample is split into three exclusive categories:

		\begin{enumerate}
			\item JetC = ``$n_c$"
			\item JetB AND NOT JetC = ``$n_b$"
			\item JetA AND NOT JetB AND NOT JetC = ``$n_a$"
		\end{enumerate}

		where A, B and C are the three thresholds, from smallest to largest.  Taking into account that threshold C is always unprescaled, the weight factors applied to stitch the sample together are:

		\begin{enumerate}
			\item $w|_{\rm JetC} = 1$
			\item $w|_{\rm JetB\ \&\&\ !JetC} = 1$
			\item $w|_{\rm JetA\ \&\&\ !JetB\ \&\&\ !JetC} = [1/PS_{\rm A} ]^{-1}$ 
		\end{enumerate}

\begin{figure}[htp]
	    	    \centering
		    \includegraphics[width=0.6\textwidth]{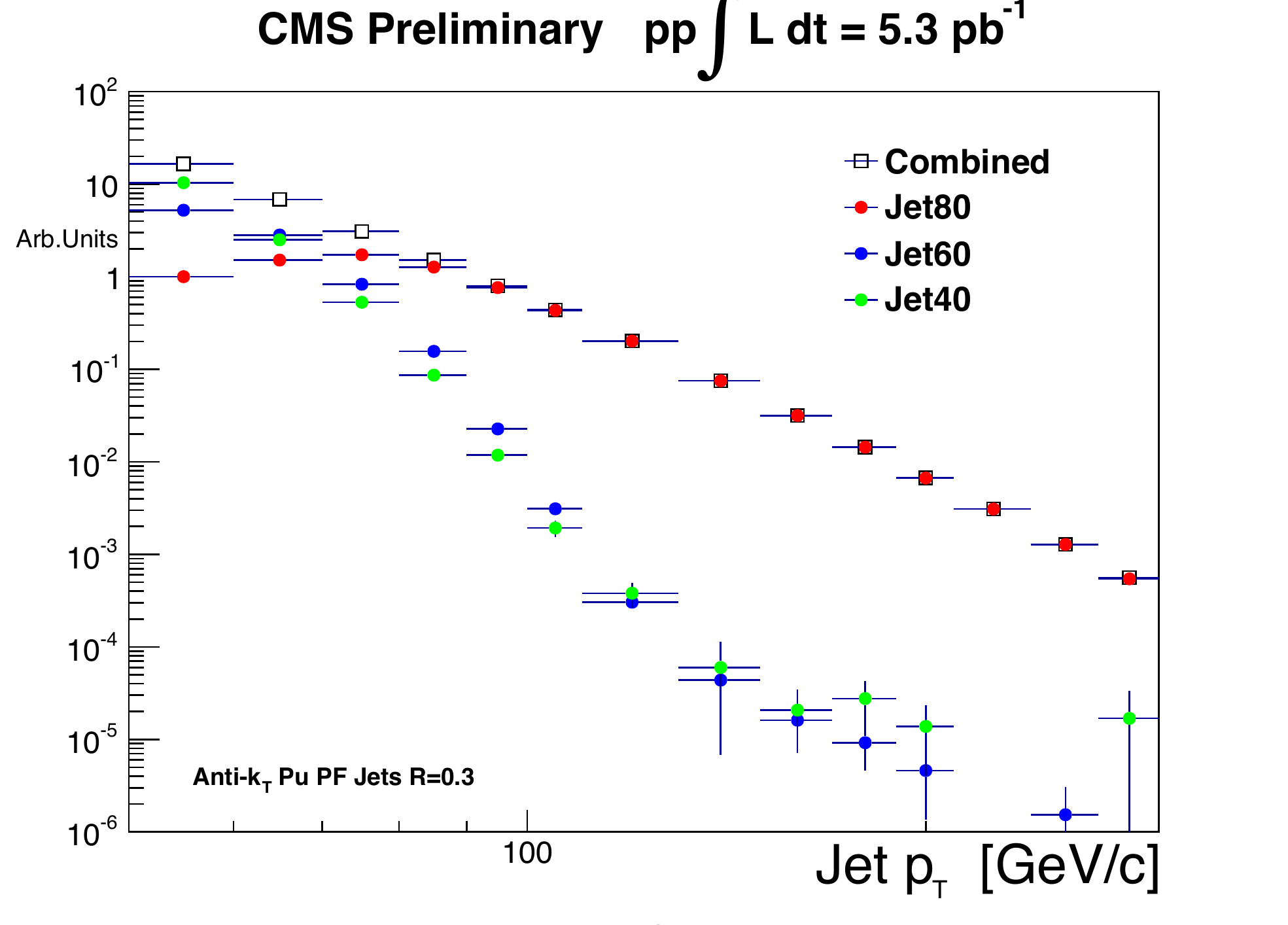}
		    \caption{Trigger combination plots of pp data for R=0.3 \akt PF jets from the different triggered datasets where Jet40 references $N_A$, Jet60 is $N_B$ and Jet80 is $N_C$.}
		    \label{fig:pp_triggercombination}
		\end{figure}

		The run-averaged pre-scale factor for a given trigger X, denoted $PS_{\rm X}$, is determined by counting the fraction of events which fire the unprescaled trigger (C), that also fire trigger X. The procedure can be described by calling the three exclusive categories $n_c$, $n_b$, and $n_a$, where the aim is to recover the original unprescaled categories $N_C$, $N_B$, and $N_A$.  Since the JetC trigger is unprescaled, this category is a trivial one: $n_c$ = $N_C$.  $N_B$, however, is the most complicated category, due to the overlap between triggers B and C, and the fact that trigger B is prescaled.  Naively, one may assume that the simplest method to obtain $N_B$ from $n_b$ is to simply scale $n_b$ by the prescale factor $PS_{\rm B}$.  However, this then leaves a difficult exercise for recovering $N_A$ from $n_a$.  Since trigger A should fire for any events that fire trigger B, there will be some ``unlucky" events that should have fired trigger B and should be counted in $N_B$, but were prescaled away by trigger B and caught by trigger A.  Therefore, simply scaling up each category by the respective trigger prescale factor will double-count events in $N_B$.  The method described here essentially relies on the trigger A to recover the majority of $N_B$, by scaling up those events by $PS_{\rm A}$.  Then,  the remaining events in $N_B$ are recovered by trigger B.  Finally, once $N_B$ is taken care of, $N_A$ is obtained by simply scaling the $n_a$ events by the prescale factor $PS_{\rm A}$, since there is no other trigger overlap.  In other words, if we break $n_a$ into two pieces, namely the events that should have only fired trigger A (``$n_{a}(A)$") and events that should have fired trigger B, but were prescaled away (``$n_{a}(B)$"), we can say:

		\begin{enumerate}
			\item $N_C = n_c$
			\item $N_B = n_b + n_a(B) * PS_{\rm A}$
			\item $N_A = n_a(A) * PS_{\rm A}$
		\end{enumerate}

		but then since $n_a(A) + n_a(B) = n_a$, we can say:

		\begin{enumerate}
			\item $N_B = n_b * 1$
			\item $N_A = n_a * PS_{\rm A}$
		\end{enumerate}

		and we recover the original algorithm. The trigger combined and individual HLT jet spectra with the aforementioned method in pp data is shown in Fig:~\ref{fig:pp_triggercombination} as a function of the jet \pt. 
	
	\subsection{Extended approach with multiple triggers}

\begin{figure}[h]
		   \centering
		   \includegraphics[width=0.45\textwidth]{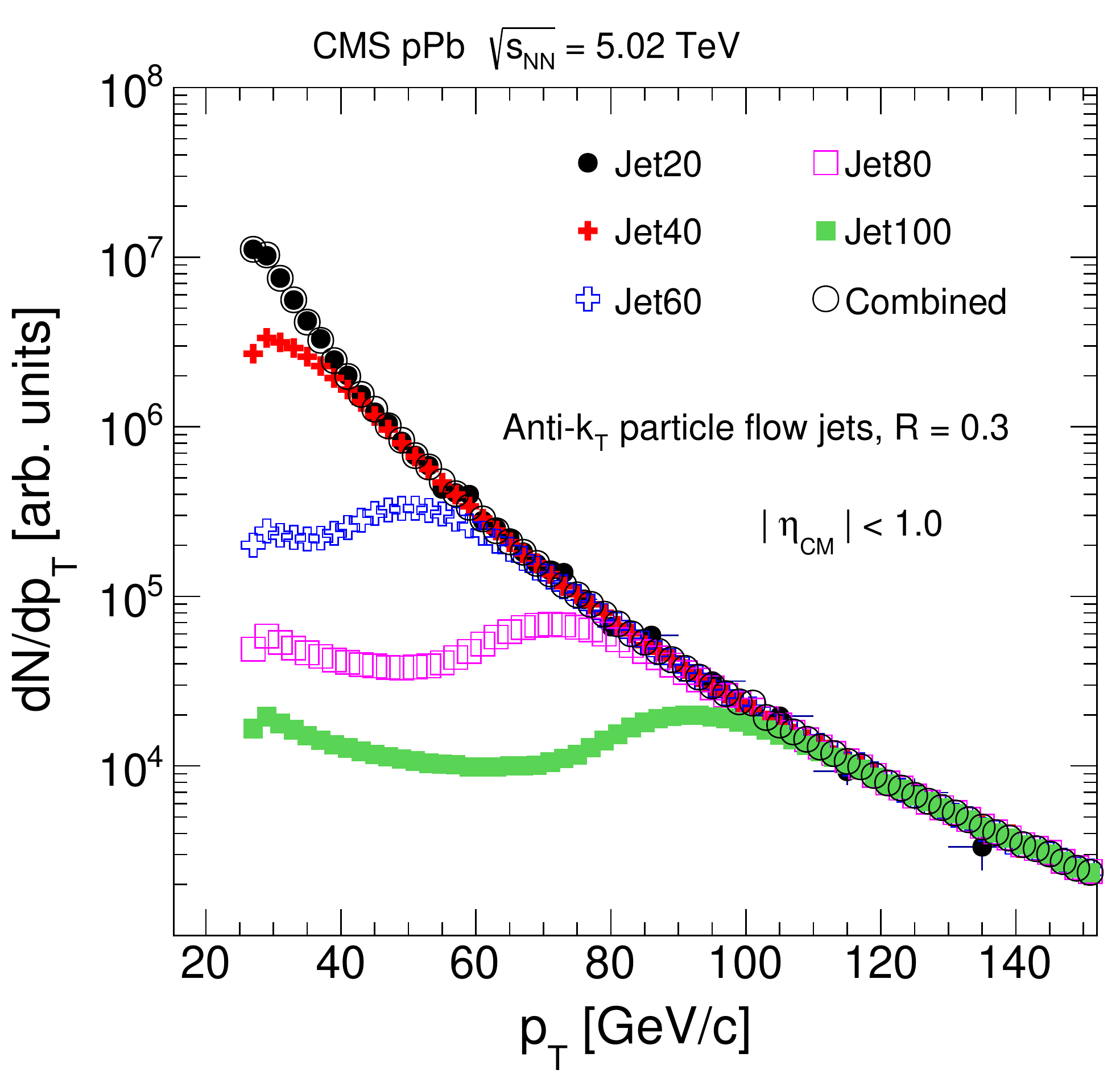} 
		   \includegraphics[width=0.45\textwidth]{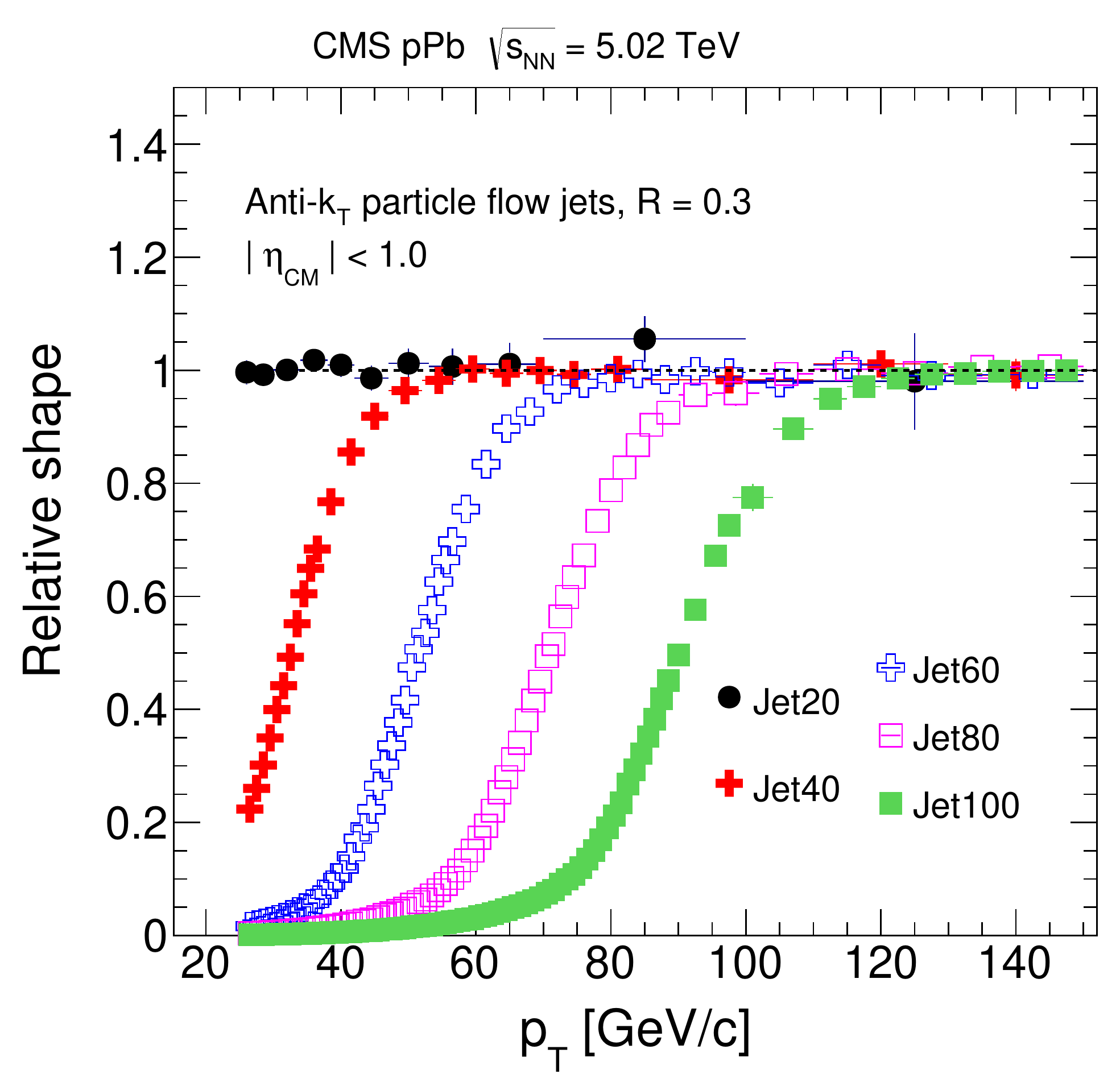} 
		   \caption{Performance of the trigger combination algorithm from various HLT single jet triggers towards a minimum bias jet yield (left) and the relative shape of the individual trigger contribution to the combined yield (right) in pPb collisions.}
		   \label{fig:pPbTriggerCombination}
		\end{figure}

		Since we increased the number of jet triggers in pPb collisions, a new different approach to combine the triggers was utilized. The events are classified according to the HLT trigger path by using the effective prescale value from each jet triggers, and then normalize to the luminosity recorded in the last trigger bin since no pre-scale is done in our highest HLT trigger path.  The sample is categorized into five groups according to its HLT trigger based on the trigger \pt~~ and then divide into different \pt~~ bins:

		\begin{enumerate}
			\item MAX(triggerPt) $>=$ threshold$_E$ = ``$n_e$"
			\item threshold$_D$ $<=$ MAX(triggerPt) $<$ threshold$_E$ = ``$n_d$"
			\item threshold$_C$ $<=$ MAX(triggerPt) $<$ threshold$_D$ = ``$n_c$"
			\item threshold$_B$ $<=$ MAX(triggerPt) $<$ threshold$_C$ = ``$n_b$"
			\item threshold$_A$ $<=$ MAX(triggerPt) $<$ threshold$_B$ = ``$n_a$"
		\end{enumerate}

		where A through E are the five \pt~ thresholds, from smallest to largest, of the five triggers used and MAX(triggerPt) refers to the maximum value of the online transverse momentum observed by the triggers and used to calculate the trigger decisions.  The weights applied to each sample are simply:

		\begin{enumerate}
			\item $w|(n_e) = PS_{\rm E}$
			\item $w|(n_d) = PS_{\rm D}$
			\item $w|(n_c) = PS_{\rm C}$
			\item $w|(n_b) = PS_{\rm B}$
			\item $w|(n_a) = PS_{\rm A}$
		\end{enumerate}

		 where ``PS" again refers to the prescale factor of a particular trigger. The left panel of Fig.~\ref{fig:pPbTriggerCombination} shows the trigger combined spectra (open black circles) as a function of the jet \pt~along with the different colored markers as individual HLT spectra. The relative contribution of the different triggers to the combined spectra is shown on the right panel of Fig.~\ref{fig:pPbTriggerCombination}. 

\section{Jet finding efficiency}

\begin{figure}[h!]
	  \begin{center}
	    \includegraphics[width=.490\textwidth]{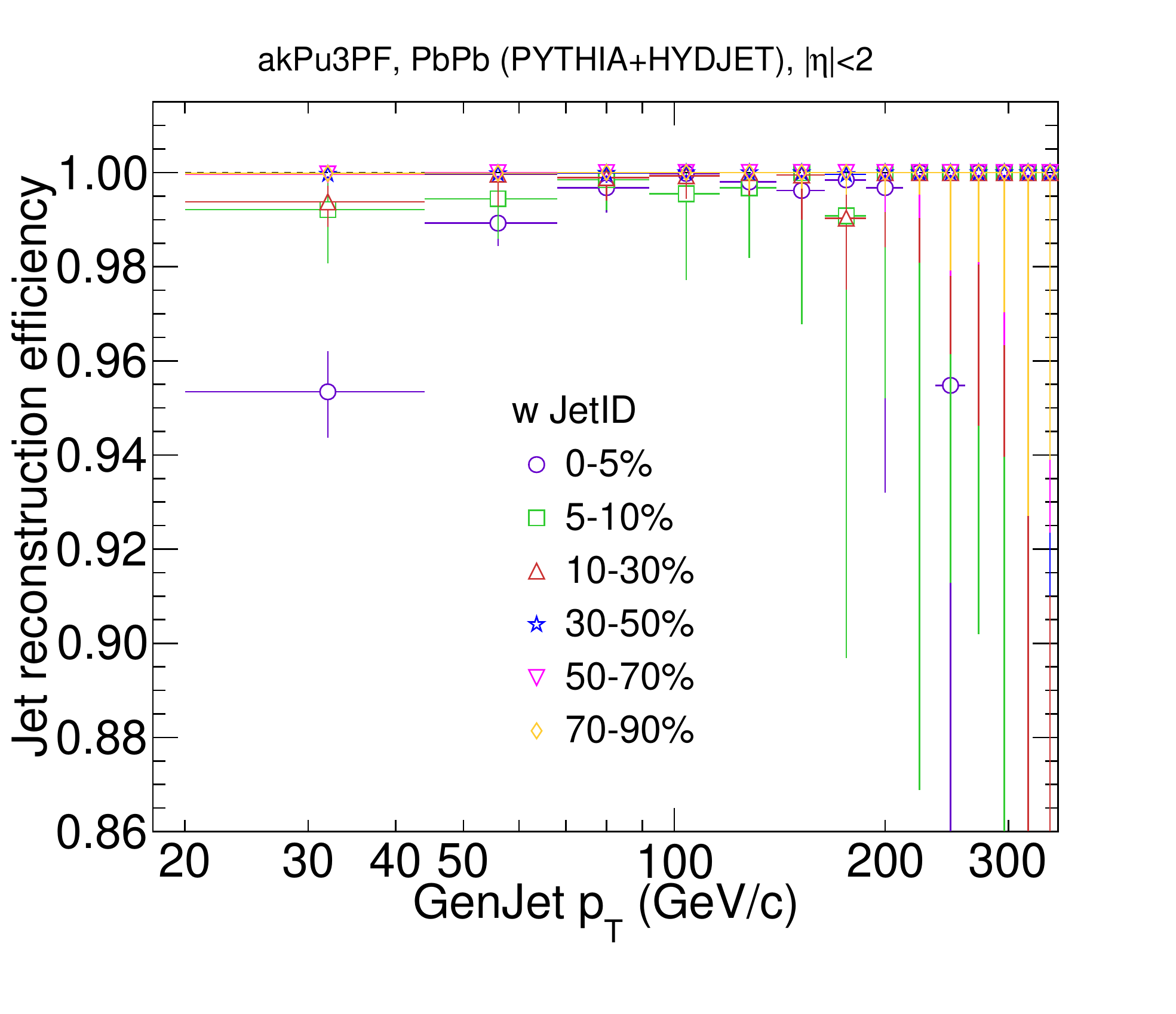}
	    \includegraphics[width=.490\textwidth]{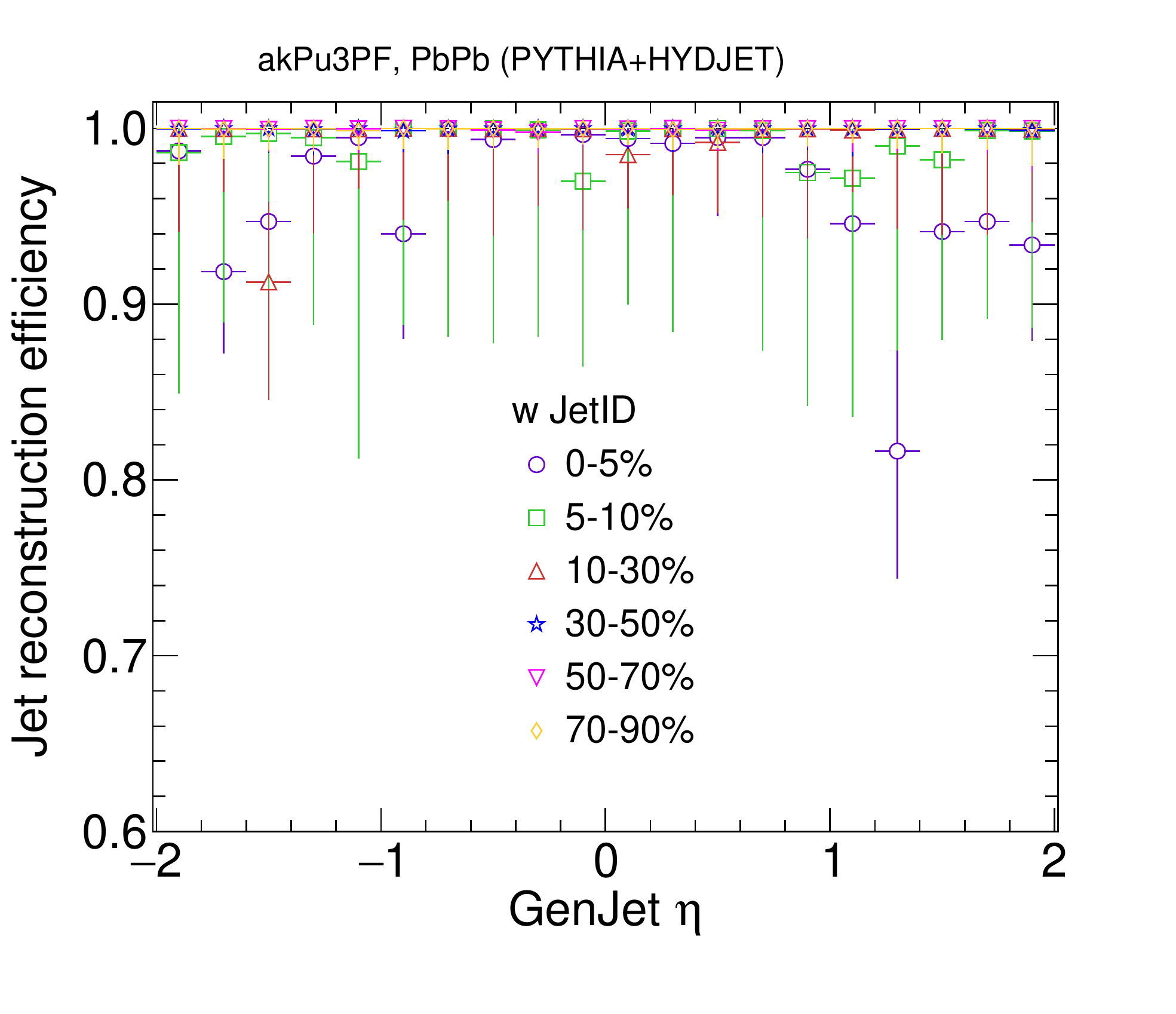}
	    \includegraphics[width=.490\textwidth]{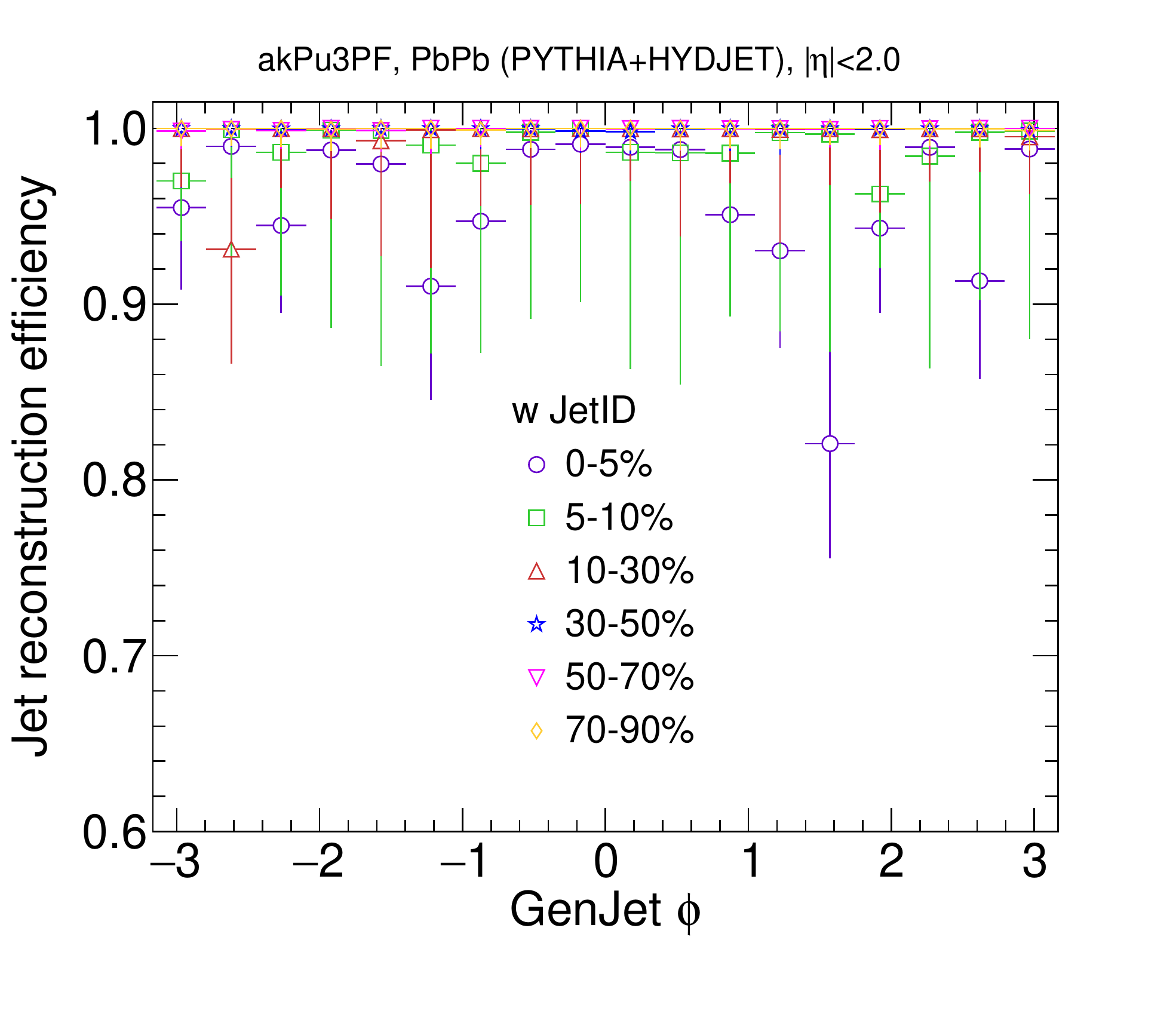}
	    \caption{Jet finding efficiency for reconstructed jets with anti-$k_{T}$ (R=0.3) pile up subtraction scheme for different centrality bins as a function of $p_T$ (top left), $\eta$ (top right) and $\phi$ (bottom) }
	    \label{fig:EffPt_1_PbPb_akPu3PF.pdf}
	  \end{center}
	\end{figure}

\begin{figure}[h!] 
	   \centering
	   \includegraphics[width=0.5\textwidth]{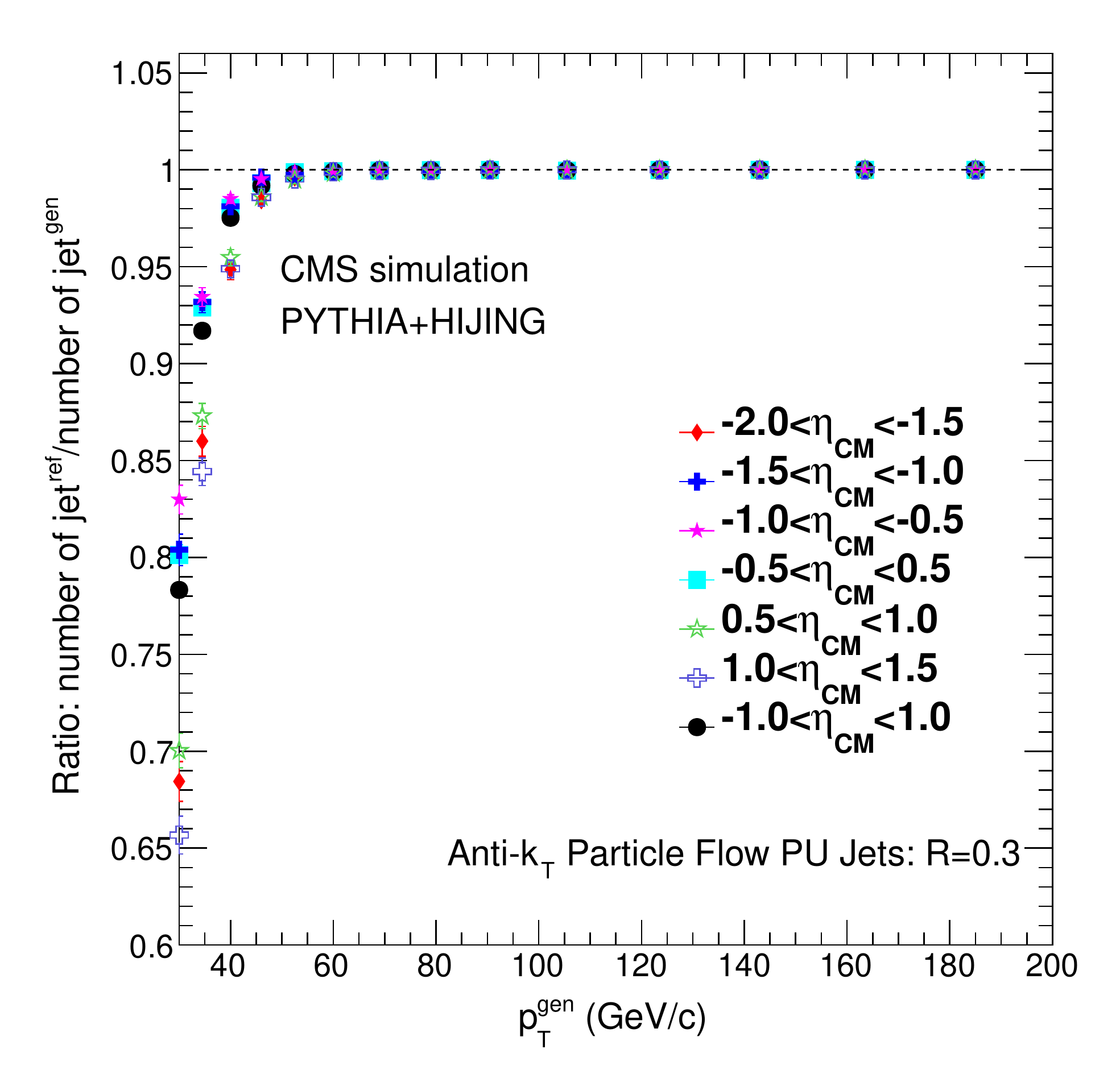} 
	   \caption{Jet finding efficiency in pPb MC as a function of gen \pt~ in seven \teta~ bins, the efficiency is approximately 100\% above 60 GeV/c.}
	   \label{fig:pPbjeteff}
	\end{figure}

	The performance of the jet reconstruction algorithm in the experiment is characterized by comparing the reconstructed jets (RecoJets) in MC simulations to the generator jets (GenJets). The RecoJets and GenJets are matched by position within $\Delta R=\sqrt{\Delta\phi^2+\Delta \eta^2} < $ jet Radius. The \pt~of the matched pairs is compared for different systems and in the case of PbPb, different centrality selections. For each selection, the ratio of RecoJet to GenJet \pt~is found. Fig:~\ref{fig:EffPt_1_PbPb_akPu3PF.pdf}-~\ref{fig:pPbjeteff} show the jet finding efficiency calculated in PbPb MC and pPb for different centralities and we find it to be less than $1\%$ for \pt~$> 65 GeV$. For the efficiency as a function of generator level jet $\eta, \phi$, the drop in efficiency are covered by the statistical error bars and also includes lower \pt jets where the efficiency is a bit lower. 

	The same procedure is also performed for pPb collisions as shown in Fig:~\ref{fig:pPbjeteff} where we see a similar performance compared to peripheral PbPb events. The different markers represent different \teta bins since the system is asymmetric (more about that in the upcoming chapters when we study the physics of pPb collisions). 
	
\section{Jet Energy Scale and Response}
	
\begin{figure}[h!]
	\begin{center}
    		\includegraphics[width=.99\textwidth]{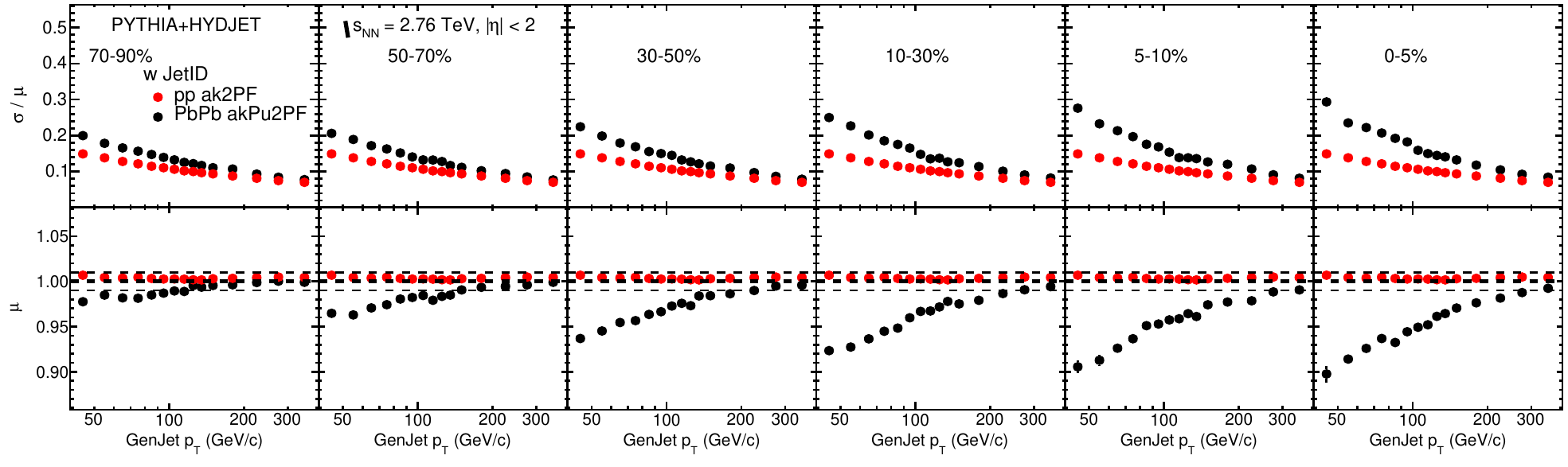}
    		\includegraphics[width=.99\textwidth]{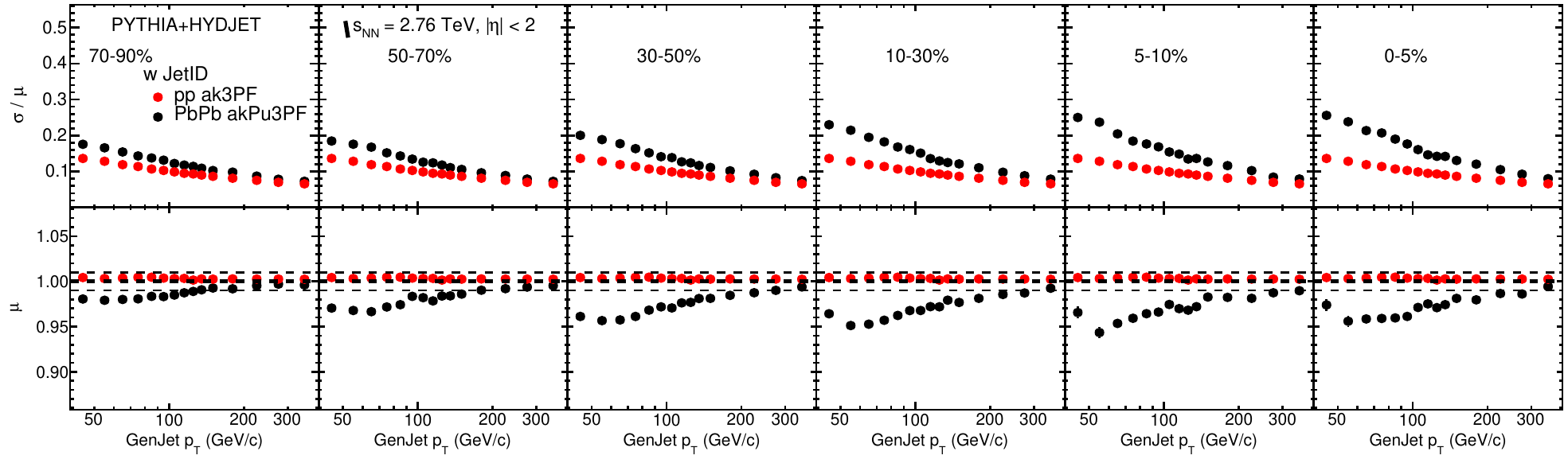}
    		\includegraphics[width=.99\textwidth]{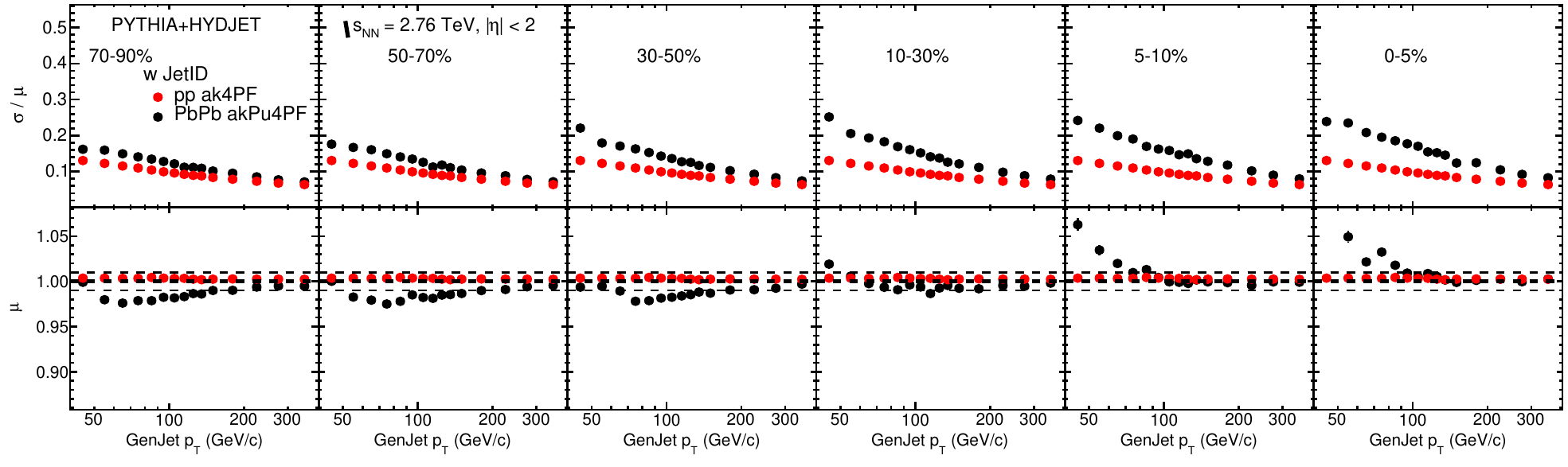}
    		\caption{Jet energy resolution and response for PF jets reconstructed with anti-$k_{T}$ for R=0.2(top), R=0.3 (middle) and R=0.4(bottom) . Black symbols shows the resolution and scale for different centralities in PYTHIA+HYDJET generated events (including pileup subtraction) and the red symbols show the jet energy resolution and response for jets from PYTHIA generated events.}
    		\label{fig:JEC_1_PbPb}
  	\end{center}
	\end{figure}

	The jet energy resolution(JER) and scale (JES) were derived from {\sc{pythia}} and  {\sc{pythia+hydjet}} generated events. The JER and JES were studied in bins of generated jets (GenJet) \pt. The pp studies have no background subtraction and the PbPb studies (Figs.~\ref{fig:JEC_1_PbPb}) include the iterative background subtraction. Jets from pp collisions are shown in red (duplicated across all centralities), and for PbPb in black. 

\begin{figure}[htbp]
	\begin{center}
		\includegraphics[width=1\textwidth]{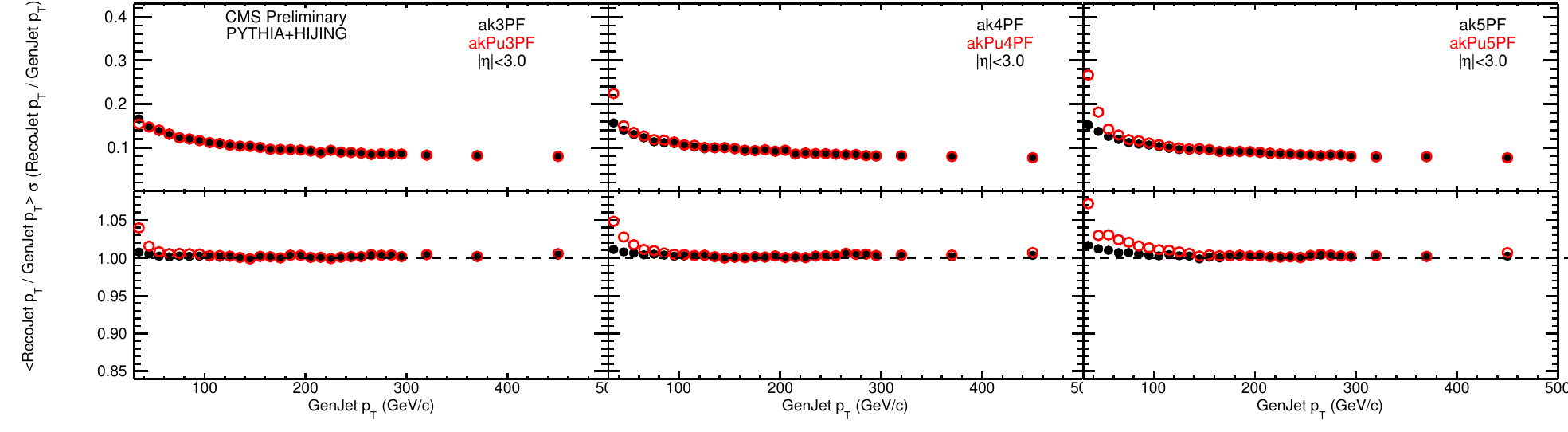}
		\includegraphics[width=1\textwidth]{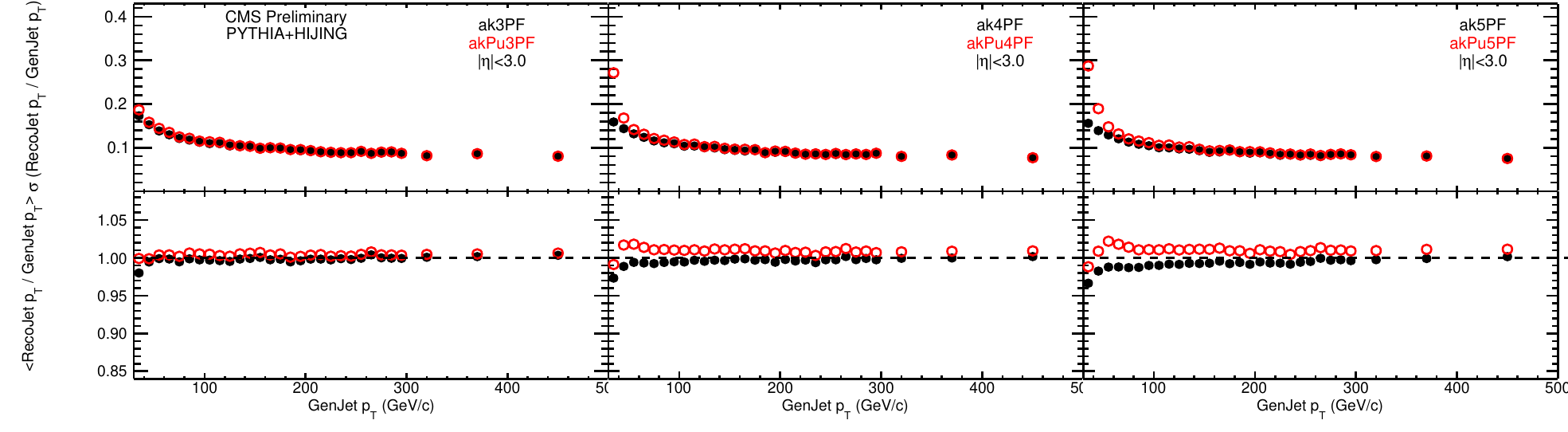}
		\caption{Jet resolution and response for anti-$k_{T}$ particle-flow jet algorithm with $R$=0.3 - 0.5 for {\sc pythia+hijing} samples (top: pPb MC with Pb going to positive direction; bottom: Pbp with Pb going to negative direction) as a function of jet \pt~for ak[3-5]PF algorithms with and without pileup subtractions after the jet energy corrections.}
		\label{fig:JESClosurePt}
	\end{center}
	\end{figure}

	The same study was performed for pPb collisions with and without the background subtraction procedure as shown in Fig:~\ref{fig:JESClosurePt} for different radii jets, except in this case, PYTHIA is embedded with HIJING events. The left and right panels correspond to p going to negative and positive direction respectively. 
	
\section{Residual correction from dijet balance}

	The dijet \pt~balance technique as described in the previous chapter can be used to correct the jet relative response in pp collisions. Two leading jets in the interval $|\eta| < 3$ are selected followed by selection of the reference jet within $|\eta^{ref}| < 1.3$. The probe jet has to be within $|\eta^{probe}| < 3.0$. If both jets are within $|\eta| < 1.3$ then one jet is chosen at random to be the reference jet. If neither jets are within $|\eta| < 1.3$ then the event is not used. To study the detector relative response, dijet balance for each event is calculated as:
	
	\begin{equation}
  		B = \frac{p_{T}^{probe} - p_{T}^{ref}}{p_{T}^{avg}}
	\end{equation}

	where $p_{T}^{avg}$ is the average \pt~of the two jets: 
	\begin{equation}
  		p_{T}^{avg} = \frac{p_{T}^{ref} + p_{T}^{probe}}{2}
	\end{equation}
	
\begin{figure}[h!]
	  \begin{center}
	    \includegraphics[width=0.9\textwidth]{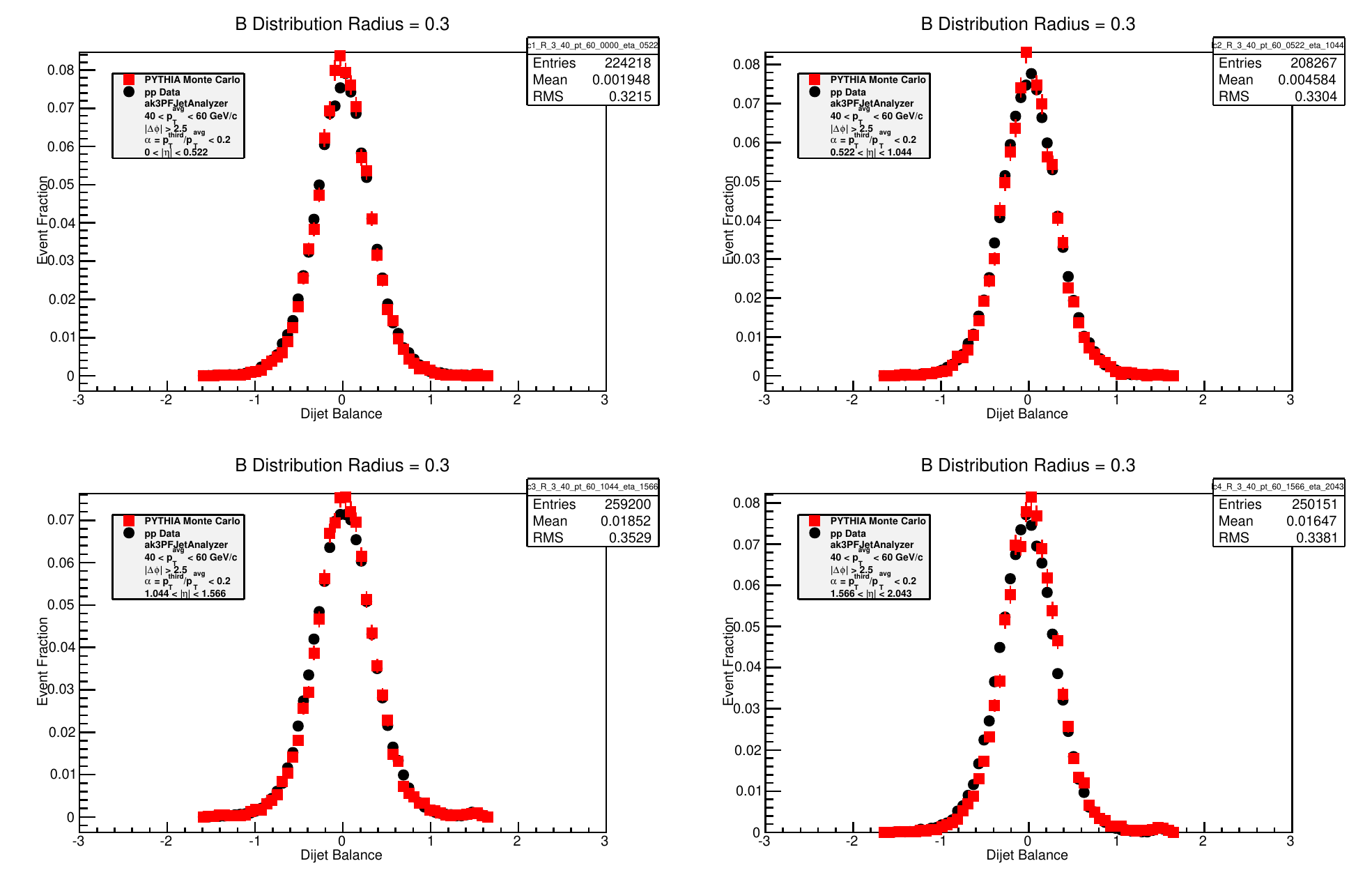}
	    \caption{Comparison of dijet balance between MC and pp data for 40 $ < p_{T}^{avg} < $ 60 \gev~and various $|\eta|$ slices. Top left is 0 $< |\eta| <$ 0.522. Top right is 0.522 $< |\eta| <$ 1.044. Bottom left is 1.044 $< |\eta| <$ 1.566. Bottom right is 1.566 $< |\eta| <$ 2.043.}
	    \label{fig:B_R_3_40_60_split}
	  \end{center}
	\end{figure}

\begin{figure}[h!]
	  \begin{center}
	    \includegraphics[width=0.45\textwidth]{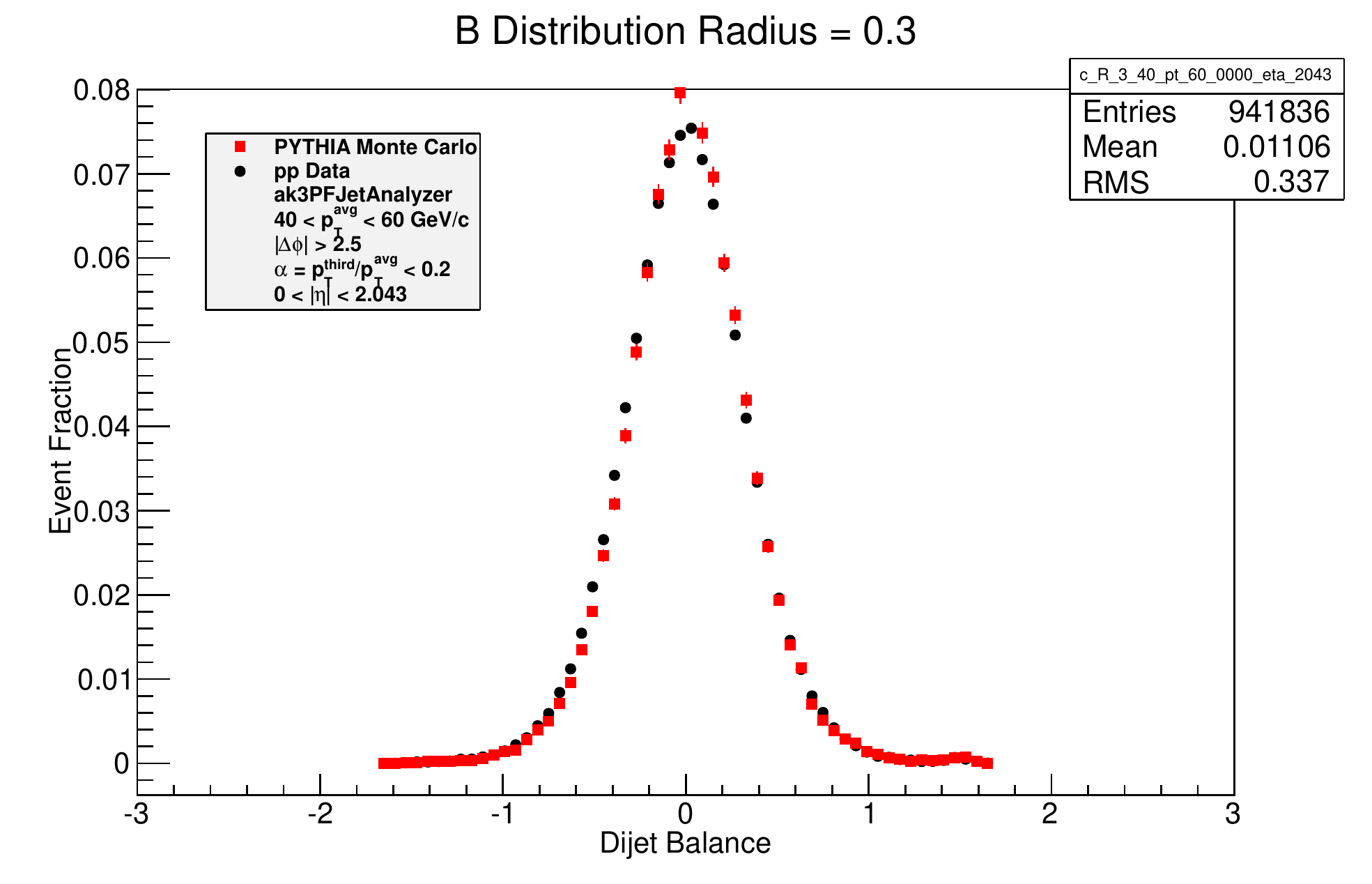}
	    \caption{Comparison of dijet balance between MC and pp data for 40 $ < p_{T}^{avg} < $ 60 \gev~and 0 $< |\eta| <$ 2.043.}
	    \label{fig:B_R_3_40_60}
	  \end{center}
	\end{figure}

	To suppress 3-jet event contributions, events were required to satisfy $\alpha = \frac{p_{T}^{third}}{p_{T}^{avg}} < 0.2$ and $|\Delta\phi| = |\phi^{probe} - \phi^{ref}| > 2.5$. Furthermore, events had to fall within one of the following $p_{T}^{avg}$ ranges to be used in the final analysis. The ranges were 40 - 60 GeV/c, 60 - 80 GeV/c, 80 - 100 GeV/c, 100 - 140 GeV/c, and 140 - 200 GeV/c. These distributions were plotted for data and MC in Fig.~\ref{fig:B_R_3_40_60_split} and Fig.\ref{fig:B_R_3_40_60} for the 40 $ < p_{T}^{avg} < $ 60.
	
	Taking the average value of B, $\langle B \rangle$, in a $\eta^{probe}$ bin, the relative response, $R_{rel}$, was computed:
	\begin{equation}
  		R_{rel}(\eta^{probe}) = \frac{2 + \langle B \rangle}{2 - \langle B \rangle}
	\end{equation}

	The corrections are estimated for pp and pPb collisions, as a function of jet \pt~and applied to the jet spectra. This particular procedure (and other residuals such as $\gamma, Z$ + jet) cannot be performed in PbPb events because of quenching. Thus the jet energy scale systematic uncertainty is always larger for PbPb when compared with pp or pPb.

\section{Dealing with mis-reconstructed low \pt~PbPb jets}
	
	As we saw in the previous sections, the background in heavy ion events is quite large and for low \pt~jets, it becomes very hard to distinguish between ``real" jets and mis reconstructed jets. We use the term real in quotes since it is not possible to accurately say in a collision if a given final state object originated from the hard scattering or not due to several non global effects. At the same time, it is a given that the background will affect some fraction of jets at that kinematic range and thus different experiments employ different techniques to remove this contamination. 
	
	The commonly used method to remove this soft background contamination is to require the presence of a hard final state object in the jet. A method to remove this contamination, used in other experiments~\cite{Aad:2014bxa,Adam:2015ewa}, is to select jets with a requirement on the leading charged-particle track or calorimeter energy deposit among the constituents of the jet. With this requirement, it is possible to remove a lot of jets made up of uncorrelated soft components, but it also introduces a fragmentation bias that is not negligible at low \pt. Some experiments try to correct for this bias by estimating its impact with a MC but since gluon fragmentation is still an open question, even in pp collisions, this correction comes with a large uncertainty that is often not quoted. 
	
	\subsection{Data driven unbiased approach}
	
		\begin{figure}[h!]
		   \centering
		   \includegraphics[width=0.8\textwidth]{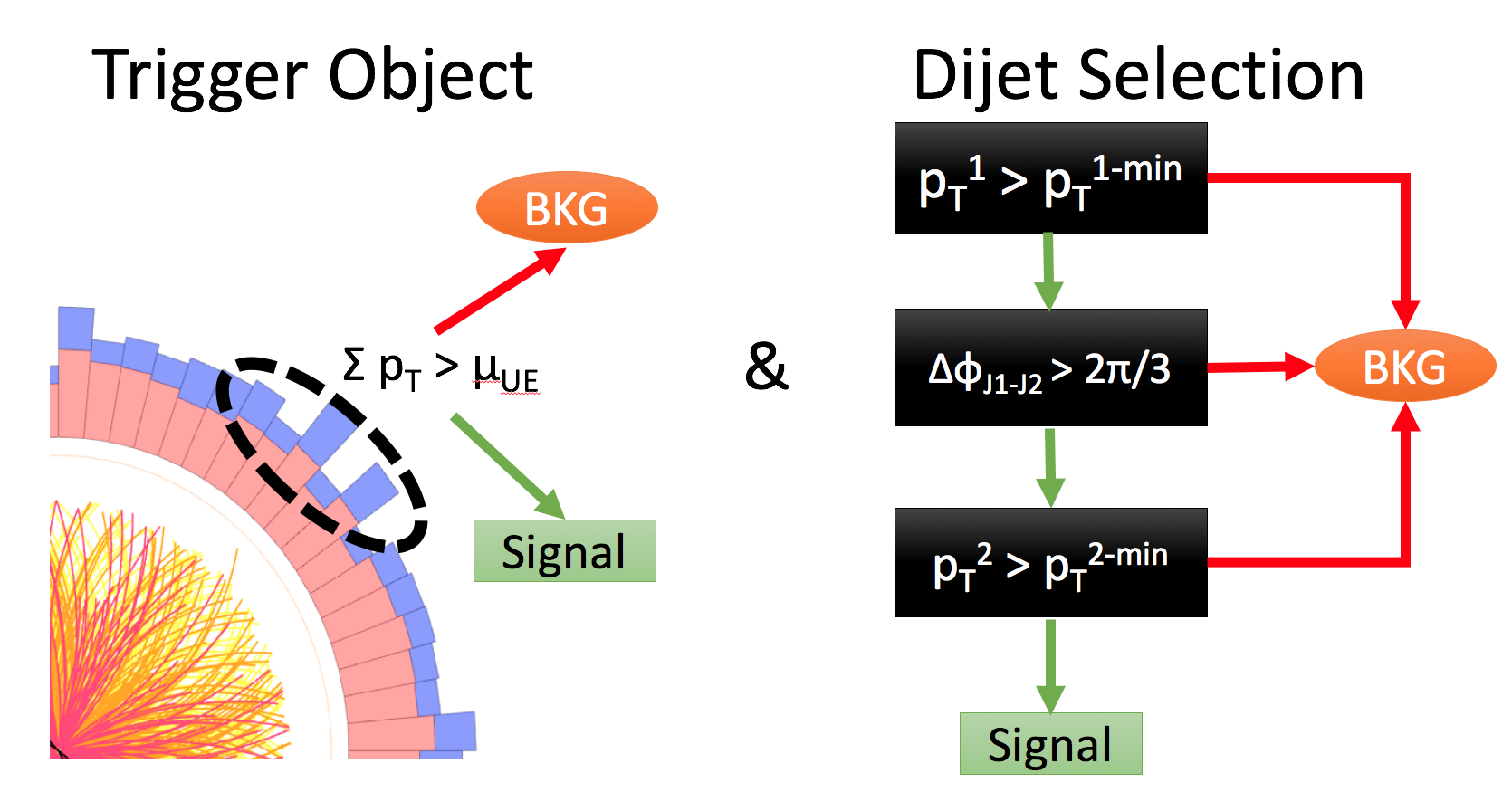} 
		   \caption{Schematic of the trigger object and dijet selection methods to correct for the mis reconstructed jet fraction.}
		   \label{fig:schematicdijet}
		\end{figure}

		In this analysis, a novel data-driven technique, based on control regions in data, is introduced to derive the spectrum of misreconstructed jets from the minimum bias sample. This spectrum is then subtracted from the jet-triggered sample. Two methods, operating in different kinematic regimes, are combined to get a correction factor. The first method (labeled the trigger object method) selects all events with a leading HLT jet \pt~of less than 60 \gev~as
a control sample potentially containing misreconstructed jets. This \pt~threshold is chosen based on analysis of random cones in minimum bias events, with the leading and subleading jets removed.

		The second method (labeled the dijet method), performed in parallel with the first method, selects minimum bias events with dijets, which can originate either from a hard scattering or fluctuating background. There are two thresholds defined in this method, one for the leading jet (\pt$^{\text{min} 1}$) and another for the subleading jet (\pt$^{\text{min} 2}$) in the reconstructed event. If an event fails any of the following selections, it is tagged as a background event. An event is tagged as a signal if it passes all of the criteria: Leading jet \pt~$ > $~\pt$^{\text{min} 1} $ and $\Delta \phi_{j1, j2} > 2\pi/3$ and subleading jet \pt~$>$ \pt$^{\text{min} 2}$.
	
		\begin{figure}[h!]
		   \centering
		   \includegraphics[width=0.8\textwidth]{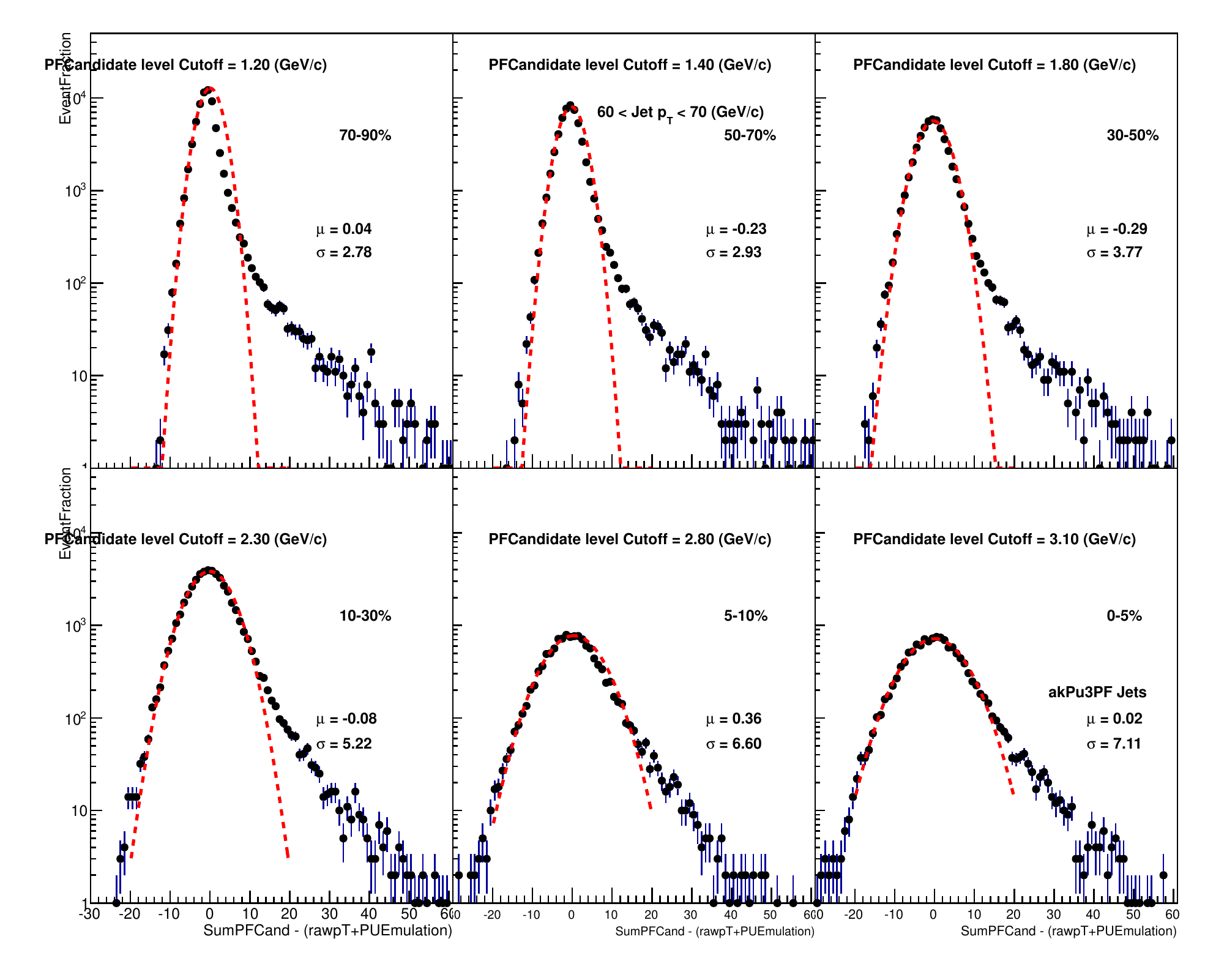} 
		   \caption{Distributions showing the efficiency of the subtraction procedure. See text for details regarding the axis definitions. The mean of all the distributions are very close to zero meaning very efficient subtraction.}
		   \label{fig:PUEmulation}
		\end{figure}
			
		To choose the thresholds for the dijet selection, the mean and RMS of the subtraction step in the iterative subtraction algorithm are mimicked by applying a cutoff on the transverse energies of the PF towers used in the random cone study as shown in Fig:~\ref{fig:PUEmulation}. The distribution plotted is essentially the subtracted event and it is represented as the difference between the amount of energy present in the region of the jet and the amount of energy subtracted plus the jet's momenta. The RMS of the background subtracted event energy distribution is used as an estimate of the fluctuation. The thresholds are set as follows: \pt$^{\text{min} 1} = 3\, \mathrm{RMS}$ for the leading jet, and \pt$^{\text{min} 2} = 1.8\, \mathrm{RMS}$ for the subleading jet, to allow for jet modification in the medium.

		\begin{figure}[h!]
		   \centering
		   \includegraphics[width=0.9\textwidth]{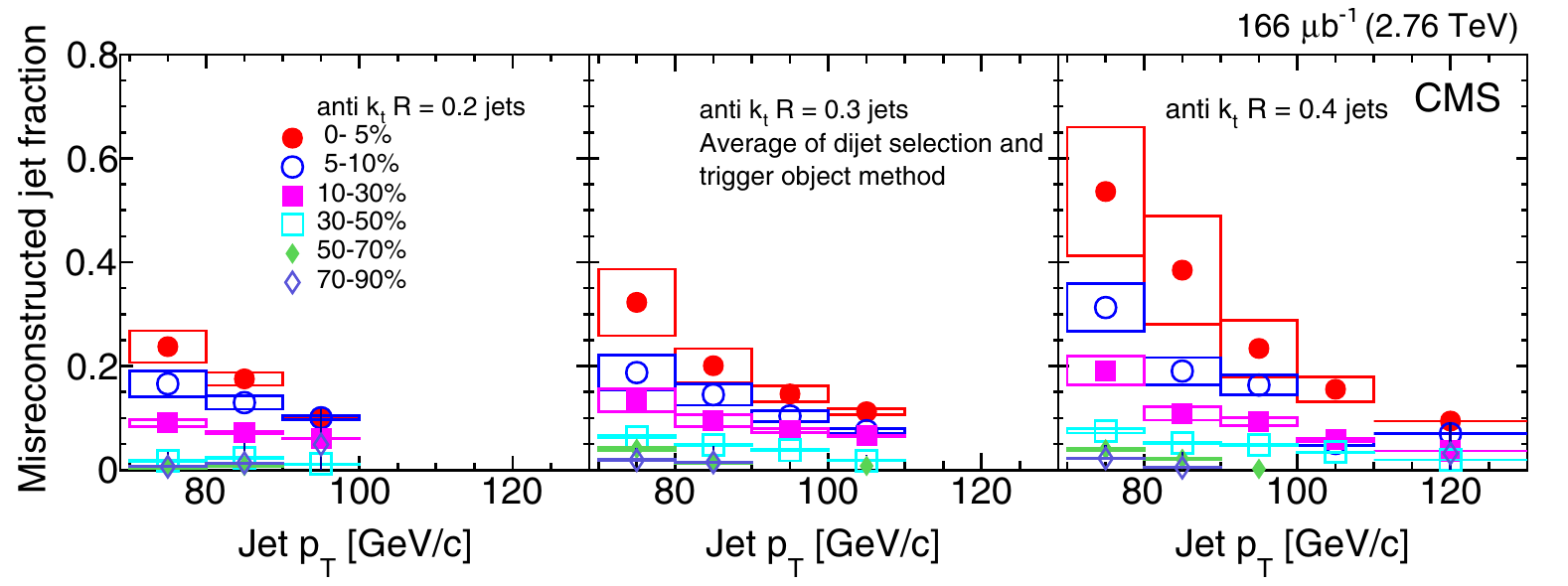} 
		   \caption{Misreconstructed jet fraction of the inclusive jet spectra~\cite{Khachatryan:2016jfl}, derived from the minimum bias sample, as a function of reconstructed jet \pt, for various centralities and three different distance parameters (left: $R = 0.2$, center: $R = 0.3$, and right: $R = 0.4$). The correction factor is the average of the dijet selection and trigger object methods discussed in the text.}
		   \label{fig:misrecocorrection}
		\end{figure}

		Since these two methods operate in different kinematic regimes, the average of the two is used to estimate the data driven correction factor for misreconstructed jet rates as can be seen in Fig.~\ref{fig:misrecocorrection}, as a function of the jet \pt. These rates for different distance parameters are shown in the different panels (left: $R = 0.2$, center: $R = 0.3$, and right: $R = 0.4$). The symbols correspond to the centrality bins in the analysis.
The minimum bias background jet spectra are then normalized to a per-event yield and the background is removed from the measured jet spectra, resulting in an inclusive jet spectrum without fragmentation bias. 

		A thorough suite of systematic studies of the data driven method was tested on heavy ion MC events. The correction, estimated in a similar way from \py~dijet events, where one does not expect any background, is added as an additional systematic uncertainty, starting from 6\% at 70 GeV to 1\% at 100 GeV. 		
		
\begin{figure}[h!]
		   \centering
		   \includegraphics[width=0.9\textwidth]{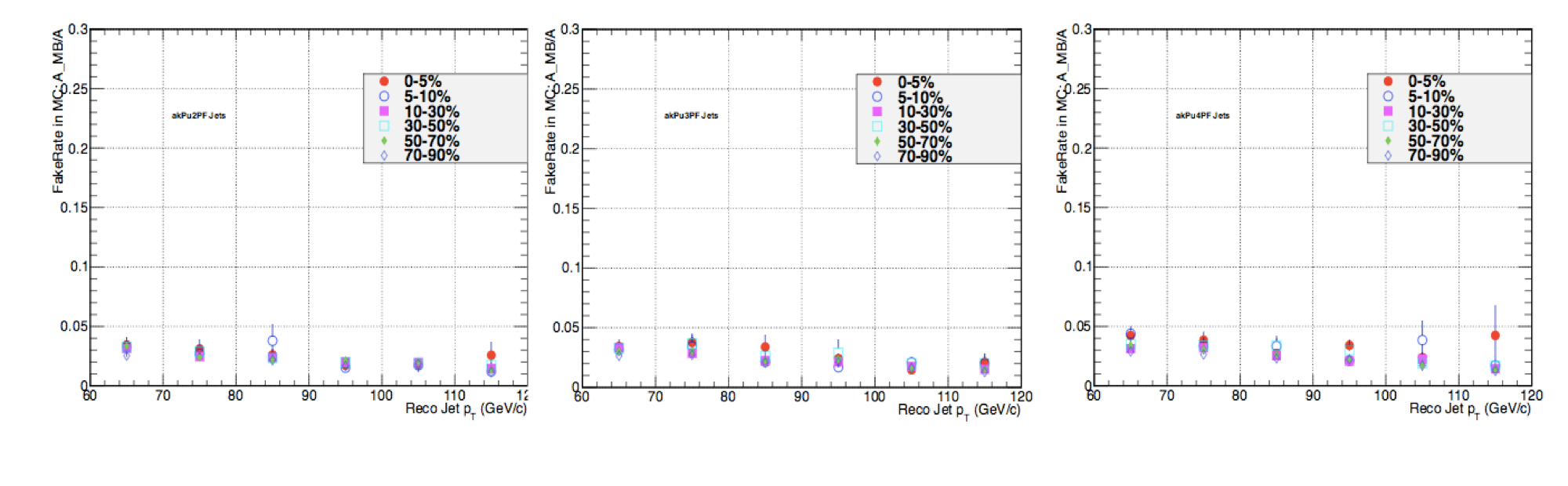} 
		   \includegraphics[width=0.3\textwidth]{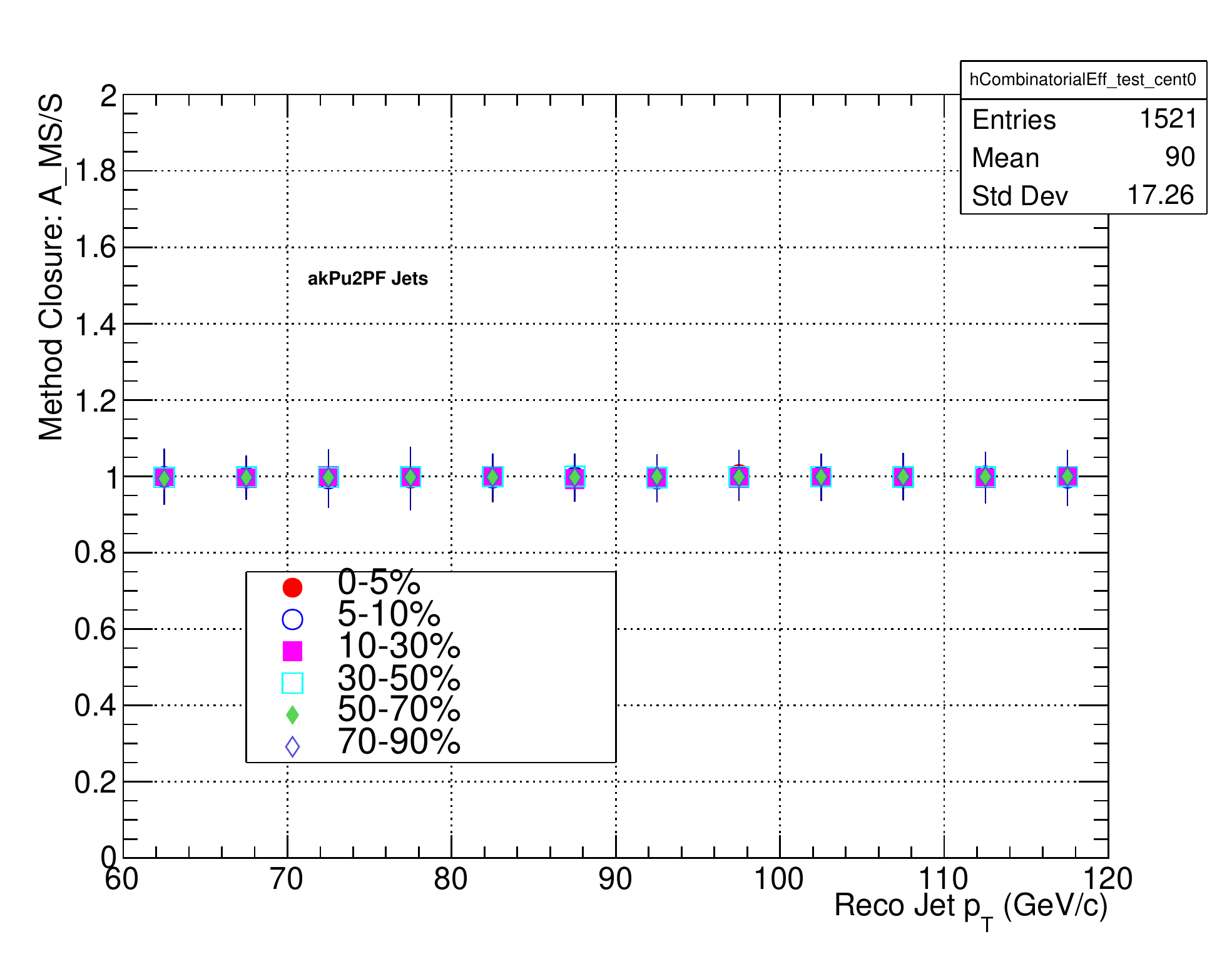} 
		   \includegraphics[width=0.3\textwidth]{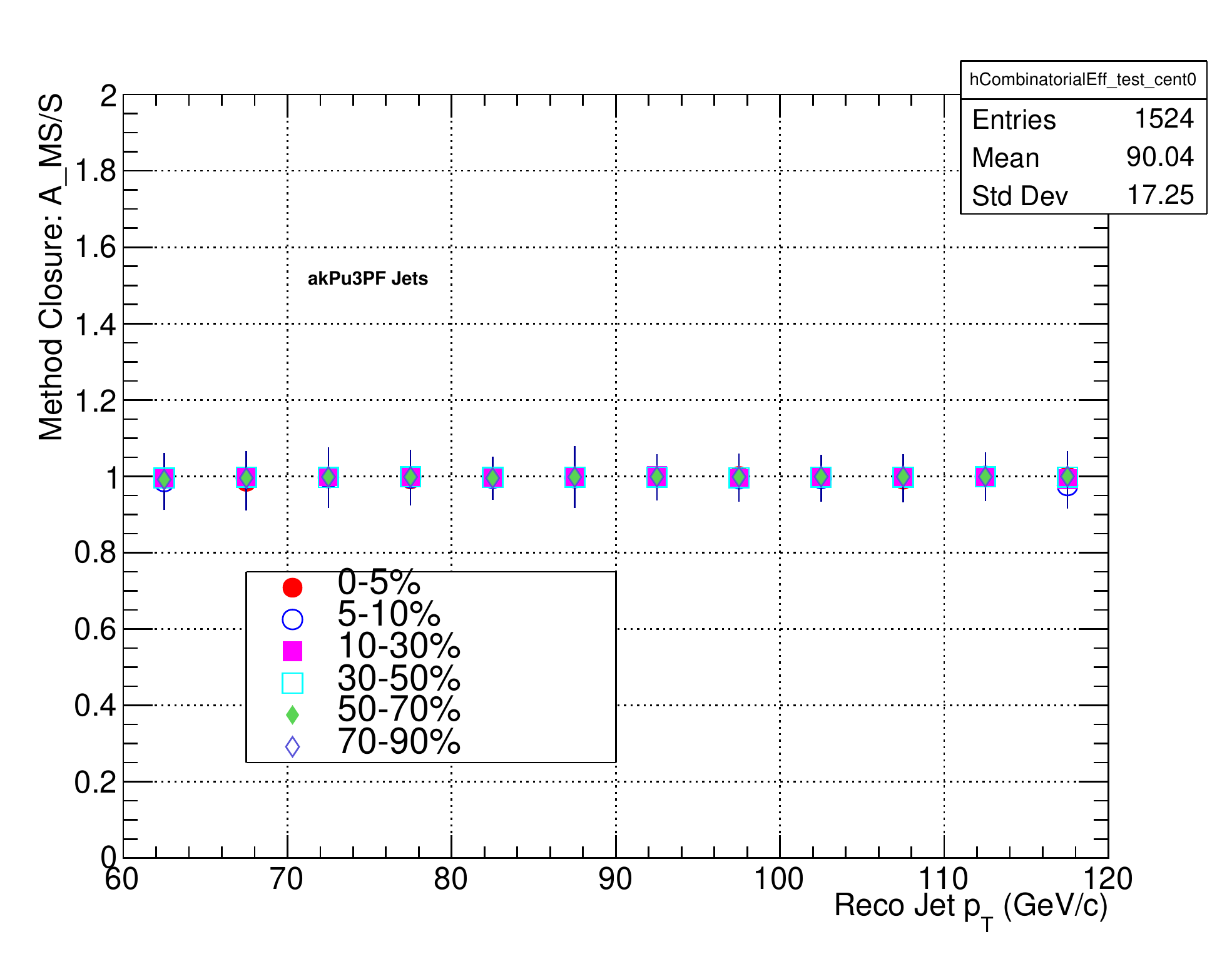} 
		   \includegraphics[width=0.3\textwidth]{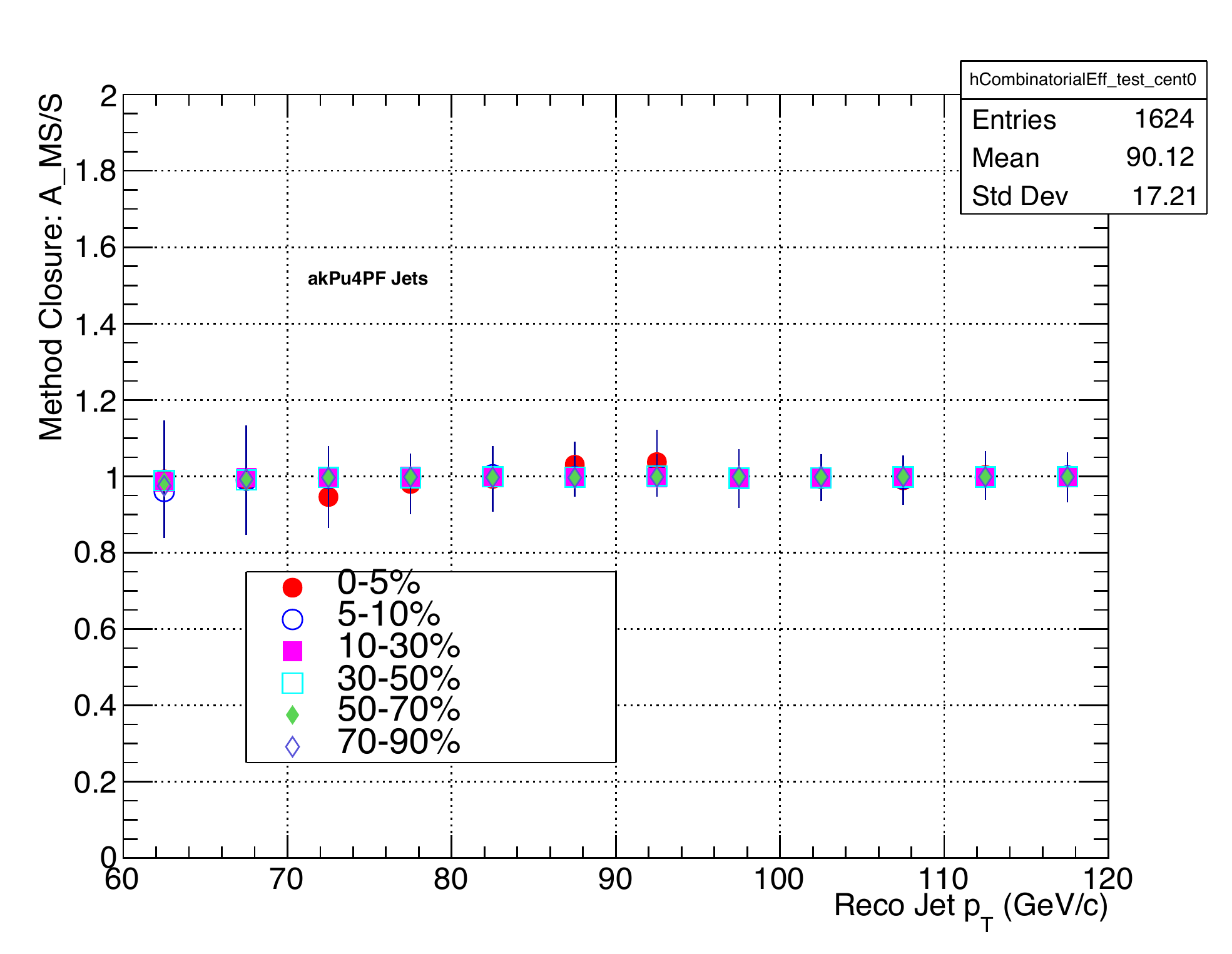} 
		   \caption{Background (top) and signal (bottom) closure of the data driven method estimated in \pyhd~for all three radii from R = 0.2 (left), R = 0.3 (middle) and R = 0.4 (right).}
		   \label{fig:methodclosure}
		\end{figure}

		The data driven method was also applied to \pyhd~simulations without quenching and, using the same \pt~threshold, this yielded a recovery efficiency of greater than 98\% for signal jets and background acceptance of less than 3\% as shown in Fig:~\ref{fig:methodclosure} as a function of jet \pt~The different markers point to different centrality bins and the panels go from R=0.2 in the left to R=0.4 on the right. Both these percentages are much smaller than the systematic uncertainty bounds of the method and thus gives us confidence in the approach undertaken.  
		
\section{Unfolding the jet spectra}


		The term ``unfolding'' is used to describe a set of techniques which are used to essentially ``invert'' the convolution of the spectra and the resolution. This is a non-trivial mathematical problem because the jet \pt~spectra are falling very steeply and the resolution is derived from MC simulations and therefore is not known exactly. Even if the transformation matrix due to the resolution were known exactly, statistical fluctuations in the measured spectrum can be  transformed into unphysical large fluctuations in the unfolded spectra.

		With binned histograms for the spectra, the unfolding problem  can be understood as the matrix inversion to solve
		
		\begin{equation}
			\mathbf{A} \mathbf{x} = \mathbf{b}
  		\label{eq:folding_matrix}
		\end{equation}
		with the correspondence

		\begin{equation}
  		\begin{split}
    			\frac{dN}{d p_{T}~^\mathrm{rec}} &\leftrightarrow \mathbf{b} \textrm{ (the measured spectrum)}\\
    			P(p_{T}~^\mathrm{rec} | p_{T}~) &\leftrightarrow \mathbf{A} \textrm{ (the response matrix)}\\
    			\frac{dN}{d p_{T}~} &\leftrightarrow \mathbf{x} \textrm{ (the true spectrum)}
  		\end{split}
		\end{equation}

\begin{figure}[hbtp]
 		\begin{center}
    			\includegraphics[width=0.8\textwidth]{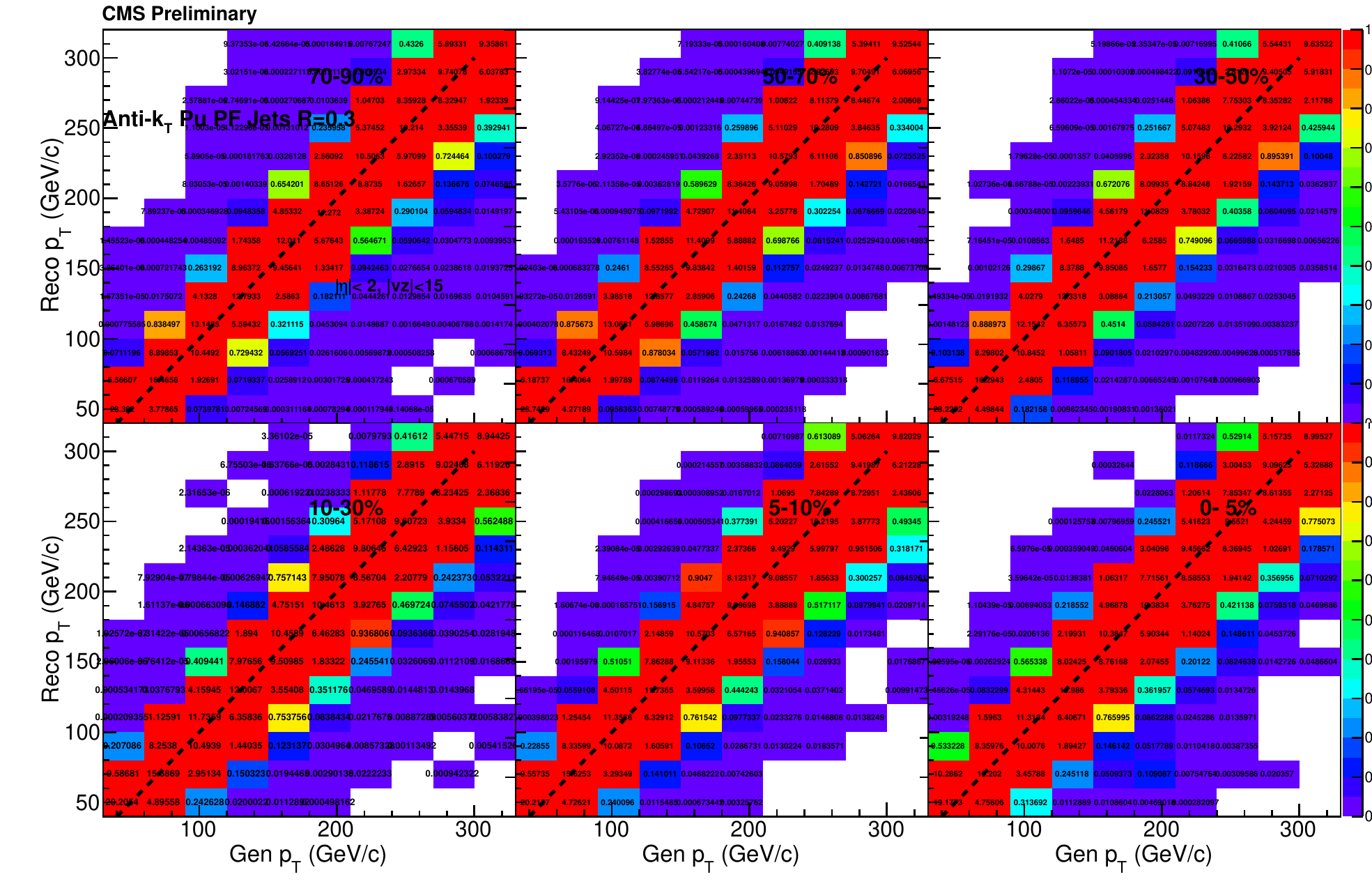}
    			\caption{PbPb Response matrix for R=0.3 akPuPF jets, with generator level \pt~on x axis and reconstructed \pt~on Y axis shown in different centrality bins.}
    			\label{fig:BayesMatrixPbPb_R3}
  		\end{center}
		\end{figure}

\begin{figure}[hbtp]
  		\begin{center}
    			\includegraphics[width=0.3\textwidth]{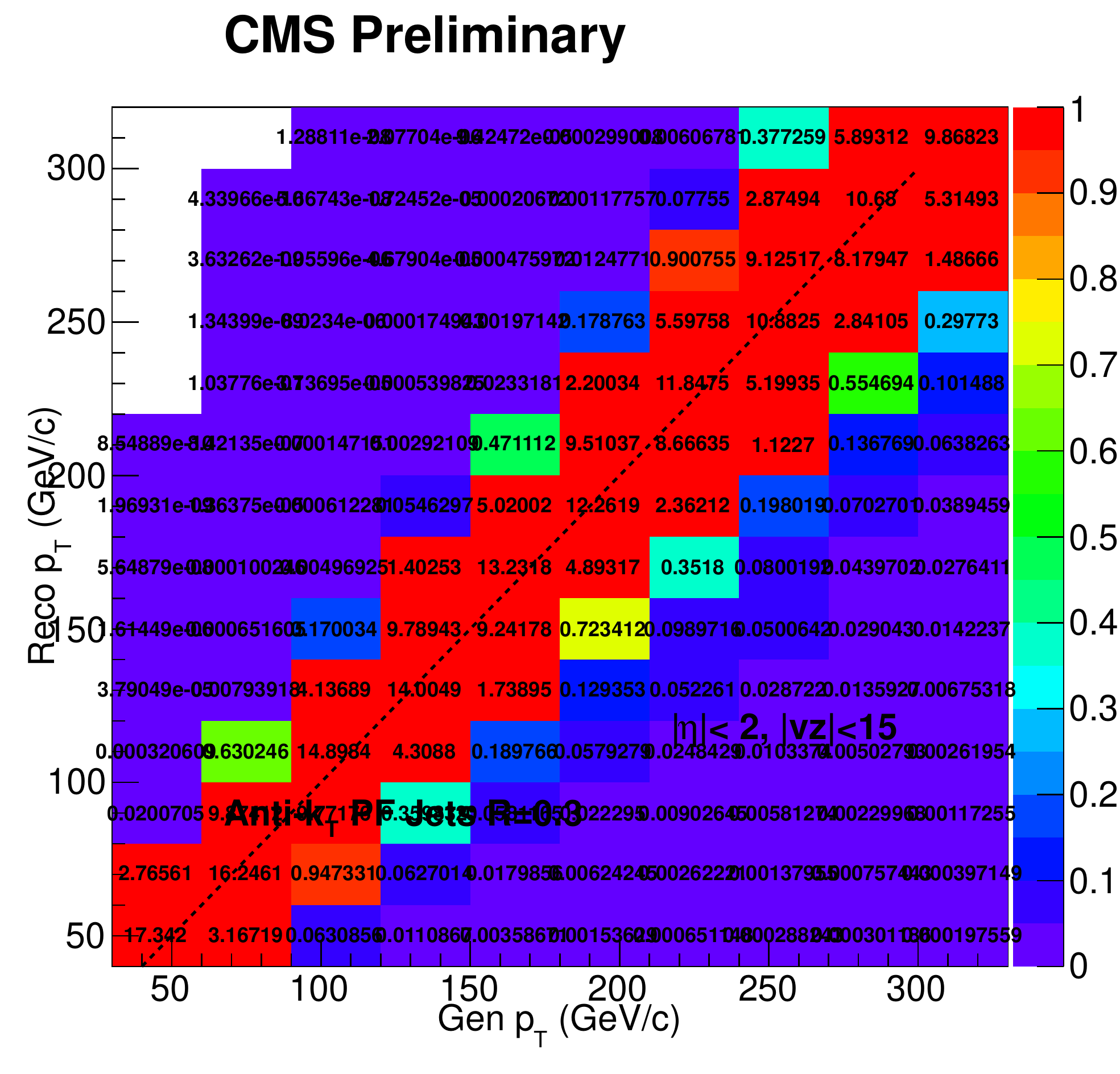}
    			\caption{PP Response matrix for R=0.3 akPF jets, with generator level \pt~on x axis and reconstructed \pt~on Y axis.}
    			\label{fig:BayesMatrixPP_R234}
  		\end{center}
		\end{figure}

		The unfolding procedures use a response matrix with Monte Carlo simulation (training), and then attempts to reconstruct the true distribution from the measured distribution  with this response matrix.  Figures~\ref{fig:BayesMatrixPbPb_R3},~\ref{fig:BayesMatrixPP_R234} show the response matrix from MC that is used in the analysis.  The response matrix shows the distribution of reconstructed jet \pt~(RecoJet) in comparison to generator level jet \pt~(GenJet), finely binned in the actual bins used in the  unfolding analysis for PbPb and pp. All the MC samples used to construct the matrix were weighted by the event cross section to properly merge different $\hat{p_{T}}$ samples. Additionally, the MC samples are weighted by the jet triggered data centrality and primary vertex distribution. 


\begin{figure}[htpb!]
		\begin{center}
			\includegraphics[width=0.5\textwidth]{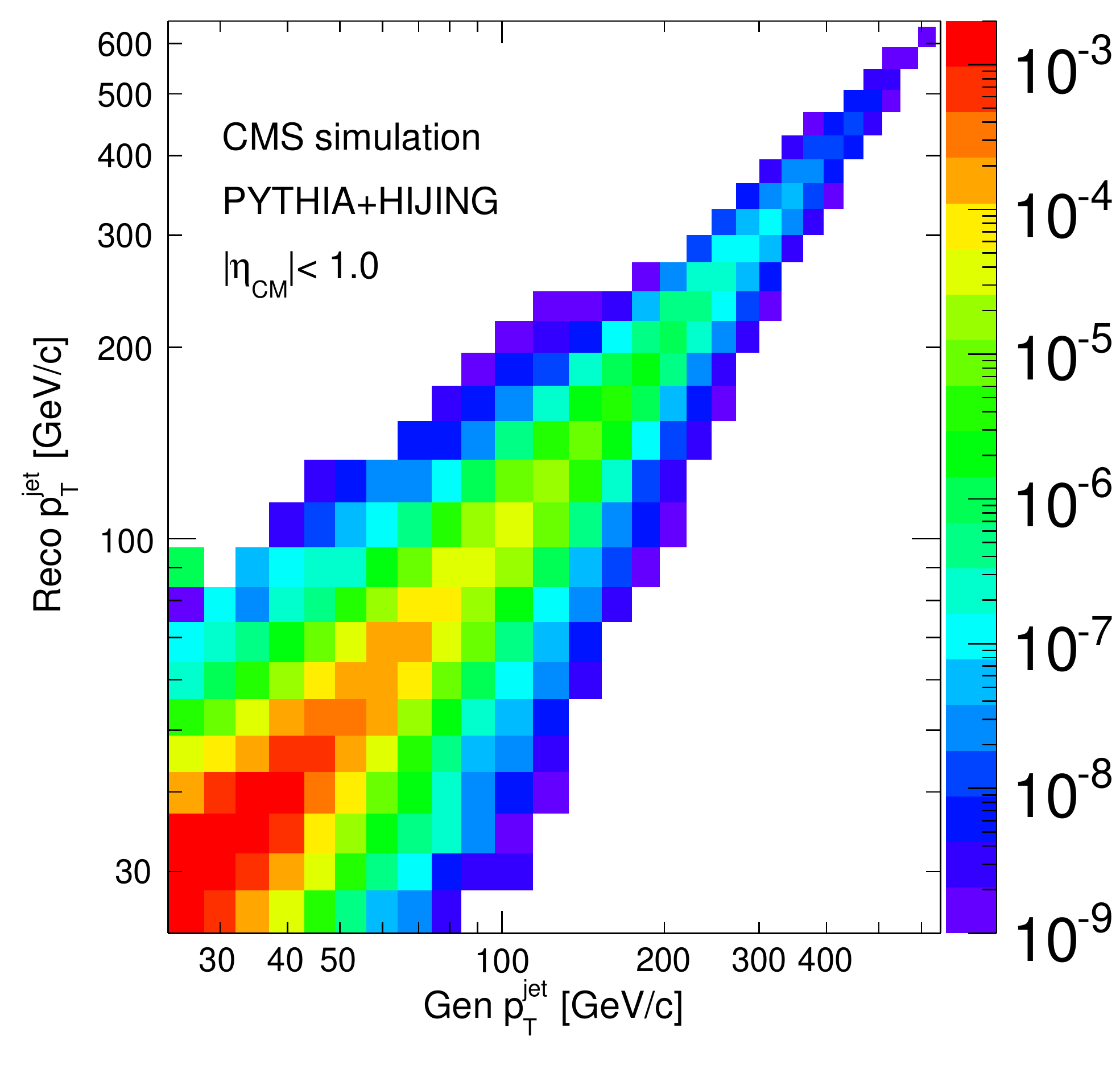}
			\caption{ Response matrix constructed from Monte Carlo sample for pPb {\sc pythia+hijing} events. }
			\label{fig:Matrix}
		\end{center}
		\end{figure}

		The pPb response matrix built with embedded  {\sc pythia+hijing} events is shown in Fig:~\ref{fig:Matrix} derived using the same procedure as described above. 

	\subsection{Kinematic efficiency}
		
\begin{figure}[h] 
   		\centering
   			\includegraphics[width=0.8\textwidth]{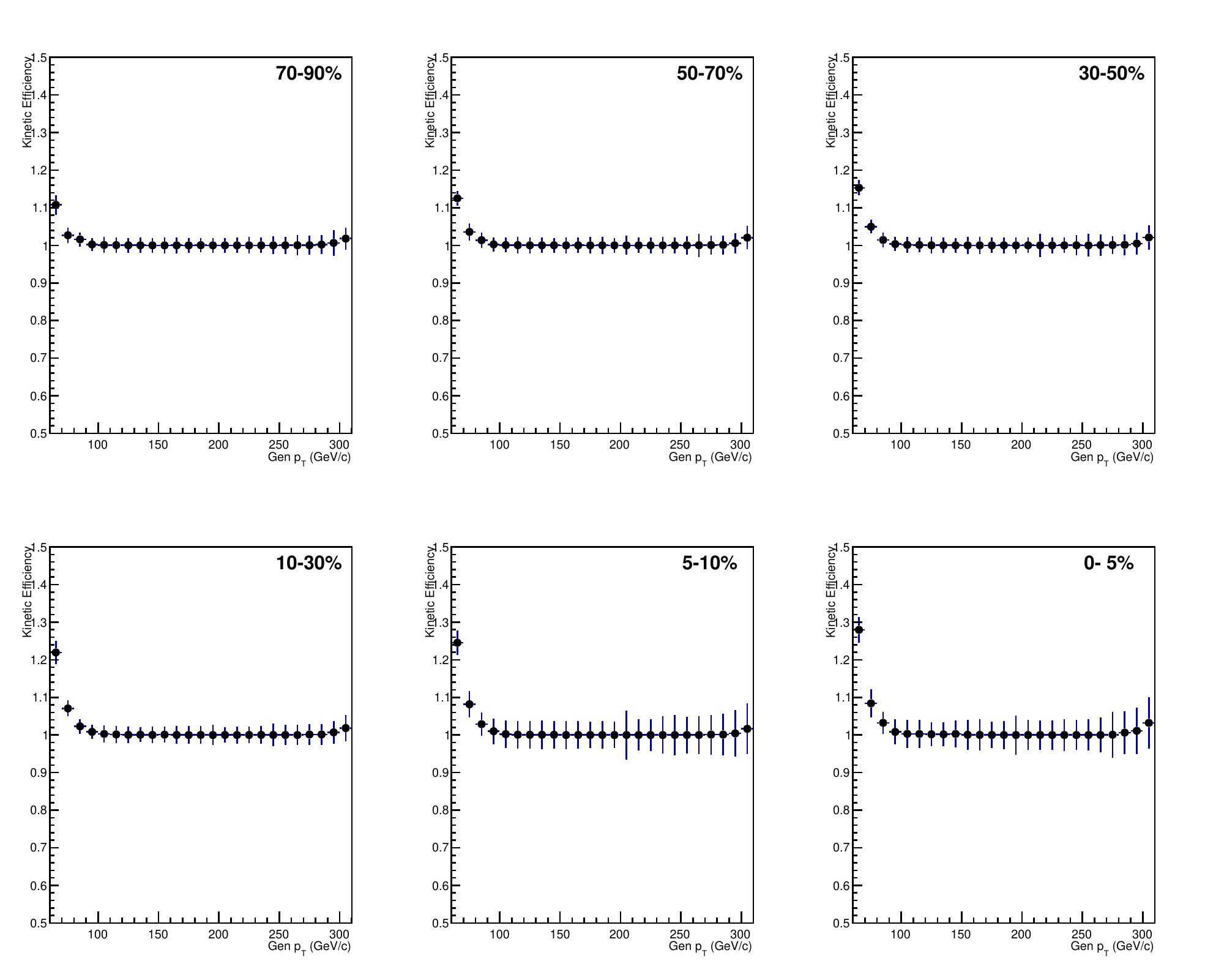} 
   			\includegraphics[width=0.4\textwidth]{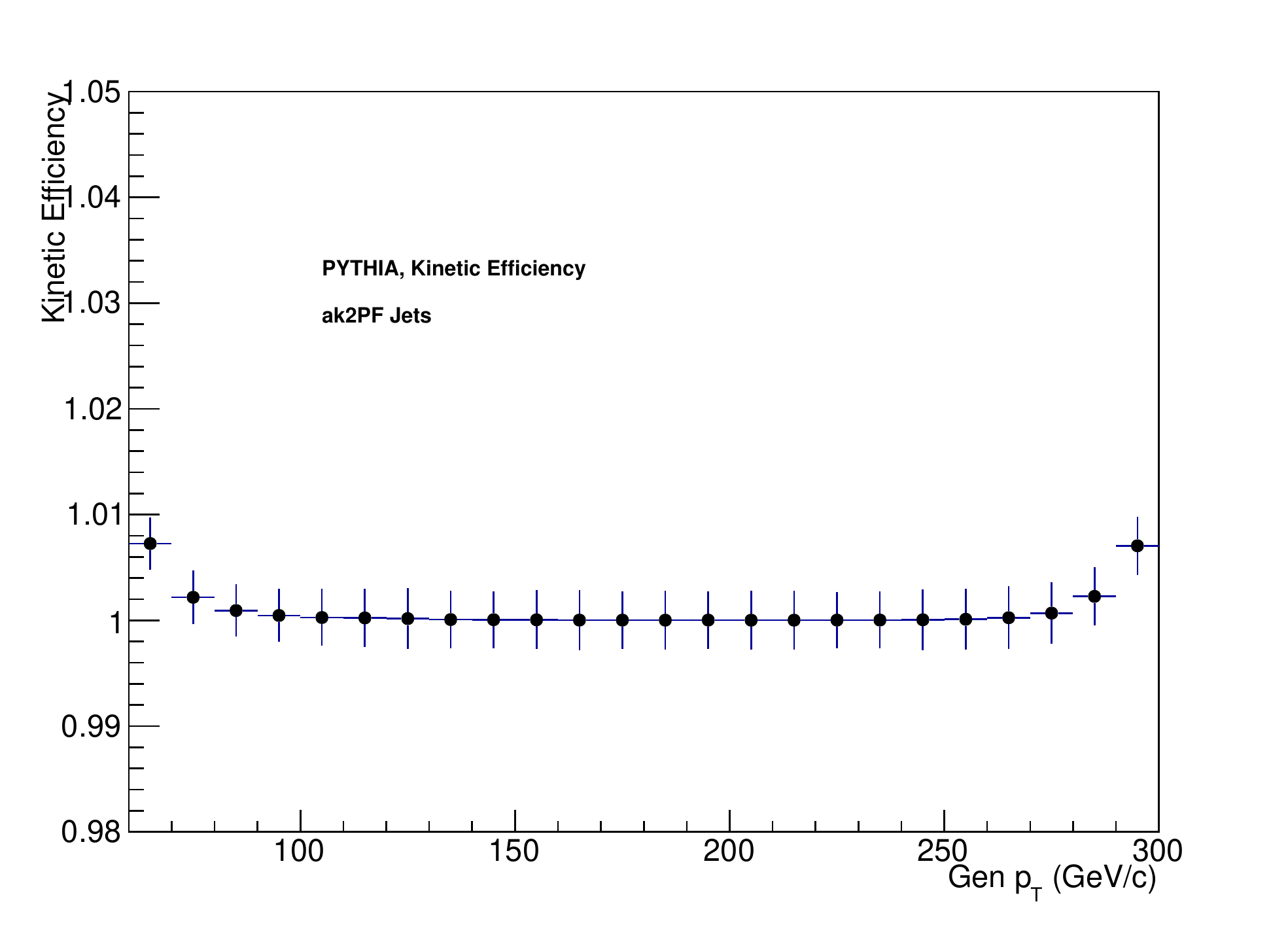} 
   			\caption{Kinematic Efficiency correction for R=0.3 Jets based on a pT cutoff }
   			\label{fig:kineticeff}
		\end{figure}

		The data kinematic range based on the trigger efficiency is chosen to be from $50 - 300$ GeV/c. Since the response matrix is built up by our MC datasets which start from a $\hat{p_T}=15$GeV/c (where $\hat{p_T}$ is the transverse momentum of the subprocess in the 2-2 scattering) we need to apply a reco level cutoff in the normalized response matrix corresponding to the same kinematic range of our data spectra. This effectively reduces the possible $p_T$ bins a given detector smeared jet can end up in the generator side. This is estimated by looking at the normalized matrix's projection onto the generator axis before and after the reco level cut and the correction factor (as shown in Figure~\ref{fig:kineticeff} for $R=0.3$ jets in PbPb and PP) is applied to the unfolded spectra. 

	\subsection{Bayesian and bin-by-bin unfolding}

		Bayesian unfolding~\cite{D'Agostini:1994zf} uses the input MC truth and reconstruction information to create a smearing matrix. Using probability theory, the physical quantity of the jet \pt~spectra is unfolded from the detector effects which modify it. Apart from the provided implementation of the root-based unfolding (\textsc{roounfold}~\cite{Adye:2011gm}), we implemented the standard Richardson--Lucy~\cite{Richardson:1972bb,Lucy:1974yx,Shepp:1982ml} method ourselves. Correlated error propagation through the Bayesian unfolding is done by taking (numerical) partial derivatives with respect to the input spectrum, and then propagated through the entire Bayesian unfolding.  Bayesian unfolding here is performed with 4 iterations. The choice of 4 iterations is  twofold: first it is the default and recommended number of iterations for Bayesian unfolding,  second, that four iterations provide reasonable closure when tested in MC.

		Unlike Bayesian unfolding, which allows for the migration of events between bins,  the bin-by-bin method assumes no migration, and thus corrects for detector effects  only in the height of each jet \pt~bin. Bin-by-bin unfolding can be a valid  technique in the case where resolution is much smaller than bin size.  
		
	\subsection{Unfolding using SVD}

		For pp and pPb collisions, we show the main physics results with the Singular Value Decomposition(SVD) unfolding methods.  For the (additive) LLS method, an ``initial guess'' $x_\mathrm{ini}$ is used to scale the problem such that the unfolding does not have to exhaust its total degrees of freedom (DOF) to purely reproduce the steeply falling spectrum shape. The standard method to perform Phillips--Tikhonov regularization is the generalized singular value decomposition (GSVD), which is available in the RooUnfold program. 

\begin{figure}[h] 
   		\centering
   			\includegraphics[width=0.8\textwidth]{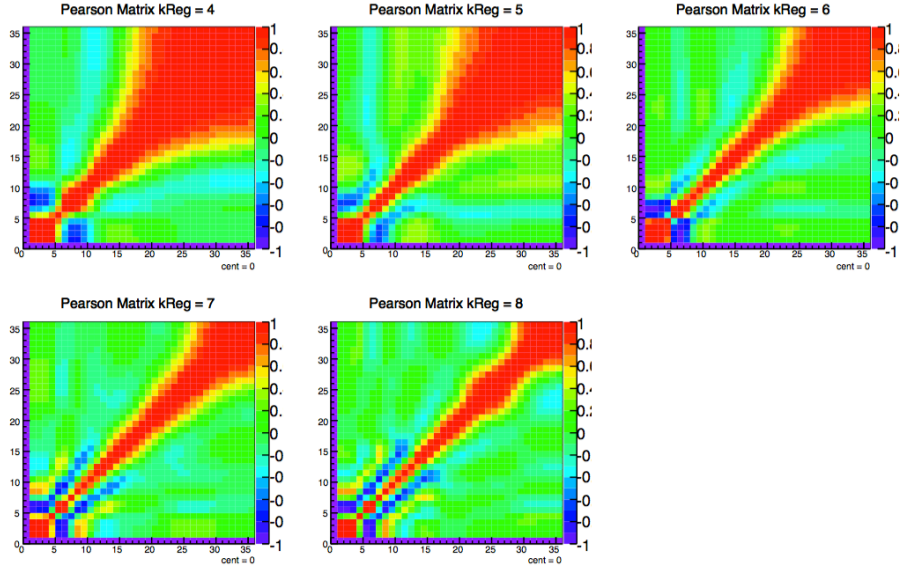} 
   			\caption{Pearson coefficients estimated for SVD unfolding procedure for $R=0.3, 0-5\%$. The X axis in these panels show the different \pt~bins (10 GeV bins) in the reconstructed jet \pt~and the Y axis are the same \pt~bin number for generator level jet \pt~for different regularization parameters The red regions symbolize strong correlation and blue, the anti-correlation.}
   			\label{pearson_coeff}
		\end{figure}

\begin{figure}[h] 
   			\centering
			\includegraphics[width=0.7\textwidth]{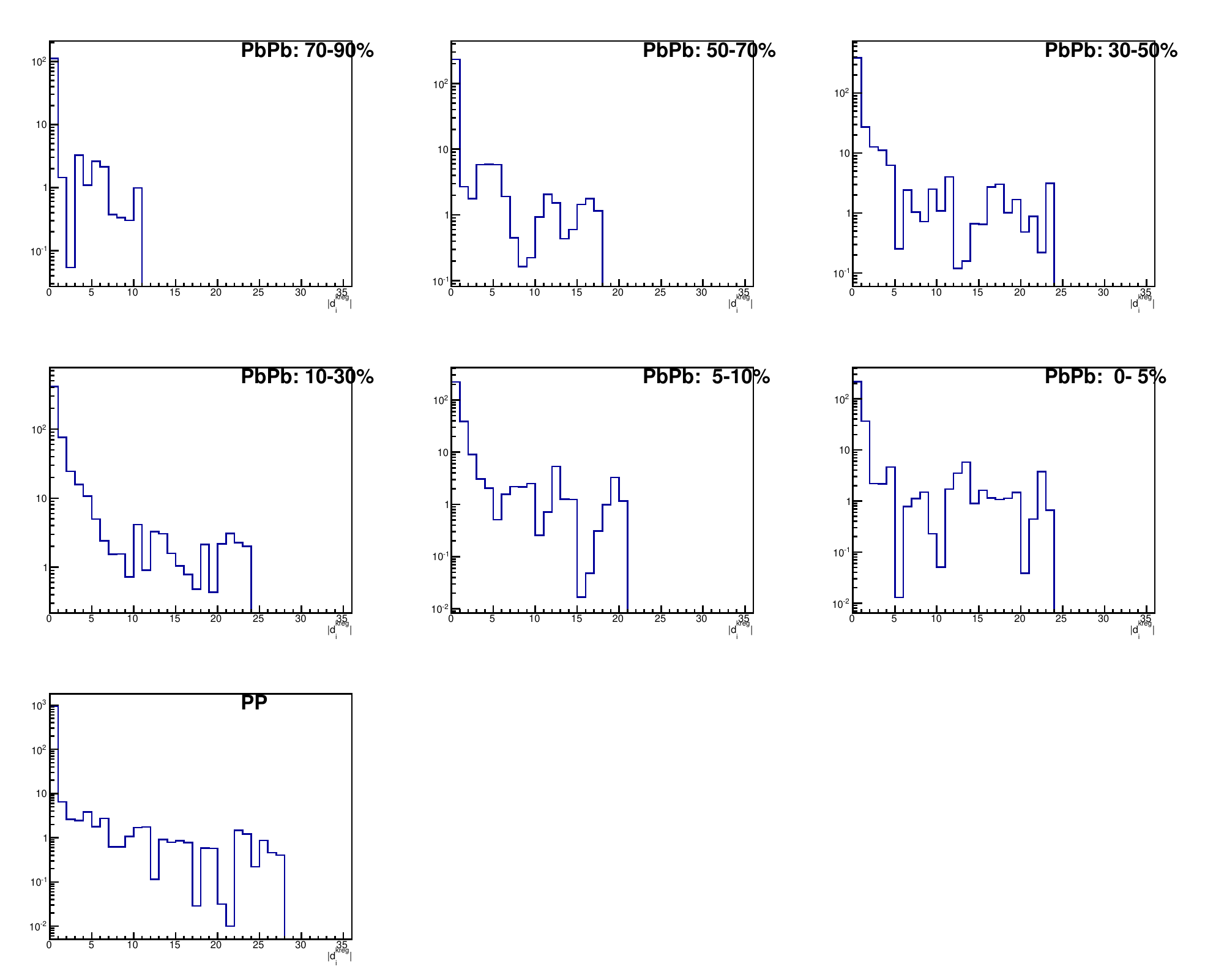} 
  			 \caption{Value of the $d_i$ vector for R=0.3 PF jets in the different centrality bins for PbPb (pileup subtracted) and pp events.}
   			\label{singularvalues_R3_allCent}
		\end{figure}

\begin{figure}[h] 
   		\centering
			\includegraphics[width=0.7\textwidth]{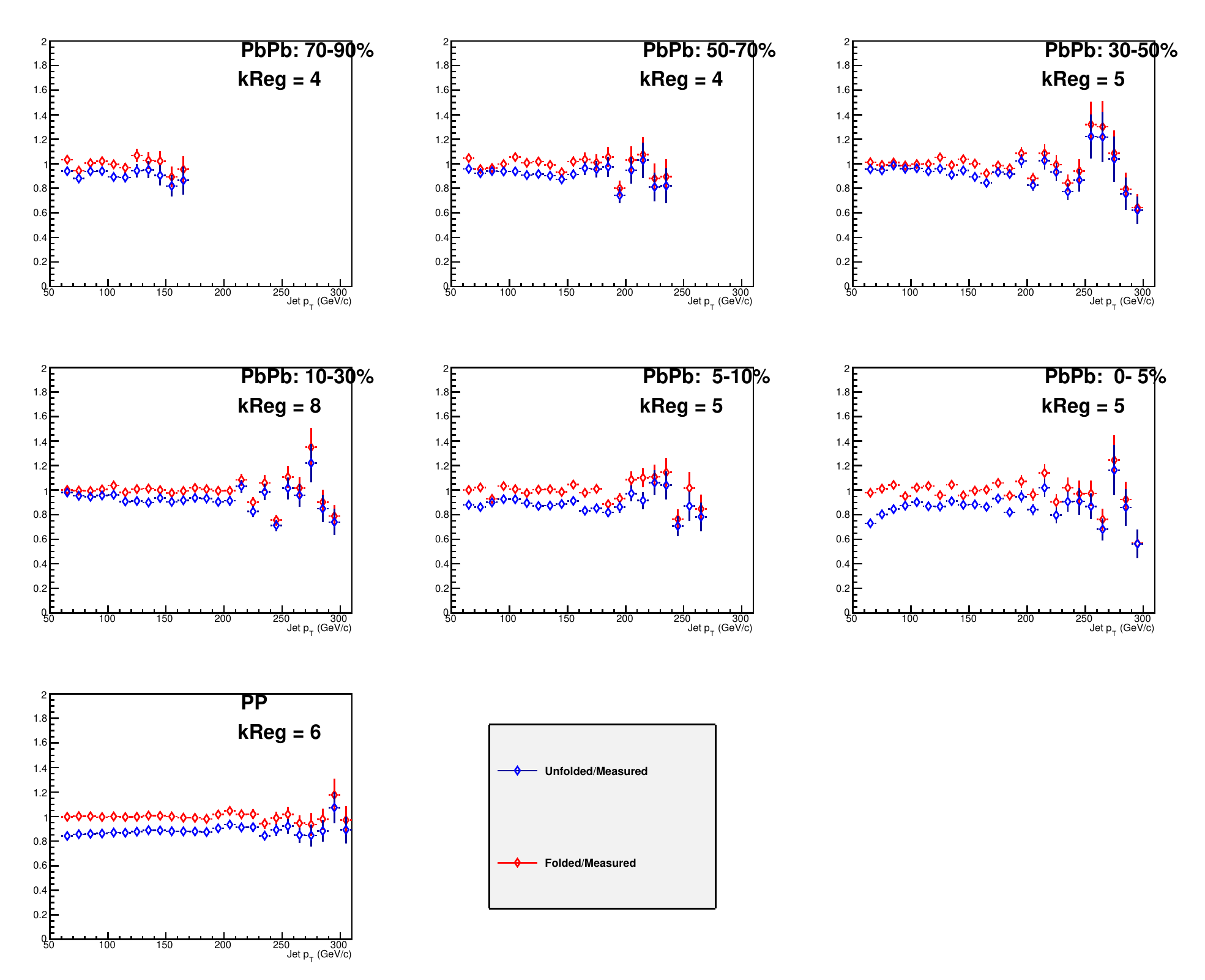} 
   			\caption{Ratio of unfolded spectra to the measured (blue) in comparison to the re-folded spectra to the measured (red) for R=0.3 PF jets in the different centrality bins for PbPb (pileup subtracted) and pp events.}
  			\label{fig:SVDratio_R3_allCent}
		\end{figure}

		The correct value for the regularization parameter is determined by looking at the Pearson coefficients and by studying the values of the $d_{i}$ vector which is the data expressed in the basis vectors of the response matrix. These $d_{i}$ are normalized by the error and statistical fluctuations of unity and the values of $i$ where $|d_{i}| >>1$ are the statistically significant equations in the linear system. The Pearson coefficients for $R=0.3$ and for the $0-5\%$ centrality bin (for example) is shown in Fig:~\ref{pearson_coeff} . For all $R=0.3$ PbPb centrality classes and pp, the $d_{i}$ vector is shown in Fig.~\ref{singularvalues_R3_allCent}. The unfolded spectra can then be re-folded to test the effectiveness of the unfolding. In Fig.~\ref{fig:SVDratio_R3_allCent}, we see that the folded spectra is similar to the measured spectra and is within the total uncertainties. Based on the studies for the different centrality classes and jet radii, we find the corresponding $kReg $ value where the $d_i$ vector gets to the value $1$. These values are collected in table~\ref{table:kReg_allradii_allcent}. Comparisons of folded and unfolded spectra are shown in Fig.~\ref{fig:SVDratio_R3_allCent} for R=0.3. 

		\begin{table}[h]
   		\centering
   		\begin{tabular}{|c|c|c|c|} 
      			\hline
      			Centrality bins    &  $R=0.2$ & $R=0.3$ & $R=0.4$ \\
      			\hline
      			$0-5\%$    & 4 & 5 & 5 \\
      			$5-10\%$   & 5 & 5 & 4 \\
      			$10-30\%$  & 5 & 8 & 5 \\
      			$30-50\%$  & 5 & 5 & 6 \\
      			$50-70\%$  & 4 & 4 & 4 \\
      			$70-90\%$  & 4 & 4 & 4 \\
      			$PP$       & 6 & 6 & 6 \\
      			\hline
   		\end{tabular}
   		\caption{Regularization parameters for SVD unfolding based on the $d_i$ vectors for each radii and centrality bin.}
   		\label{table:kReg_allradii_allcent}
		\end{table}

\begin{figure}
  		\begin{center}
    			\includegraphics[width=0.4\textwidth]{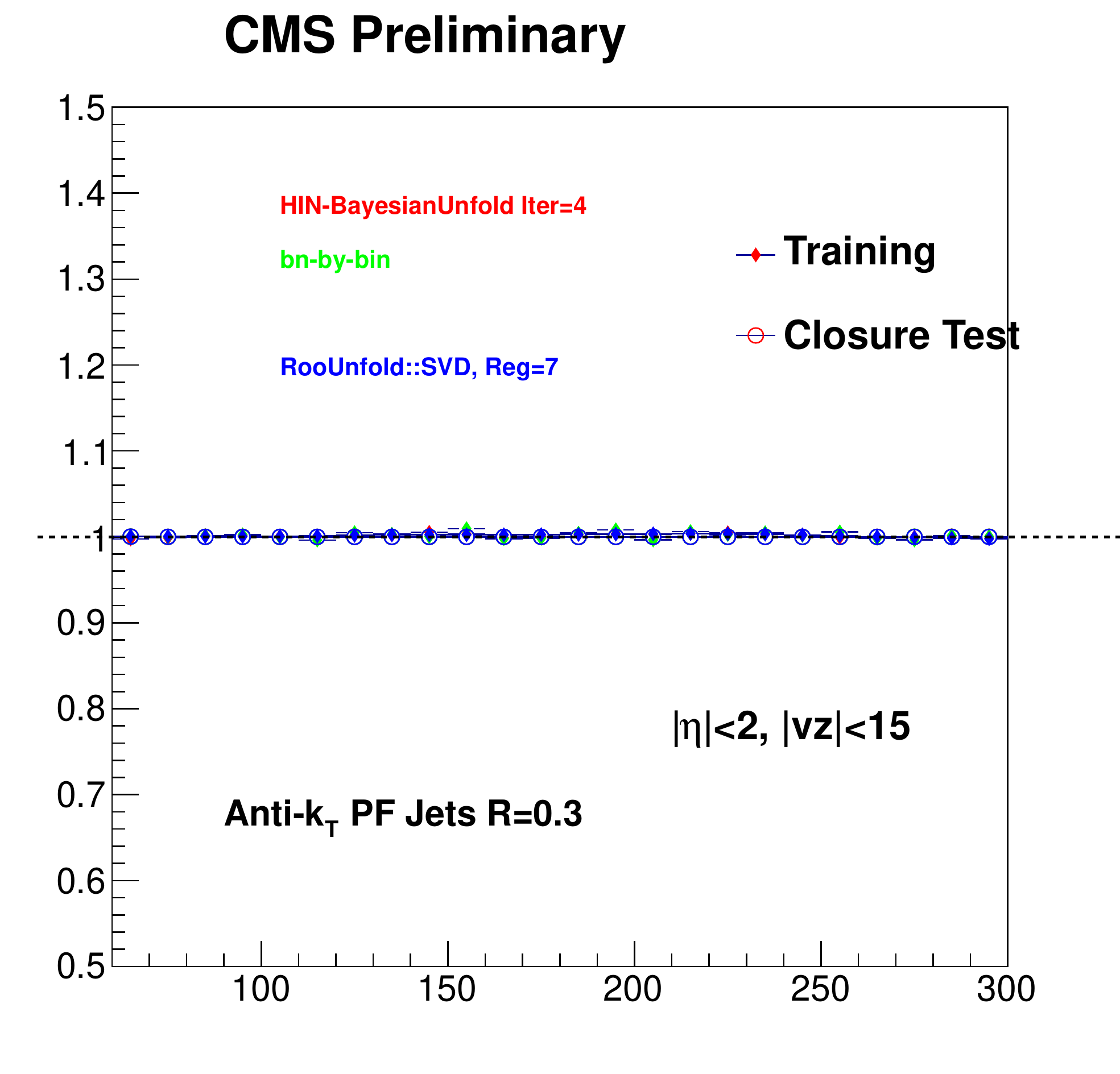}
			\includegraphics[width=0.4\textwidth]{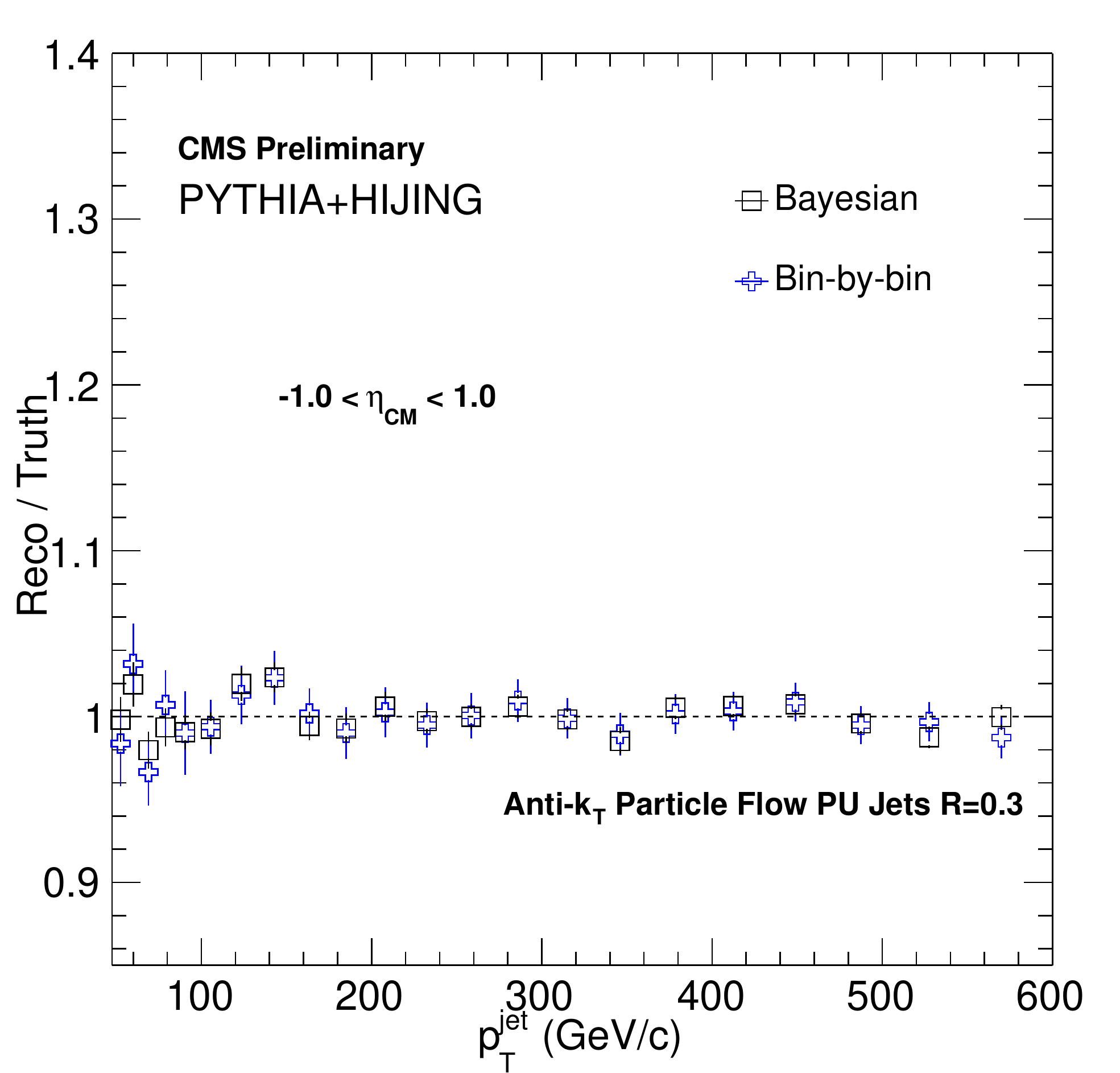}
    			\caption{(left )PP MC closure test showing the performance of bin-by-bin and unfolding methods using one half of the statistics as the truth and other half as data for R=0.3 akPF jets. (Right) A comparison of Monte Carlo sample unfolded by Bayesian and  Bin-By-Bin unfolding methods to the generator truth information for jet \pt~spectra ratio of unfolded jet spectra to the generator truth.}
    			\label{fig:MCClosureTestPP}
 		\end{center}
		\end{figure}

\begin{figure}
                  \begin{center}
                    \includegraphics[width=0.7\textwidth]{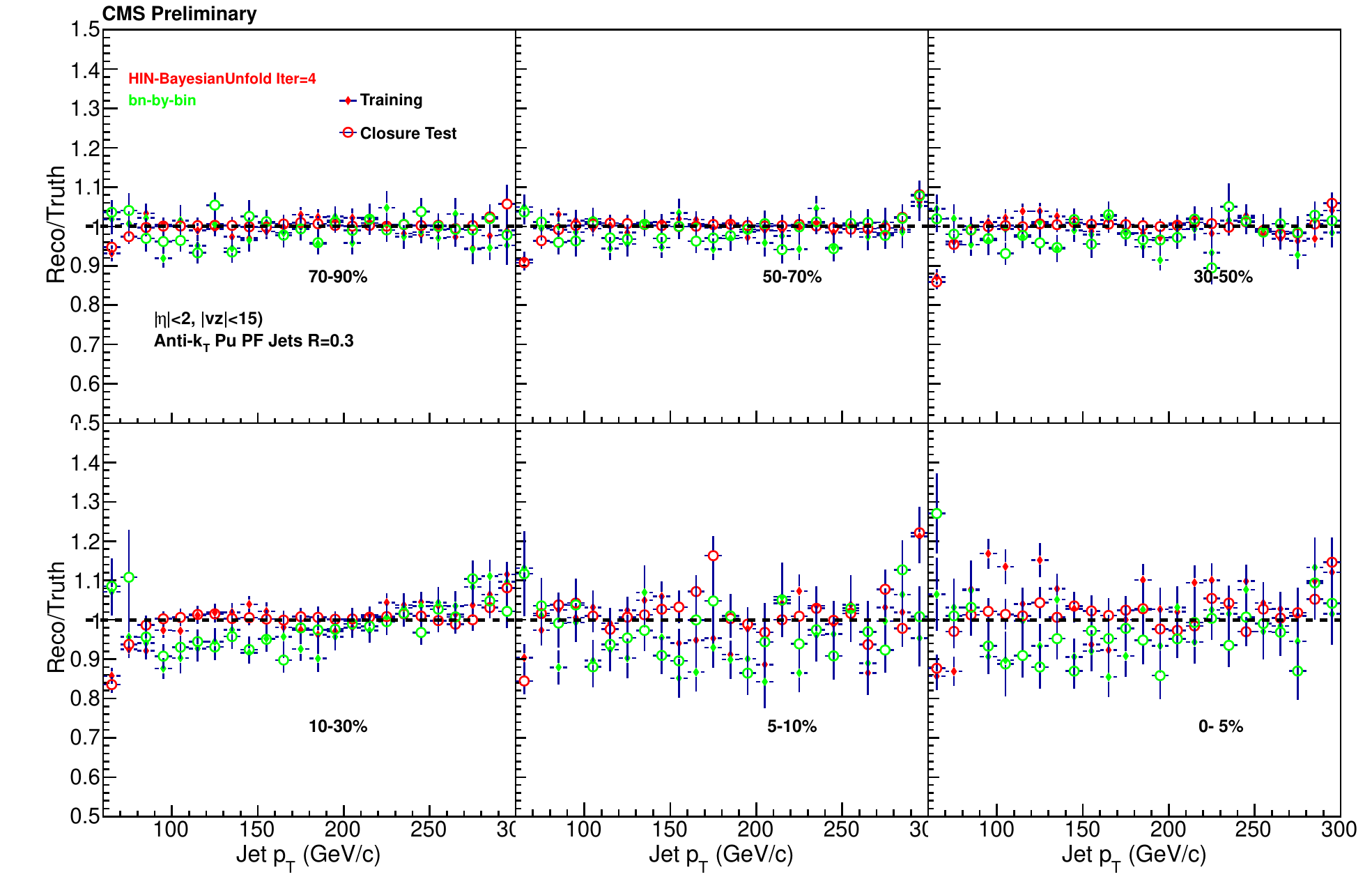}
                    \includegraphics[width=0.7\textwidth]{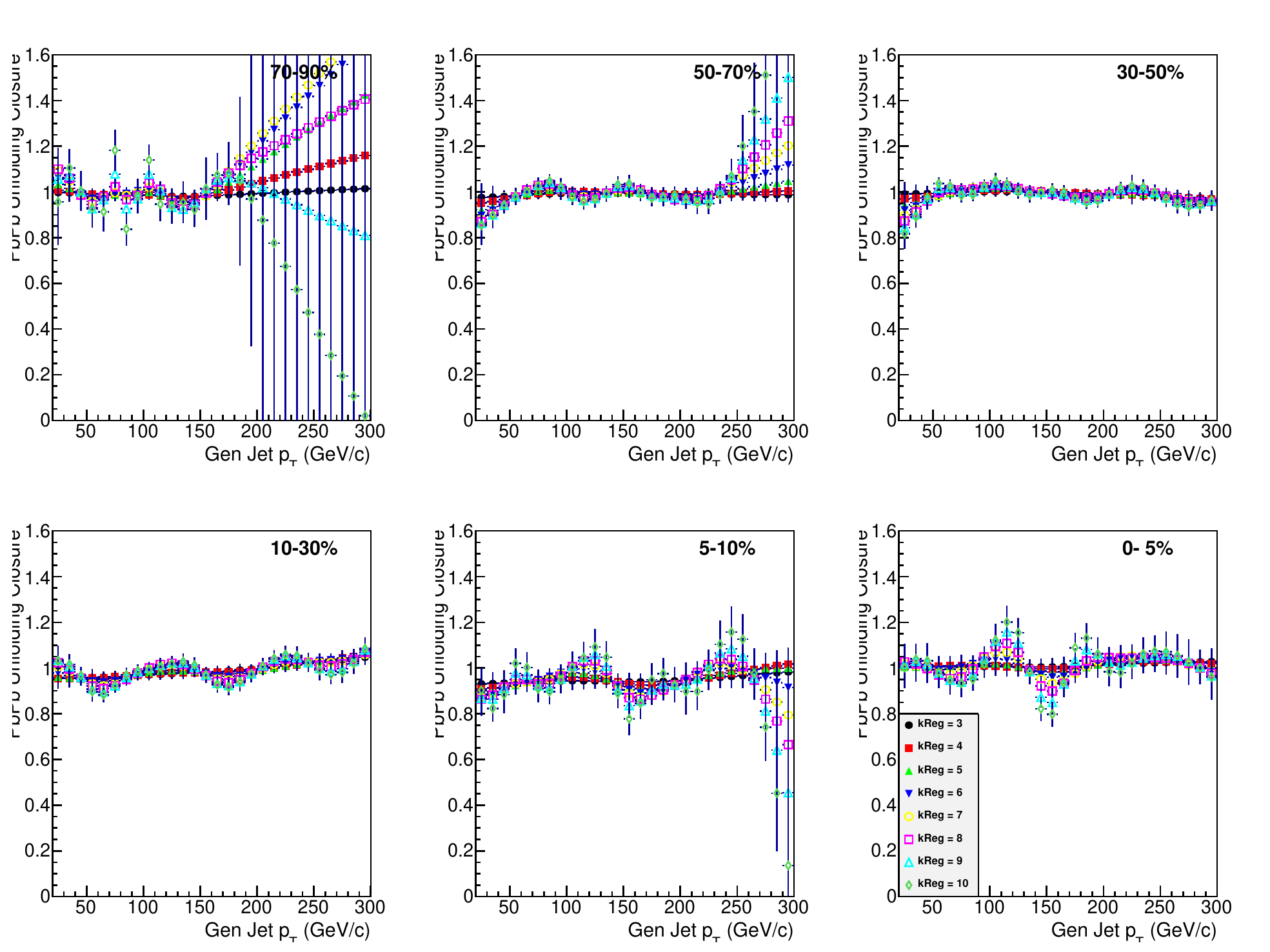}
                    \caption{(Top)PbPb MC closure test showing the performance of Bayesian and bin-by-bin unfolding methods using one half of the statistics as the truth and other half as data for R=0.3  akPuPF jets shown in different centrality bins. The Black circles correspond to the training which takes the data and response matrix from the same half of the statistics and the check (in red circles) show the result of the unfolding when they are taken in opposite halves. (Bottom) MCClosure for SVD method, R=0.3. The black points are the training method where the one half of the statistics is used as both the response matrix and the data whereas the red points show the real effect of the closure when the data is from the opposite half of the statistics.}
                    \label{fig:MCClosureTestPbPb}
                  \end{center}
                \end{figure}

		Monte Carlo closure has been performed to verify the validity of the unfolding methods.  Fig.~\ref{fig:MCClosureTestPP}~-~Fig.~\ref{fig:MCClosureTestPbPb} show the closure test for the Bayesian, bin-by-bin and SVD unfolding methods with pp, pPb and PbPb MC samples. This is done by taking half of the MC sample as "data", and utilizing the other  half as the prior to unfold the "data". The jet spectra for unfolded, reconstructed MC jets  from the "data" are compared to the generator level jets. The ratio of unfolded jet spectra to generator level jet spectra and the ratio of measured jet spectra by generator level jet spectra are compared to show the closure of the Bayesian and bin-by-bin methods. 

	\subsection{Statistical error propagation}

\begin{figure}[htbp] 
		   \centering
		   \includegraphics[width=0.6\textwidth]{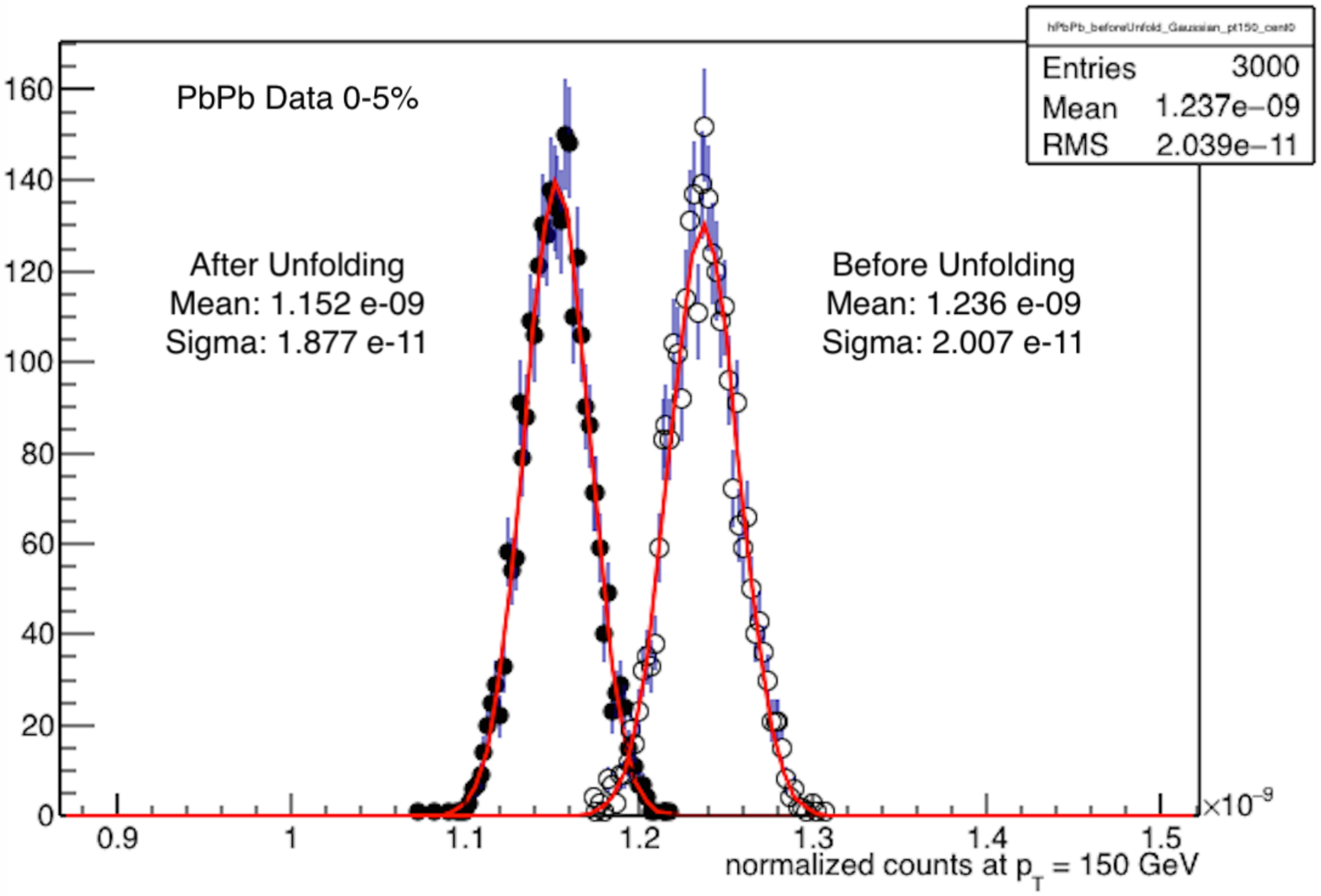}
		   \caption{Before (empty circles) and after (black filled circles) distributions of a large set of spectra at a specific \pt~bin of 150 $GeV/c$ in the most central PbPb data bin. The distributions are fitted by a standard gaussian and the RMS value of the unfolded fit is taken as the final statistical error of the respective \pt~bin.}
		   \label{fig:gausErrorCorr}
		\end{figure}

                	While for SVD method we use the kCovToy option and propagate the estimated statistics uncertainties, in Bayesian unfolding, the unfolded jet spectra's statistical error information is lost. We can perform a data driven approach where we randomly generate a large set of jet spectra using a gaussian distribution for each \pt~bin which upon unfolding will give us another gaussian distribution (as shown in Figure~\ref{fig:gausErrorCorr}) of the bin content per \pt~bin in our final spectra. 
	
\section{Estimating systematic uncertainties}

	An experimental measurement is incomplete without an estimate of systematic uncertainties and how they affect the result. For a measurement of the jet spectra, the systematic uncertainty due to a particular sources is estimated as overall change in the final spectra due to finite variations in the source. 

	\begin{figure}[h!] 
	   \centering
	   \includegraphics[width=0.8\textwidth]{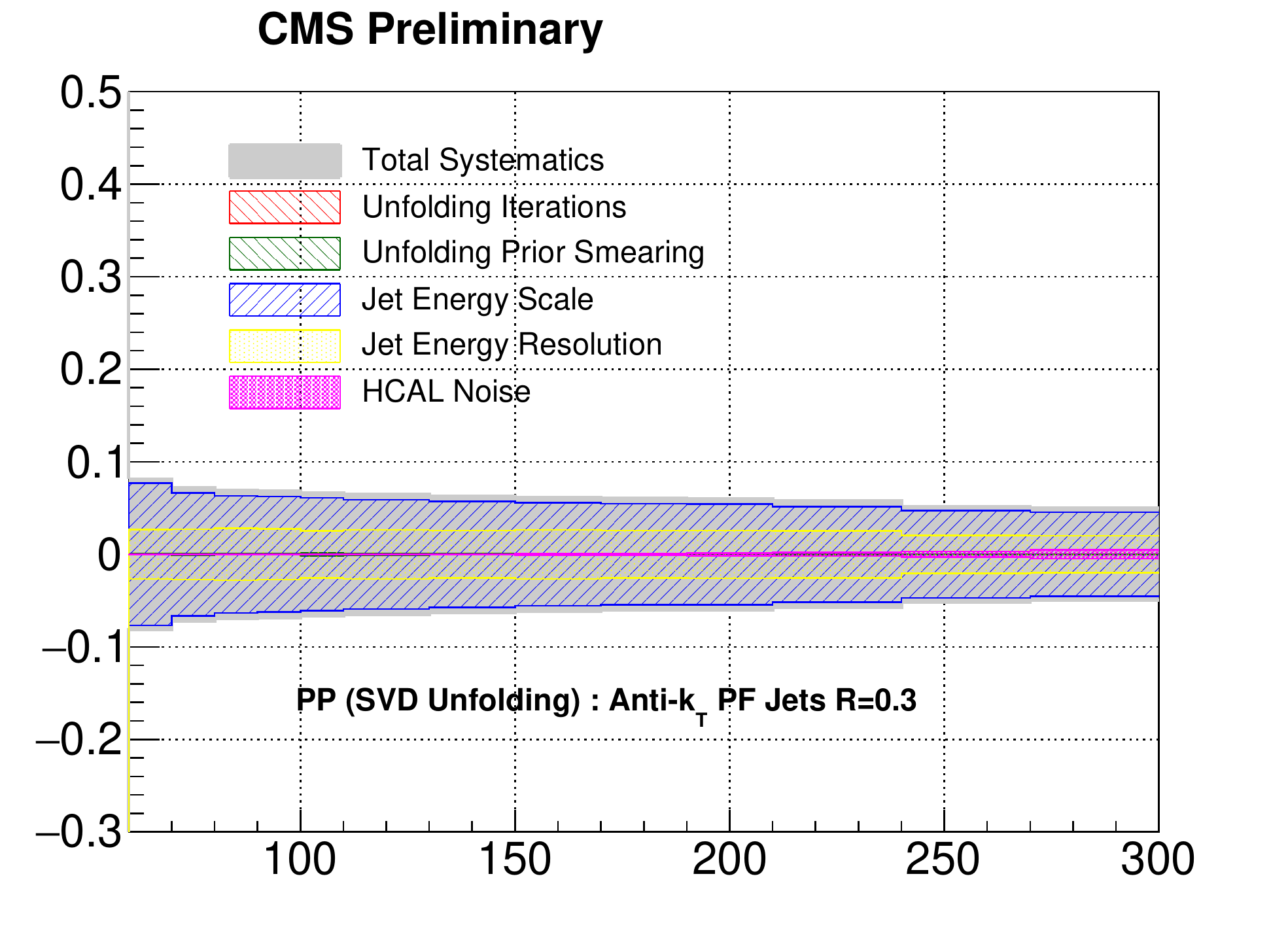} 
	   \caption{Systematic uncertainty from a variety of sources on the pp \akt R=0.3 jet spectra shown in different panels corresponding to the centrality classes.}
	   \label{fig:ppsysuncert}
	\end{figure}

	The unfolded jet spectra for R=0.3 jets has an overall systematic uncertainty of $7.6\% - 5.3\%$. We have clumped the unfolding with the JES systematic uncertainty together since they both affect the overall energy scale. A residual jet energy correction, using the dijet balance method~\cite{Chatrchyan:2011ds}, is derived and applied to the jets from pp collisions. It corresponds to less than 1\% correction to the jet \pt. There is an additional uncertainty on the integrated luminosity of $3.7\%$ which is estimated by running van der merve scans during the same data collection period. The systematic uncertainties from the different sources are shown in Fig:~\ref{fig:ppsysuncert} as a function of the jet \pt~for R=0.3 jets.  

	A summary of the systematic uncertainties in the jet spectra in pPb collisions, the jet yield asymmetry measurements in pPb collisions, the reference pp spectra, and the nuclear modification factors $R^{\ast}_{pPb}$  are listed in Table~\ref{tab:sys}.  The uncertainties depend on the jet \pt~and pseudorapidity, and the table shows representative values in two jet \pt~and $\eta_\mathrm{CM}$ ranges. The uncertainties vary smoothly between these ranges. The total systematic uncertainties listed for the nuclear modification factors $R^{\ast}_{pPb}$ do not include the scale uncertainty of 4.3\%  from the integrated luminosity measurements in pPb (3.5\%) and pp (2.4\%) collisions. The luminosity uncertainties cancel in the measurements of the jet yield asymmetry. The remaining uncertainties are partially correlated in jet \pt, with the unfolding uncertainty dominating at low jet \pt~and the JES uncertainty dominating at high jet \pt.

	\begin{table*}[htbp]
	\centering
	\topcaption{Systematic uncertainties in the measurement of the jet spectra in pPb collisions are shown in the first four lines. The sources and corresponding systematic uncertainties in the extrapolated pp reference are presented in the next four lines. The total uncertainties in the jet spectra in pPb collisions, the reference pp spectra, the jet yield asymmetry in pPb collisions, and $R^{\ast}_{pPb}$ are shown in the bottom four lines. The uncertainties depend on the jet pt and pseudorapidity, and the table shows representative values in two jet \pt~and $\eta_\mathrm{CM}$ ranges. The uncertainties vary smoothly between these two ranges. Total systematic uncertainties listed for the nuclear modification factors $R^{\ast}_{pPb}$ do not include the scale uncertainty of 4.3\% due to the uncertainty in the integrated luminosity measurements in pPb (3.5\%) and pp (2.4\%) collisions.}
	\label{tab:sys}
	\begin{tabular}{llrr{c}@{\hspace*{5pt}}rr}
	&\multirow{2}{*} {Source}
	&\multicolumn{2}{c} {Jet \pt$ < 80$\gev } && \multicolumn{2}{c} {Jet \pt$ > 150$\gev} \\ \cline{3-4}\cline{6-7}
	&& $|\eta_\mathrm{CM}|< 1$  & $|\eta_\mathrm{CM}|> 1.5$ && $|\eta_\mathrm{CM}|< 1$  & $	|\eta_\mathrm{CM}|> 1.5$\\
	\hline
	pPb: &JES \& unfolding     & 5\%  & 8\% &&7\%  & 10\%\\
	& Misreconstructed jet contribution& 1\% & 1\% &&1\%  & 1\%\\
	& Jet pointing resolution    & 1\%  & 2\% &&1\%  & 2\%\\
 	& Integrated luminosity     & 3.5\%  & 3.5\% &&3.5\% &  3.5\%  \\
	\hline
	pp: &Input data            & 6\%  & 8\% &&5\%  & 7\%\\	
	& Cone-size dependence       & 5\%  & 5\% &&5\%  & 5\%\\
	& Collision-energy dependence& 4\%  & 5\% &&6\%  & 7\%\\
 	& Integrated luminosity     & 2.4\%& 2.4\% &&2.4\% &2.4\%  \\
	\end{tabular}
	\end{table*}

	\begin{figure}[h!] 
	   \centering
	   \includegraphics[width=0.8\textwidth]{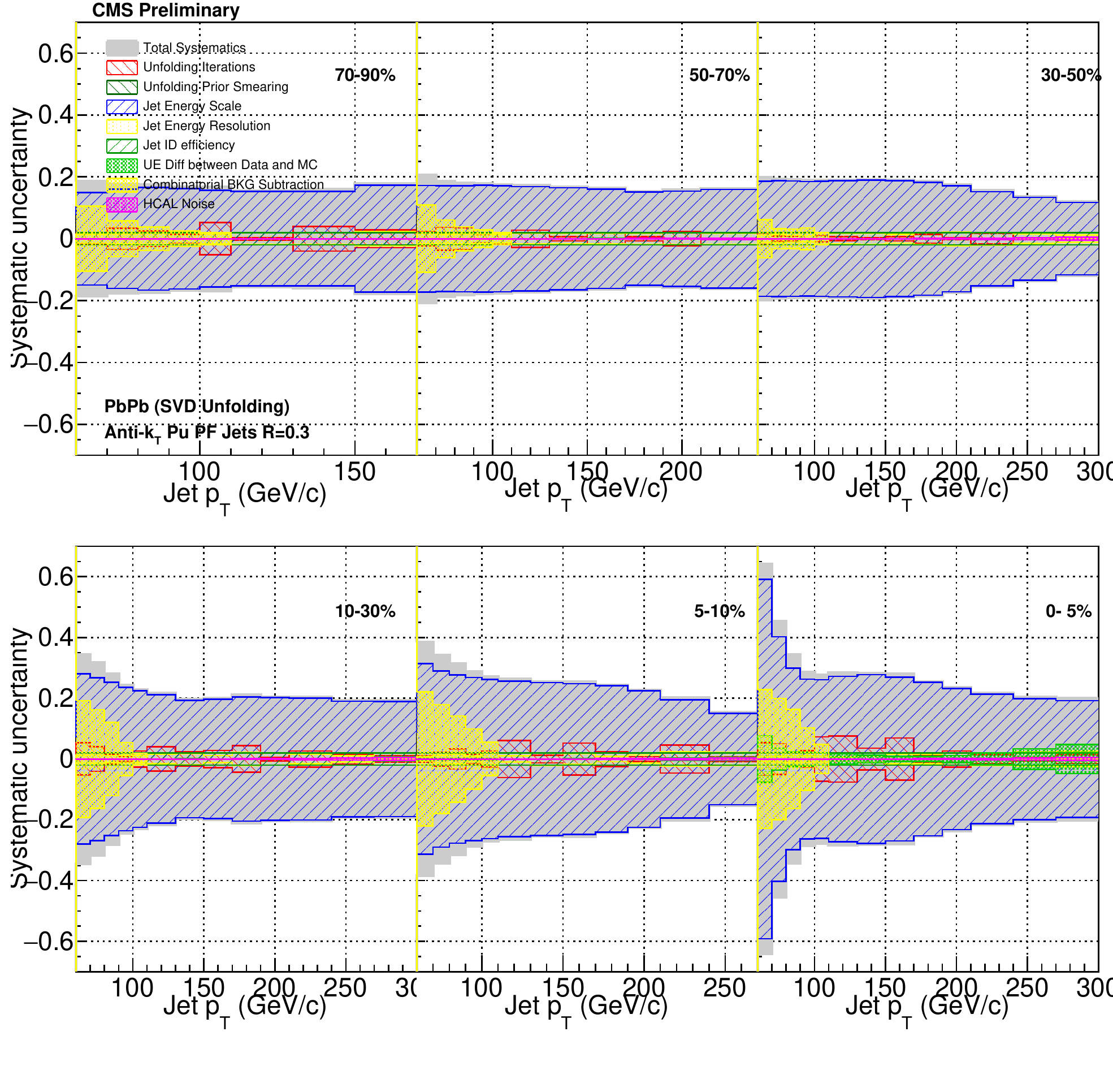} 
	   \caption{Systematic uncertainty from a variety of sources on the PbPb \akt R=0.3 jet spectra shown in different panels corresponding to the centrality classes.}
	   \label{fig:pbpbsysuncert}
	\end{figure}
	
	The jet response matrix is smeared by 1\%, at both the generator and reconstructed levels to account for variations in the simulations. Separately the regularization parameter used for the unfolding is varied between 4 and 8 resulting in at most 8\% systematic uncertainty for the PbPb jet yield. The JER uncertainty is estimated for each \pt~bin in the analysis and is found to be at most 3\%, for both pp and PbPb. Studies of the underlying event fluctuations in jet-triggered and minimum bias events show a contribution of up to 5\% to the uncertainty of reconstructed jet yields based on differences between data and \pyhd~quantified in the right side of Fig.~\ref{fig:bkgmeanrms}. The contributions due to jet reconstruction efficiency, detector noise, and unfolding response matrix smearing are about 1\% each. Since in PbPb, the per-event jet yield is being measured, there is a 3\% uncertainty on the number of minimum bias events and there is no uncertainty quoted for the luminosity.  The overall systematic uncertainties from the different sources are shown in Fig:~\ref{fig:pbpbsysuncert} for R=0.3 jets. 

\begin{table*}[htb]
  \centering
  \topcaption{Summary of the systematic uncertainties in the PbPb jet yield for the central (0--5\%), peripheral (70--90\%) bins, and the pp jet cross section. Each column showcases the total systematic uncertainties for the corresponding source for the different radii and two jet \pt~ranges i.e,  $70 < $ \pt~$< 80$ and $250 < $ \pt$ < 300$ \gev~. The \taa uncertainties are not shown in the table. Other sources mentioned in the text that are smaller than 1\% are not listed explicitly below.}
  \label{table:systematics}
	\begin{tabular}{llrrr{c}@{\hspace*{5pt}}rrr}
    & \multirow{2}{*} {Source}
    & \multicolumn{3}{c} {$70 < $\pt$< 80$} && \multicolumn{3}{c} {$250 < $\pt$< 300$} \\ \cline{3-5}\cline{7-9}
    && R = 0.2 & R = 0.3 & R = 0.4 && R = 0.2 & R = 0.3 & R = 0.4 \\
    \hline
    PbPb: & Data driven correction & 13\% & 20\% & 27\% && \NA & \NA & \NA \\
    (0-5\%)& JES \& unfolding & 32\% & 32\% & 48\% && 19\% & 19\% & 21\% \\
    & JER & 3\% & 3\% & 3\% && 3\% & 3\% & 3\% \\
    & Underlying event & 5\% & 5\% & 5\% && \NA & \NA & \NA \\
    \hline
    PbPb: & Data driven correction & 8\% & 10\% & 12\% && \NA & \NA & \NA \\
    (70-90\%)& JES \& unfolding & 16\% & 16\% & 18\% && \NA & \NA & \NA \\
    & JER & 3\% & 3\% & 3\% && \NA & \NA & \NA \\
    & Underlying event & 5\% & 5\% & 5\% && \NA & \NA & \NA \\
    \hline
    pp: & JES \& unfolding & 7\% & 7\% & 6\% && 5\% & 4\% & 5\% \\
    & JER & 3\% & 3\% & 3\% && 2\% & 2\% & 2\% \\
    & Integrated luminosity & 3.7\% & 3.7\% & 3.7\% && 3.7\% & 3.7\% & 3.7\% \\
	\end{tabular}
\end{table*}

\clearpage

\chapter{Baseline Measurements in p-p and p-Pb Collisions}
\label{ch_pppPbcoll}

\begin{chapquote}{Max Planck}
``An experiment is a question which science poses to Nature, and a measurement is the recording of Nature's answer." 
\end{chapquote}

\section{Baseline measurements}

	As we have introduced in the previous chapters, measuring the inclusive jet cross section in an important fundamental test of QCD. In line with excellent CMS jet measurements~\cite{Chatrchyan:2014gia, Khachatryan:2015luy, Khachatryan:2016wdh}, we focused on the inclusive jet cross section at 2.76 TeV for small radii jets, such as 0.2, 0.3 and 0.4 so that we can highlight the maximal deviation from present state of the art theory calculations. We then proceed to pPb collisions at the LHC and look at measurements of the inclusive jet cross sections and study any possible effects cold nuclear matter effects by comparing with an extrapolated pp reference spectra at 5 TeV. 

\section{Jet spectra in pp collisions}

\begin{figure}[h] 
	   \centering
	   \includegraphics[width=0.9\textwidth]{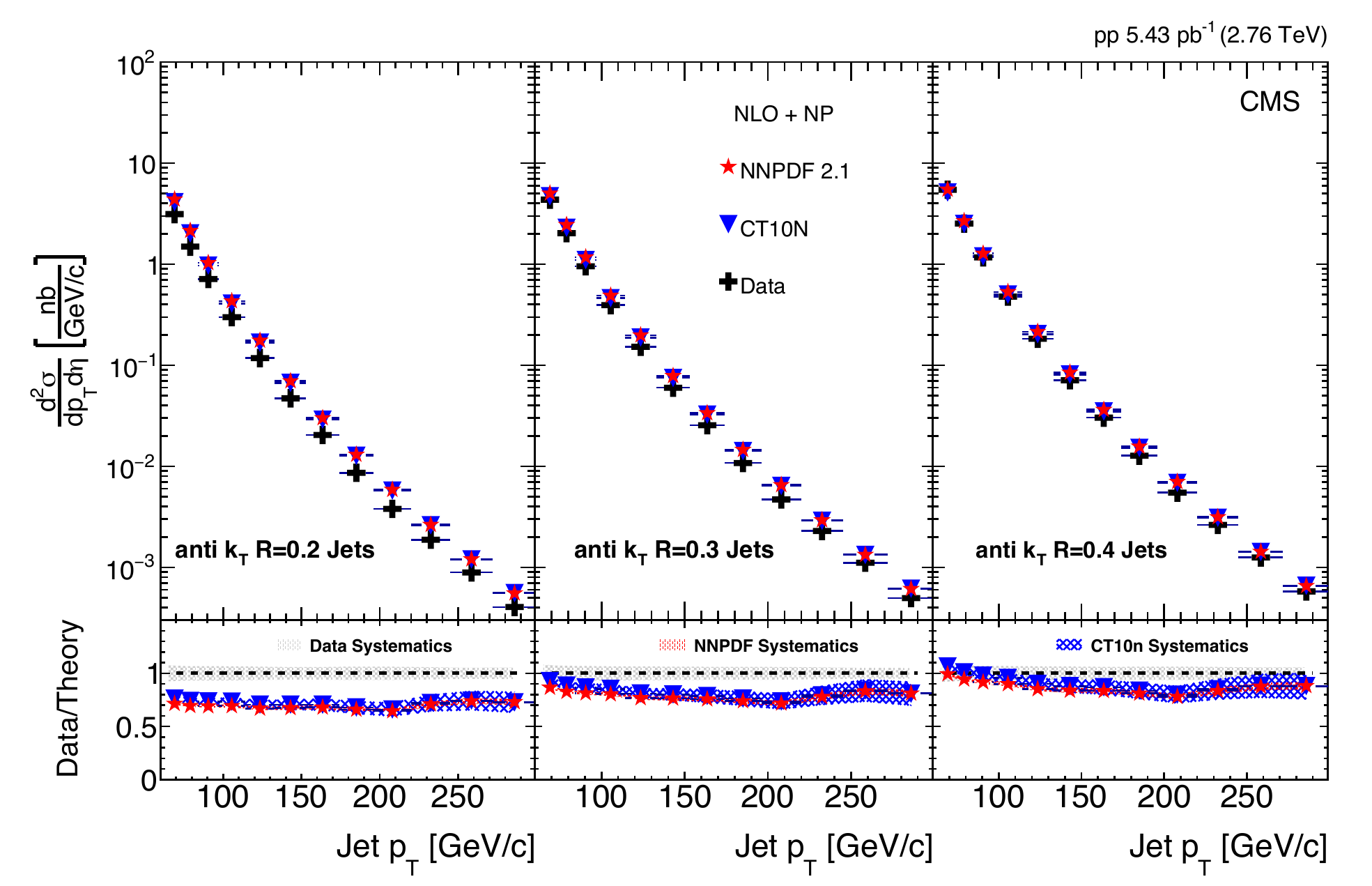} 
	   \caption{Double differential inclusive jet cross sections in pp collisions at 2.76 TeV for \akt R = 0.2 (left), R = 0.3 (middle) and R = 0.4 (right) PF jets~\cite{Khachatryan:2016jfl}. The data is compared to NLO calculations utilizing NNPDF2.1 (red star) and CT10N (purple triangle) PDFs with the bottom panels showing the ratio of data over theory. The shaded regions show the systematic uncertainty for the data and theory systematic uncertainties.}
	   \label{fig:ppJetSpectra}
	\end{figure}

	The inclusive differential jet cross sections in pp collisions at 2.76 TeV are shown in Fig.~\ref{fig:ppJetSpectra} for three different distance parameters. A comparison is made to NLO pQCD~\cite{Wobisch:2011ij} calculations with the fastNLO framework including non-perturbative (NP) corrections. These calculations are shown for two parton distribution functions (PDF) sets: NNPDF 2.1~\cite{Ball:2012cx} (red stars), and CT10N~\cite{Gao:2013xoa} (purple triangles) including  such as multi-parton interactions and hadronization. The bottom panel of Fig.~\ref{fig:ppJetSpectra} shows the ratio of the data for jet cross sections in pp collisions to theoretical calculations. The agreement with data gets better at larger distance parameters. In Ref.~\cite{Khachatryan:2015luy} the ratio tends closer to unity for jets with R = 0.7. The theoretical uncertainties shown are due to variations of the strong coupling constant and the parton shower, factorization scales involved in the NLO calculations for the different PDF sets.
		
	\subsection{Theoretical progress for small R jets}
	
		In our landmark result, we essentially showed an evolution of the jet cross-section as a function of the jet radii and how theory calculations at NLO+NP over estimated the data at small radii. There has been recent interest in this field of small radii jets in p-p collisions since phenomenologically, radii dependent corrections were always assumed to be small at NLO but they turn out to be quite large at NNLO. We will briefly go over two of the latest developments from our theory colleagues including matching NNLO corrections with Leading Logarithmic resummations (LL$_{R}$) for small radii jets. 
	
		\subsubsection{NNLO matched with LL$_{R}$}
		
			The overall jet cross section is sensitive to the perturbative series involving terms such as $\alpha^{n}_{s} \ln^{n} 1/R^{2}$ and at high $Q^{2}$ where $\alpha_{s}$ is small, the terms are under compensated by large logarithms in the $\ln^{n} 1/R^{2}$ when we take the small radii limit.  Calculations for the first order terms are available in the literature~\cite{Dasgupta:2016bnd} and NLL$_{R}$ is still in progress since it involves several additional complications that we will not discuss here.  At the leading log, there is an additional contribution to the micro-jet cross section 
			\begin{equation}
				\sigma^{LL_{R}} (p_{t}, R) \equiv \frac{d\sigma^{LL_{R}}_{jet}}{dp_{t}} = \sum_{k} \int_{p_{t}} \frac{d\prime{p_{t}}}{\prime{p_{t}}} f^{incl}_{jet/k} \left(\frac{p_{t}}{\prime{p_{t}}}, t(R, R_0, \mu_R)\right)\frac{d\sigma^{k}}{d\prime{p_{t}}}
			\end{equation} 
			where the micro-jet fragmentation function ($f^{incl}_{jet/k} (p_{t}/\prime{p_{t}}, t)$) for micro jets with a fraction of the parton's energy. The additional term in the small R limit is built into the $t$ and can be written as
			\begin{equation}
				t(R, R_{0}, \mu_{R}) = \int^{R^{2}_{0}}_{R^{2}} \frac{d\theta^{2}}{\theta^{2}} \frac{\alpha_{s}(\mu_{R} \theta/R_{0})}{2\pi} = \frac{1}{b_{0}} \ln \frac{1}{1-\frac{\alpha_{s} (\mu_{R})}{2 \pi} b_{0} ln\frac{R^{2}_{0}}{R^{2}}},
			\end{equation}
			where we have $b_{0}$ proportional to the casimir factor $C_{A}$. There is an angular scale $R_{0}$ which is set to 1 in this calculation and when we set $R = R_{0}$ this functional $t$ turns into the regular fragmentation function with a delta function at $1-z$, where $z$ is the fractional energy of the splitting. 

			\begin{figure}[h] 
			   \centering
			   \includegraphics[width=0.45\textwidth]{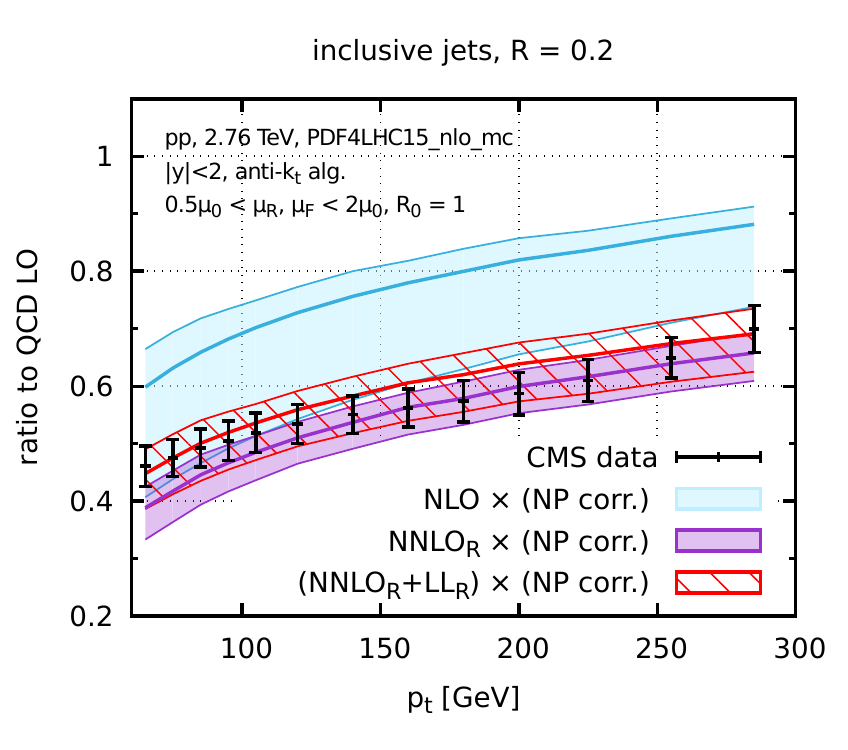} 
			   \includegraphics[width=0.45\textwidth]{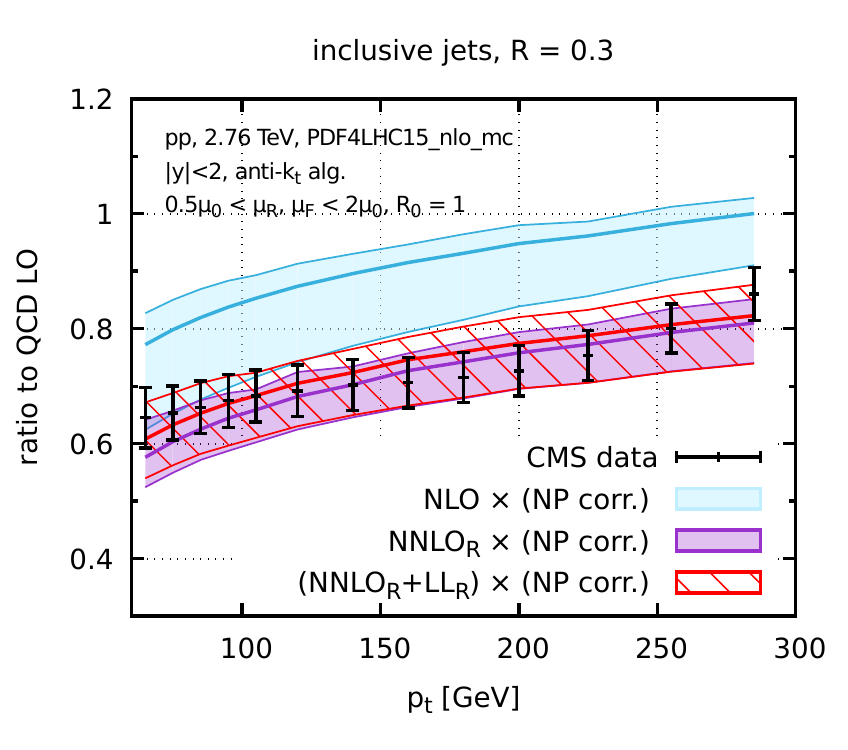} 
			   \includegraphics[width=0.45\textwidth]{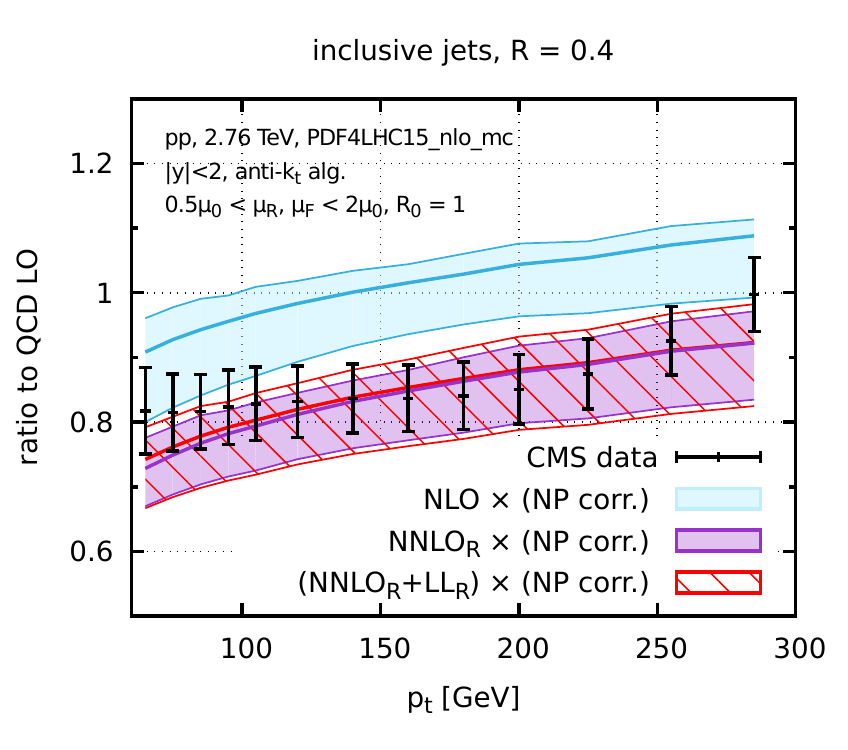} 
			   \caption{Ratio of CMS data (black points)  NLO (blue shaded region), NNLO$_{R}$ and NNLO$_{R}$ + LL$_{R}$, all with NP corrections, to LO perturbative calculations of the inclusive jet spectra for three radii as a function of the jet \pt. Figure courtesy Frédéric Dreyer}
			   \label{fig:ppjettheorynnlo}
			\end{figure}

			This additional correction for small radii jets is actually large, up to $40\%$ for R=0.2, in the framework. A matching procedure is also introduced in this calculation to combine the cross section at the NLO with LL$_{R}$ 
			\begin{equation}
				\sigma^{NLO+LL_{R}} = (\sigma_{0} + \sigma_{1} (R_{0})) \times \left[\frac{\sigma^{LL_{R}} (R)}{\sigma_{0}} \times \left(1 + \frac{\sigma_{1}(R) - \sigma_{1}(R_{0}) - \sigma_{1}^{LL_{R}}(R)}{\sigma_{0}}\right)\right], 
			\end{equation}	
			where again, $R_{0}$ is the arbitrary radius of order 1, $\sigma_{1}(R)$ denotes the pure NLO contribution, without the LO, to the inclusive jet spectra and $\sigma_{1}^{LL_{R}}(R)$ is the pure NLO within the LL$_{R}$ resummation. This matching procedure is conceptually handy when extending the calculation to NNLO+LL$_{R}$ as was done in Fig:~\ref{fig:ppjettheorynnlo}. Both NNLO$_{R}$ and NNLO$_{R} +$ LL$_{R}$ very nicely match data for all the three jet radii across the full \pt~range measured. An interesting feature to note is that both data and theory systematic uncertainties are of a similar magnitude. In order to estimate, which of NNLO$_{R}$ vs NNLO$_{R}+ $LL$_{R}$ captures the data performance better, we need to reduce the uncertainties in both data/theory and come up with novel observables such as taking ratios of cross-sections in different kinematic ranges etc.
		
		\subsubsection{LL$_{R}$ resummation in SCET}
		
\begin{figure}[h] 
			   \centering
			   \includegraphics[width=0.3\textwidth]{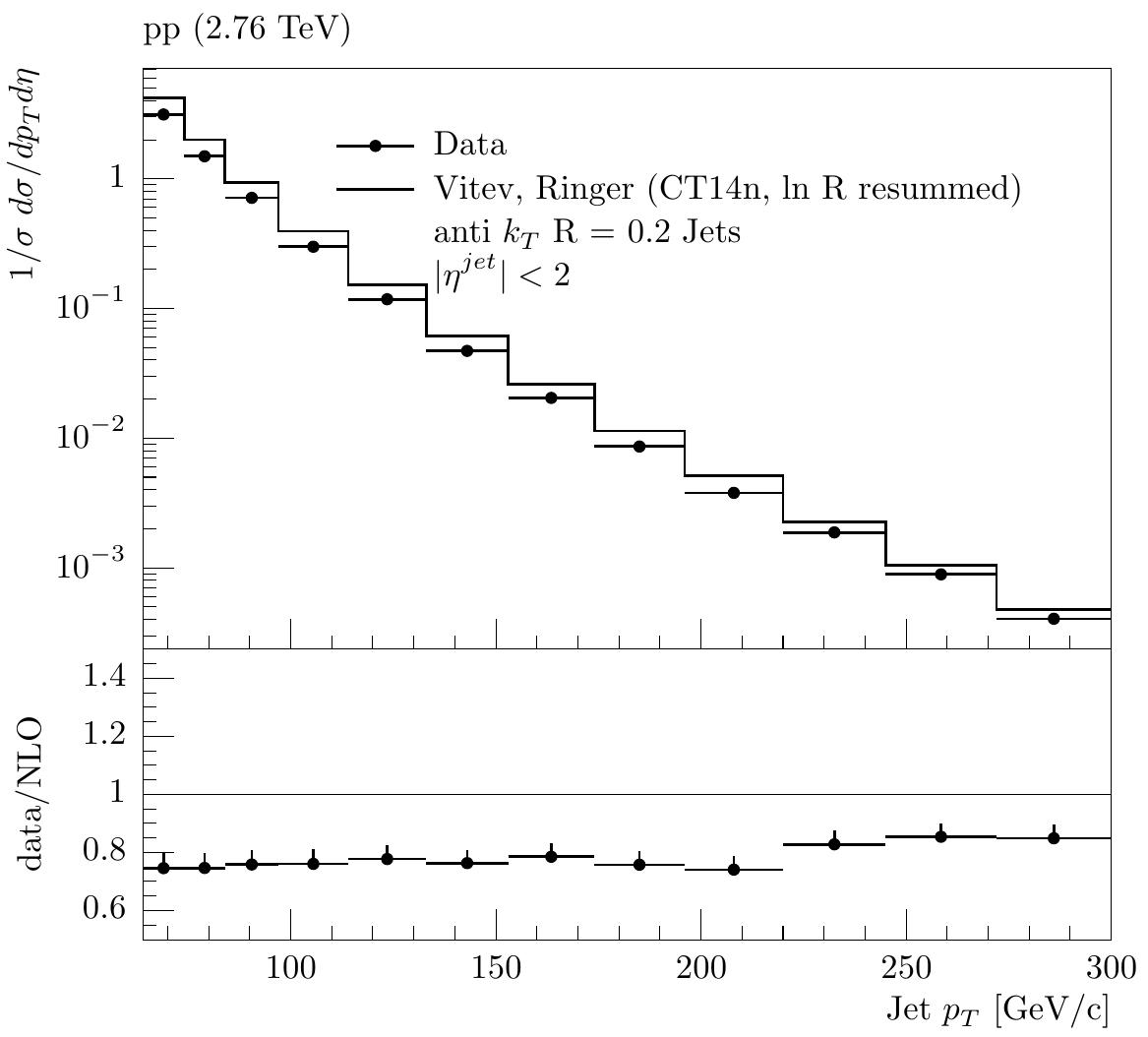} 
			   \includegraphics[width=0.3\textwidth]{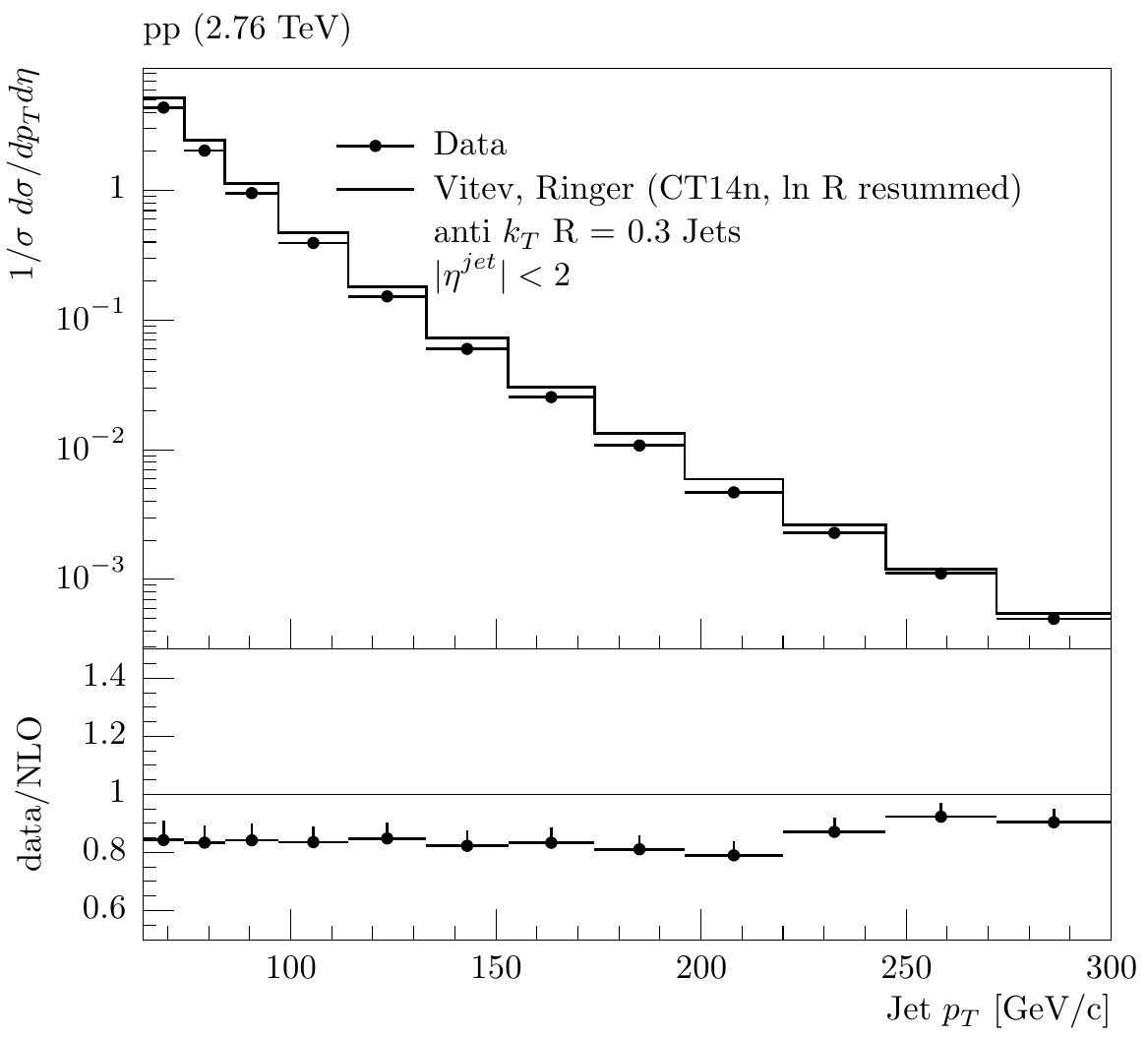} 
			   \includegraphics[width=0.3\textwidth]{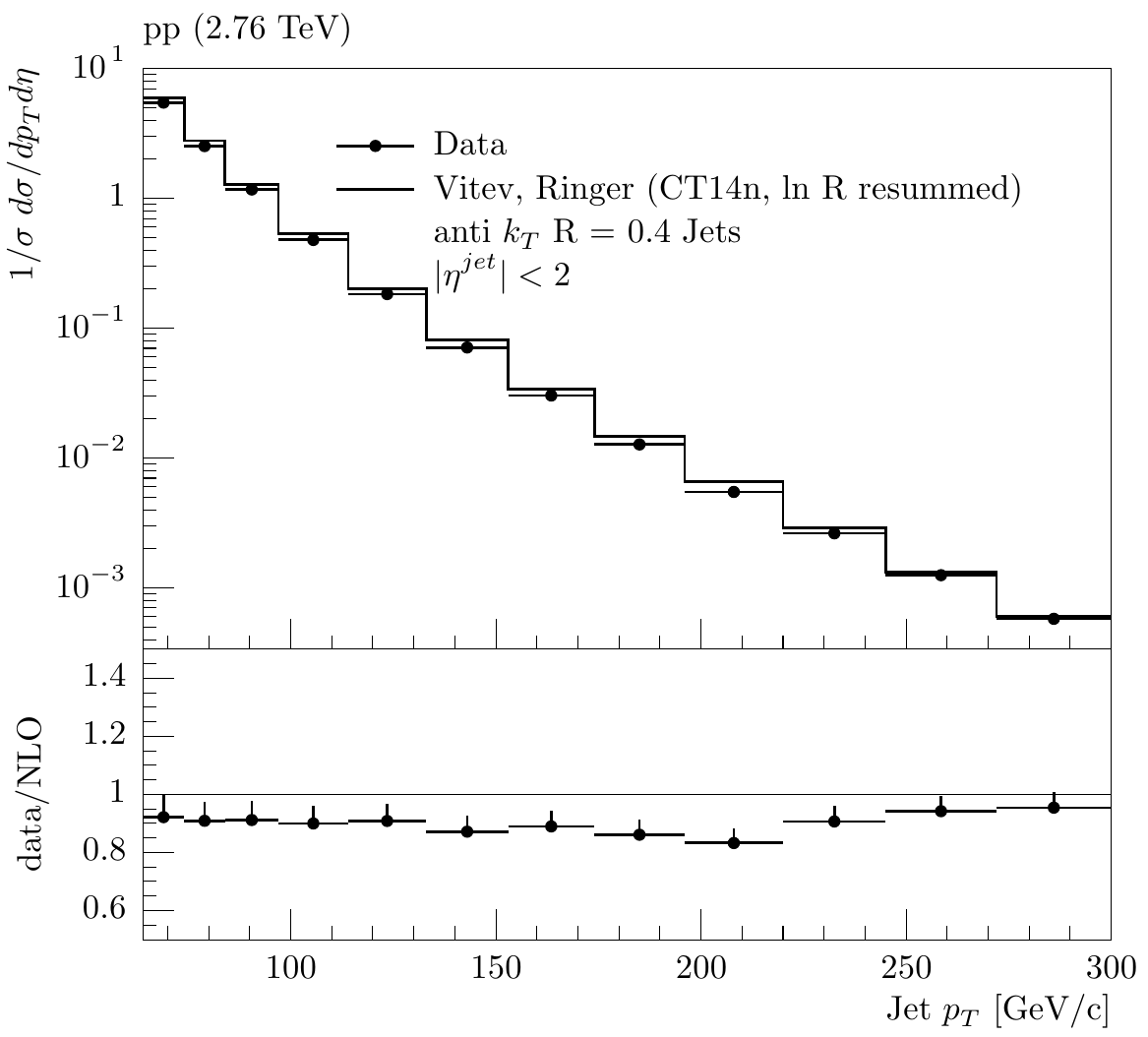} 
			   \caption{Comparison of data for inclusive jet cross section with SCET framework, w/o NP corrections for three jet radii. Bottom panels show the ratio data/NLO for mid rapidity jets. Figure courtesy Felix Ringer.}
			   \label{fig:ppjettheoryscet}
			\end{figure}

			Within the soft collinear effective theory (SCET) framework, there has also been recent work to estimate the radius dependence at fixed order~\cite{Kang:2017frl}. Similar to the resummation we saw above, a perturbatively calculable function is introduced that describes the formation of an observed jet coming from the fragmenting parent parton as a function of the radii.  In a recent paper, this function was outlined for a quark jet as the sum of the leading order born diagram, modified splitting with the AP and additional corrections including virtual partons, including when the final state partons exists the jet. In this framework, the inclusive differential jet cross section is compared with our CMS data as shown in Fig:~\ref{fig:ppjettheoryscet}. These theory curves do not have the NP corrections added but still are able to perform much better than the NLO calculation. 
			

\section{Asymmetric pPb collisions at the LHC}

	Proton lead collisions at the LHC offer an intermediary stage between pp and PbPb. One does not expect the formation of the QGP in pPb collisions and thus we can test any additional nuclear effects. We continue with the same goal of studying inclusive jet production in the natively asymmetric pPb collisions and comparing them with pp cross sections at the same center of mass energy. Since there was no pp data at 5.02 TeV in Run1, we compare it to an extrapolated reference spectra as will be discussed in detail in the following sections.  
		
	\subsection{Jet Yields and asymmetries in production}
	
		The observation of inclusive charged hadron yield modification in pPb collisions~\cite{Khachatryan:2015xaa} resulted in the \rpa~rises above unity in the high \pt~region, which was explained by anti-shadowing. It is worthwhile to repeat the same analysis with an independent observable to see if the same effect can be observed in the jet production yield in pPb collisions since jets can be better constrain our partonic kinematics in the hard scattering process. We study the inclusive jet production in different pseudorapidity ranges to measure possible modification of the nPDF.  
	
\begin{figure}[h] 
		   \centering
		   \includegraphics[width=0.6\textwidth]{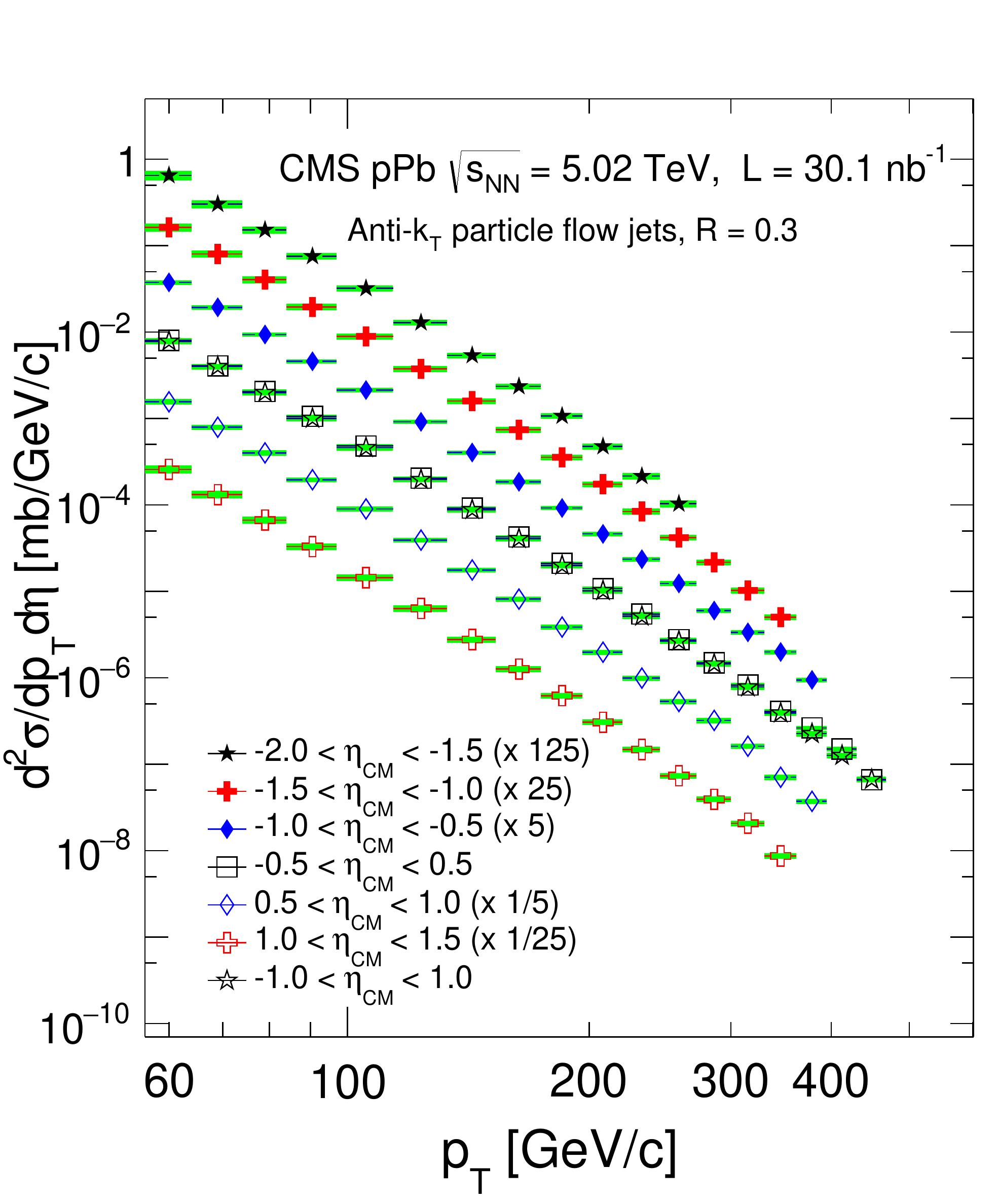} 
		   \caption{The unfolded inclusive jet yields in $\sqrt{s_{NN}}=5.02$~TeV pPb collisions in several \teta\ bins. Positive pseudorapidity values correspond to the proton going side~\cite{Khachatryan:2016xdg}.}
		   \label{fig:pPbSpectra}
		\end{figure}

		The unfolded inclusive jet spectra in pPb collisions at $\sqrt{s_{NN}}=5.02$~TeV are shown in Fig.~\ref{fig:pPbSpectra} for the following \teta\ intervals: $-1.0<\eta_{\rm CM}<1.0$, $-2.0<\eta_{\rm CM}<-1.5$, $-1.5<\eta_{\rm CM}<-1.0$, $-1.0<\eta_{\rm CM}<-0.5$, $-0.5<\eta_{\rm CM}<0.5$, $0.5<\eta_{\rm CM}<1.0$ and $1.0<\eta_{\rm CM}<1.5$. The jet spectra from different \teta\ ranges are scaled by arbitrary factors described in the legend to enhance visibility. 
		
		The pPb jet spectra are obtained in several pseudorapidity intervals. The \teta\ dependent jet cross section spectrum is studied by measuring the jet yield asymmetry, $Y_{\rm asym}$, in various $\eta_{\rm CM}$ ranges, where $Y_{\rm asym}$ is defined as follows: 
		
		\begin{equation}
			Y_{\rm asym}(p_{T}) = \frac{d^{2}N_{jet}/dp_{T}d\eta |_{\eta_{\rm CM}\in[c,d]}} {d^{2}N_{jet}/dp_{T}d\eta |_{\eta_{\rm CM}\in[a,b]}}
		\end{equation}
		
		where $a$, $b$, $c$ and $d$ are the boundary of two different \teta\ intervals.

\begin{figure}[h] 
			   \centering
			   \includegraphics[width=0.6\textwidth]{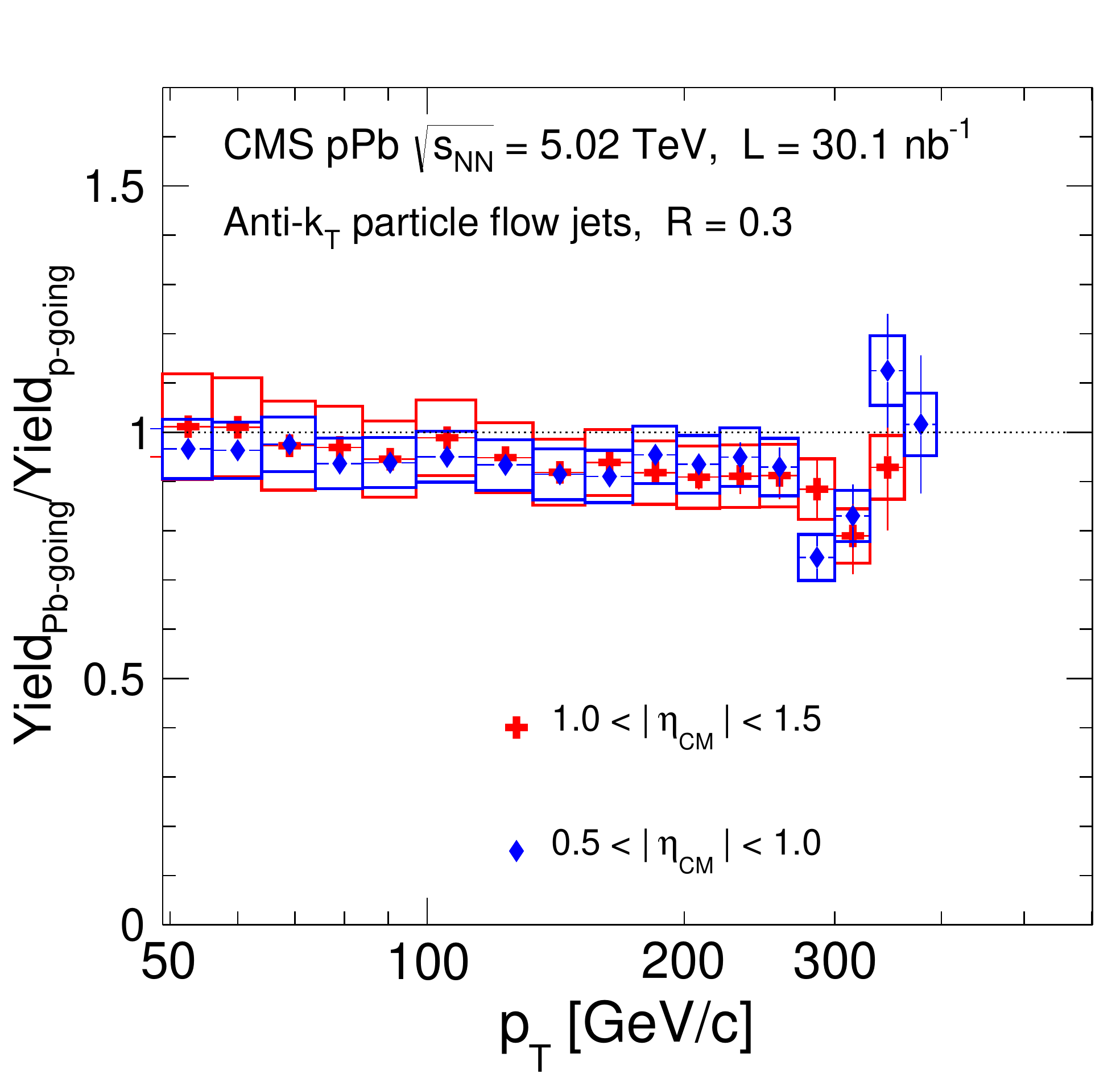} 
			   \caption{Inclusive jet asymmetry as a function of jet \pt~for $0.5<|\eta_{\rm CM}|<1.0$ and $1.0<|\eta_{\rm CM}|<1.5$. The asymmetry is computed as the jet yields on the Pb-going side divided by the those on the p-going side~\cite{Khachatryan:2016xdg}. }
			   \label{fig:Asymmetry}
			\end{figure}

		To study the \teta\ dependent jet production, the backward-forward asymmetry of the jet cross section ratio, $Y_{\rm asym}$, is calculated in Pb going direction with respect to proton going direction  in various $\eta_{\rm CM}$ ranges. Figure~\ref{fig:Asymmetry} shows $Y_{\rm asym}$ as a function of jet \pt~for $1.0<|\eta_{\rm CM}|<1.5$, and $0.5<|\eta_{\rm CM}|<1.0$ intervals.

		Figure~\ref{fig:Asymmetry} shows $Y_{\rm asym}$ as a function of jet \pt~for $1.2<|\eta_{\rm CM}|<2.2$, $0.7<|\eta_{\rm CM}|<1.2$, and $0.3<|\eta_{\rm CM}|<0.7$ regions.

\begin{figure}[h] 
			   \centering
			   \includegraphics[width=0.4\textwidth]{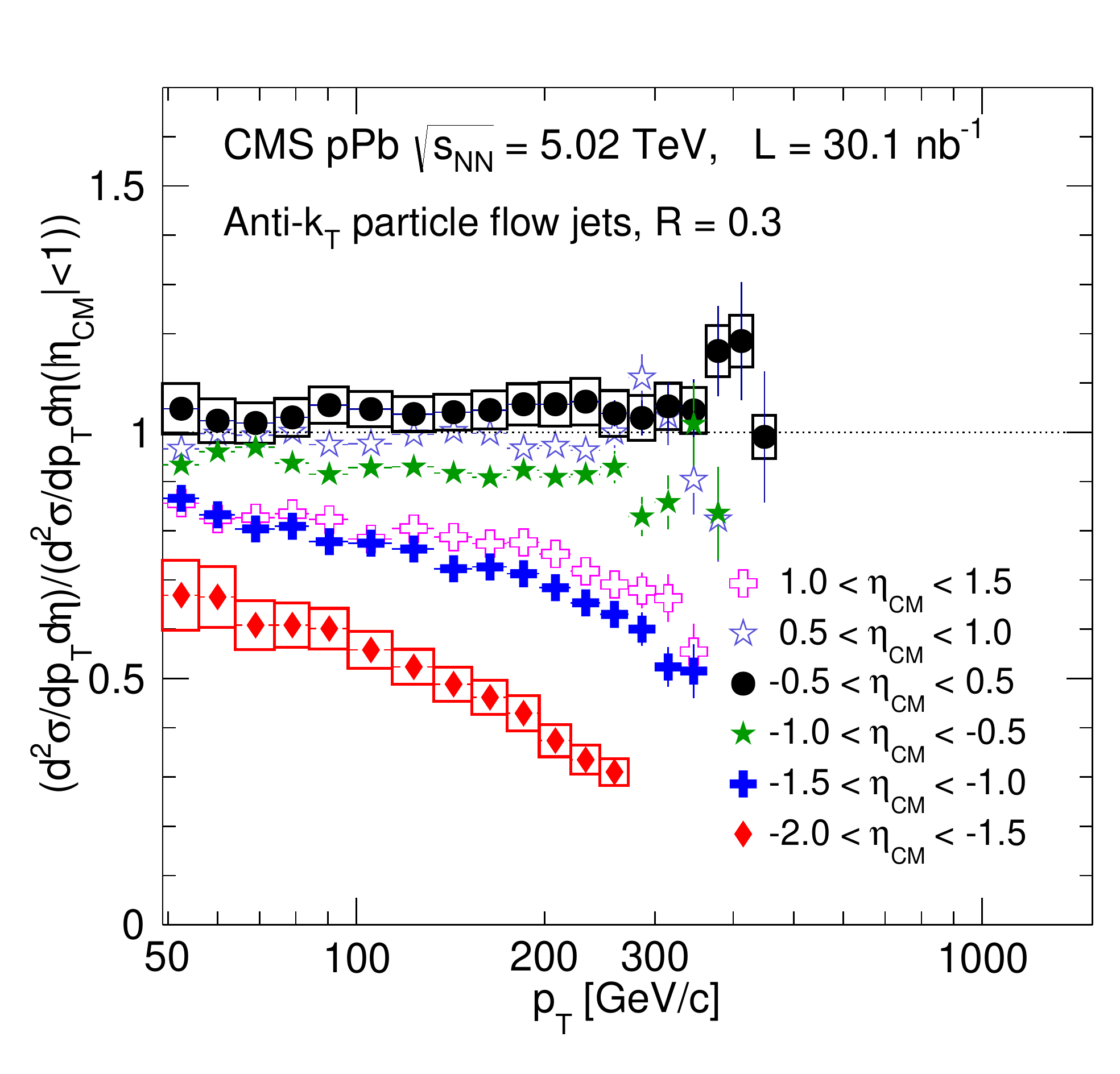} 
			   \includegraphics[width=0.4\textwidth]{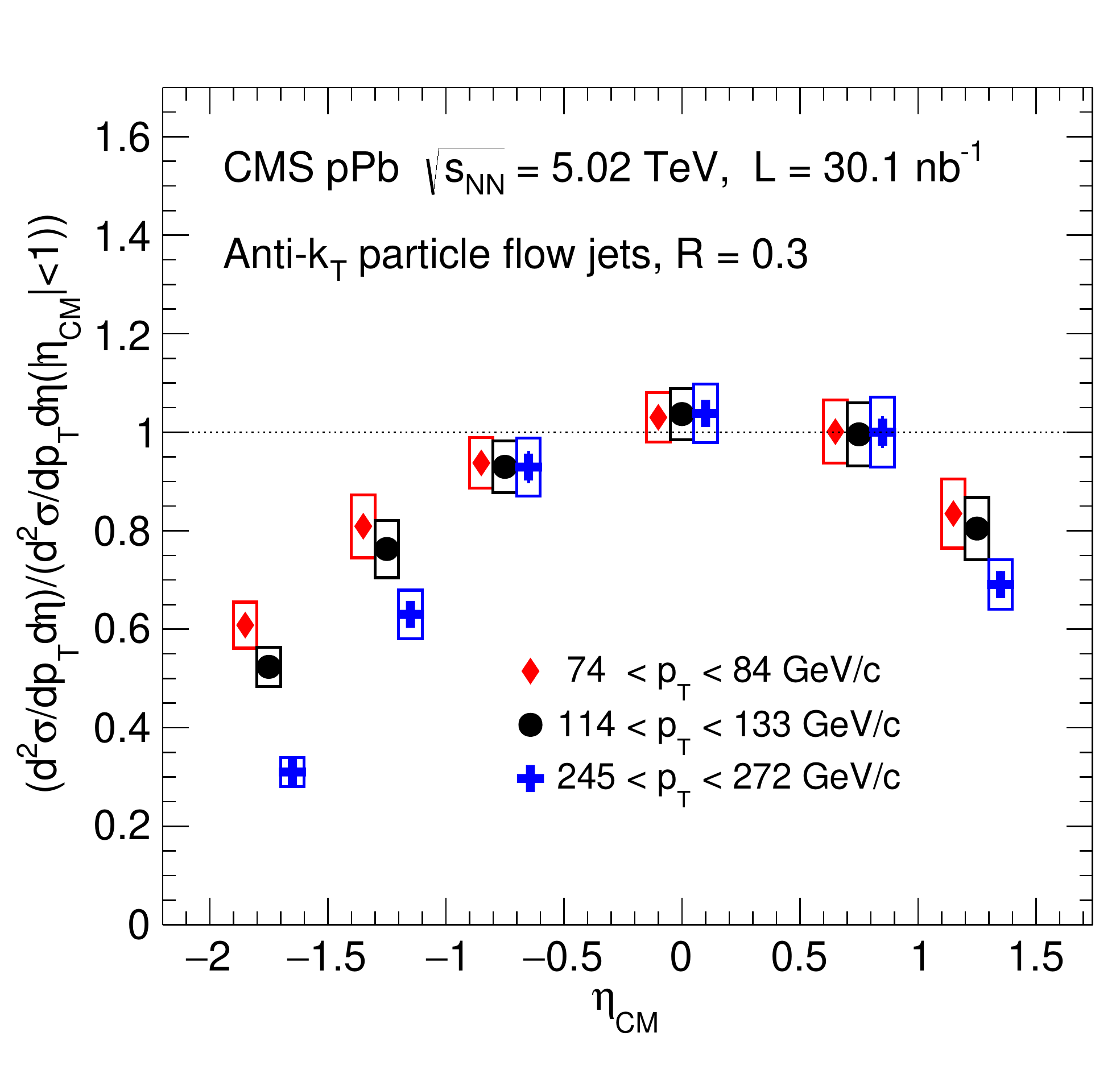} 
			   \caption{Inclusive jet cross section ratios of various $\eta$ intervals to $|\eta_{\rm CM}|<1$ (left) as a function of jet \pt~and for three selected jet \pt\ ranges as a function of the $\eta$ (right)~\cite{Khachatryan:2016xdg}.}
			   \label{fig:ForwardMid}
			\end{figure}

		Since the $Y_{\rm asym}$ strongly fluctuates due to the statistics limitation at large \teta\ range, one can then study the \teta\ dependence by dividing the jet yields in different \teta\ bins to the most central one ($|\eta_{\rm CM}|<1$), Fig.~\ref{fig:ForwardMid} shows the jet production ratio in different \teta\ bin to the one in $|\eta_{\rm CM}|<1$, it shows clearly \teta\ dependent jet production yield in pPb collisions.     

		Inclusive jet cross section ratios of various $\eta$ intervals to $|\eta_{\rm CM}|<1$ for three selected jet \pt\ ranges on the right panel of Fig:~\ref{fig:ForwardMid}. The error bars on the data points are statistical uncertainties and open boxes represent the systematic uncertainties in pPb collisions. The data points are shifted in \teta\ to enhance visibility of the uncertainty boxes.

	\subsection{Extrapolating a pp reference at 5 TeV}
	
		The pp reference at $5.02$ TeV is extrapolated from the published  jet cross section measurement in pp collisions at $\sqrt{s} = 7$~TeV~\cite{Chatrchyan:2014gia} by using the scaling factors calculated based on PYTHIA Z2 simulation. The challenge of such an extrapolation is only one collision energy data is available and also a difference cone size is used in pPb analysis compared to pp published data. 
	
		A schematic of the extrapolation method used is given below: 
		\begin{enumerate}
			\item Compare data	with	PYTHIA and POWHEG	for	$R=0.7$	and	$R=0.5$	
			\item Calculate	scaling factor	of	collision	energy	dependence 	($7 \rightarrow 5.02	TeV$)	by	using  PYTHIA and POWHEG	calculation for $R=0.7$	and	$R=0.5$ respectively
			\item Scaling the published data in pp collisions at $\sqrt{s} = 7$~TeV from $R=0.5$	or $R=0.7$ to	obtain the inclusive	jet	spectrum	for the same cone size	at $\sqrt{s} = 5.02$~TeV
			\item Make the cone-size dependent cross section ratio at $\sqrt{s} = 5.02$~TeV using PYTHIA and POWHEG generators	 and available data points
			\item Calculate the cone-size dependent cross section ratio from PYTHIA and POWHEG
			\item Scale the jet spectra obtained in (3) in pp collisions at $\sqrt{s} = 5.02$~TeV from $R=0.5$	or $R=0.7$ to $R=0.3$ using the factors from (5) to obtain the jet	spectra	at	$R=0.3$	 
			\item Study the collision energy dependence for a fixed cone size jet from PYTHIA 
			\item Estimate systematics during this extrapolation using NLO calculation as scaling factors
			\item Cross check with ${x_T}$ based interpolation using CMS published data~\cite{Khachatryan:2015luy}.
		\end{enumerate}
		
\begin{figure}[h] 
		   \centering
		   \includegraphics[width=0.6\textwidth]{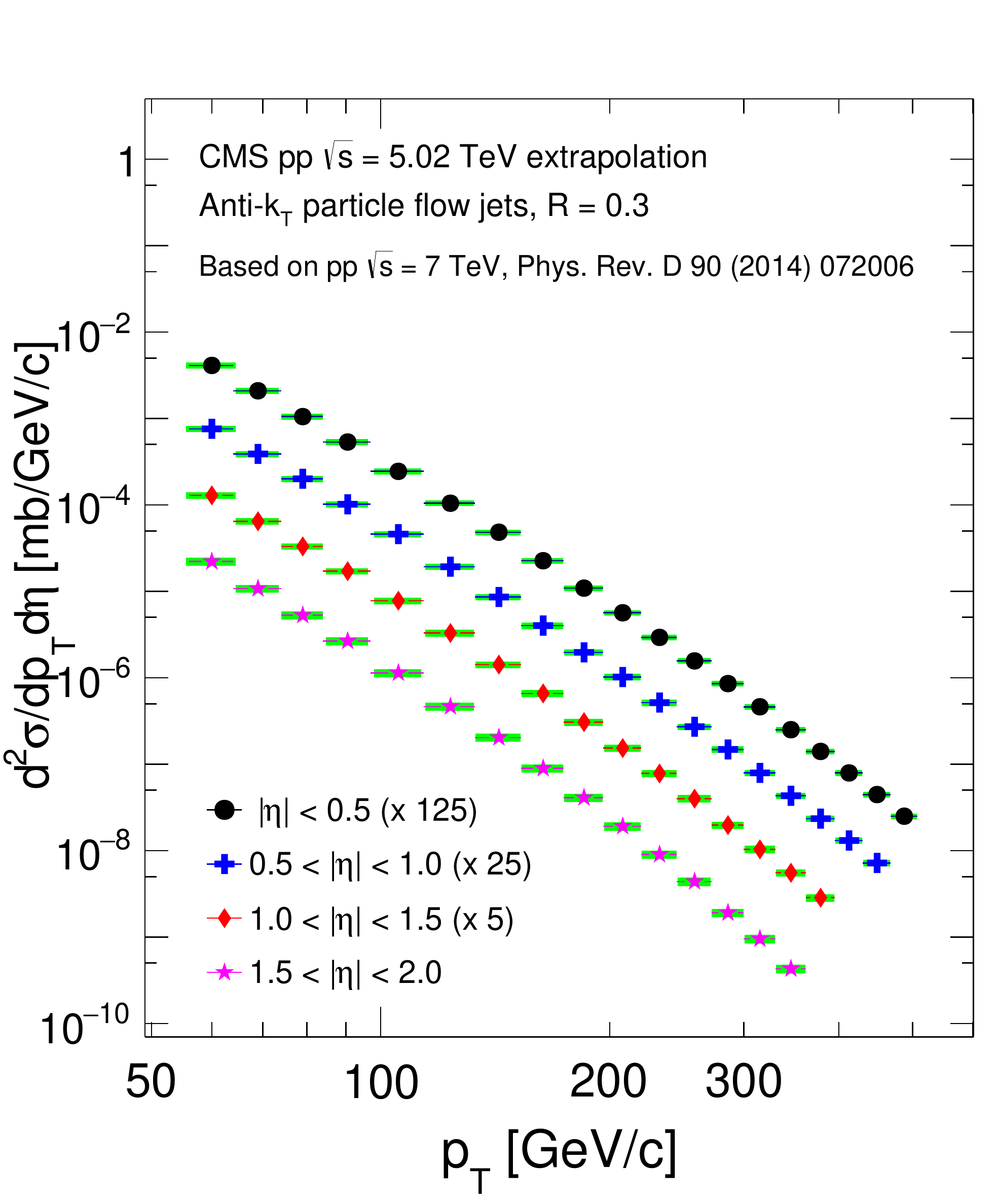} 
		   \caption{The extrapolated jet spectra in four absolute rapidity intervals with an artificial scaling factor applied to 
enhance the visibility. The horizontal error bars represent the bin size and the points are plotted in the center of the bin. The shaded boxes denote the systematic uncertainties in the extrapolation procedure~\cite{Khachatryan:2016xdg}.}
		   \label{fig:ppRef5TeV}
		\end{figure}

		Figure~\ref{fig:ppRef5TeV} shows the extrapolated jet spectra in four absolute rapidity intervals with an artificial scaling factor applied to enhance the visibility.
		
	\subsection{Nuclear modification factors - \rpa}	
	
\begin{figure}[h] 
		   \centering
		   \includegraphics[width=0.9\textwidth]{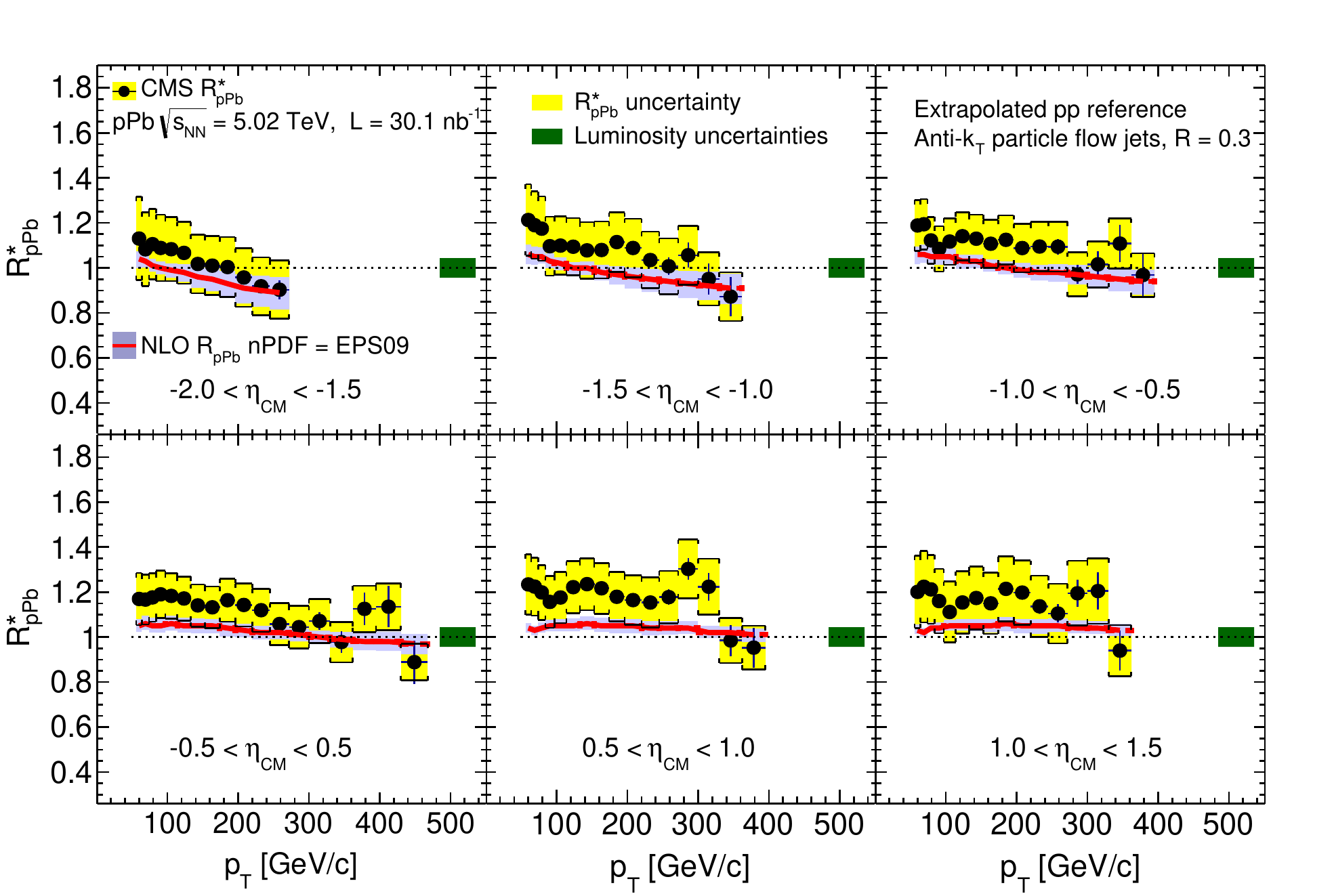} 
		   \caption{The inclusive jet nuclear modification factor \rpa~as a function of jet \pt\ in $\sqrt{s_{NN}}=5.02$~TeV pp collisions with the interpolated pp reference. The error bars on the data points are the statistical uncertainties and the open boxes represent the systematic uncertainties. The filled color boxes are the systematic uncertainties due to the pp reference interpolation~\cite{Khachatryan:2016xdg}. }
		   \label{fig:JetRpPb}
		\end{figure}

		These extrapolated pp spectra at $\sqrt{s} = 5.02$~TeV are then used to calculate nuclear modification factor \rpa~via:
		\begin{equation}
		\begin{split}
			R_{pA} &=\frac{1}{N_{coll}}\frac{d^{2}N_{jet}^{pA}/dp_{T}d\eta}{d^{2}N_{jet}^{pp}/dp_{T}d\eta}  \\
   			              &=\frac{1}{A}\frac{d^{2}\sigma_{jet}^{pA}/dp_{T}d\eta}{d^{2}\sigma_{jet}^{pp}/dp_{T}d\eta} \\
  			              &=\frac{1}{A}\frac{1}{L}\frac{d^{2}N_{jet}^{pA}/dp_{T}d\eta}{d^{2}\sigma_{jet}^{pp}/dp_{T}d\eta}
		\end{split}
		\end{equation}
		
		where $N_{coll}$ is the number of binary collisions,  the $L = 35$~nb$^{-1}$ is the recorded luminosity in pPb collisions as determined in Ref.~\cite{CMS:2013rta} and $A$ is the mass number of the lead nucleus. The inclusive jet nuclear modification factor \rpa~as a function of jet \pt\ for six pseudorapidity bins are shown in  Fig.~\ref{fig:JetRpPb} using PYTHIA based extrapolated pp reference. No dependence on the jet \pt~is observed in most of the rapidity bins, except the backward (Pb going direction) ones where a weak decrease in the \rpa~is observed.   In the mid-pseudorapidity bins, the nuclear modification is quite similar. While at large rapidity, a clear separation of the \rpa~between forward and backward range, with proton beam direction has slightly larger nuclear modification factors compared to Pb beam direction.  

		\begin{figure}[h] Fi
		   \centering
		   \includegraphics[width=0.5\textwidth]{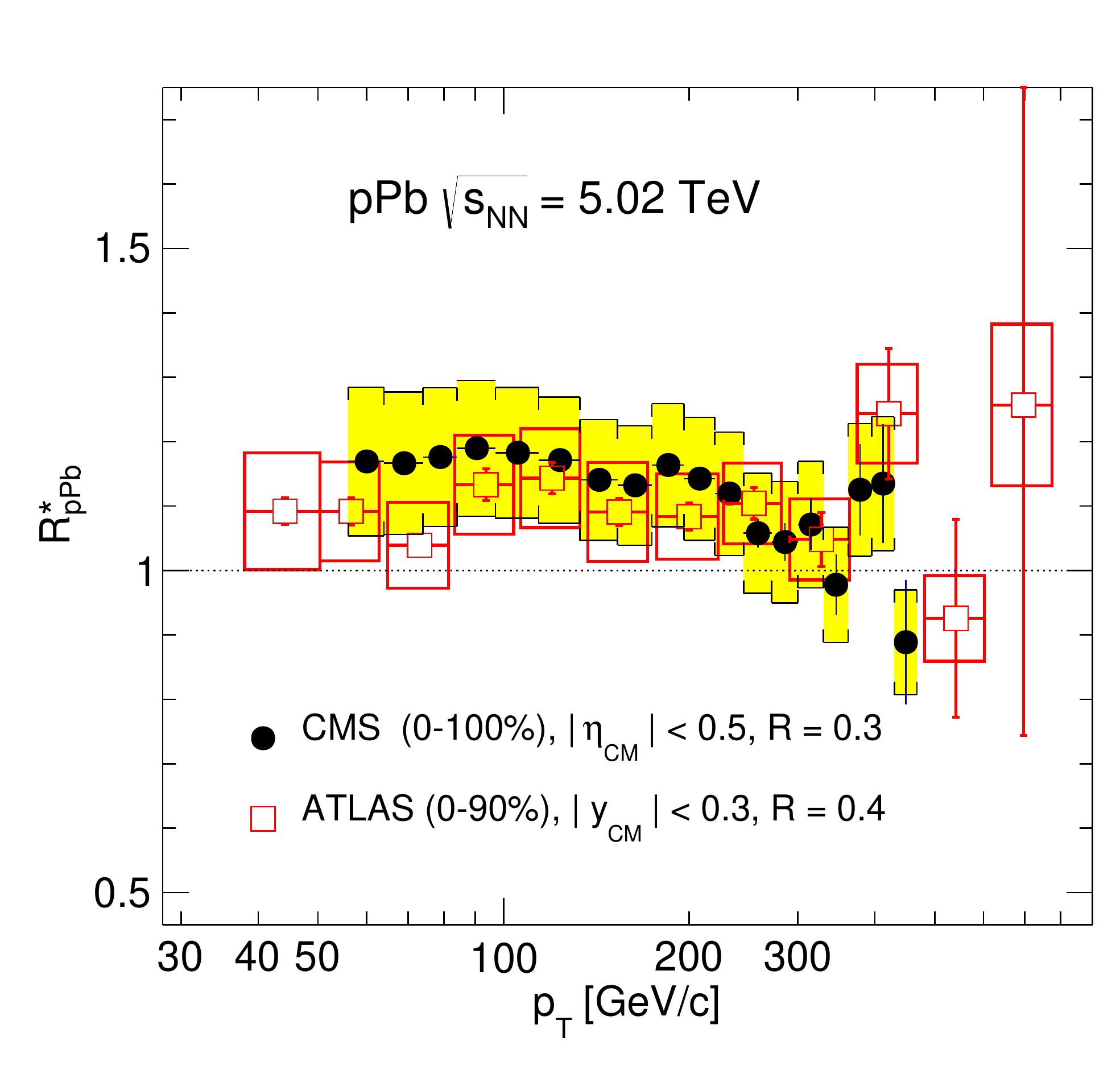} 
		   \caption{Comparing high \pt~jet \rpa~at CMS with ALICE for mid rapidity jets. Note: CMS and ATLAS have different jet radii and the references are extrapolated versus interpolated respectively~\cite{Khachatryan:2016xdg}.}
		   \label{fig:comparingrpaATLASCMS}
		\end{figure}

		We also compared the CMS measurement with the ATLAS result except it was for a larger jet radii ($R = 0.4$ vs $R = 0.3$) in Fig:~\ref{fig:comparingrpaATLASCMS} with a different method of extracting a pp reference. In both cases the results are comparable and show the \rpa~$\approx 1.15$ across the measured jet \pt~range. The larger systematic uncertainties are mostly due to the pp reference extrapolation. 
		
		Inclusive jets at high \pt~are predominantly composed of gluons and light quarks, due to the low x region we end up probing. A natural experimentalist question that follows is to ask if there is any dependence on the mass of the quark that initiates the jet. To answer this question, we measured the \rpa~for jets that were selected as b-jets.

\section{Probing initial parton flavor dependence}

	 The nuclear modification of jets in heavy ion collisions should depend on the flavor of the fragmenting parton~\cite{CasalderreySolana:2007zz}. For example, under the assumption that radiative energy loss is the dominant mechanism, gluon jets are expected to quench more strongly than quark jets, owing to the larger color factor for gluon emission from gluons than from quarks~\cite{djordjevicGluon}.  There are also theoretical predictions that radiative energy loss may not be dominant for heavy quarks, including models based on collisional energy loss of quarks within the medium and models favoring an interpretation based on mesonic recombination and disassociation within the medium, e.g. Refs.~\cite{vanHees:2005aa,teaney:2005aa}.  It is expected that there should be some mass-dependence of partonic energy loss at low momentum, and therefore, b quark jet (b jet) energy loss might be different from that of light quark jets~\cite{buzzatti:2011,djordevic:2015}. 
	
	We will briefly go over the measurement of b-jet \rpa~and please refer to my friend and colleague Kurt Jung's PhD thesis~\cite{KurtJungPhDThesis} for more details regarding b-jets in CMS. 

	\subsection{Heavy flavor jet \rpa}

\begin{figure}[h] 
		   \centering
		   \includegraphics[width=0.6\textwidth]{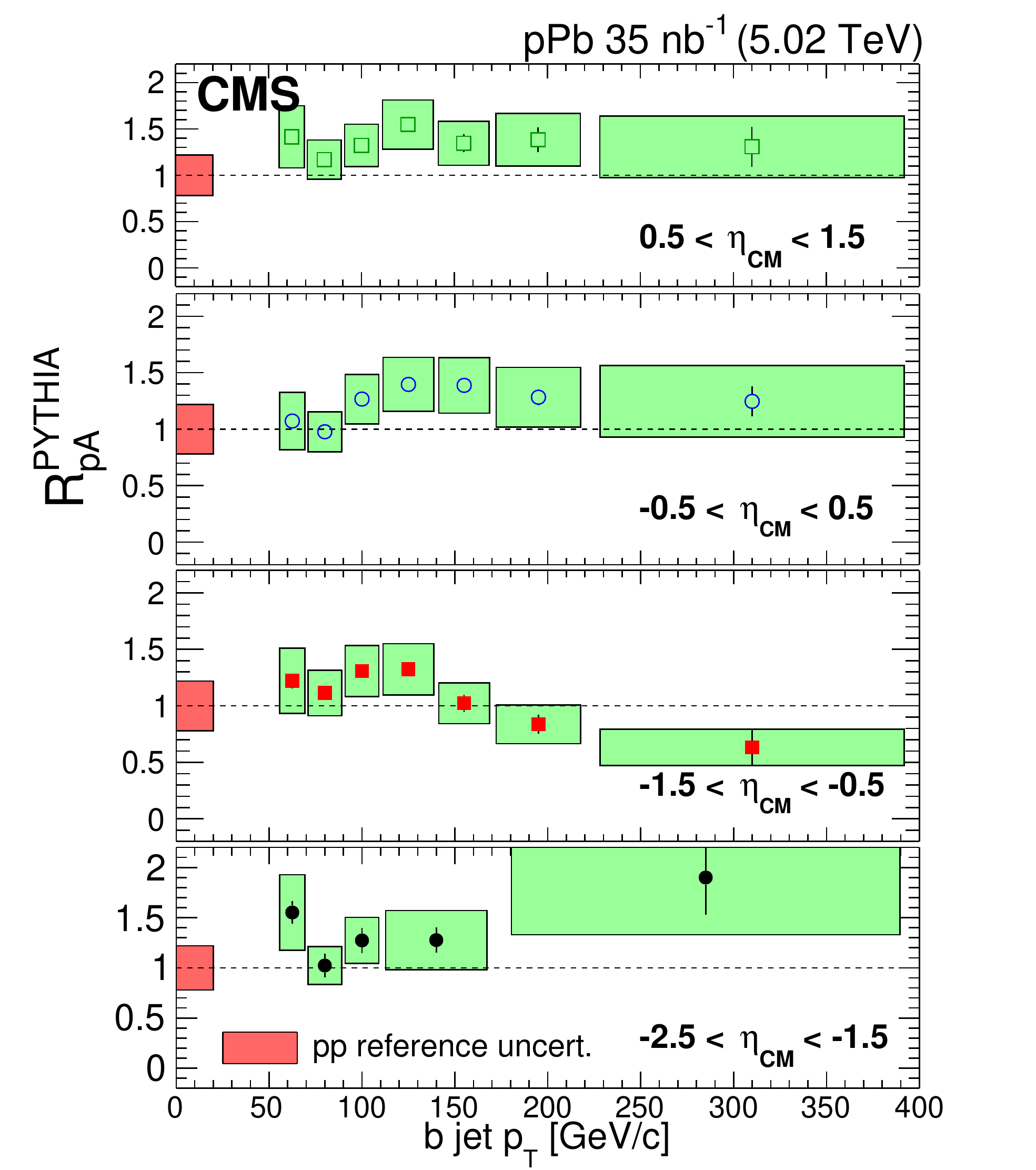} 
		   \caption{\RpAPythia measurements for the four \teta\ ranges are shown. Positive $\eta$ corresponds to the direction of the proton beam. Statistical uncertainties are represented using vertical bars, while systematic uncertainties are shown as colored bands on the left and filled boxes on the right. The pp reference uncertainties are shown separately as red boxes around unity on the right panel~\cite{Chatrchyan:2014hqa}.}
		   \label{fig:bjetrpa}
		\end{figure}

		When the b-jet \rpa~is measured with respect to a PYTHIA based reference, we see consistency between the pPb data and the PYTHIA pp reference, indicating a lack of $\eta$-dependent effects for each \teta selection. Fig:~\ref{fig:bjetrpa} shows the \RpAPythia~measurements for the same four \teta selections. The average values are consistent with unity within uncertainties.
	
\begin{figure}[h] 
		   \centering
		   \includegraphics[width=0.6\textwidth]{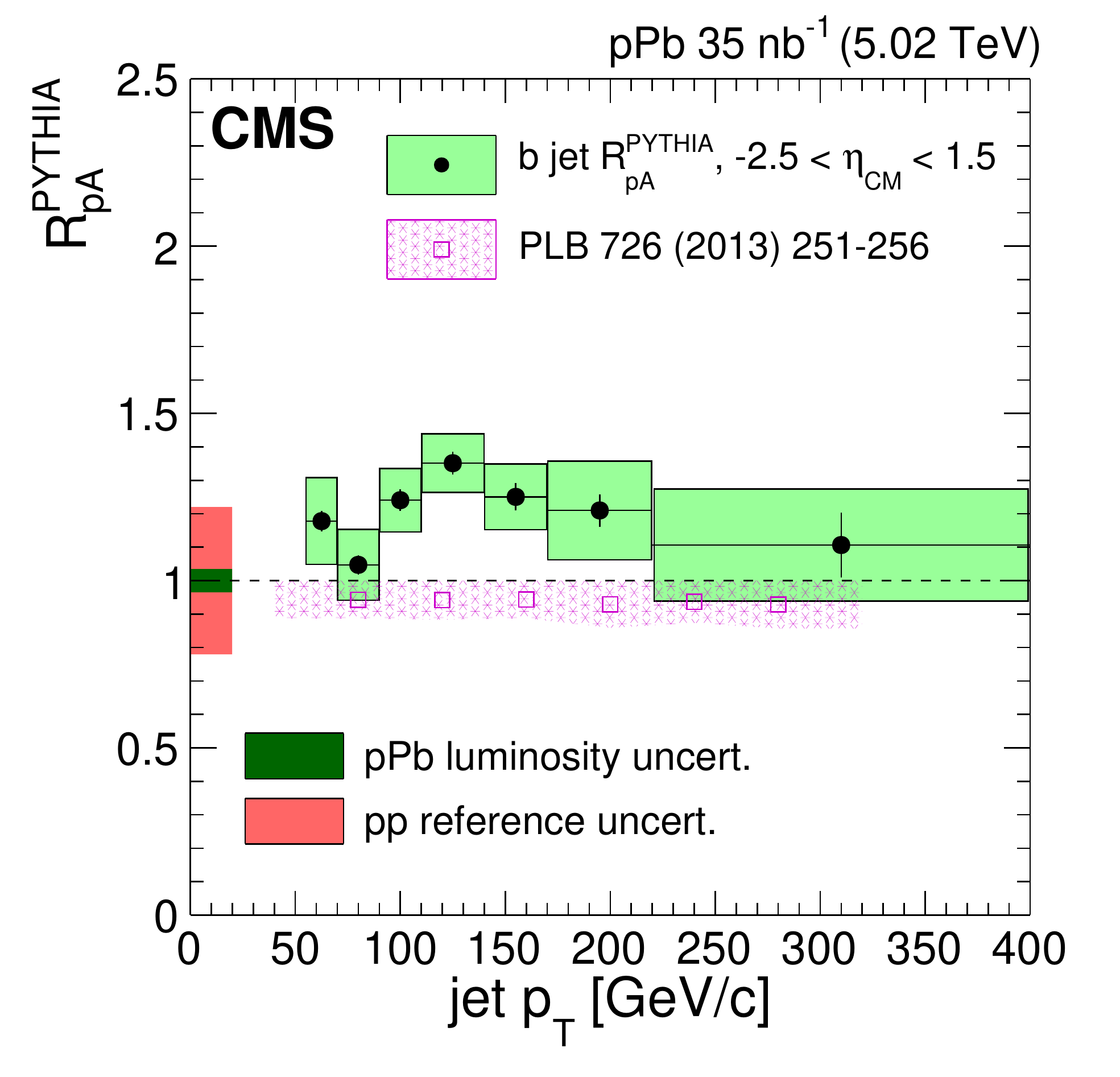} 
		   \caption{The b jet \RpAPythia~as a function of jet \pt~is shown as points with filled boxes for systematic uncertainties~\cite{Chatrchyan:2014hqa}.  The pp reference and integrated luminosity uncertainties are shown as red and green bands around unity, respectively. A pQCD prediction from Huang et. al.~\cite{vitev:2013aa} is also shown.}
		   \label{fig:rpavspt}
		\end{figure}

		Fig:~\ref{fig:rpavspt} shows the pseudorapidity-integrated \RpAPythia. Fitting a constant to this distribution returns a value of \RpAPythia~$= 1.22 \pm 0.15\,(\text{stat}+\text{syst pPb})\pm0.27\,(\text{syst PYTHIA})$, which indicates that the b jet yield in pPb is consistent with the pp PYTHIA simulation, especially considering the 22\% uncertainty on just the PYTHIA reference.  In addition, Fig.~\ref{fig:rpavspt} shows the comparison of the measured \RpAPythia~to predictions from a pQCD model that includes modest initial-state energy-loss effects \cite{vitev:2013aa}. The model and data are roughly consistent within the total systematic uncertainties from both PYTHIA and the pPb data.

	In this chapter we saw a selection of measurements in pp and pPb collisions geared towards studying how well we can comprehend the baseline from a theoretical standpoint. When we take into account the latest theoretical developments at the NLO and LL$_{R}$, we find a very good agreement for the inclusive jet cross sections. But that is not the end of the story since  the important parts of a jet measurement, such as fragmentation and hadronization, are still model dependent and we still do not understand them intuitively. There is a renewed effort to go back to basic jet structure measurements, but this time, looking at gluon jets\footnote{What is a gluon jet? Is it based on the originating parton or the hardest fragmented object or the initial hard scattering. This question is starting to take prominence in the literature again and how to distinguish gluon from quark jets without a known/unknown bias.} at the LHC in order to tune our models and calculations.

\clearpage

\chapter{Quantifying Jet Quenching In Pb-Pb Collisions}
\label{ch_quantJetQuenching}
\begin{chapquote}{Sir Issac Newton}
``If I have seen further it is by standing on the shoulders of Giants."
\end{chapquote}

In the last chapter, we discussed the physics of the nuclear modification factor in pPb offering evidence of no cold nuclear effects on jet production and jet energy. In search of the hot nuclear effects such as jet quenching, we report the measurement of the jet yield in PbPb collisions for three jet radii and a variety of centrality classes. 

\section{Invariant jet yield}

\begin{sidewaysfigure}[p]
	   \centering
	   \includegraphics[width=0.33\textwidth]{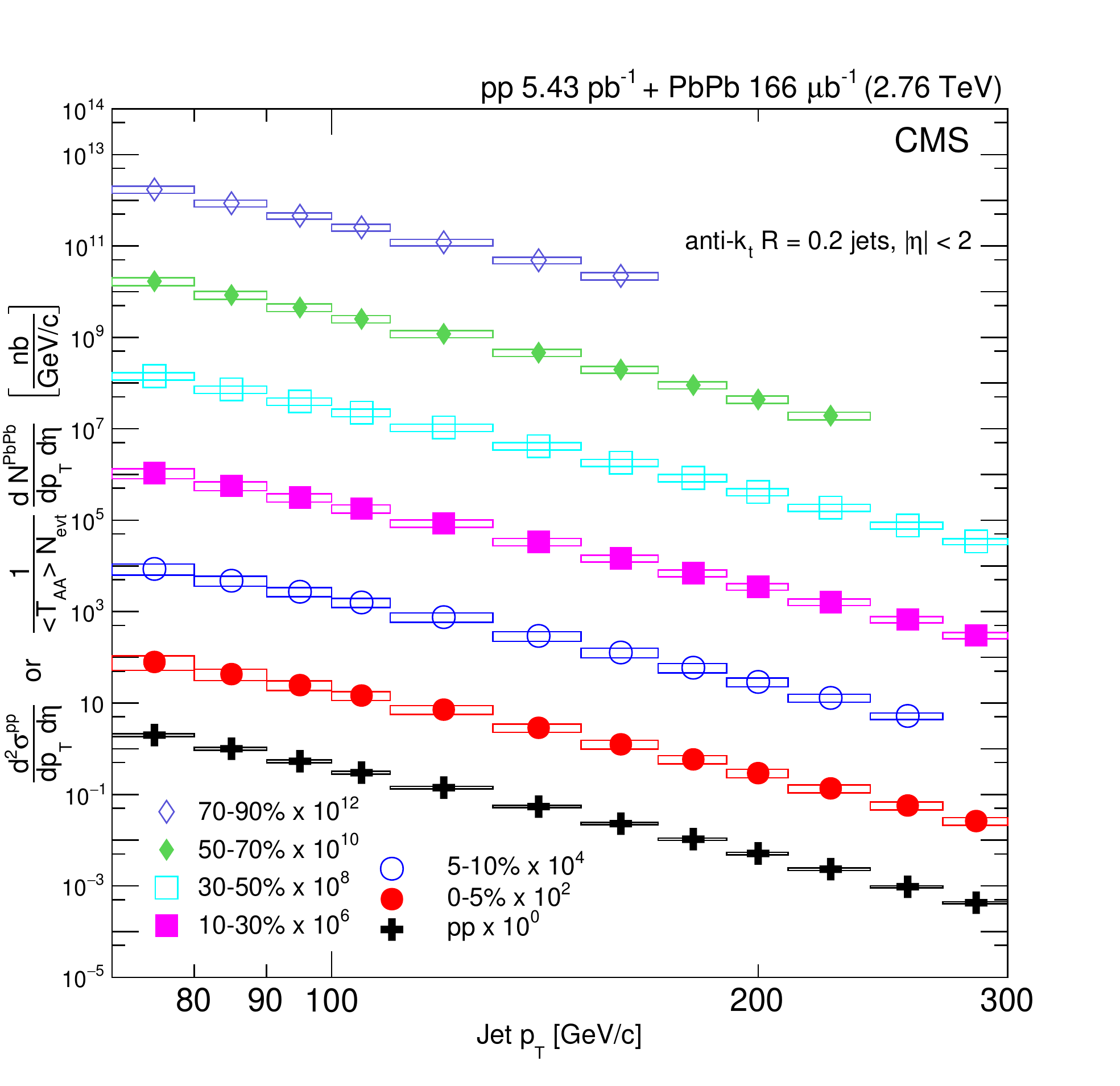} 
	   \includegraphics[width=0.33\textwidth]{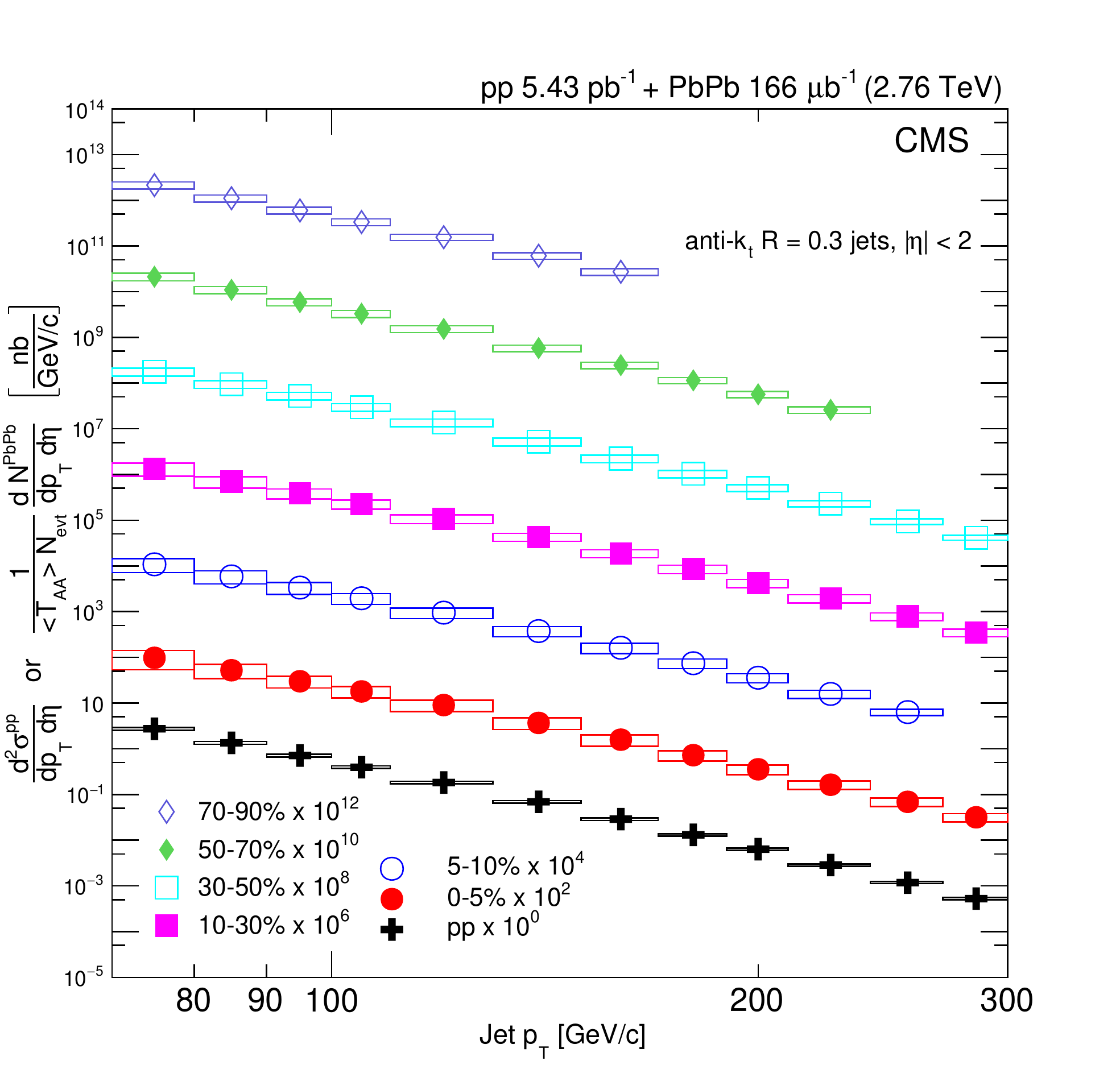} 
	   \includegraphics[width=0.33\textwidth]{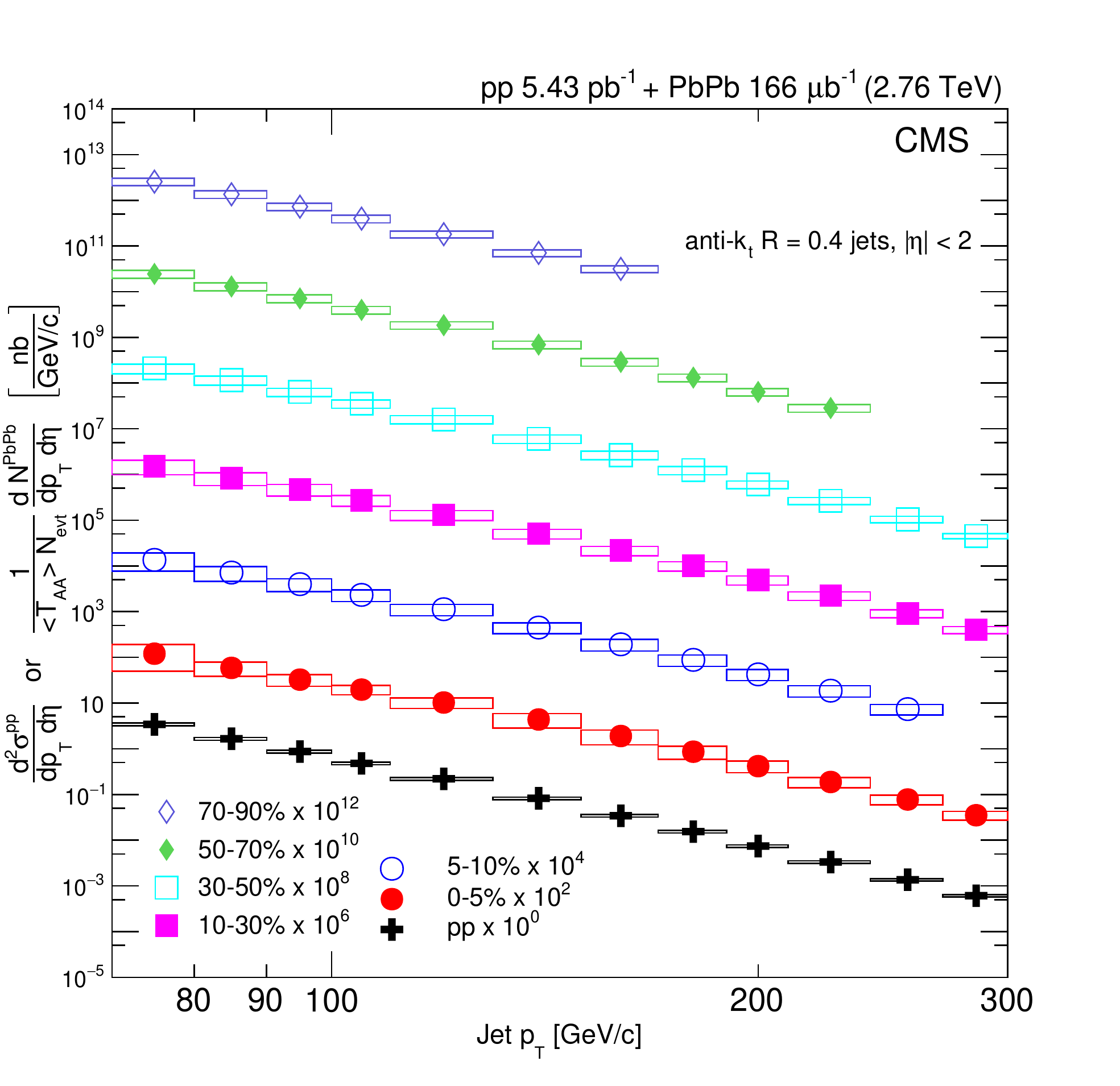} 
	   \caption{Inclusive jet spectra for PbPb jets of distance parameter $R  =  0.2, 0.3$ and $0.4$, in different centrality bins, and pp reference data~\cite{Khachatryan:2016jfl}. The PbPb jet spectra for different centrality classes are scaled by $\left<T_{AA}\right>$ and multiplied by a different factor for better visualization. Vertical bars represent statistical uncertainty (too small to see on this scale) with the systematical uncertainty in the colored boxes around the data points.}
	   \label{fig:jetYieldPbPb}
	\end{sidewaysfigure}

	The unfolded jet cross sections for PbPb and pp events are shown in Fig.~\ref{fig:jetYieldPbPb} for different distance parameters. The PbPb spectra are normalized by the number of minimum bias events, and are scaled by $\left<T_{AA}\right>$, with each centrality multiplied by a different factor, to separate the spectra for better visualization. The pp reference data is normalized to the integrated luminosity of the analyzed data set. The high \pt~cutoffs for the spectra (hence also the \raa) are dictated by statistical limitations.

\begin{figure}[h!] 
	   \centering
	   \includegraphics[width=0.7\textwidth]{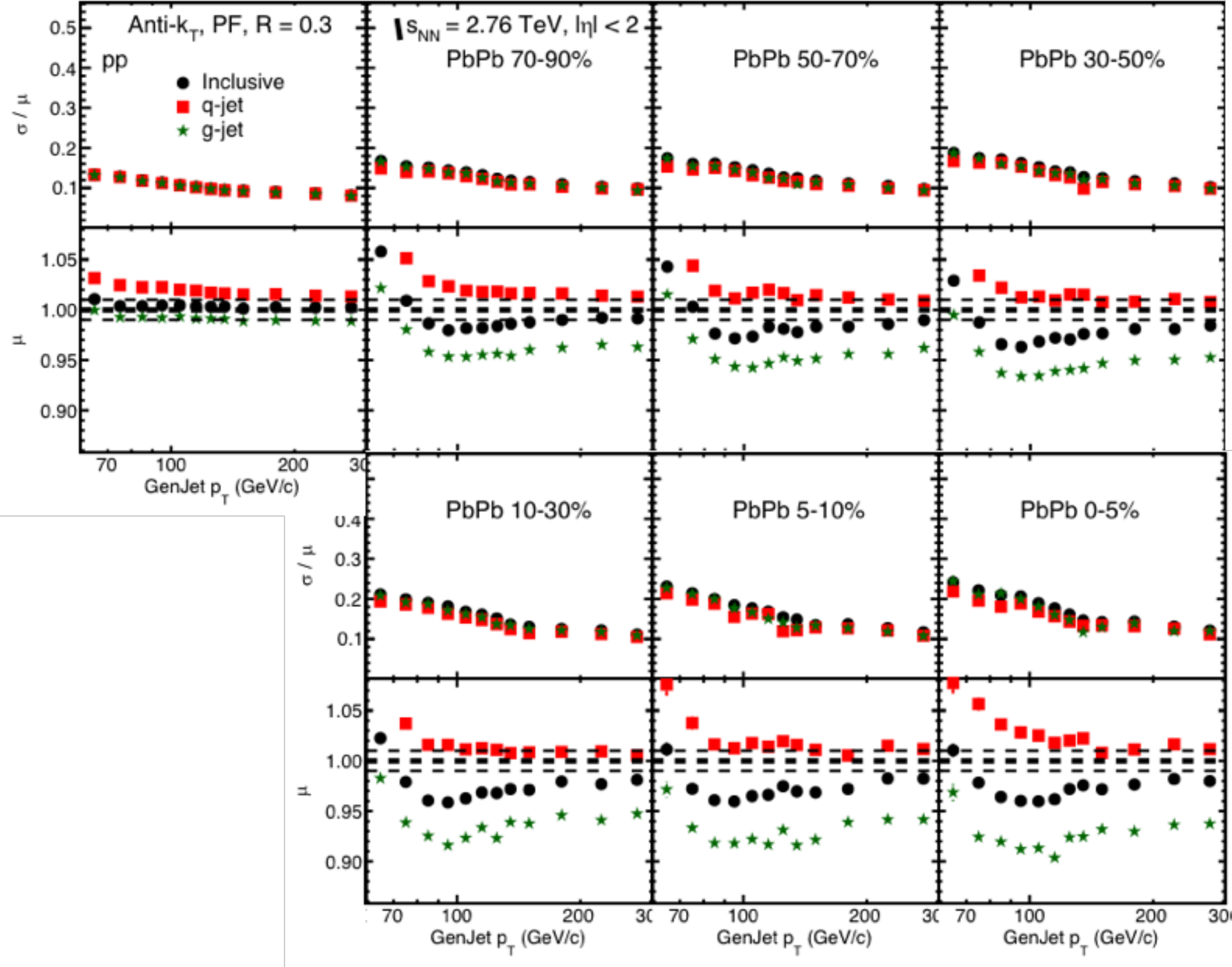} 
	   \caption{Jet Energy resolution (top) and response (bottom) for PbPb PU subtracted inclusive and split into quark/gluon jets with \pyhd. Left most panels show pp, followed by peripheral collision to central collisions.}
	   \label{fig:jecqg}
	\end{figure}

	The JES uncertainty ranges from 6--32\% (from peripheral to central events), varying due to the uncertainty in the heavy ion tracking and the quark/gluon fragmentation. The fragmentation difference is extended for PbPb jets due to expected asymmetric jet quenching effects for quark and gluon jets. They are estimated in a MC sample by separating jet collections as those originating from quarks or gluons and separately estimating the JES and JER as shown in Fig:~\ref{fig:jecqg}. The pp is shown on the top left panel and the other panels correspond to different centrality classes. The black markers are for inclusive jets and the green and red correspond to gluon and quark jet. Half of the spread between the quarks and gluons is chosen as the additional JES systematic uncertainty on the PbPb jet yield.  
	
\section{Inclusive jet nuclear modification factors}

\begin{figure}[h!]
	   \centering
	   \includegraphics[width=0.9\textwidth]{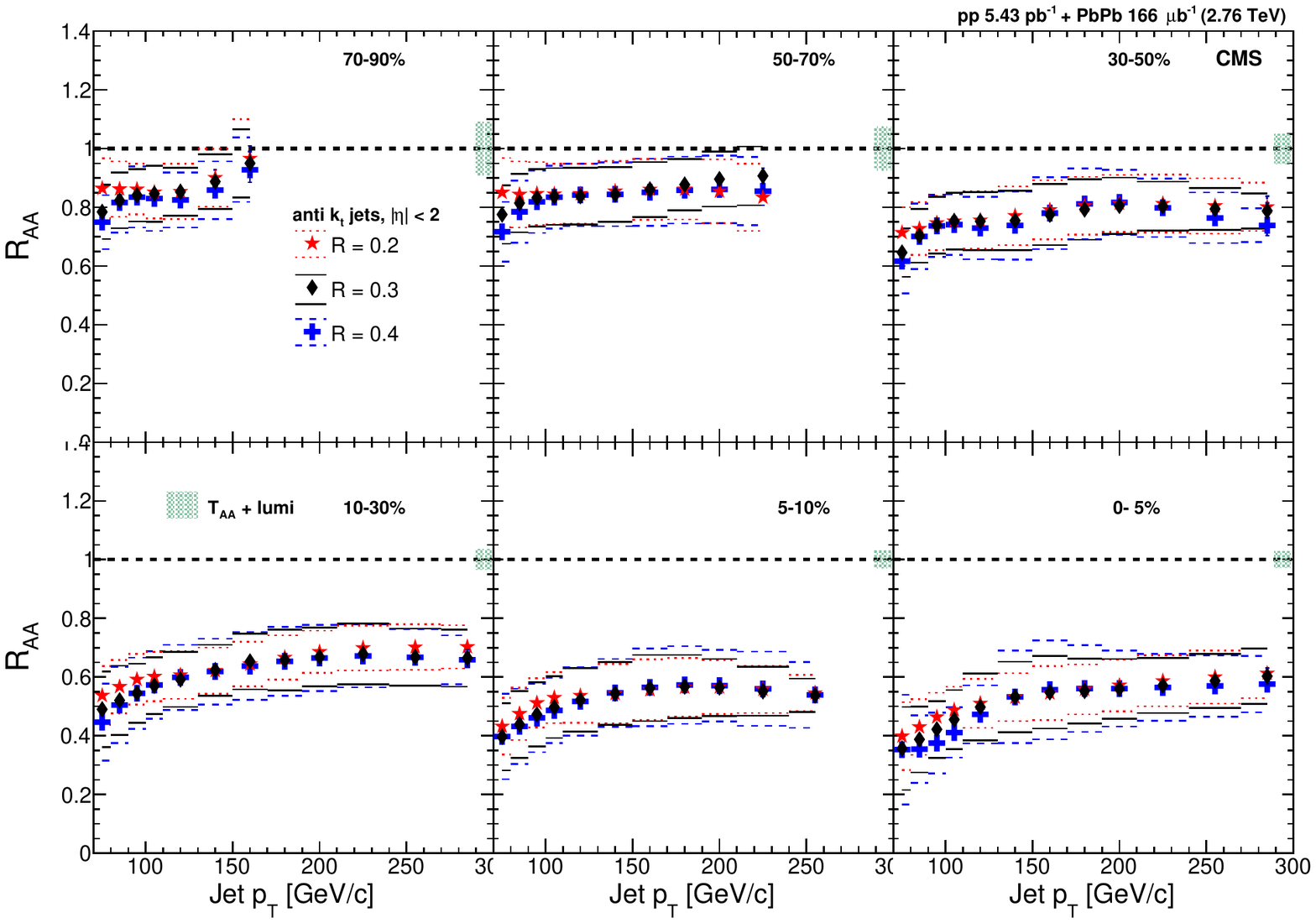} 
	   \caption{Inclusive jet \raa~as a function of the jet \pt, for \akt jets with distance parameters $R  =  0.2$ (red stars), 0.3 (black diamonds), and 0.4 (blue crosses) for different centrality bins~\cite{Khachatryan:2016jfl}. The vertical bars (smaller than the markers) indicate the statistical uncertainty and the systematic uncertainty represented by the bounds of the dotted, solid, and dashed horizontal lines. The uncertainty boxes at unity represent the \taa~and luminosity uncertainty.}
	   \label{fig:jetraa}
	\end{figure}

	The jet \raa, calculated from the PbPb and pp spectra after all corrections including SVD unfolding, are shown for different distance parameters in Fig.~\ref{fig:jetraa}. The jet \raa~decreases with increasing collision centrality in the range of the measured jet \pt. Within the systematic uncertainty, the jet \raa~shows the same level of suppression for the three distance parameters. Systematic uncertainties, from different contributions to the jet \raa~from the individual spectra in pp and PbPb collisions, are summed in quadrature with an overall uncertainty of 19--40\%, from peripheral to central collisions for $R  =  0.3$  jets. 

\begin{figure}[h!]
	   \centering
	   \includegraphics[width=0.45\textwidth]{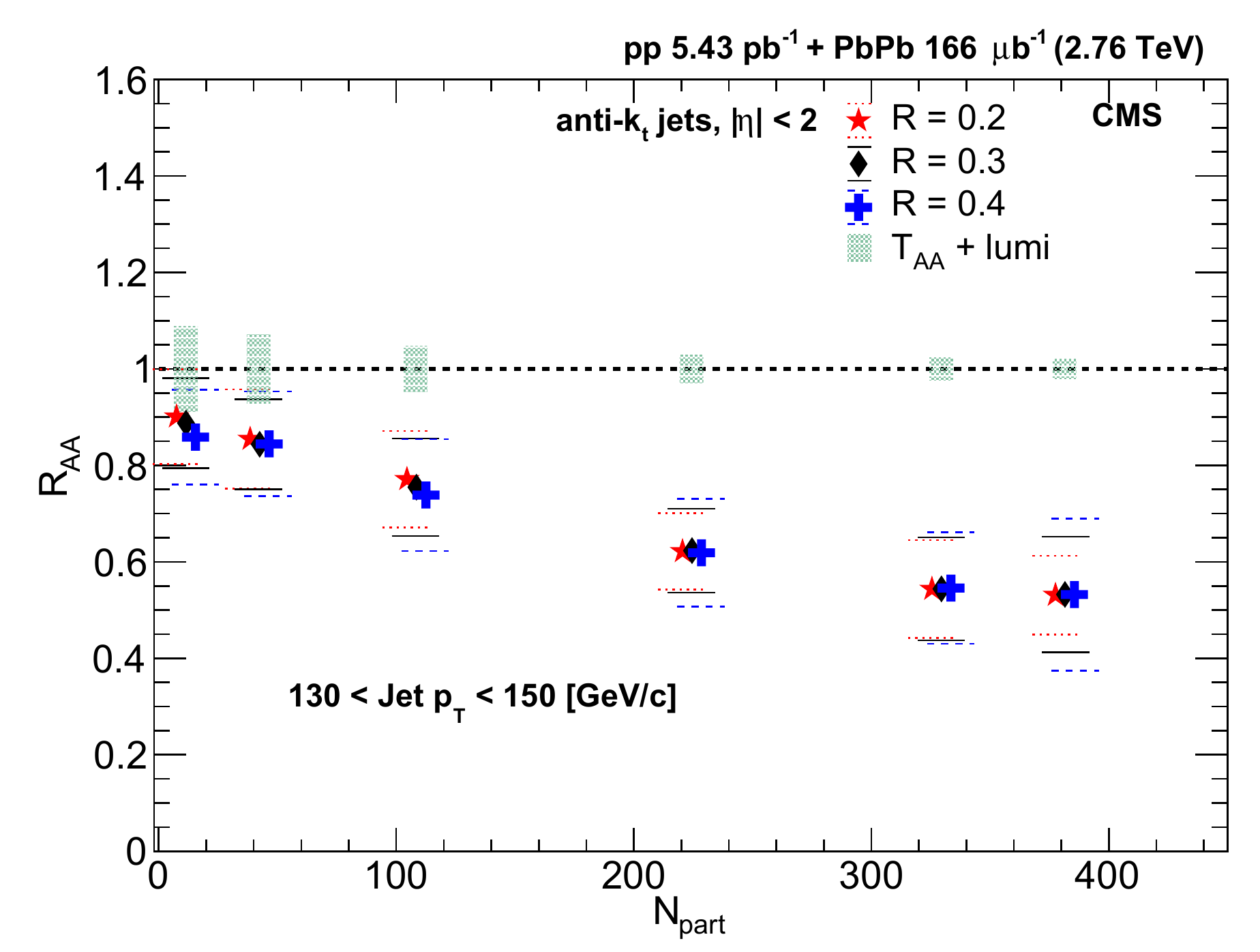} 
	   \includegraphics[width=0.45\textwidth]{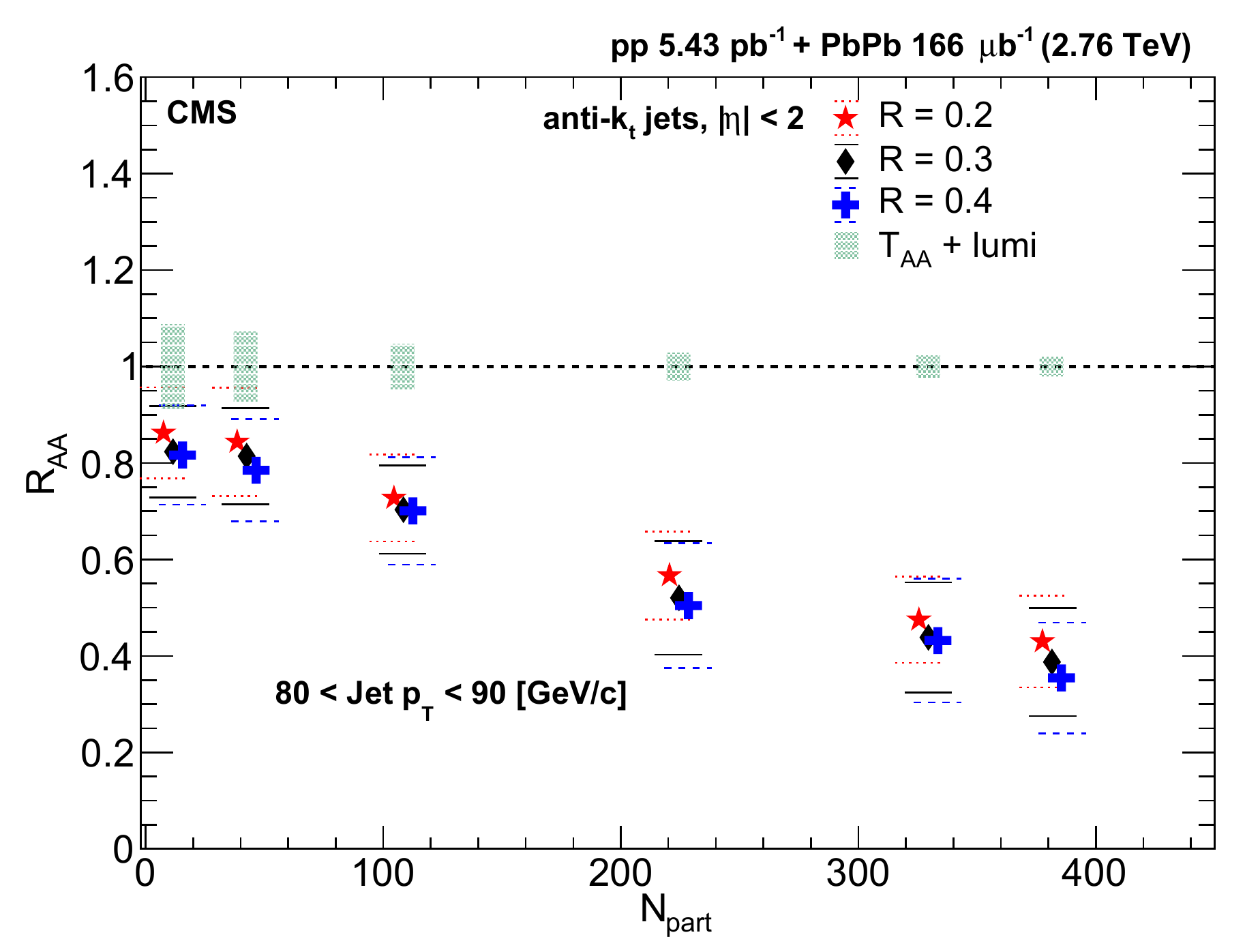} 
	   \caption{Inclusive jet \raa~for \akt jets with distance parameters $R  =  0.2$ (red stars), 0.3 (black diamonds), and 0.4 (blue crosses), as a function of the average $N_{\text{part}}$ for each collision centrality, for jets of $80 < $\pt$ < 90$ and $130 < $\pt$ < 150$ \gev, in the left and right panels respectively~\cite{Khachatryan:2016jfl}. Points are shifted to the left ($R  = 0.2$) and right ($R  = 0.4$) for clarity. The statistical uncertainty is indicated by colored vertical lines (smaller than the markers). The systematic uncertainty is represented by the bounds of the dotted, solid, and dashed horizontal lines for the corresponding distance parameters. The uncertainty boxes at unity represent the \taa~and luminosity uncertainty.}
	   \label{fig:raavsnpart}
	\end{figure}

	To focus on the centrality dependence of the jet \raa, two ranges of jet \pt~are selected and the corresponding jet \raa~values are plotted as a function of the average number of participants ($N_\text{part}$) in Fig.~\ref{fig:raavsnpart}, for jets of $80 < $\pt$ <90$ and $130 < $\pt$ < 150$ \gev. The systematic uncertainty is shown in the three bounds of lines for $R  =  0.2$ (dotted),  0.3 (solid), and 0.4 (dashed) jets. The jet \raa~shows a clear trend of increasing suppression as the number of participants in the PbPb collision increases. Overall, in the kinematic range explored, the \raa~show the same level of suppression across the three distance parameters.

\section{Comparisons with theory}

	Comparing data with predictions is an important theme of my work so far and thus in this section, we shall go over several different models of jet energy loss and see if they can accurately predict the jet \raa, a hallmark heavy ion measurement.     
	
	\subsubsection{Toy Model of Energy Loss}
	
\begin{figure}[h!]
		   \centering
		   \includegraphics[width=0.8\textwidth]{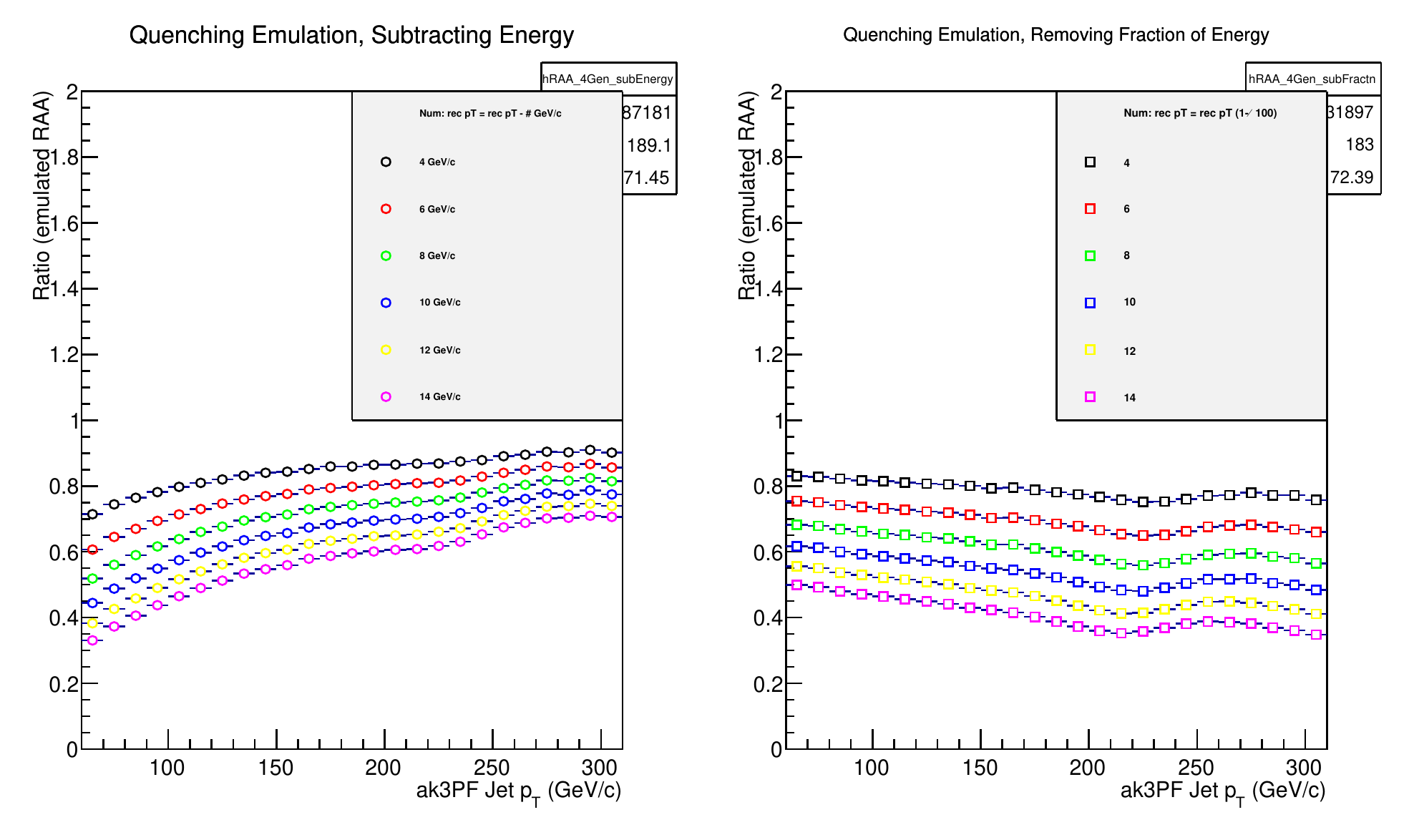} 
		   \caption{Expectation of the Jet \raa~from a toy model of energy loss in \py~with removal of certain amount (left) or fraction (right) of jet's energy for R = 0.3 jets.}
		   \label{fig:raatoymodel}
		\end{figure}

		Lets begin by understanding the impact of energy loss on the jet \raa. Fig:~\ref{fig:raatoymodel} shows the \raa~for inclusive jets with two different methods for energy loss; losing an fixed amount of \pt~per jet on the left and losing a fraction of the jet's \pt~with different colored markers pointing to different energy/fractions. The main difference between the two emulations, as performed in \py~for R=0.3 \akt jets, we see that when a jet loses a standard amount of energy due to quenching regardless of the jet's \pt, the \raa~has a increasing behavior as one would expect. On the other hand, the \raa~has a slight decreasing tendency when one loses a fraction of the jet's \pt~and in both cases, when one increases the amount of energy/fraction, the \raa~decreases. By looking at the trend in data, one can naively estimate the behavior of energy loss. 
	
	\subsubsection{\jw}
	
\begin{figure}[h!]
		   \centering
		   \includegraphics[width=0.5\textwidth]{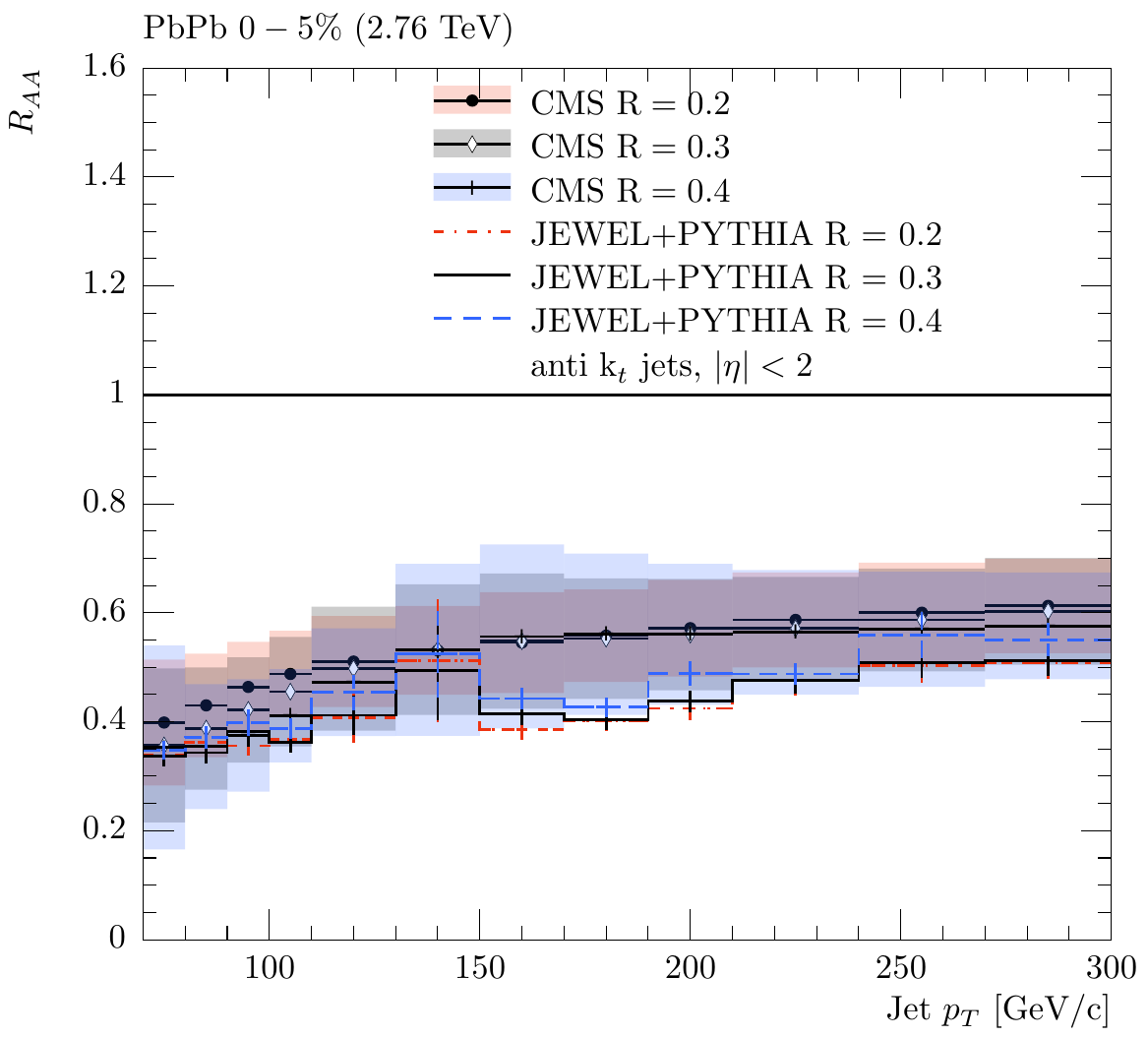} 
		   \caption{Comparison of the CMS Jet \raa~with the \jw~MC based on pQCD for three different jet radii.}
		   \label{fig:raavsjewel}
		\end{figure}

		\jw~\cite{Zapp:2012ak} is a pQCD based model of energy loss and we will discuss it in further detail in the next chapter. For now, all we need to know is that for the jet \raa observable, \jw~does a good job of reproducing the data as shown in Fig:~\ref{fig:raavsjewel} for the different radii. 

	\subsubsection{SCET-g}
	
\begin{figure}[h!]
		   \centering
		   \includegraphics[width=0.8\textwidth]{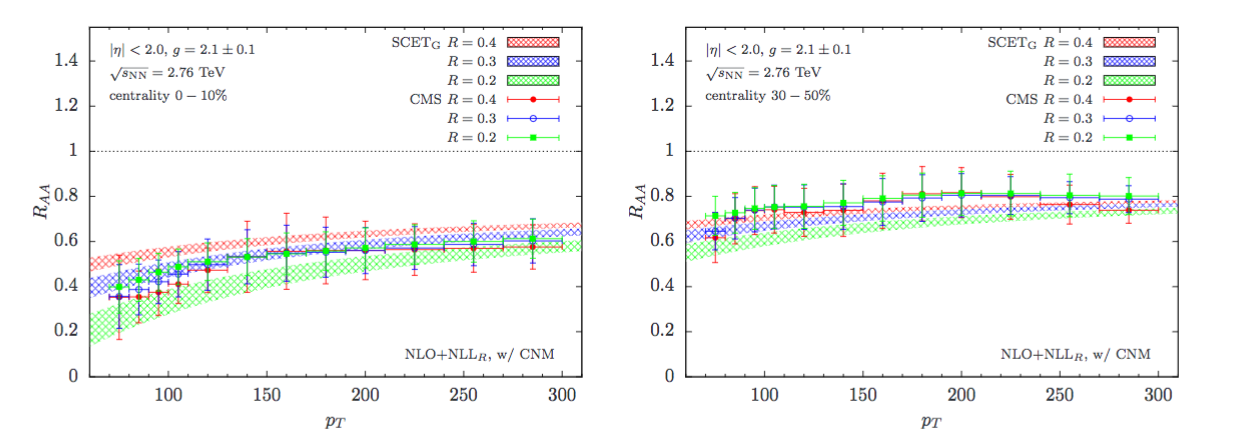} 
		   \caption{Comparison of the CMS Jet \raa~with the SCET-g theoretical framework for three different jet radii.}
		   \label{fig:raavsscet}
		\end{figure}

		The SCET-g framework~\cite{Chien:2015hda} is a heavy ion extension of the framework that was introduced in an earlier chapter for small radii pp jets. In this framework, they estimate the effect of a single gluon coupling with the medium in the different fragmentation legs and thus are able to emulate energy loss with the medium. Their result for the inclusive jet \raa~is shown in Fig:~\ref{fig:raavsscet} for two different centrality bins (central on the left and peripheral on the right) for three different radii. It is important to note the radial ordering here with R = 0.4 jets having a smaller quenching than R=0.3 jets followed by R=0.2. This picture is consistent with energy loss at angles away from the jet axis with going to larger radii tending to recover the energy lost rather quickly.  

	\subsubsection{Hydro+Shower}
	
\begin{figure}[h!]
		   \centering
		   \includegraphics[width=0.5\textwidth]{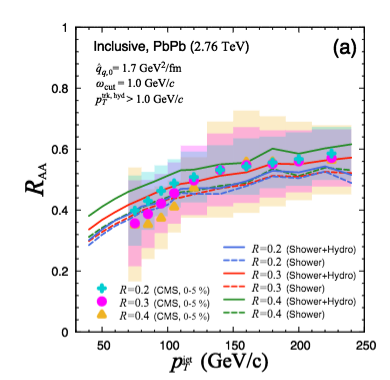} 
		   \caption{Comparison of the CMS Jet \raa~with the CCNU model including shower and hydro model for three different jet radii. Figure taken from~\cite{Tachibana:2017syd}.}
		   \label{fig:raavsshowerhydro}
		\end{figure}

		This is one of the more recent models on the market and it uses both a hydro component and a particle shower component in its energy loss calculations~\cite{Tachibana:2017syd}. It has certain parameters such as $\hat{q}$ (which was introduced earlier) and a soft radiation cutoff for emissions and tracks to stay in the perturbative regime. Their calculations for the \raa~as shown in Fig:~\ref{fig:raavsshowerhydro} have solid lines for shower + hydro model and the dotted lines for only the shower part. Within the data systematic uncertainties, the model performs admirably and the shower+hydro mode showcases the same radii dependence as with a certain amount of energy lost from a jet independent of its momenta. 

	\subsubsection{Holographic HYBRID}
	
\begin{figure}[h!]
		   \centering
		   \includegraphics[width=0.5\textwidth]{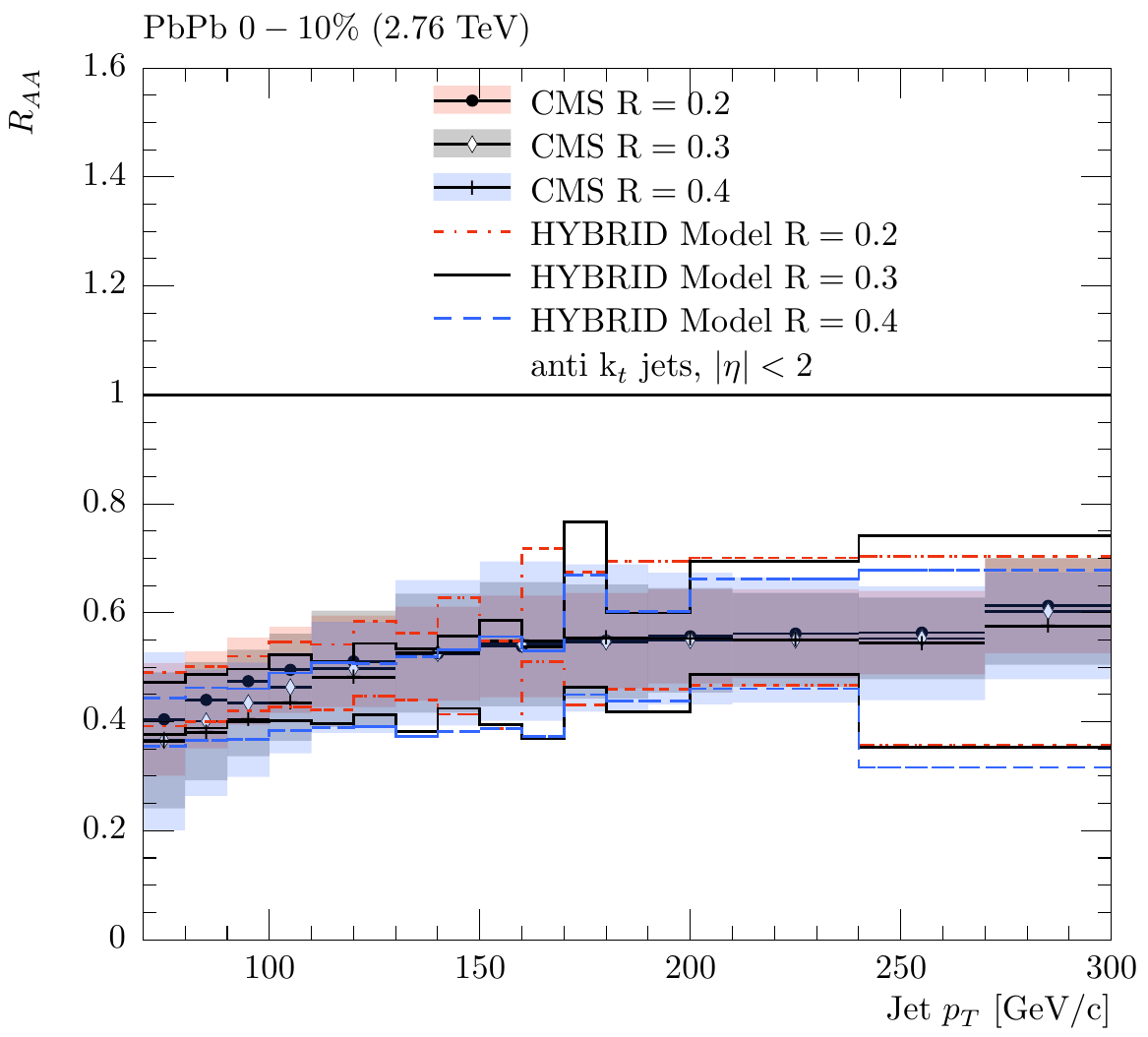} 
		   \caption{Comparison of the CMS Jet \raa~with the holographic hybrid model for three different jet radii. Figure courtesy Dani Pablos.}
		   \label{fig:raavshybrid}
		\end{figure}

		The HYBRID~\cite{Casalderrey-Solana:2015vaa} model uses holographic techniques to perform parton energy loss. The AdS dual of a parton losing energy is a string falling into a black hole. For the inclusive jet \raa, we can see their comparison with the published data points in Fig:~\ref{fig:raavshybrid} which happens to line up very nicely ableit with large theoretical systematic uncertainties. It is called HYBRID since the jets are described at weak coupling and the parton energy loss at strong coupling. It is interesting to note the radii dependence in this theoretical calculation is opposite to other models that we have seen for a given jet \pt. A physical interpretation for this particular ordering is partly due to the medium kick being at quite large angles away from the jet axis with the recovery happening beyond the $\Delta R \approx 0.4$ limit.  

\section{What we learn from the jet \raa}
	
\begin{figure}[h!]
	   \centering
	   \includegraphics[trim={4cm 1cm 0 0},clip]{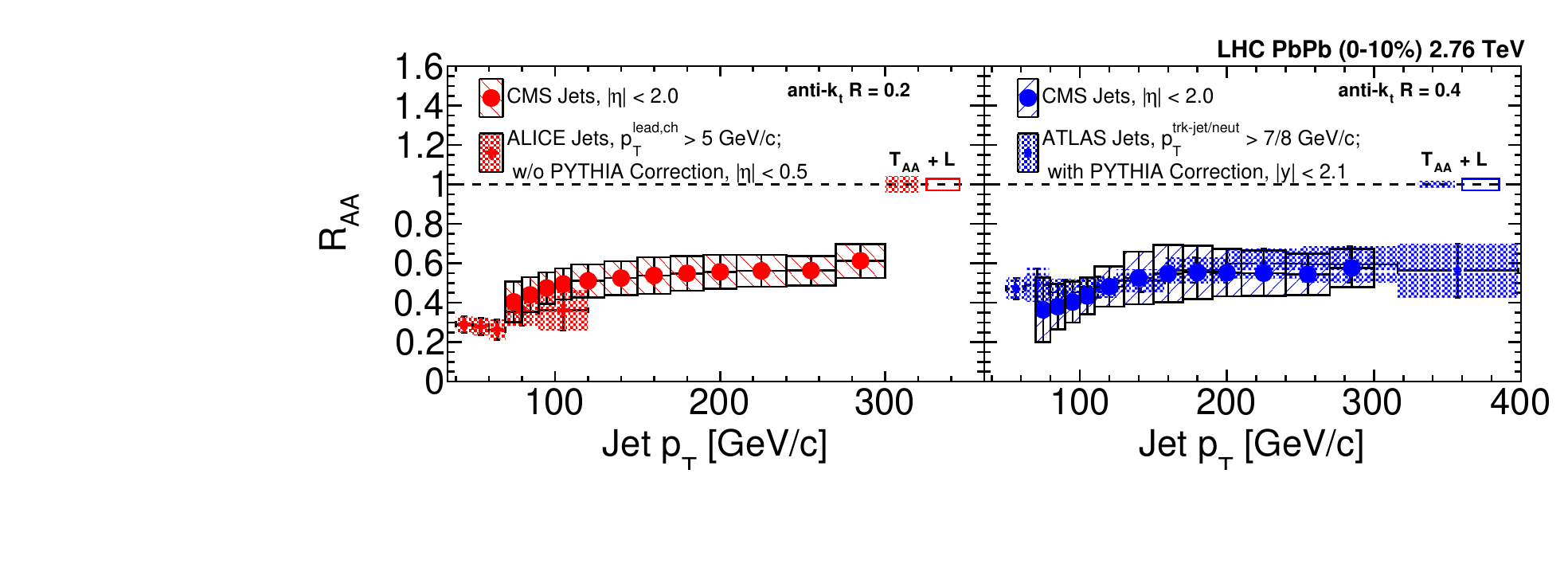} 
	   \caption{Inclusive Jet \raa~from CMS compared with ALICE (R = 0.2, left) and ATLAS (R = 0.4, right), both including fragmentation based cuts to remove the misreconstructed jet fraction~\cite{Khachatryan:2016jfl}.}
	   \label{fig:raavsaliceatlas}
	\end{figure}

	One can think of the \raa~as an ensemble observable elucidating bulk/overall properties but failing for finer details. In the previous section, we see that a whole host of theory calculations and models can predict the jet \raa~with reasonable accuracy. The interesting point is that even amongst experiments at the LHC, there is very good consistency in the measurement as shown in Fig:~\ref{fig:raavsaliceatlas}. The panels show CMS compared with ALICE (left) for R = 0.2 jets and ATLAS (right) for R = 0.4 jets and even though ALICE and ATLAS employ somewhat of a fragmentation biased jet selection, overall the agreement is pretty remarkable. This is why the CMS result is a legacy measurement and one that bridges the gap between ALICE and ATLAS in terms of both kinematic reach and with utilizing unbiased inclusive jets.  
	
	From the data, we see the following 
	\begin{itemize}
		\item Jet quenching is a final state effect - due to lack of modification in \rpa
		\item Clear centrality dependence in PbPb - more overlap leads to larger quenching
		\item slight dependence on the jet \pt
		\item independence on the jet radii within the measured systematic uncertainties 
	\end{itemize}
	
	From our CMS measurement, the inclusive jet \raa for the most central collisions, ranges $\approx 0.4 - 0.6$ as a function of jet \pt. Since these are ratios of jet spectra, a quick back of the envelope calculation tells us that an \raa of 0.4 means a $60\%$ reduction in the spectra which in turn corresponds to energy loss of $\approx 10.5$ GeV for a 70 GeV jet, assuming a power law spectra with an exponent of 4. Similarly, \raa of 0.6 corresponds to energy loss of $\approx 30$ GeV for a 300 GeV jet. With the latest data from Run2 at the LHC, ATLAS and ALICE have both presented their preliminary \raa~and this time even at 5.02 TeV, the results lie in the same scale i.e the \raa~appears to be flat around $0.6$ even for 1 TeV jets. The reasons for a 1 TeV jet to lose roughly $\approx 100$ GeV, with our back of the envelope calculation are still unknown and thus provides an exciting new question for both experiments to quantify this energy loss and for theorists to model such a behavior. Thus, we need to move towards more differential measurements on the jet structure/profile compared to a jet quenching MC which we will look at in the next chapter.

\clearpage

\chapter{Jet Quenching with JEWEL}
\label{ch_jewel}

\begin{chapquote}{Richard Feynman}
`` If you make a theory, for example, and advertise it, or put it out, then you must also put down all the facts that disagree with it, as well as those that agree with it."
\end{chapquote}

\section{Introducing \jw}

	\jw~or Jet Evolution With Energy Loss, is a fully dynamical perturbative framework developed to simulate jet quenching in heavy ion collisions. It includes a simultaneous scale evolution of hard partons into jets and their subsequent re-scatterings with the medium. The parton shower in \jw~is virtuality ordered and all partons in the shower in addition to the jet evolution, undergo re-scattering in the background. These interactions are described by $2-2$ pQCD elastic or inelastic matrix elements at LO+LL. When the we consider re-scattering in the QGP,  \jw~goes beyond factorization theorems and relies on a few assumptions, 
	\begin{itemize} 
		\item re-scattering resolves the partonic structure of the QGP for sufficiently hard interactions
		\item infra-red continuation to regularize the pQCD matrix elements and include the dominant effect of soft scattering
		\item interplay of different sources of radiation governed by the formation times  
		\item the physical picture of the LPM interference via eikonal kinematics is also valid in the non-eikonal regime. 
	\end{itemize}
	
	For a full discussion of the \jw~framework and its implementation the reader is referred to~\cite{Zapp:2012ak}. We begin this chapter, as is our preferred style throughout, by summarizing the most important features. The vacuum case of jet evolution in \jw~reduces to a standard virtuality ordered final state parton shower in based on \py~6.4~\cite{Sjostrand:2006za} which generates the initial state parton showers, hard jet production matrix elements, hadronization and hadron decays. The strong coupling $\alpha_{s}$ runs at one loop evaluated according to the standard perturbative scale choices and $\Lambda_\text{QCD}$ is adjusted to fit LEP data and doesn't change for subsequent generations.

	\subsection{MonteCarlo implementation of in-medium energy loss}

		In \jw~the jet-medium interactions give rise to energy loss. The knowledge about the energy-momentum transferred from the jet to the medium can be used for detailed studies of the medium response to jets~\cite{Floerchinger:2014yqa}. \jw~has the option to retain recoiling medium partons in the event, but this requires special analysis techniques~\cite{KunnawalkamElayavalli:2016dda, KunnawalkamElayavalli:2016lzw} (paper in preparation). For inclusive jet observables like the jet \pt, dijet asymmetry, angular correlations etc. \jw~can be run without storing the recoils. To study jet-substructure observables and their sensitivity to the medium response, \jw~can store the recoiling partons which  requires a background subtraction procedure to be discussed in detail in the upcoming sections.  

		All scattering processes within the formation time of a medium-induced emission act coherently, which means that only the vectorial sum of the momentum transfers matters for the gluon emission. This is the QCD analogue of the Landau-Pomerantchuk-Migdal effect, which is implemented according to a generalization of the algorithm derived in~\cite{Zapp:2011ya}. The emissions due to the scale evolution of the jet get dynamically interleaved with radiation associated to re-scatterings in such a way that re-scattering can only induce radiation if its formation time is shorter than the lifetime of the hard parton. This implies that only a hard re-scattering can perturb the hard parton shower emissions related to the initial jet production process, so that the hard jet structure is protected from medium modifications. This principle shares important features with color coherence (cf. e.g.~\cite{CasalderreySolana:2012ef}), but is not a dynamical implementation of color coherence. It is missing, for instance, soft and large angle emissions from coherent sub-systems. 	
		
		The full list of parameters are available in the \jw~manual but the main ones that change typically as the beam energy, initial temperature and the formation time of the QGP, along with the weighted exponent in the power law of jet spectra. The initial conditions are provided by hydrodynamic calculations. There are also additional parameters which involve the debye mass screening factor which was set based on pion quenching at RHIC. This means that for a given collision system and center of mass energy, \jw~can provide predictions for several jet observables simultaneously without additional tunings or including new parameters. All one needs is to generated events in \jw~and produce predictions for observables which can then be directly compared to measurements.  
		
		We generate events in the standard setup~\cite{Zapp:2013vla} at $\sqrt{s_\text{NN}} = 2.76$ TeV and $\sqrt{s_\text{NN}} = 5.02$ TeV with the simple parametrisation of the background discussed in detail in~\cite{Zapp:2013zya}. This background model describes a thermal quark-gluon gas undergoing Bjorken expansion with a superimposed transverse profile obtained from an optical Glauber model.  The initial conditions for the background model are initial time $\tau_\text{i}=0.6 $ fm and temperature $T_\text{i}=485$ MeV for $\sqrt{s_\text{NN}} = 2.76$ TeV~\cite{Shen:2012vn} and $\tau_\text{i}=0.4$ fm and $T_\text{i}=590$ MeV for $\sqrt{s_\text{NN}} = 5.02$ TeV~\cite{Shen:2014vra}. They are taken from a hydrodynamic calculation describing soft particle production. The proton PDF set is \textsc{Cteq6LL}~\cite{Pumplin:2002vw} and for the Pb+Pb sample the \textsc{Eps09}~\cite{Eskola:2009uj} nuclear PDF set is used in addition, both are provided by \textsc{Lhapdf}~\cite{Whalley:2005nh}. The only parameter in \jw~that can be fitted to jet quenching data is the scaling factor of the Debye mass. It was adjusted once to describe the single-inclusive hadron suppression at RHIC and has remained the same since. We use the \textsc{Rivet} analysis framework~\cite{Buckley:2010ar} for all our studies. Jets are reconstructed using the same jet algorithm as the experiments (\akt~\cite{bib_antikt}) from the fastjet package~\cite{Cacciari:2011ma}.

\section{Electroweak bosons and the QGP}

	Since the QGP interacts via the strong force, it is transparent to electro-weak bosons which propagate without energy loss. Thus for events where we have a quark scattering of a boson, say a photon in the final state, reconstructing the photon's energy can provide a handle on how much the jet in the back to back direction got quenched.  For this reason such events are called as standard candles of the QGP by some aficionados in the field. In general they are also used in pp collisions to apply corrections to the jet energy since there is no jet quenching. Thus for a heavy ion MC generator, it is important to have the ability to generate such events to establish confidence in the implementation of energy loss.  

	\subsection{Adding V+Jet to \jw}
	
\begin{figure}[h] 
		   \centering
		   \includegraphics[width=0.5\textwidth]{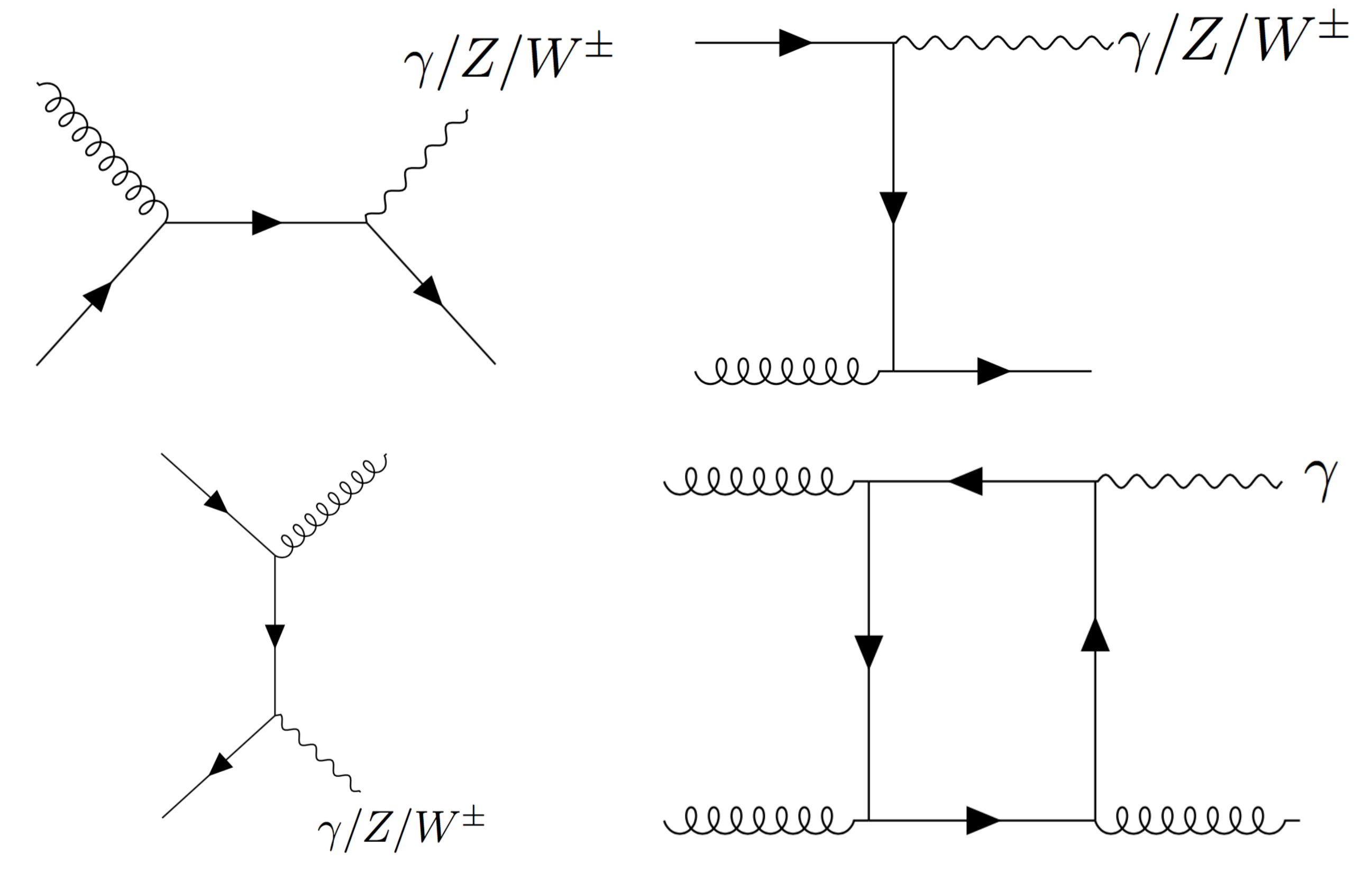} 
		   \caption{Feynman diagrams for $V$+jet processes included in \jw~\cite{KunnawalkamElayavalli:2016ttl}}
		   \label{fig:production}
		\end{figure}

		We have included the lowest order processes with a jet recoiling against a vector boson~\cite{KunnawalkamElayavalli:2016ttl} with the corresponding  diagrams shown in Fig.~\ref{fig:production}. These correspond to either a quark scattering off a gluon (Compton scattering) or a quark--anti-quark pair annihilating to produce a boson and a gluon. For photons, the box diagram $gg \rightarrow \gamma g$ is also included. This process is of higher order than the others, but is included as it can be numerically important in certain phase space regions. The leptonic decays of the heavy boson $Z$ and $W$ are simulated as well. 

		Hard photons can also be radiated off quarks during jet evolution. These fragmentation photons are typically accompanied by hadronic activity and are  suppressed by requiring the photon to be isolated. However, it is still possible that fragmentation photons pass the isolation criterion. The probability for this to happen is small and depends on the cuts. It has to the best of our knowledge not been quantified in a heavy ion environment in the presence of jet quenching. In the current \jw~version  fragmentation photons are also not included. For the analyses shown here the fragmentation component is expected to be small due to the applied photon isolation.

		As we mentioned before, \jw~is a leading-order framework. While NLO corrections to $V$+jet processes can be sizable, in the observables shown here corrections affecting only the cross section largely cancel due to the normalization to number of bosons or number of boson-jet pairs. The corrections to differential distributions remain, but are typically smaller.
	
	\subsection{Predictions and comparisons with data}
	
		\begin{figure}[h] 
		   \centering
		   \includegraphics[width=0.45\textwidth]{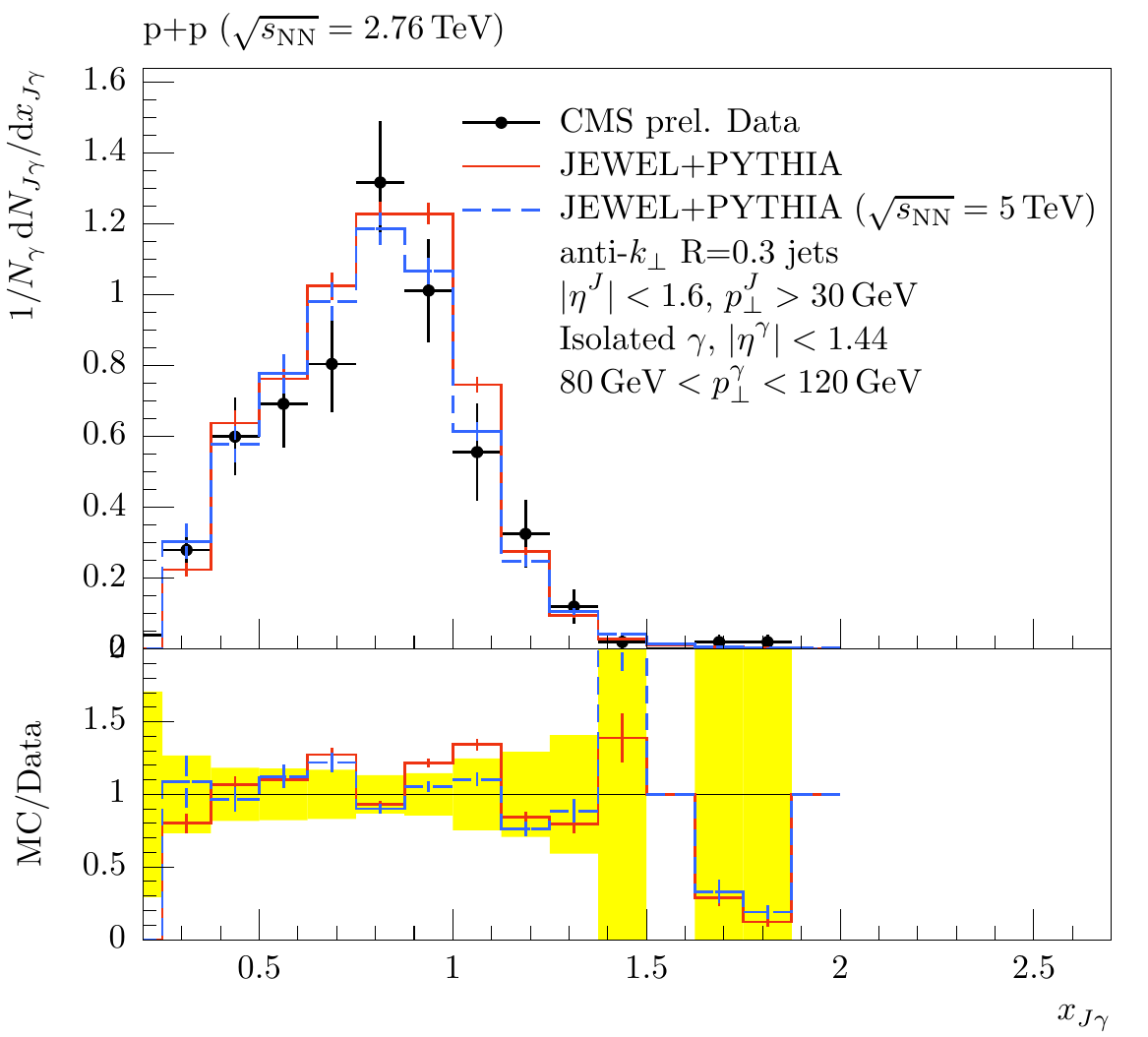} 
		   \includegraphics[width=0.45\textwidth]{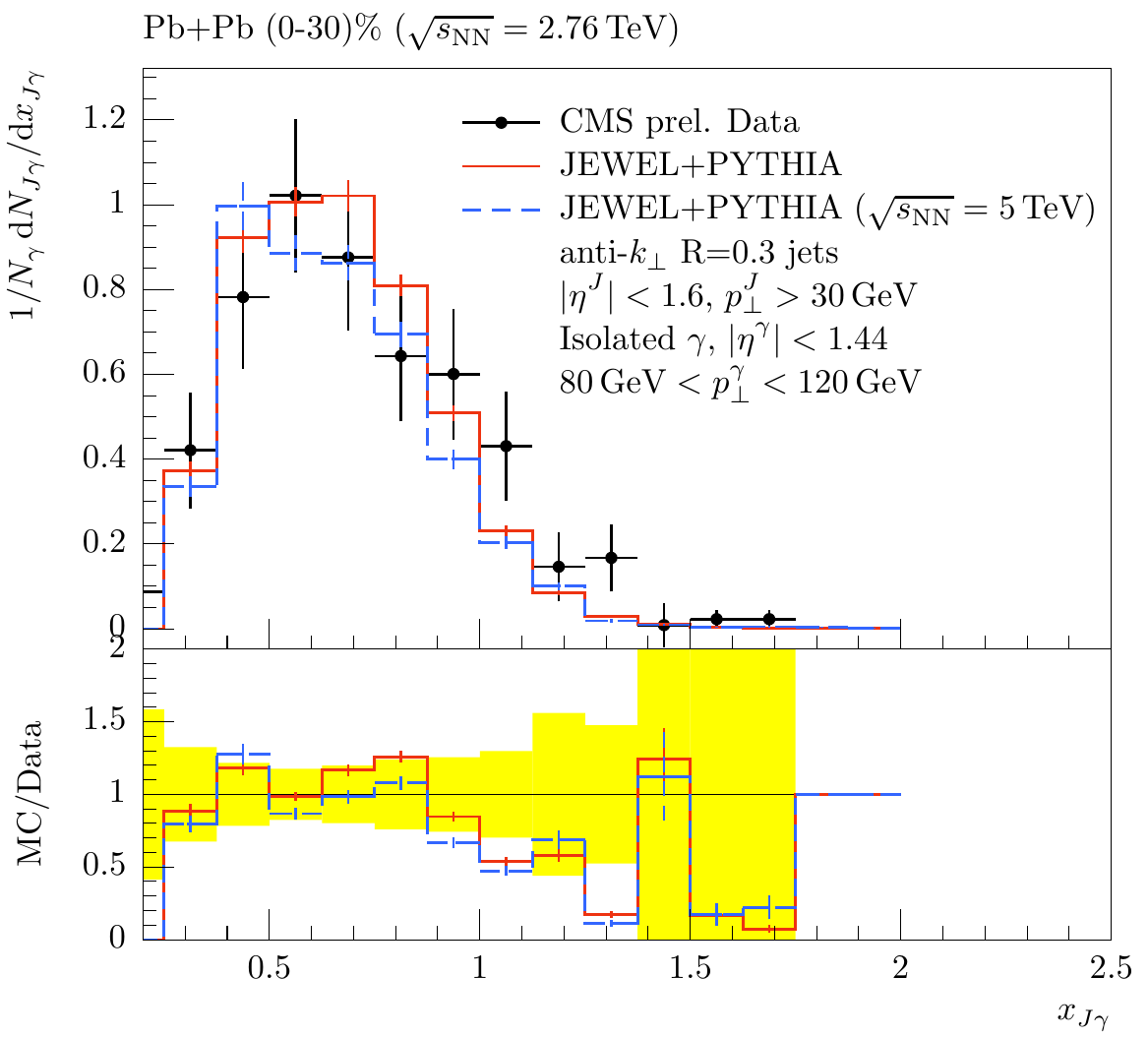} 
		   \caption{Momentum imbalance $x_{J \gamma} = p_{T}^J/p_{T}^\gamma$ in $\gamma$+jet events for photon transverse momentum $80 < p_{T}^\gamma < 120$ \gev compared to preliminary CMS data~\cite{cmsypjet} in p+p (left) and central Pb+Pb events (right) at $\sqrt{s_\text{NN}} = 2.76$ TeV~\cite{KunnawalkamElayavalli:2016ttl}. The \jwpy~prediction for  $\sqrt{s_\text{NN}} = 5.02$ TeV is also shown. The CMS data are not unfolded for jet energy resolution, therefore the jet \pt~was smeared in the Monte Carlo sample using the parametrization from~\cite{Chatrchyan:2012gt}. The data points have been read off the plots and error bars correspond to statistical errors only. The yellow band in the ratio plot indicates the errors on the data points.}
		   \label{fig:ypjet_xjy}
		\end{figure}

		\begin{figure}
		 \includegraphics[width=0.5\textwidth]{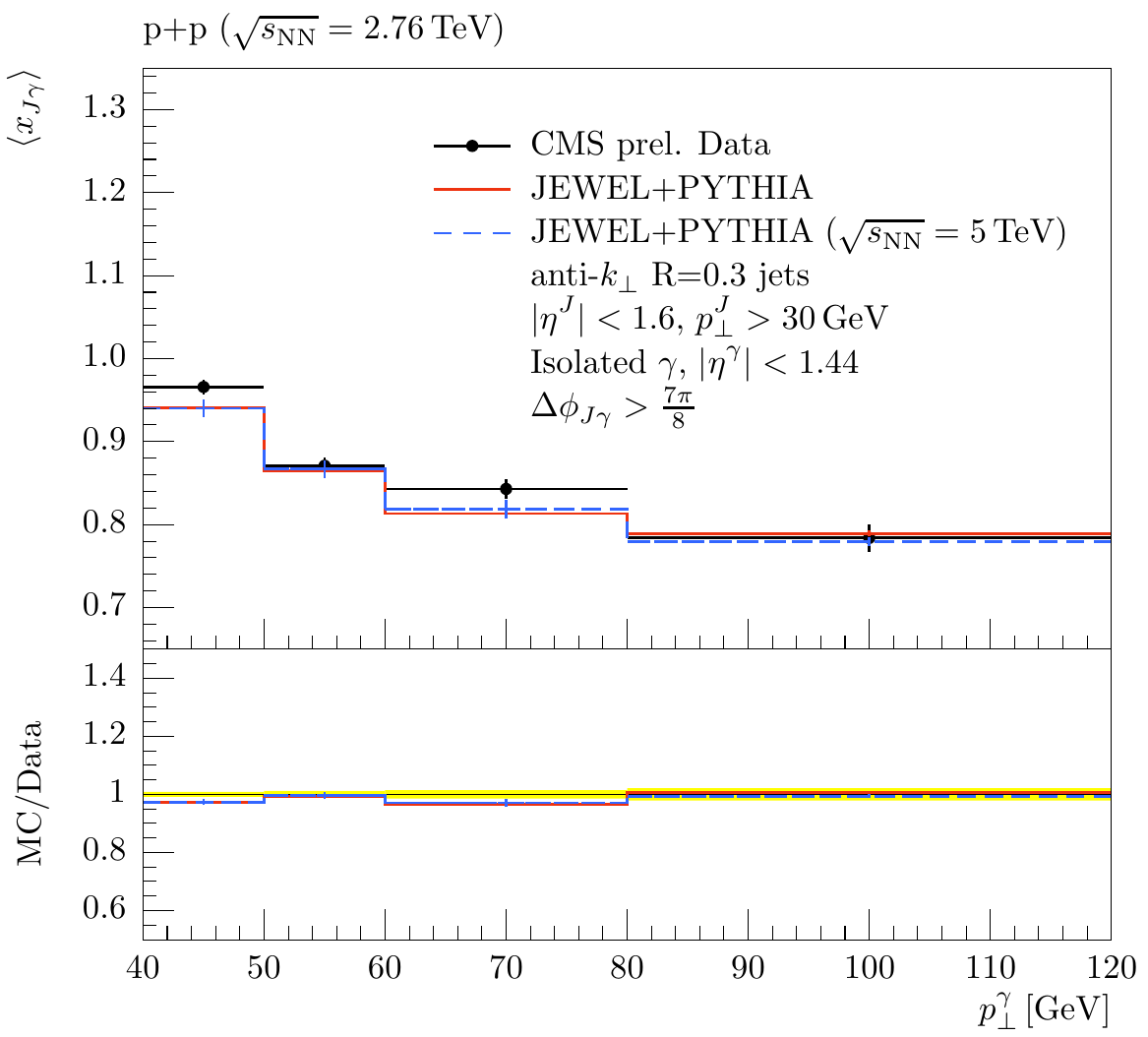}
		 \includegraphics[width=0.5\textwidth]{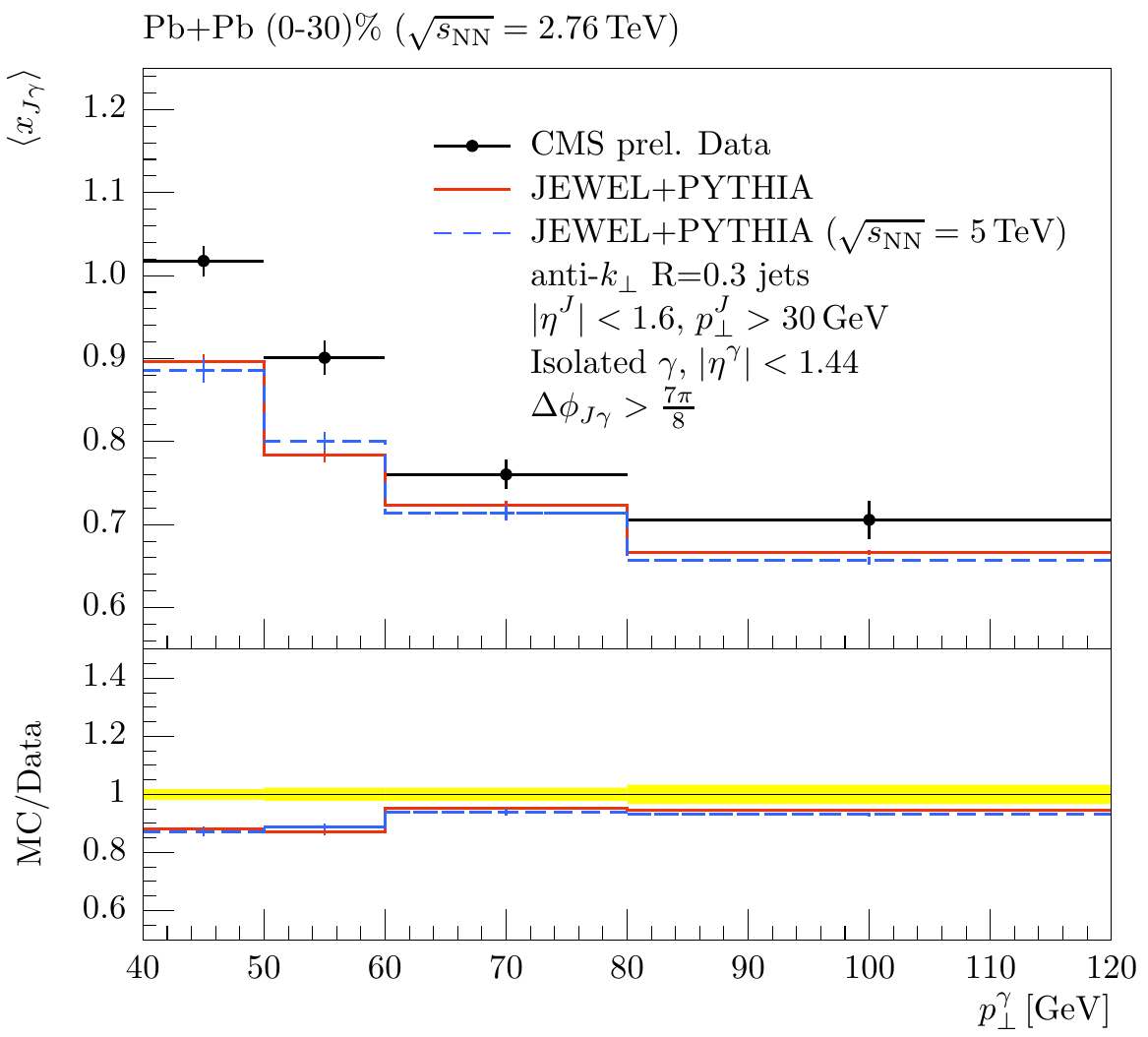}
		 \caption{Average value of the $x_{J \gamma}$ shown as a function of the photon's \pt~compared to preliminary CMS data~\cite{cmsypjet} in p+p (left) and central Pb+Pb events (right) at $\sqrt{s_\text{NN}} = 2.76$ TeV. The \jwpy~prediction for  $\sqrt{s_\text{NN}} = 5.02$ TeV is also shown. The CMS data are not unfolded for jet energy resolution, therefore the jet \pt~was smeared in the Monte Carlo sample using the parametrization from~\cite{Chatrchyan:2012gt}.  The data points have been read off the plots and error bars correspond to statistical errors only. The yellow band in the ratio plot indicates the errors on the data points.}
		 \label{fig:ypjet_average}
		\end{figure}

		\begin{figure}[h] 
		   \centering
		   \includegraphics[width=0.45\textwidth]{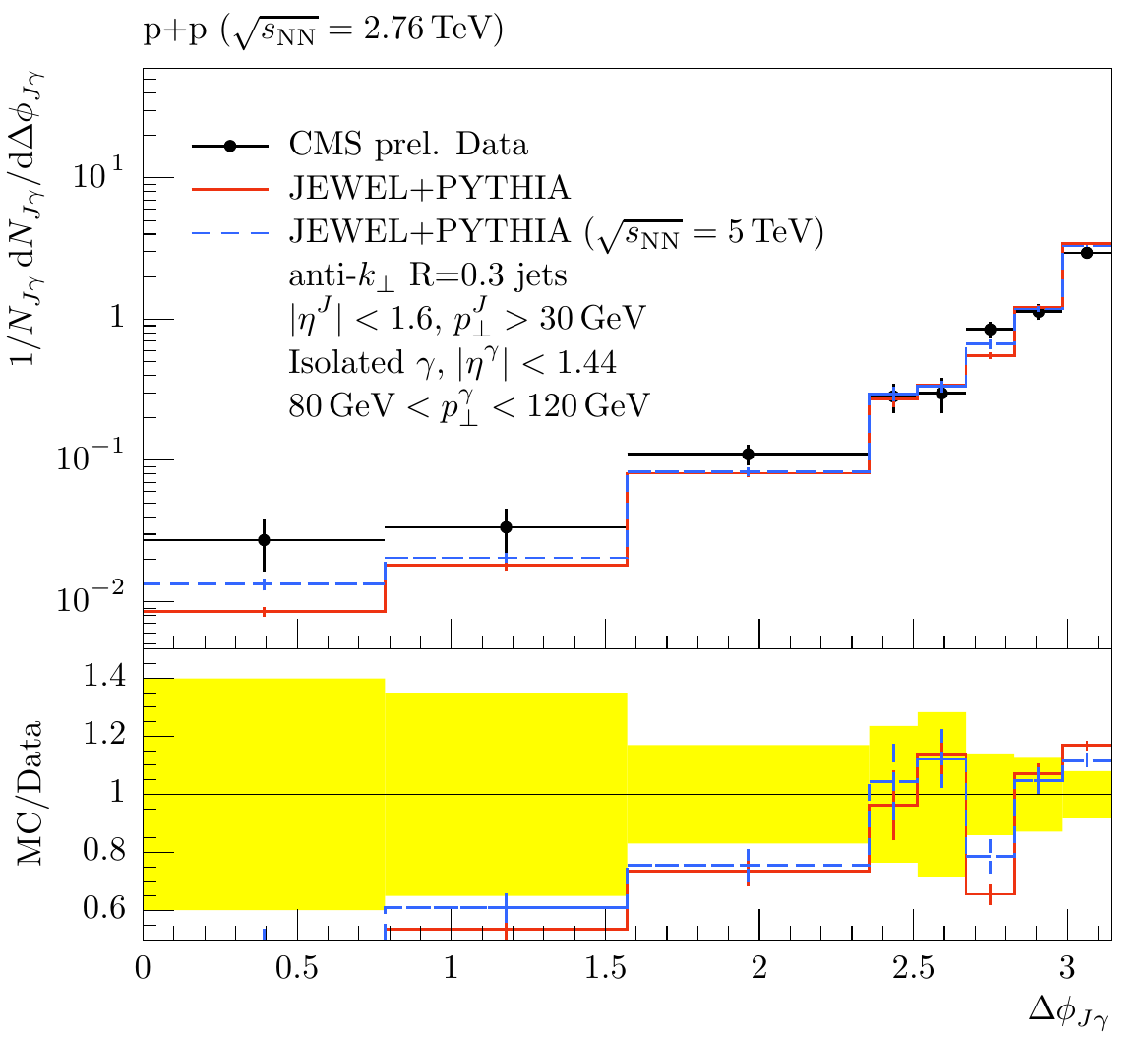} 
		   \includegraphics[width=0.45\textwidth]{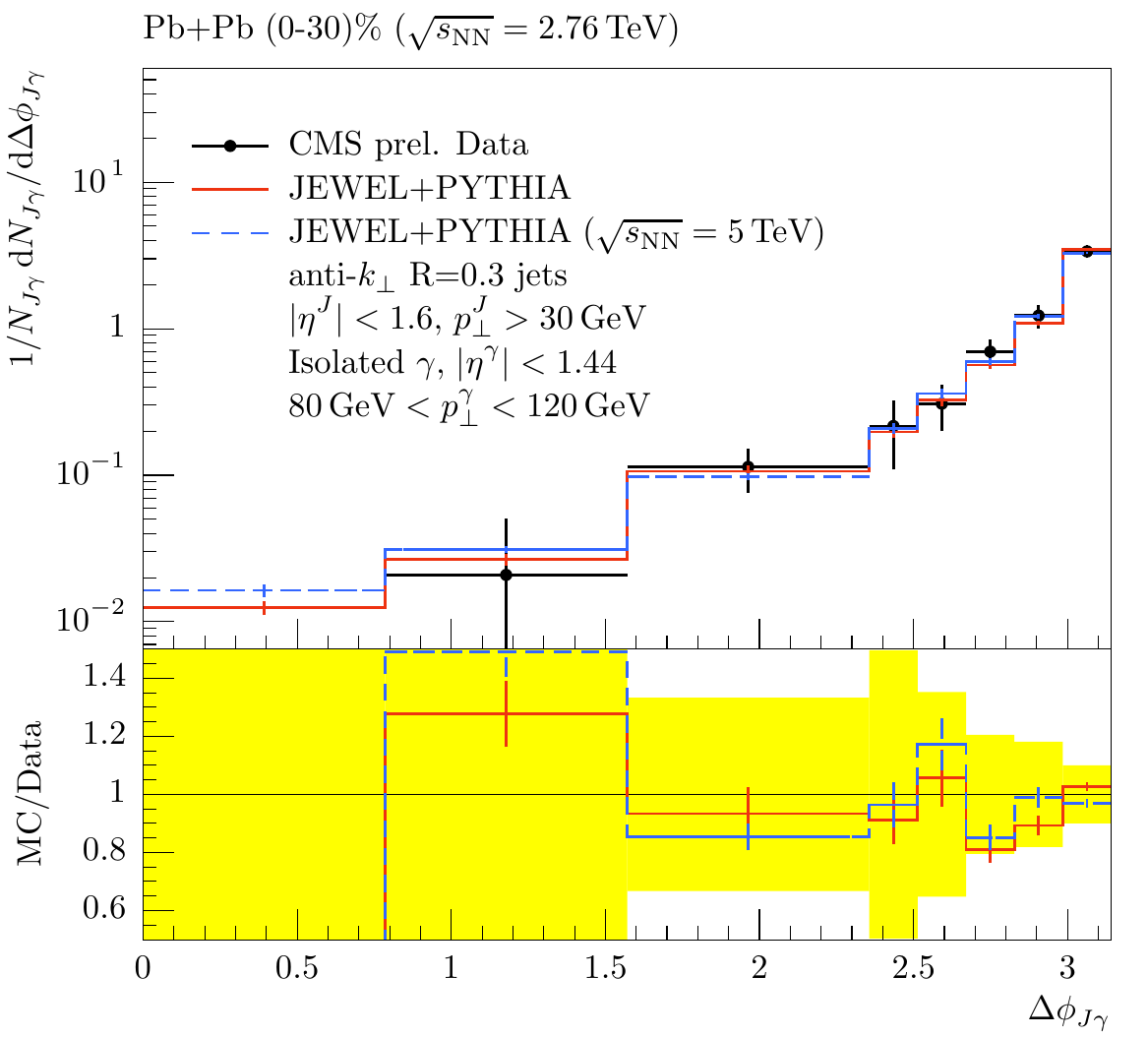} 
		   \caption{Azimuthal angle $\Delta \phi_{J\gamma}$ between the photon and the jet for photon transverse momentum $80 < p_{T}^\gamma < 120$\gev compared to preliminary CMS data~\cite{cmsypjet} in p+p (left) and central Pb+Pb events (right) at $\sqrt{s_\text{NN}} = 2.76$ TeV~\cite{KunnawalkamElayavalli:2016ttl}. The \jwpy~prediction for  $\sqrt{s_\text{NN}} = 5.02$ TeV is also shown. The CMS data are not unfolded for jet energy resolution, therefore the jet \pt~was smeared in the Monte Carlo sample using the parametrization from~\cite{Chatrchyan:2012gt}. The data points have been read off the plots and error bars correspond to statistical errors only. The yellow band in the ratio plot indicates the errors on the data points.}
		   \label{fig:ypjet_deltaphijy}
		\end{figure}

		In order to suppress the background from fragmentation and decay photons, isolation cuts are applied by requiring the sum of energy in a cone of radius $0.4$ (in the $\eta-\phi$ phase space) around the photon to be less than $7\%$ of the photon's energy. In addition, the photon has to be within $|\eta^\gamma| < 1.44$ and have a transverse momentum \pt$^\gamma > 40$\gev. The jets are reconstructed with the \akt algorithm with a resolution parameter of $R=0.3$. Jets are required to have a \pt$^{J}>30$\gev and to be in the barrel region ($|\eta^J|<1.6$). Furthermore, only jets that are back-to-back with the photon ($\Delta \phi_{J\gamma} > 7\pi/8$) are selected.

		Fig.~\ref{fig:ypjet_xjy} shows our results for the transverse momentum asymmetry in $\gamma+$jet pairs ($x_{J\gamma} = p_{T}^J/p_{T}^\gamma$) compared with preliminary CMS~\cite{cmsypjet} data points for p+p and central ($0-30\%$) Pb+Pb collisions at $\sqrt{s_\text{NN}} = 2.76$ TeV.  Fig~\ref{fig:ypjet_xjy} shows the average value of the $x_{J\gamma}$ as a function of the photon transverse momenta in four \pt~bins, again for p+p and central Pb+Pb collisions. \jwpy~is able to reproduce the effect of the \pt~imbalance for $\gamma+$jets events very nicely for both p+p and Pb+Pb events. In central Pb+Pb collisions $\langle x_{J\gamma}\rangle$ is slightly lower in \jwpy~than in the data indicating stronger medium modifications in \jw, particularly at relatively low photon \pt. In Fig.~\ref{fig:ypjet_deltaphijy} the azimuthal angle ($\Delta \phi_{J \gamma}$) between the photon and the jet is shown. We again find a very reasonable agreement with \jwpy~for pp collisions slightly more peaked. In all three figures we also show the \jwpy~predictions for $\sqrt{s_\text{NN}} = 5.02$ TeV, which turn out to be very similar to the $\sqrt{s_\text{NN}} = 2.76$ TeV results. The agreement with the ATLAS measurement~\cite{atlasypjet} is of a very similar quality.

\begin{figure}[h] 
		   \centering
		   \includegraphics[width=0.45\textwidth]{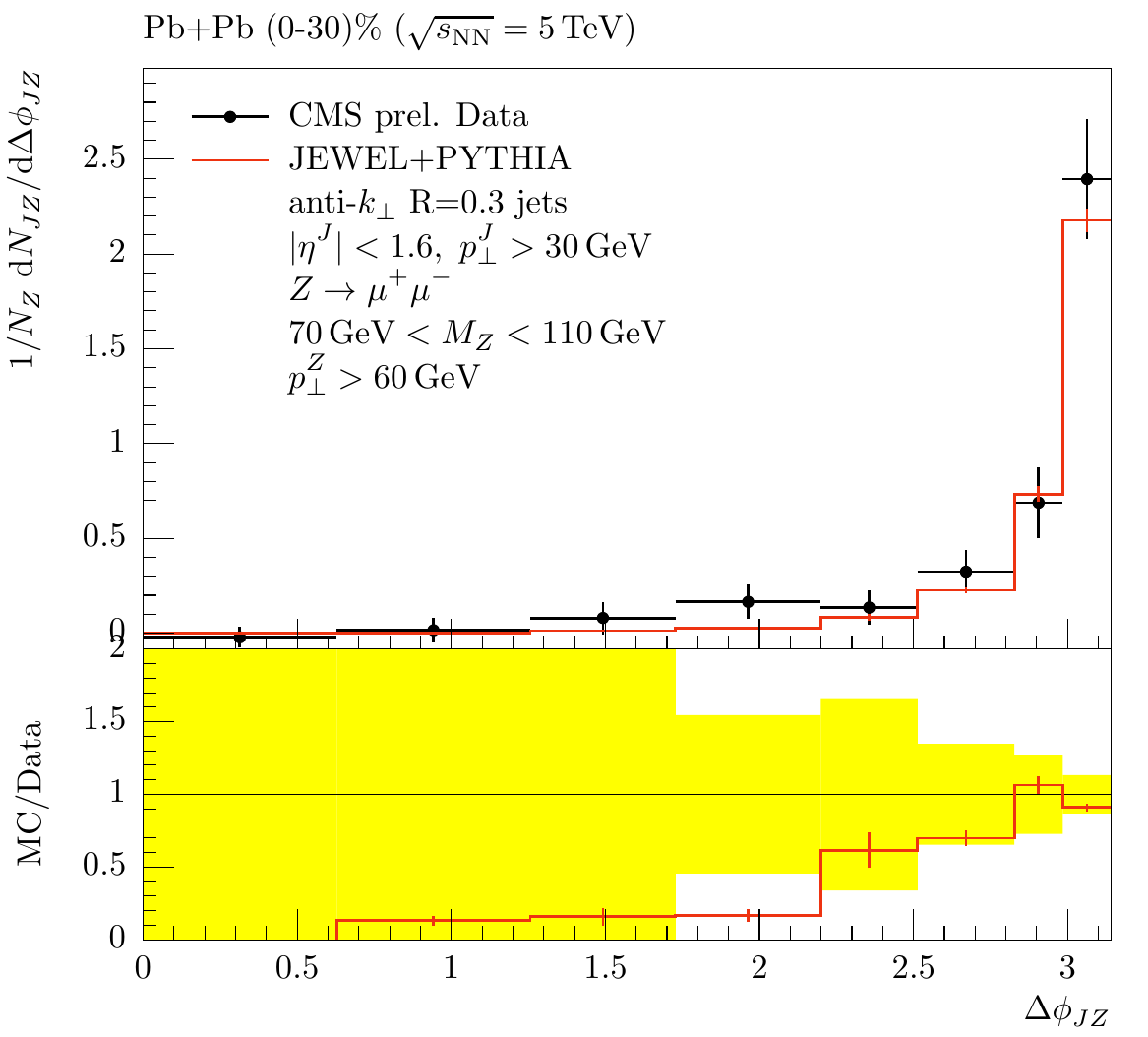} 
		   \includegraphics[width=0.45\textwidth]{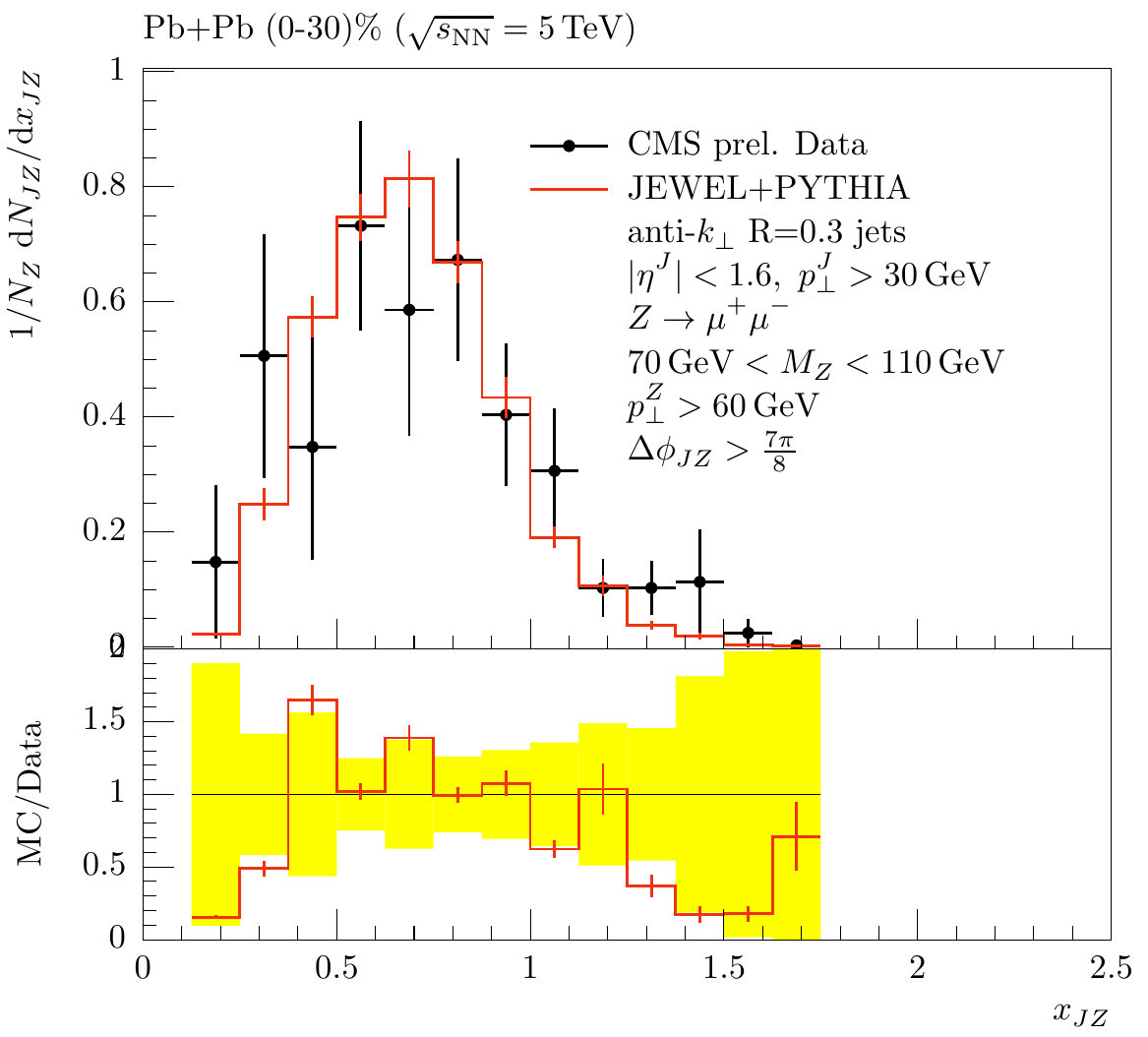} 
		   \caption{Azimuthal angle $\Delta \phi_{JZ}$ between the $Z$ and the jet (left) and momentum imbalance $x_{JZ}$ (right) in $Z$+jet events compared to preliminary CMS data~\cite{Sirunyan:2017jic} in central Pb+Pb events at $\sqrt{s_\text{NN}} = 5.02$ TeV~\cite{KunnawalkamElayavalli:2016ttl}. The CMS data are not unfolded for jet energy resolution, therefore the jet \pt~was smeared in the Monte Carlo sample using the parametrisation from~\cite{Chatrchyan:2012gt}. The data points have been read off the plots and error bars correspond to statistical errors only. The yellow band in the ratio plot indicates the errors on the data points.}
		   \label{fig:zpjet_atlas_r4}
		\end{figure}

		In the case of $Z$ and $W$ production we utilize the muon decay channel in our simulations (this is purely convenience, the electron channel can be simulated as well). The $Z$ candidate's momentum is reconstructed from the di-muon pairs. For comparison to the ATLAS measurement we require its reconstructed mass in the window $66 < M_Z < 102$ \gevc and \pt$^Z>60$ \gev. The jets are reconstructed with the same \akt algorithm with resolution parameter $R=0.4$, with the kinematic cut on its \pt$^J>25$ \gev and it is required to be found in the barrel region $|\eta^J|<2.1$. Similar to the $\gamma+$jet case, we impose $\Delta \phi_{JZ} > \pi/2$ to select the back to back pairs. 
Fig.~\ref{fig:zpjet_atlas_r4} shows the ATLAS~\cite{atlaszpjet} preliminary result for the \pt~imbalance compared to \jwpy~for central (0-20\%) Pb+Pb collisions at  $\sqrt{s_\text{NN}} = 2.76$ TeV. For comparison we also show the \jwpy~result for p+p. In central Pb+Pb events we observe a clear shift of the distribution towards smaller $x_{JZ}$ compared to p+p and a reasonable agreement between the MC and data.

		Reconstructing a $W$ boson candidate in the heavy ion environment is difficult due to the ambiguous nature of the missing transverse energy (MET) in the event. Due to in-medium energy loss, the MET in such events does not accurately represent the neutrino, required to reconstruct the $W$. We therefore investigate the possibility of using the charged decay lepton instead of a reconstructed $W$. In both cases we require the lepton to have a high \pt$^\mu > 60$ \gev and $|\eta^\mu| < 2.5$, for reconstructed $W$'s the mass window is $60 < M_W < 100$ \gevc. Jets are reconstructed with $R=0.4$ and kinematic cuts \pt$^J > 25$ \gev and $|\eta^J| < 2.1$. We also impost a $\Delta R_{J\mu}>0.6$ to ensure no overlap between our reconstructed jet and lepton collections. 

\begin{figure}[h] 
		   \centering
		   \includegraphics[width=0.45\textwidth]{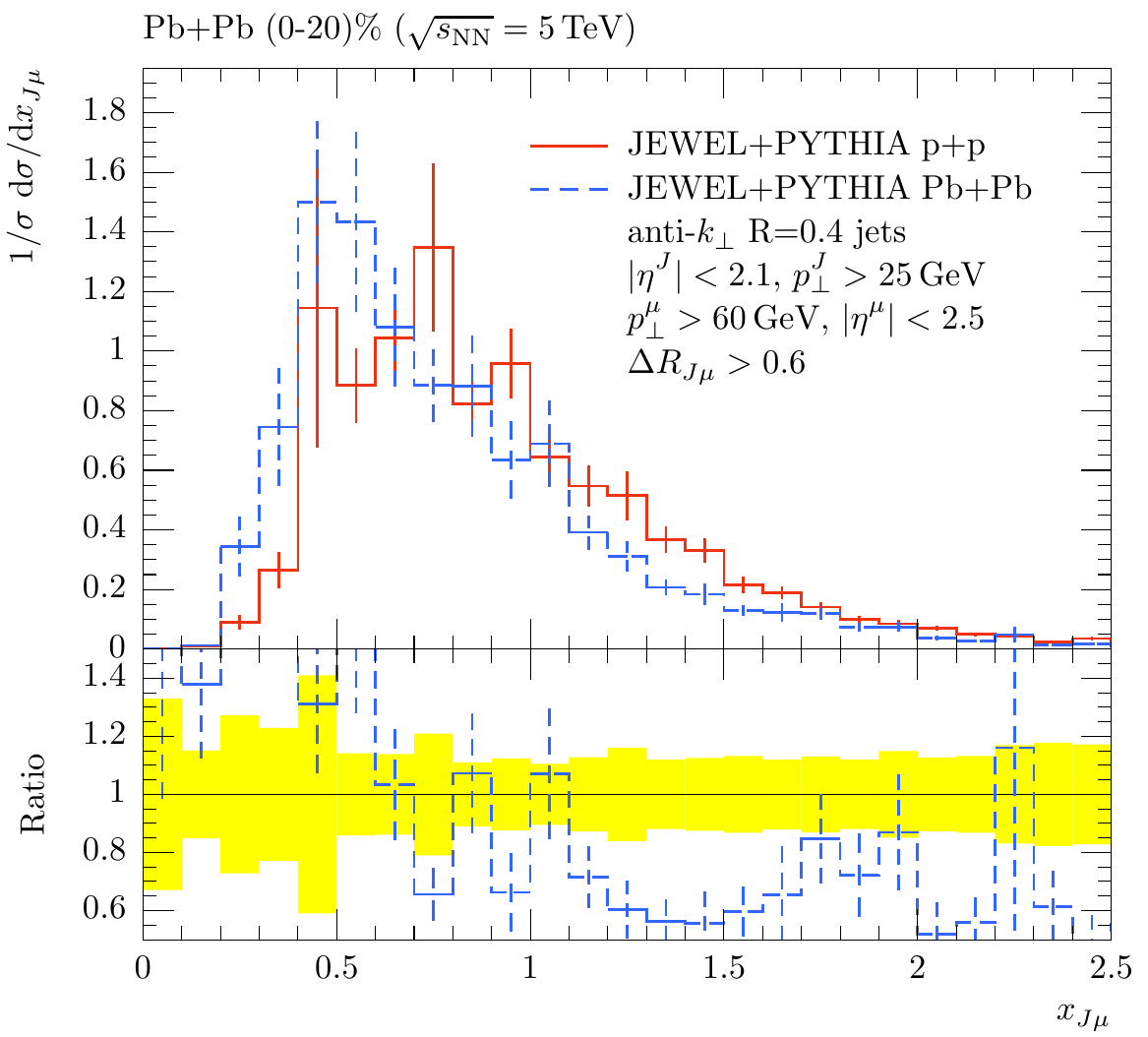} 
		   \includegraphics[width=0.45\textwidth]{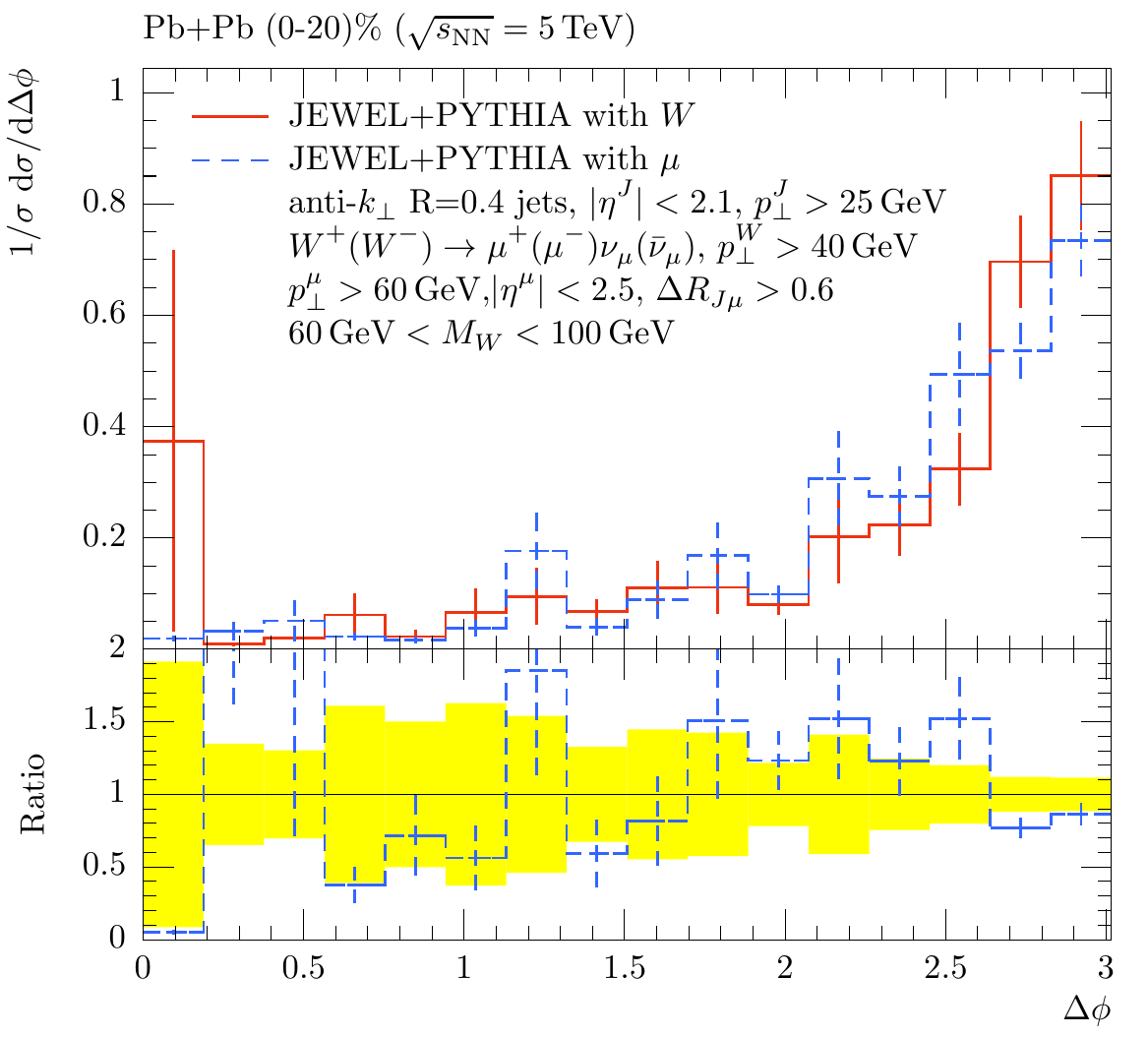} 
		   \caption{Left: Azimuthal angle $\Delta \phi$ between the generator level $W$ and the jet in $W$+jet events compared to the azimuthal angle between the decay muon and the jet in central Pb+Pb events at $\sqrt{s_\text{NN}} = 5.02$ TeV~\cite{KunnawalkamElayavalli:2016ttl}. Right: Momentum imbalance $x_{J\mu}$ with respect to the decay muon in $W$+jet events in p+p and central Pb+Pb collisions at $\sqrt{s_\text{NN}} = 5.02$ TeV. In the ratio plots the dashed blue histogram is divided by the solid red one and the yellow band indicates the uncertainty on the latter.}
		   \label{fig:wpjet_r4}
		\end{figure}

		The left panel of Fig.~\ref{fig:wpjet_r4} shows the $\Delta \phi$ distributions in central ($0-20\%$) Pb+Pb events for the reconstructed jets with the generator level $W^{\pm}$ in the red line and with the leading lepton ($\mu$) in the event in the blue dotted line. We see that the $\Delta \phi$ distribution are similar for the $W^{\pm}$ and leading lepton and therefore we show the transverse momentum imbalance with the leptons. This is shown in the right panel of Fig.~\ref{fig:wpjet_r4} for p+p and central Pb+Pb collisions. Again, there is a clear shift towards larger asymmetries in central Pb+Pb events. 

\begin{figure}[h] 
		   \centering
		   \includegraphics[width=0.7\textwidth]{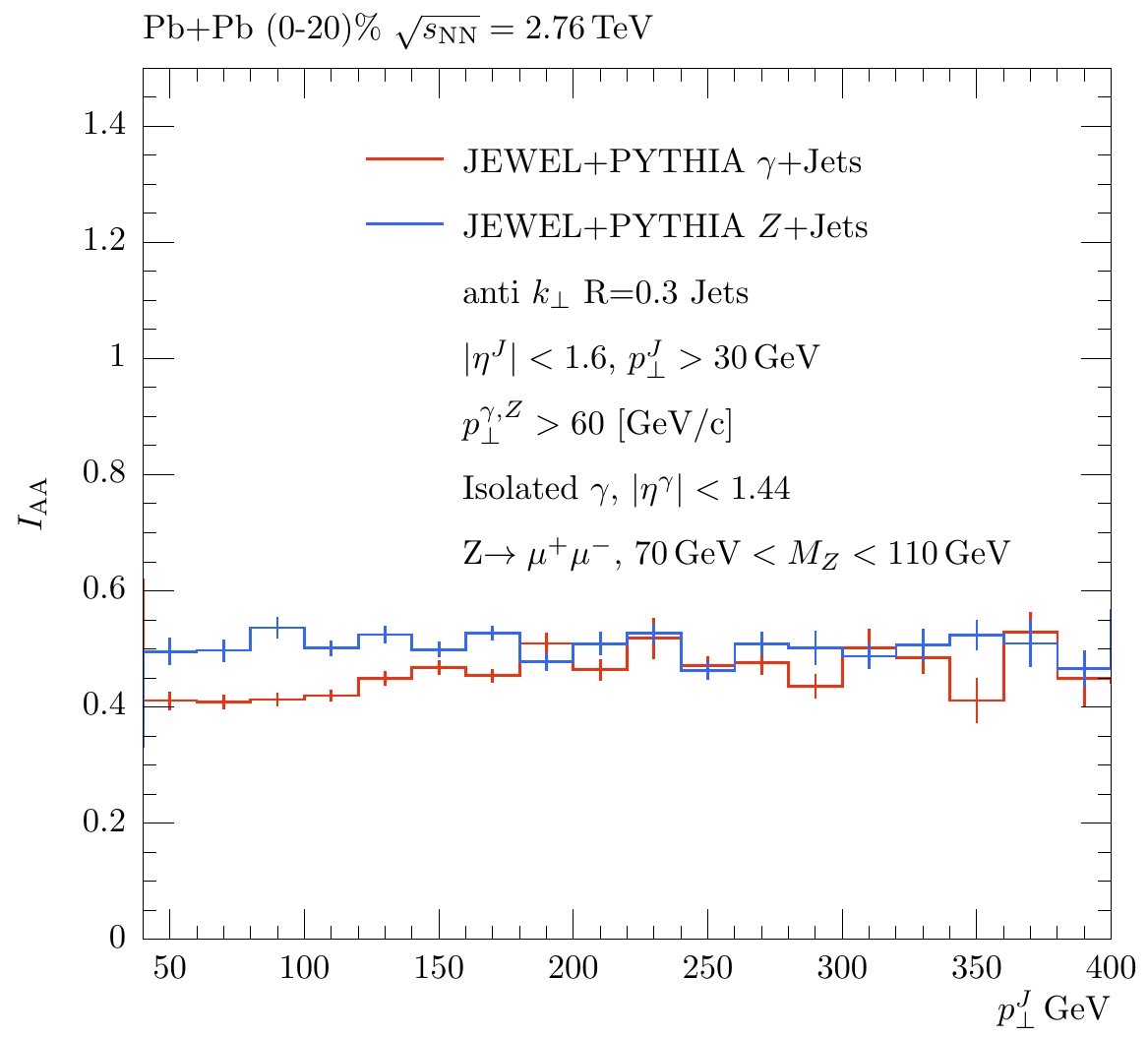} 
		   \caption{Nuclear modification factor of the jet in $Z$+jet (blue) and $\gamma$+jet (red) events in central Pb+Pb events at $\sqrt{s_\text{NN}} = 2.76$ TeV~\cite{KunnawalkamElayavalli:2016ttl}.}
		   \label{fig:zyjet_raa}
		\end{figure}

		It is also informative to look at the nuclear modification factors ($I_{AA}$) of jets in events recoiling against a $\gamma$ or a $Z$. Due to the large mass of the $Z$ boson, the jet spectrum is harder than for jets recoiling off a $\gamma$. This influences the $I_{AA}$ for $Z$+jets to be less suppressed at the low \pt~range as shown in Fig.~\ref{fig:zyjet_raa}. 

\section{Background subtraction in \jw}

	In \jw~the background medium is assumed to consist of an ensemble of partons, the phase space distribution and flavor composition of which have to be provided by an external medium model. Partons belonging to a jet may interact with these background partons through $2\to 2$ scattering processes described by perturbative matrix elements, with associated gluon emission generated by the parton shower. Further details of the inner workings and Monte Carlo implementation of \jw~are available in~\cite{Zapp:2012ak,Zapp:2011ya}.

	As mentioned in the introduction, there are two operational modes for event generation with \jw~concerning the treatment of background partons recoiling against a scattering with the jet (so called ``recoils'' or ``recoiling partons''). Events can be generated with or without storing the recoil information. When run without recoils, the recoiling partons do not show up in the event. In this case no medium response is considered and inclusive and inter-jet observables can be compared to (background subtracted) experimental data. So far this was the recommended mode for jet observables. 

	However, jet structure observables are sensitive to medium response and hence it is desirable to include these effects in \jw~by keeping the recoiling partons in the event. After the scattering these recoiling partons do not interact further with the medium and free-stream towards hadronization. This represents a limiting case for the recoil behavior, that can be regarded as being the limit opposite the assumption of immediate thermalization of recoil energy and momentum made by hydrodynamic frameworks. The truth is expected to be between these two extreme cases, since one would expect that these partons interact further with the medium, but do not necessarily fully thermalize. 

	So far the background partons could be either (anti-)quarks or gluons. For hadronization, however, all recoiling partons are converted to gluons. It is assumed that the recoiling parton is a color neighbor of the hard parton it interacted with. The recoil gluons are thus inserted in the strings connecting the partons forming the jet. Therefore, the hadronic final state including recoiling partons is not an incoherent superposition of jets and activity arising from recoils. At hadron level, it is impossible to assign a certain hadron to the jet or medium response.

	The four-momentum of the recoiling partons has two components: the thermal momentum it had before the interaction with the jet, and the four-momentum transferred from the jet in the scattering process. Only the latter is interesting for investigating medium response, the former is part of the uncorrelated thermal background that is subtracted from the jet. As \jw~generates only the jets and not full heavy ions events, it is not possible to use the experimental background subtraction techniques for the Monte Carlo events. Instead, a dedicated procedure for removing the thermal four-momentum components from the jets when running with recoils has to be devised. Therefore, along with the recoiling partons, we are also storing the thermal four-momenta, which constitute our background\footnote{Technically, this is done by adding one line labeled as comment for each thermal momentum to be subtracted to the HepMC event record.}. These will be systematically removed from the jets during the analysis step, as detailed in the following section.

\section{Subtraction of the thermal component}
\label{sec::submethods}

\begin{figure}[h] 
	   \centering
	   \includegraphics[width=0.9\textwidth]{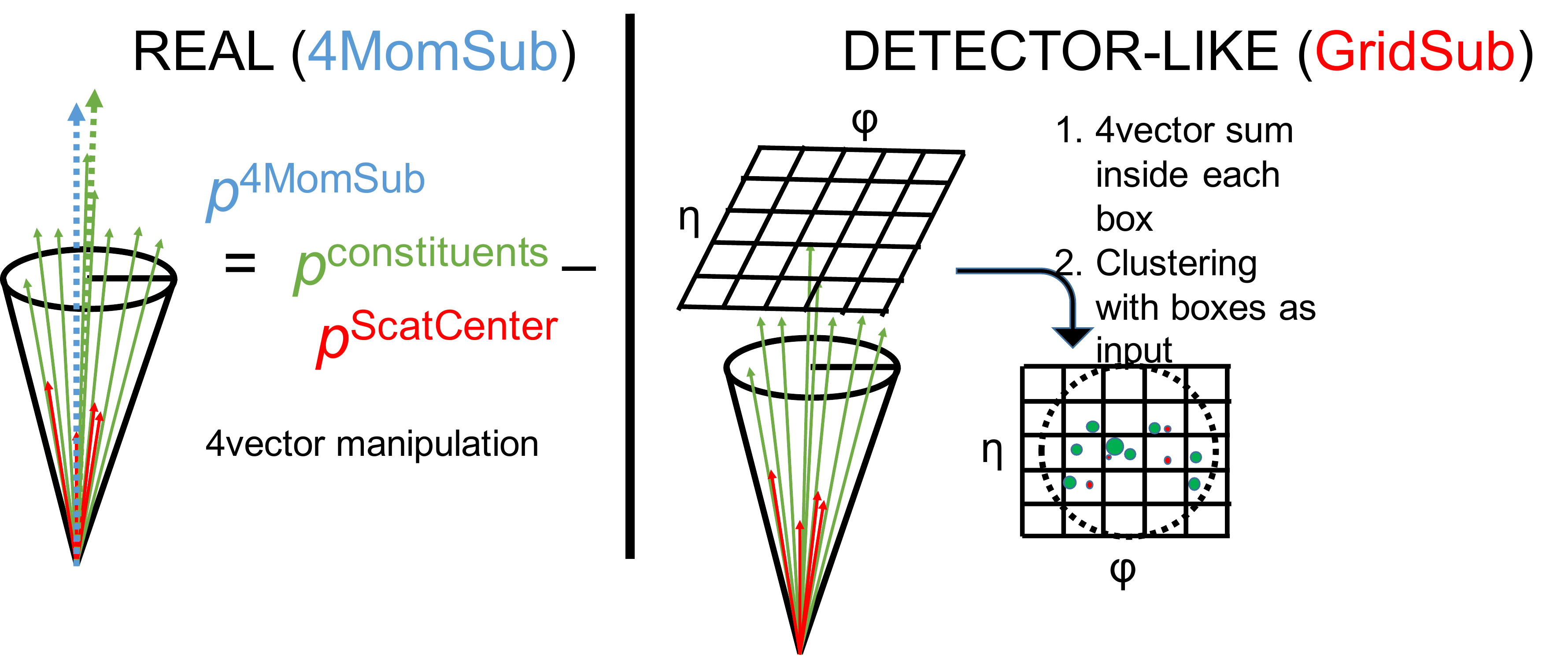} 
	   \caption{Schematic of the two different background subtraction methods utilized in \jw~with 4MomSub on the left and GridSub on the right.}
	   \label{fig:jewelbcksubdiagram}
	\end{figure}

	As discussed in the previous section, in order to compare predictions for jet  observables with data, it is imperative to perform a background subtraction, as shown diagrammatically in Fig:~\ref{fig:jewelbcksubdiagram} on \jw~events when running with the recoils. This is to avoid a mismatch between the prediction and data, since the jets in data have the fluctuating underlying event subtracted. In this section, we present two independent subtraction methods for \jw, that can be employed at the analysis level\footnote{Example \textsc{Rivet}~\cite{Buckley:2010ar} analyses are available for download on the \jw~homepage \url{http://jewel.hepforge.org/}}. 

	\subsection{4MomSub}

		This method removes the thermal momenta exactly from the jet's four-momentum. In order to determine which thermal momenta should be subtracted, an additional set of neutral particles with very small energy and momenta and pointing in the direction of the thermal momenta are added to the final state particles list. These ``dummy" particles are effectively the same as ghosts that \textsc{FastJet}~\cite{Cacciari:2011ma} uses during its clustering to determine the jet area. They can get clustered into jets without affecting the jet's momentum or structure. Thermal momenta, that are matched to a dummy (in the azimuthal angle - pseudorapidity plane) inside a jet, are subtracted from the jet's momentum. The resulting four vector constitutes the subtracted jet momentum. An algorithmic implementation of the procedure is detailed below: 

		\begin{enumerate}
			\item Cluster the initial jet collection from the final state particles (including dummies). 
			\item Compile a list of the thermal momenta (particles in the HepMC event record with status code 3).
			\item For each jet, get the list of thermal momenta that have $\Delta R < 1\cdot 10^{-5}$ with one of the jet constituents, i.e a dummy particle.
			\item Sum up the four-momenta of the matched thermal momenta. This constitutes the background.
			\item For each jet subtract the background four-momentum from the jet's four momentum, this provides the corrected jet collection.
			\item Calculate jet observables from corrected jet four-momenta. 
		\end{enumerate}

This method is easily generalized to subtraction of sub-systems of jets, such as sub-jets or annuli used for the jet profile.

	\subsection{GridSub}
	\label{sec::gridsub}

		This is a generic, observable independent subtraction method. A finite resolution grid (in the $\phi-\eta$ plane) is superimposed on the jet and its constituents. The four-momenta of particles in each cell in the grid are then vectorially summed and thermal momenta subtracted, yielding the cell four-momentum. Finally, we re-cluster the jet with the cell four-momenta as input to the jet clustering algorithm. This method does not require dummy particles. It is also possible to first discretize the entire event, subtract thermal four-momenta cell-by-cell, and then cluster jets. The algorithms for the two variants are given below.

		\textbf{Jet clustering before discretization (GridSub1):}
		\begin{enumerate}
			\item Cluster the initial jet collection from the final state particles. 
			\item Compile a list of the thermal momenta (particles in the HepMC event record with status code 3).
			\item Define the grid resolution and place grid over jets.
			\item Inside each cell sum the jet constituents' four-momenta and subtract the thermal four-momenta that fall into the cell (note: no matching is required, thermal four-momenta with distance $\Delta R < $ jet radius are considered\footnote{Alternatively, when dummy particles are written to the event record, one can also match thermal momenta and dummies to decide which momenta should be included.}), providing a single four momentum for each cell.	
			\item In case a cell contains more thermal momentum than jet constituents, the cells is set to have zero four-momentum. This is deemed to be the case when the (scalar) \pt~of the thermal component is larger than the \pt~of the particle component.
			\item Re-cluster the jets with the cell four-momenta as input to get the final, subtracted jets.
			\item Calculate jet observables from re-clustered jets. 	
		\end{enumerate}
		This version is the default.

		\textbf{Discretization before jet clustering (GridSub2):}
		\begin{enumerate}
			\item Compile a list of the thermal momenta (particles in the HepMC event record with status code 3).
			\item Define the grid resolution and place grid over the entire event.
			\item Inside each cell sum the final state particles' four-momenta and subtract the thermal four-momenta that fall into the cell (note: no matching is required), providing a single four momentum for each cell.
			\item In case a cell contains more thermal momentum than particle momentum, the cells is set to have zero four-momentum. This is deemed to be the case when the (scalar) \pt~of the thermal component is larger than the \pt~of the particle component.
			\item Cluster the jets with the cell four-momenta as input to get the final, subtracted jet.
			\item Calculate jet observables. 	
		\end{enumerate}

		Due to the finite size of the grid, it is possible to have certain cells with more thermal momentum than particle momentum, resulting in a total negative four-momentum, which in our case is set to zero before clustering. Thus, the GridSub method systematically removes less background from the jet than 4MomSub. The smearing introduced by the GridSub method will be quantified systematically in the following section.

		The use of the 4MomSub method is recommended when possible, since it does not introduce finite-resolution effects and is consequently more accurate.

	\subsection{Limitations of the subtraction and the issue of track jets}
	\label{sec:limitations}

		Since the subtraction techniques introduced above subtract the thermal momenta, which are at parton level, from the hadronic final state, they only yield meaningful results for observables that are insensitive to hadronization effects. This is the case for most infra-red safe observables based on calorimetric jets. Examples for observables that do not fall into this category are fragmentation functions and charged jet observables. In general, all cuts on the final state particles, also \pt~cuts, are problematic.

		A few of the recent experimental results involve the use of charged or track jets~\cite{Abelev:2013kqa,Cunqueiro:2015dmx,Acharya:2017goa}, i.e\ jets reconstructed using only tracks. When the subtraction is naively applied,  the techniques end up overestimating the contribution of the four-momenta to subtract. Thus, in order to compare with such experimental results, a heuristic procedure is applied. The observable of interest is calculated for full jets and re-scaled. The re-scaling between the full and the charged jet distribution is extracted from the corresponding \jw~simulation for p+p collisions.  If it is larger than the resolution of the observable, it is applied to the full jet subtracted distribution in Pb+Pb. In this way an estimate of the charged jet distribution is derived. For example, a naive way of estimating the charged jet four-momentum is by re-scaling the full jet quantity with the fraction of charged particles in the jet. The charged jet mass distribution discussed in section~\ref{sec::jetshapes} is estimated using this technique and compared with data. In other cases, for instance the jet radial moment girth (also shown in  section~\ref{sec::jetshapes}), the distributions for charged and full jets are the same in p+p collisions. In this case we compute the observable for full jets in Pb+Pb collisions as well and do not apply any re-scaling.

		Obviously, this method comes with an additional uncontrolled systematic uncertainty, since it is not guaranteed that the same relation between full and charged jet distribution holds in Pb+Pb and p+p.

\section{Systematic studies}
\label{sec::systematics}

	The background subtraction techniques introduced in the previous section and their effects on jets are studied henceforth in a systematic fashion. 

	\subsection{Smearing due to finite resolution of the grid}

		\begin{figure}[h!] 
			\centering
			\includegraphics[width=0.47\textwidth]{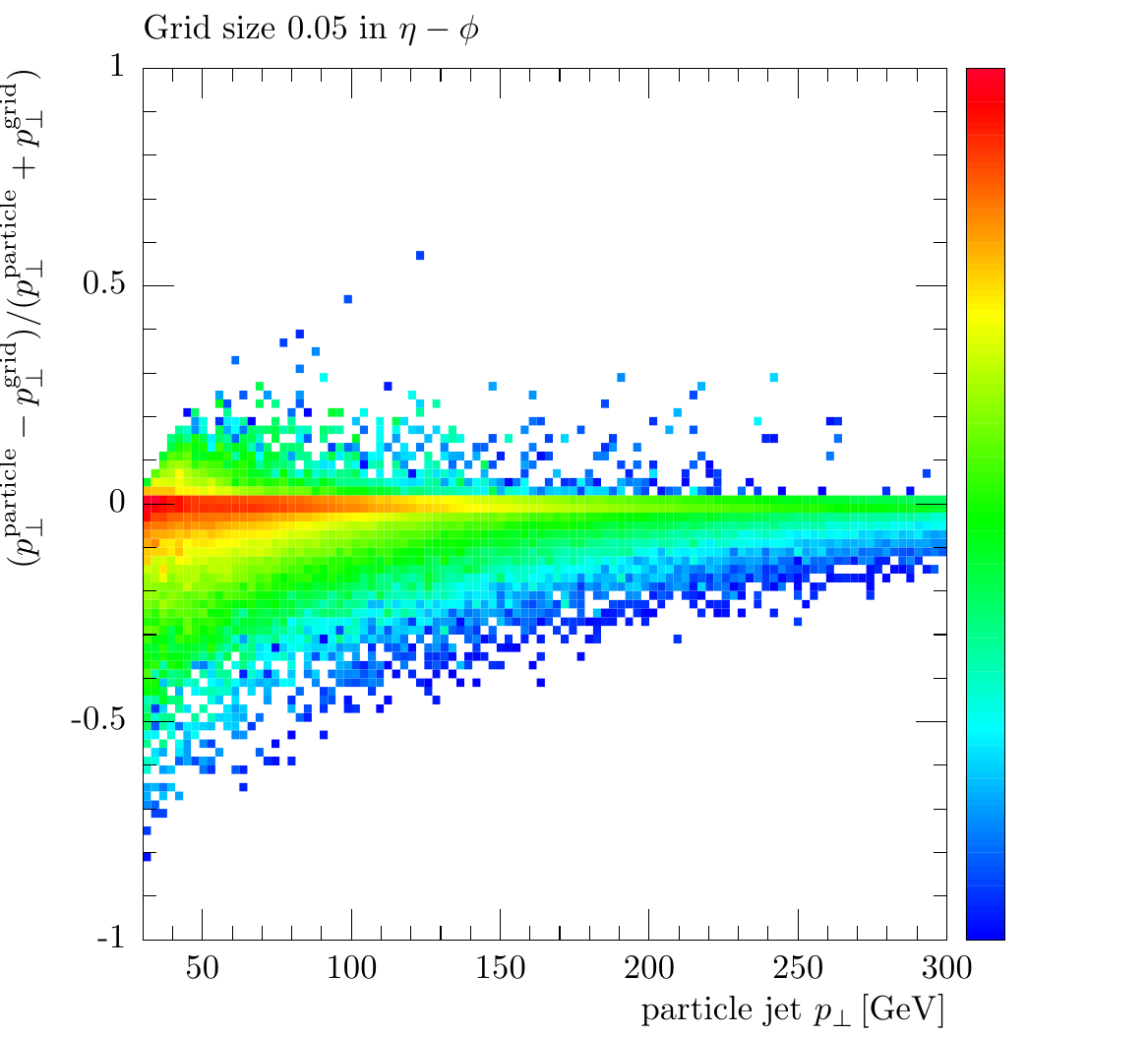} 
			\includegraphics[width=0.47\textwidth]{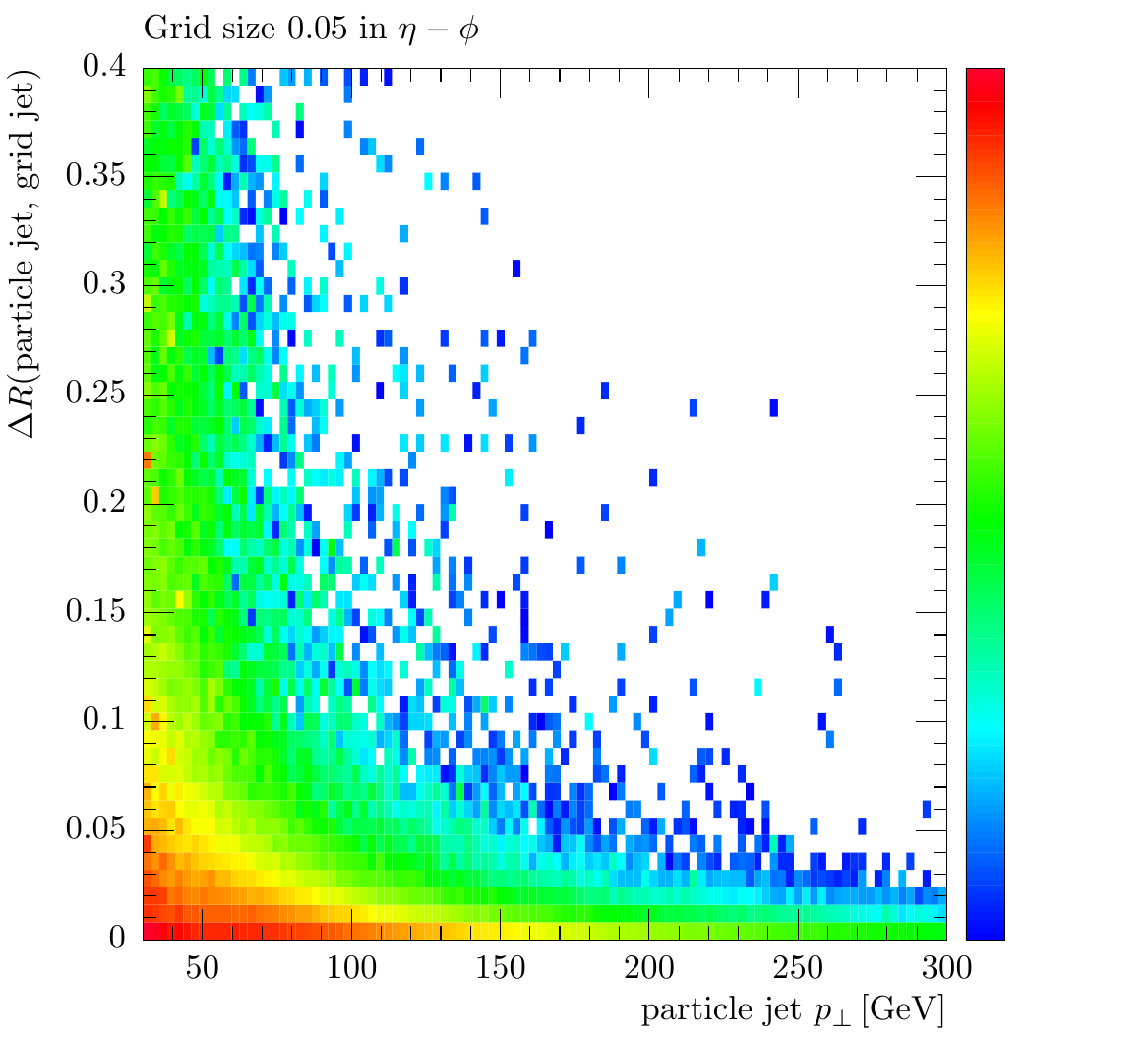} 
			\includegraphics[width=0.47\textwidth]{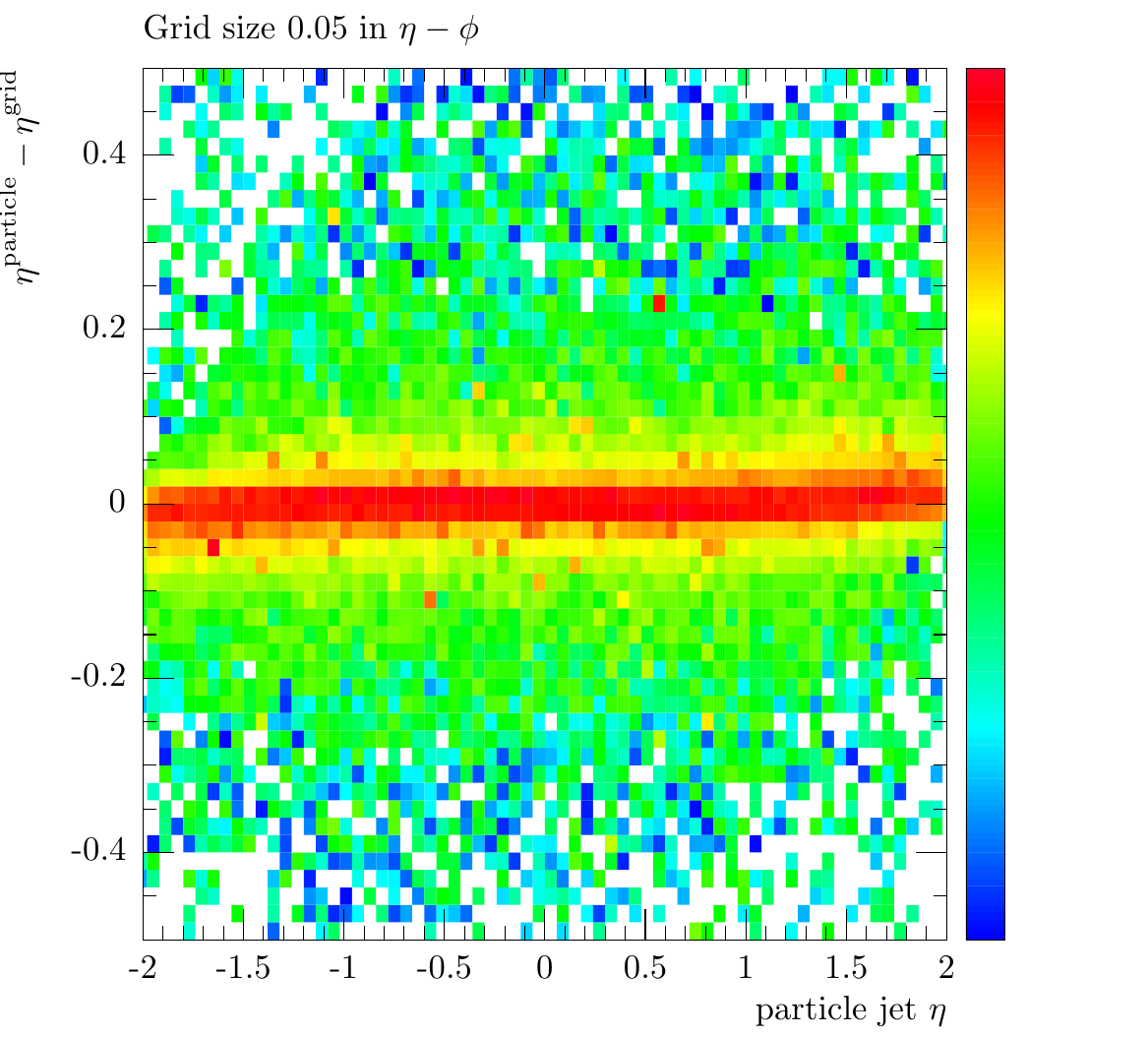} 
			\includegraphics[width=0.47\textwidth]{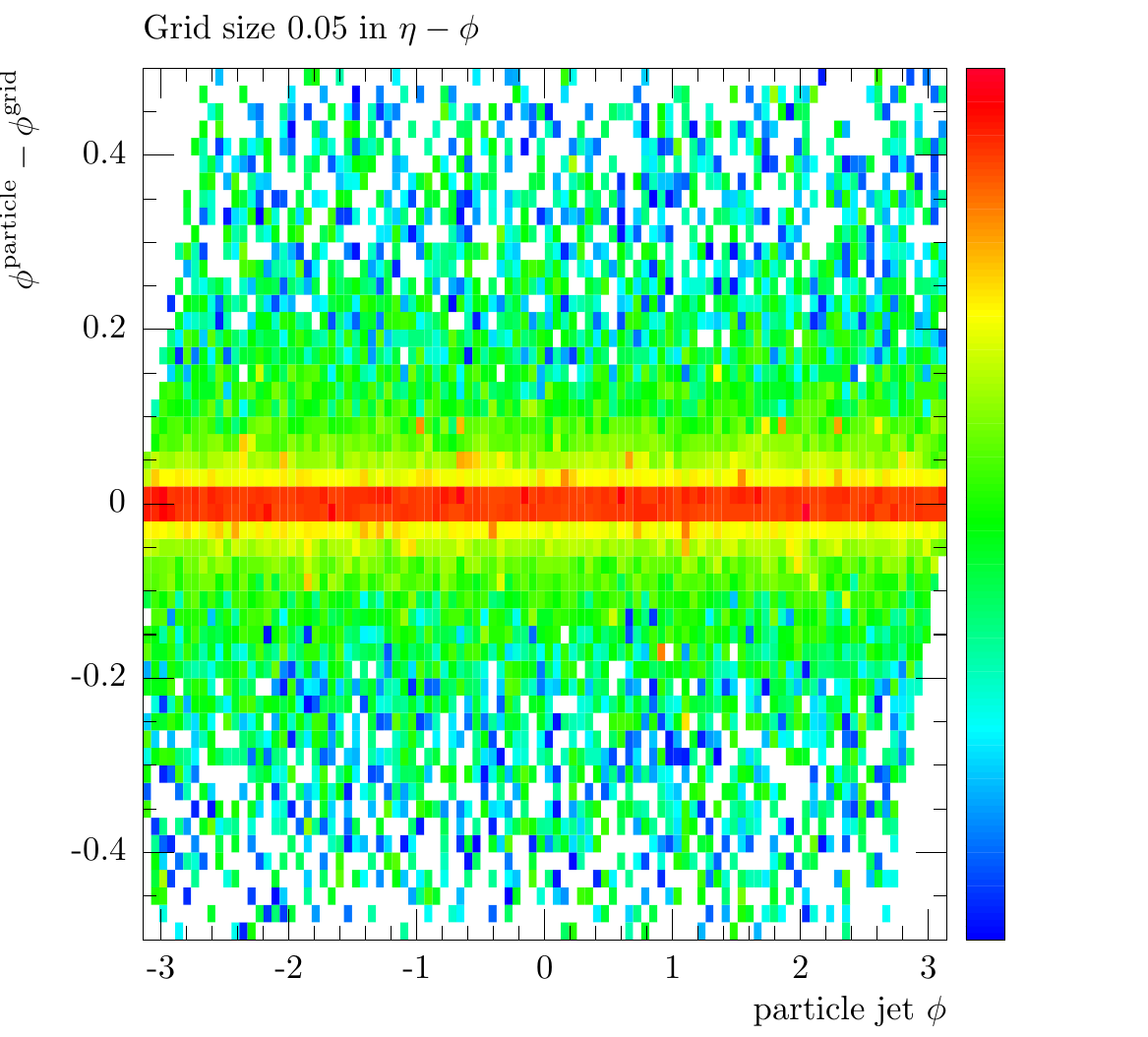} 
			\caption{Smearing introduced by the grid on the jets, quantized by the smearing in jet \pt~(top left) and the absolute shift of the jet axis in the $\eta-\phi$ plane (top right) shown as a function of the particle jet \pt. Bottom plots show the relative shifts in jet $\eta$ (left) and $\phi$ (right), shown as a function of the respective particle jet $\eta$ and $\phi$. (Note that the log scales on the z axis span six orders of magnitude.)}
			\label{fig:gridjetsmearing}
		\end{figure}

		An immediate consequence of the grid, before any subtraction is introduced, is that both the jet \pt~and the position of the jet in the $\eta-\phi$ plane are smeared. This effect is studied in \jw~with p+p events generated at hadron level to highlight the inherent behavior. All our studies of the grid are shown for a nominal grid size of $0.05$ in $\eta-\phi$ plane, which we find to be a good compromise between resolution and under-subtraction (which is more severe for smaller cell sizes). The systematic uncertainties are estimated by varying the grid size by a factor of two and most final observables are shown to be quite insensitive to the grid size within these limits.

		In each event, jets are first reconstructed from the final state hadrons. Then the event is discretized using a grid and jets are reconstructed based on the grid cells. Finally, each jet of the smaller of the two collections is matched to the one from the other set that is closest in $\Delta R$, with the constraint that $\Delta R$ is smaller than the reconstruction radius (this is the standard CMS procedure for comparing generator level jets to jets after detector simulation). The smearing is quantified in Fig.~\ref{fig:gridjetsmearing} with the top panels showing the smearing in jet \pt~(on the left) and jet axis (on the right) as a function of the particle jet \pt. The latter is broken down into the respective shifts in $\eta$ and $\phi$, which are shown in the bottom panels. The deviations are observed to be small in the \pt~range studied here. There is a clear trend for the grid jet \pt~to be larger than the corresponding particle jet \pt, which is due to the fact that the effective area of the grid jets can be larger due to the discretization. As one would expect, increasing the jet \pt~reduces the smearing introduced by the grid. 

		\begin{figure}[h!] 
			\centering
			\includegraphics[width=0.47\textwidth]{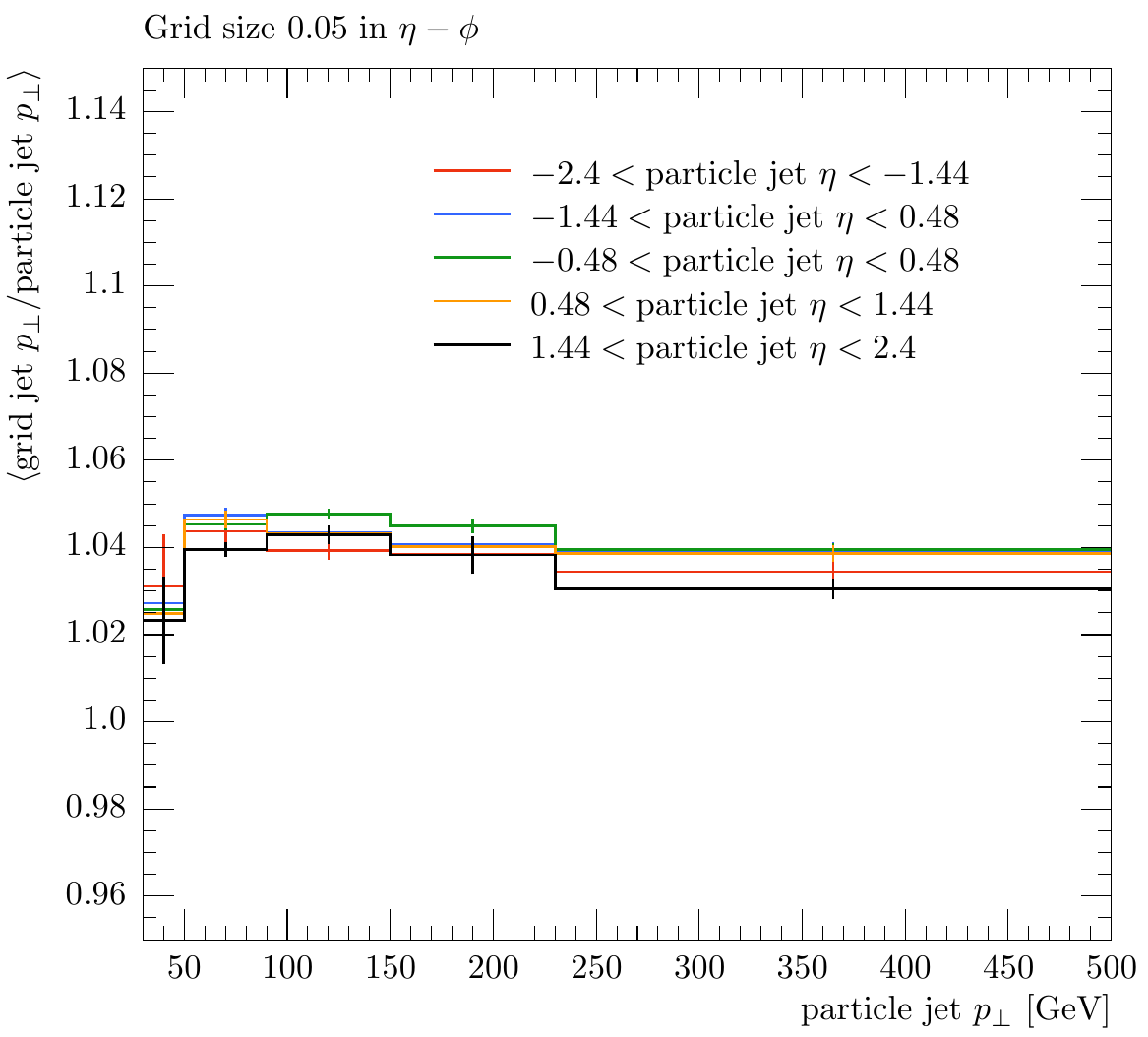} 
			\includegraphics[width=0.47\textwidth]{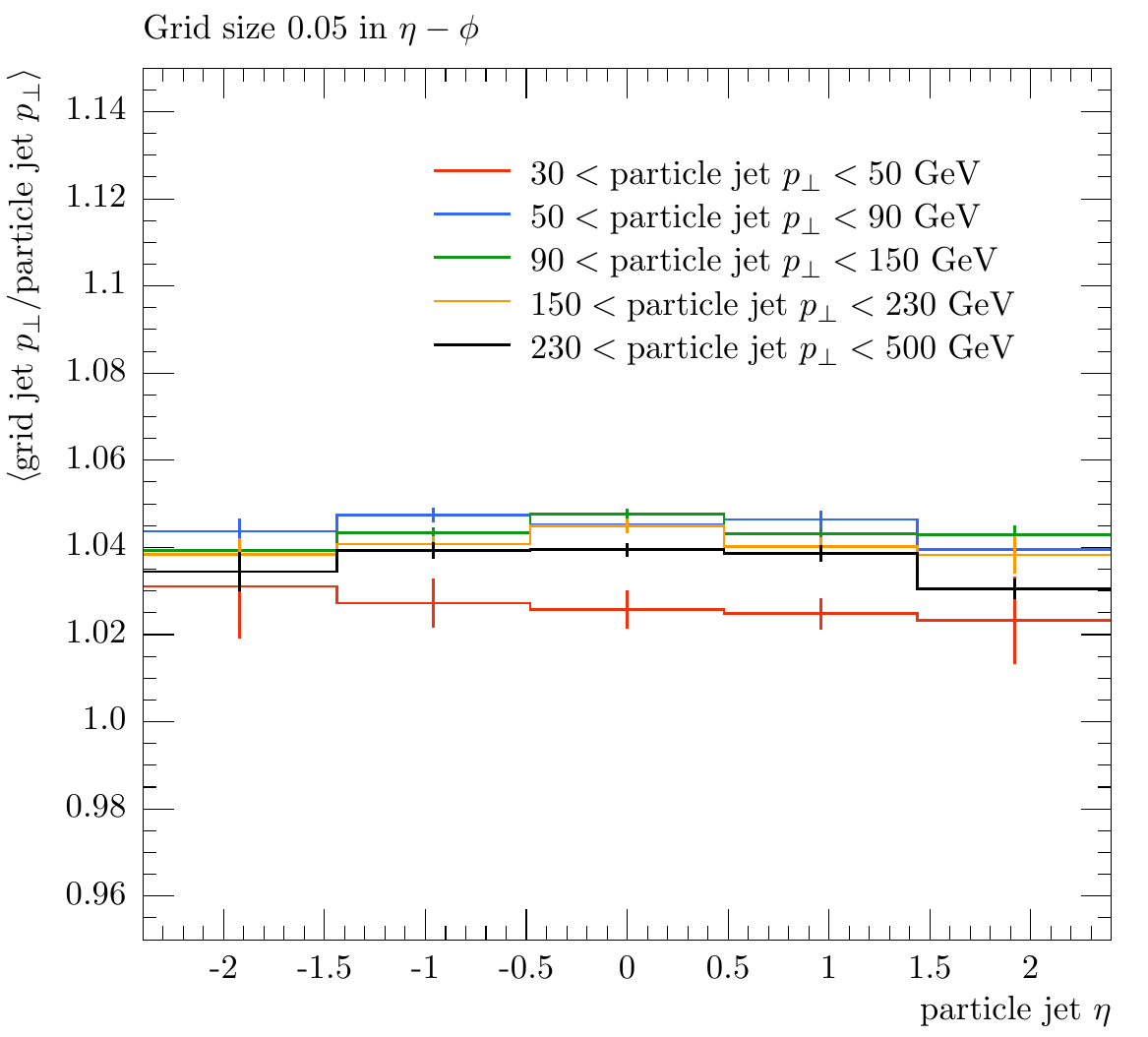} 
			\caption{Average shift in the jet \pt~introduced by the discretization as a function of the particle jet \pt~and $\eta$, respectively.}
			\label{fig:jes}
		\end{figure}		

		In Fig.~\ref{fig:jes} the \pt~shift seen in Fig.~\ref{fig:gridjetsmearing} is quantified. The ratio between grid jet \pt~and the particle jet \pt~is seen to be around 1.04 and thus reasonably close to unity, and largely independent of jet \pt~and $\eta$ for \pt$^\text{jet} > 50$ \gev. Such shifts usually are corrected in experiments~\cite{Chatrchyan:2011ds,Berta:2016ukt} by introducing detector level correction factors as a function of the jet \pt~and $\eta$. In this paper, GridSub jets are not corrected for this shift in their \pt, since it is reasonably small. Also, it partially cancels when looking at ratios of Pb+Pb with p+p due to its independence on jet kinematics. Furthermore, since the mismatch is related to nearby jets, increasing the jet \pt~cut leads to a reduction of the effect.

	\subsection{Under-subtraction due to cells with negative energy}

		\begin{figure}[h!] 
			\centering
			\includegraphics[width=0.47\textwidth]{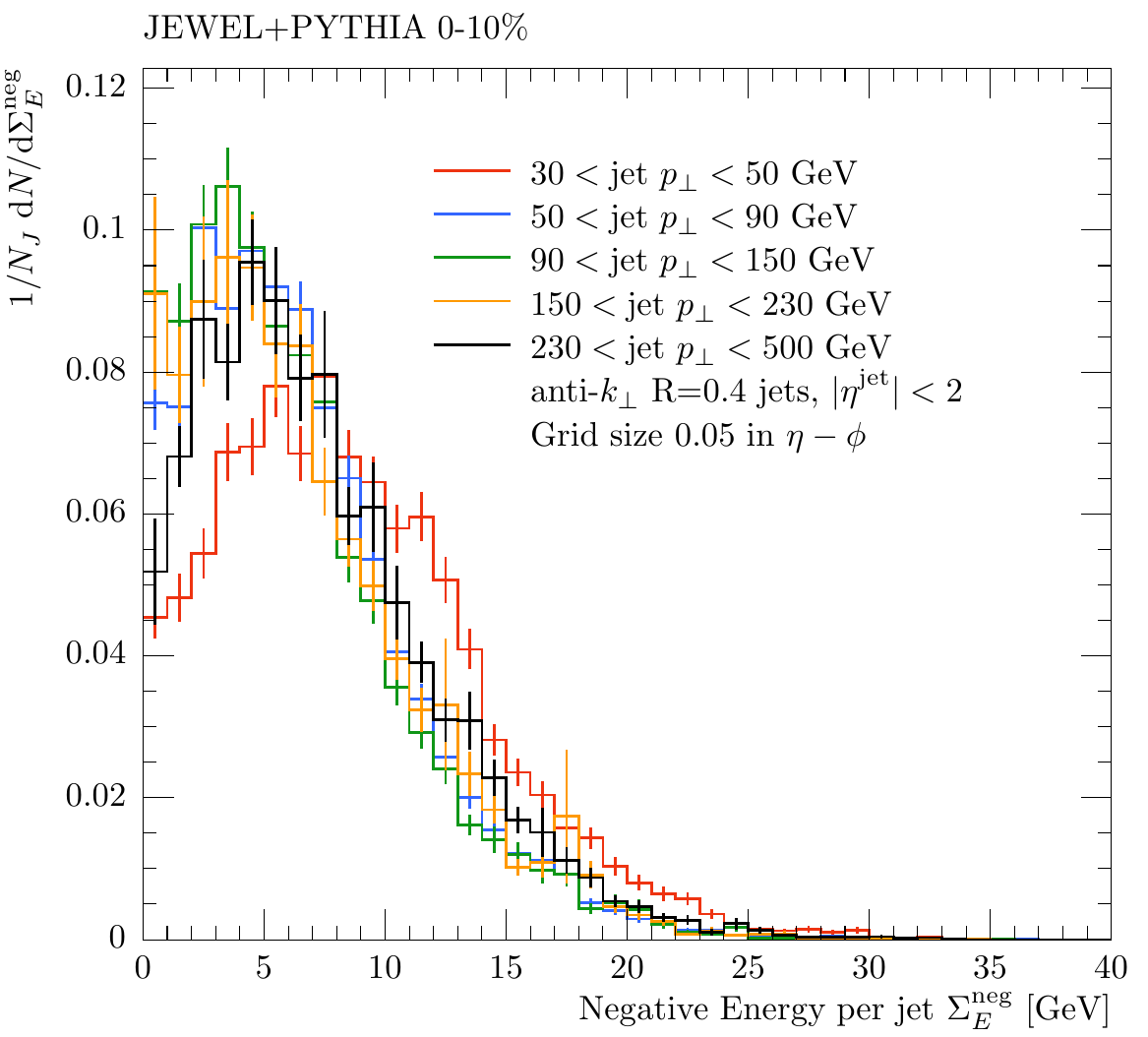} 
			\caption{Total negative energy per jet introduced by the GridSub technique shown for different jet \pt~bins  for central Pb+Pb ($0-10\%$) \jwpy~events generated with recoils.}
			\label{fig:gridnegativeenergy}
		\end{figure}		

		As previously mentioned, the GridSub technique sets the cell's four momentum to zero if it contains more thermal than particle momentum. This leads to a systematic under-subtraction, that increases with decreasing cell size. We quantify this effect using the event sample with medium response included. Jets are reconstructed and subtracted using the default grid subtraction, but here we keep track of the energy of cells whose four-momentum is set to zero. For each jet we then check if it contains such cells and sum the (negative) energy that these cells originally had. The sum of the negative energy per grid jet is shown in Fig.~\ref{fig:gridnegativeenergy} for different jet \pt~ranges. The contribution of negative energy, i.e the amount of thermal energy that remains un-subtracted from the jet, is largely independent of the jet \pt~(except for the lowest \pt~bin) and small compared to the jet \pt~over most of the covered \pt~range. 

	\subsection{Comparison of two GridSub versions}

		\begin{figure}[h!] 
			\centering
			\includegraphics[width=0.47\textwidth]{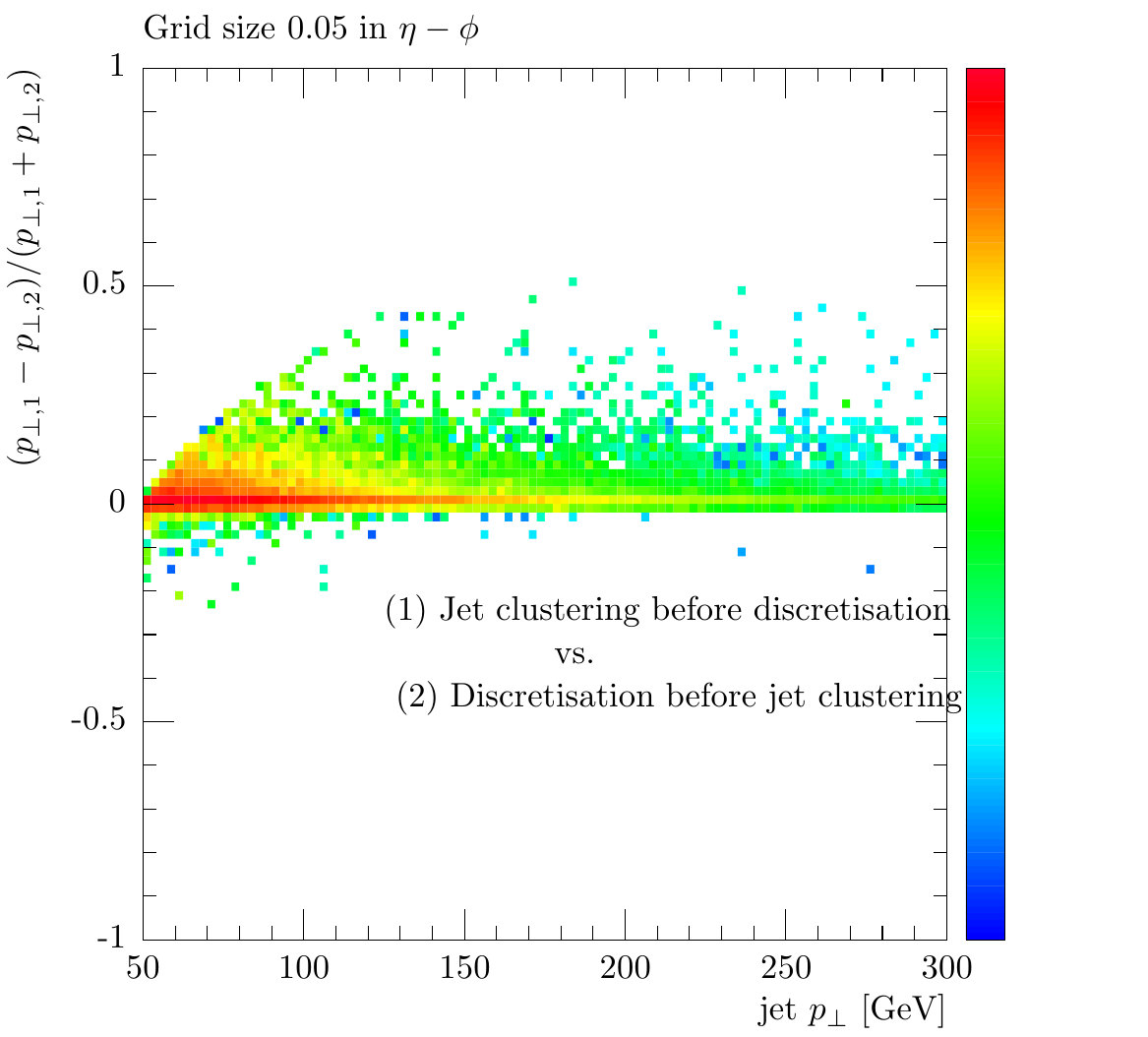} 
			\includegraphics[width=0.47\textwidth]{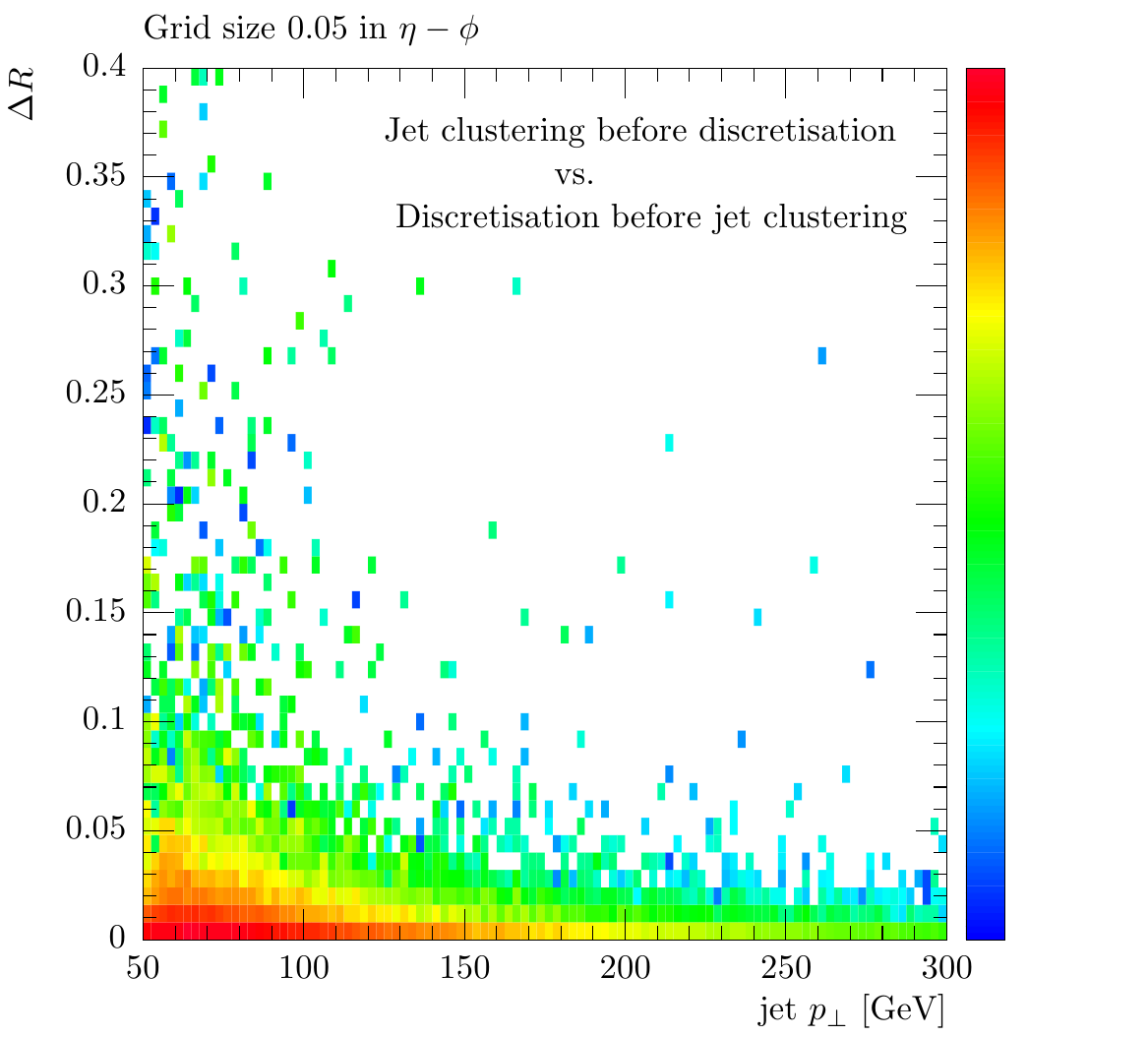} 
			\caption{Relative \pt~difference (left) and shift of jet axis (right) between the two GridSub versions ((1) jet clustering before discretisation and (2) discretisation before jet clustering).}
			\label{fig:asybkgsubjetfinding}
		\end{figure}

		As discussed in section~\ref{sec::gridsub} we have implemented two versions of the grid based subtraction, that differ in the order of jet clustering and discretization. It is to be expected that the two versions yield different results, as there is no reason why the two operations should commute. Using again the hadron level event sample with medium response included we quantify the differences between the versions. To this end, we find and subtract jets with both versions and event-by-event match the jets following the procedure detailed above. The relative difference in jet \pt~and shift of the axis due to the different ordering of operations is shown in Fig.~\ref{fig:asybkgsubjetfinding}. Both these effects are determined to be quite small, but the jet \pt~is consistently larger, when the initial jet clustering is performed before the discretization of the event.

	\subsection{Effects on jet \pt~with 4MomSub and GridSub subtraction}

		\begin{figure}[h!] 
			\centering
			\includegraphics[width=0.47\textwidth]{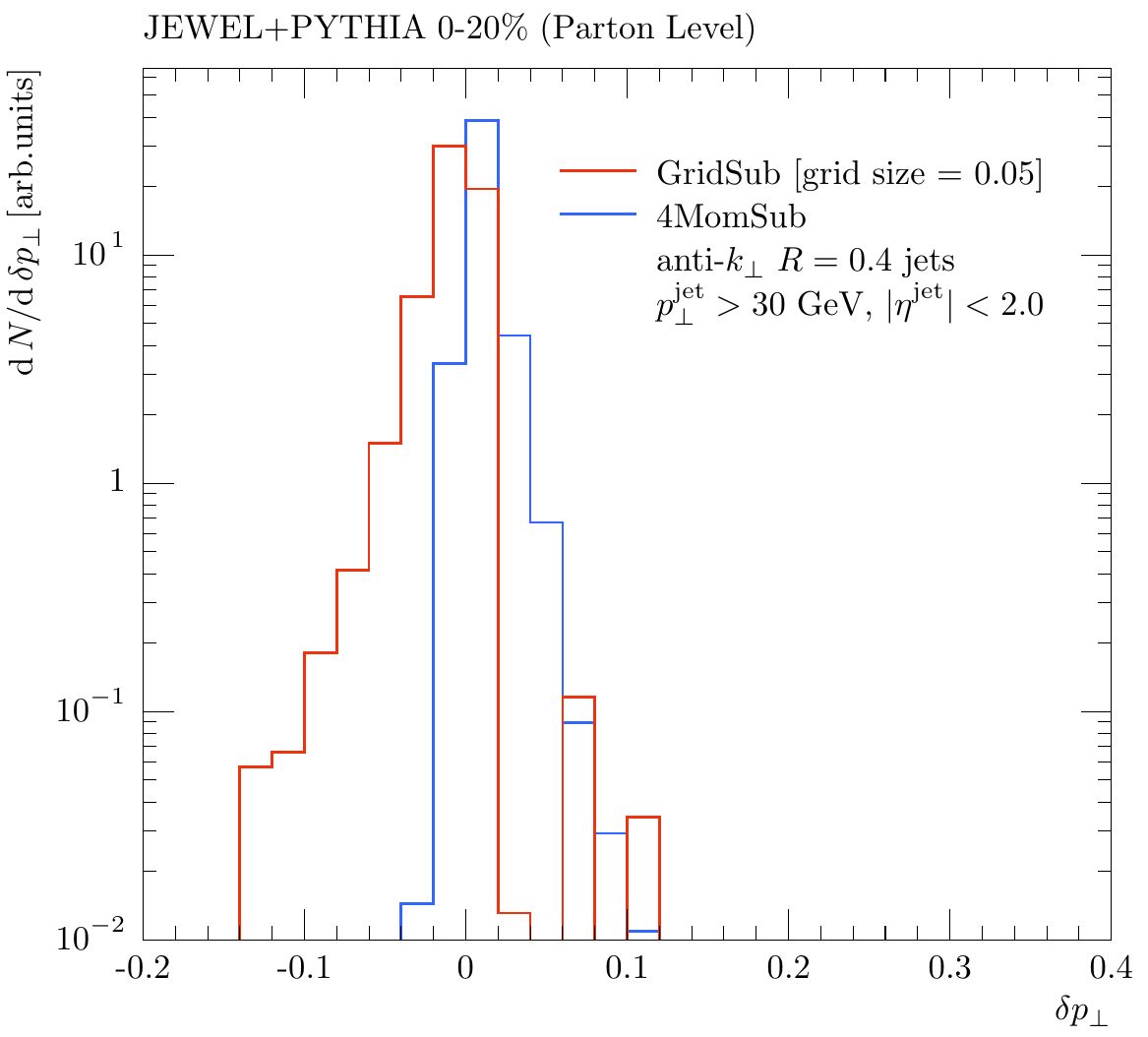} 
			\caption{Relative \pt~difference $\delta$ \pt~$= ($\pt$^\text{w/o\ rec} - $\pt$^\text{w\ rec})/($\pt$^\text{w/o\ rec} + $\pt$^\text{w\ rec})$ between parton level jets reconstructed with and without recoiling partons for the two subtraction methods.}
			\label{fig:subAsym_partonLevel}
		\end{figure}

		A final check of the two subtraction methods (4MomSub and GridSub1) is done at parton level, where the same jets can be reconstructed with and without recoiling partons. The subtraction is performed with either of the two methods and the matching procedure is again the same as before. 

		Fig.~\ref{fig:subAsym_partonLevel} shows the relative \pt~difference between jets reconstructed with and without recoiling partons. As expected, the 4MomSub distribution is narrower compared to GridSub1, due to additional jet smearing introduced by the discretization of the event into cells of a finite size. Additionally, the 4MomSub distribution has a tail on the positive side. This is a momentum conservation effect: the thermal distribution is isotropic (except for the longitudinal boost), while the recoiling partons have a net momentum in direction of the jet due to momentum conservation. Therefore, when including medium response more momentum is added to the jet than is subtracted. This is a physical effect that is independent of the subtraction method, but for the GridSub method the shoulder is towards the negative side. This is due to the aforementioned nature of the GridSub to under-subtract the jets, which overcompensates the momentum conservation effect.  

\section{Application to traditional jet quenching observables}	
\label{sec::tradobs}	

	\begin{figure}[h!] 
		\centering
		\includegraphics[width=0.47\textwidth]{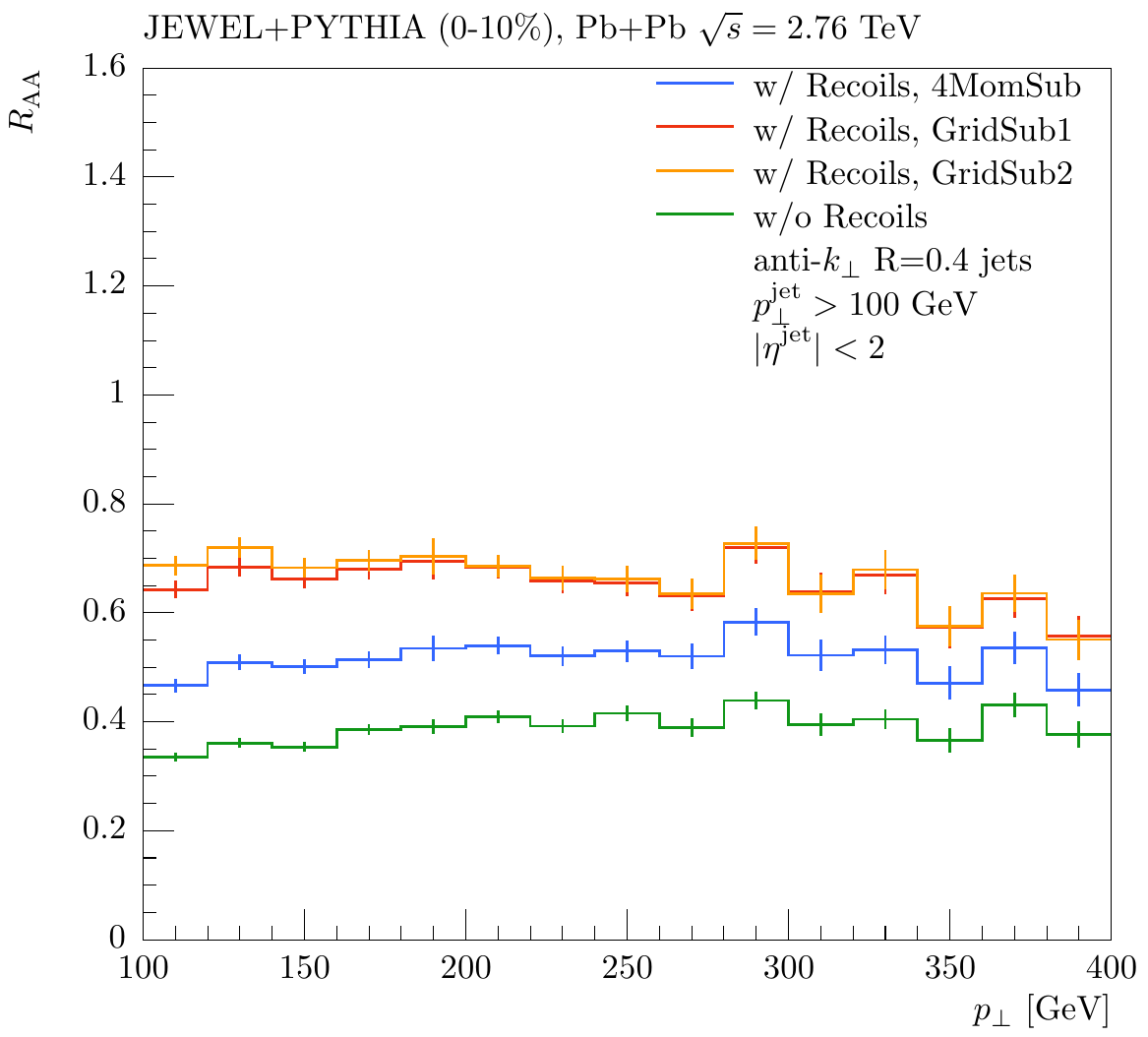} 
		\caption{Inclusive jet nuclear modification factor \raa for Pb+Pb central events in \jwpy. The green line represents \jwpy~without medium response while the blue, red and orange lines show the result including medium response with 4MomSub, GridSub1 and GridSub2 respectively.}
		\label{fig:effect_bkgsub}
	\end{figure}		

	Observables built from the jet \pt~and axis, such as jet \raa or the di-jet asymmetry $A_J$, for smaller radii jets typically show a rather mild sensitivity to medium response. The jet axis is dominated by the hard jet components and for the jet \pt~the only effect of medium response is a partial recovery of lost energy. For small reconstruction radii, this is at best a moderate effect, while for very large radii, such as $R \approx 1.0$, the effect becomes sizable. For such large radii also the systematic uncertainties related to the subtraction become large. Experimentally, the study of such large jets in a heavy ion environment constitutes an almost impossible task of discriminating between underlying event and the jets. For small radii jets at small momenta the same problem persists, which is why different experiments utilize different procedures to remove the effect of the underlying event in the jet collection of interest~\cite{Adam:2015ewa,Aaboud:2017bzv,Kodolova:2007hd}.  

	As our primary validation, Fig.~\ref{fig:effect_bkgsub} shows the nuclear modification factor \raa of jets, i.e the ratio of jet yield in Pb+Pb over binary collisions scaled p+p, for a moderate radius of $R=0.4$. As expected, including medium response leads to a small increase of \raa over the entire jet \pt~range. The grid based subtraction leads to a significantly larger increase. This reflects the under-subtraction of the GridSub method discussed in section~\ref{sec::systematics}. Increasing the cell size leads to a reduction of \raa. There is good agreement between the two versions of the grid subtraction.

	\begin{figure}[h!]
		\centering
		\includegraphics[width=0.47\textwidth]{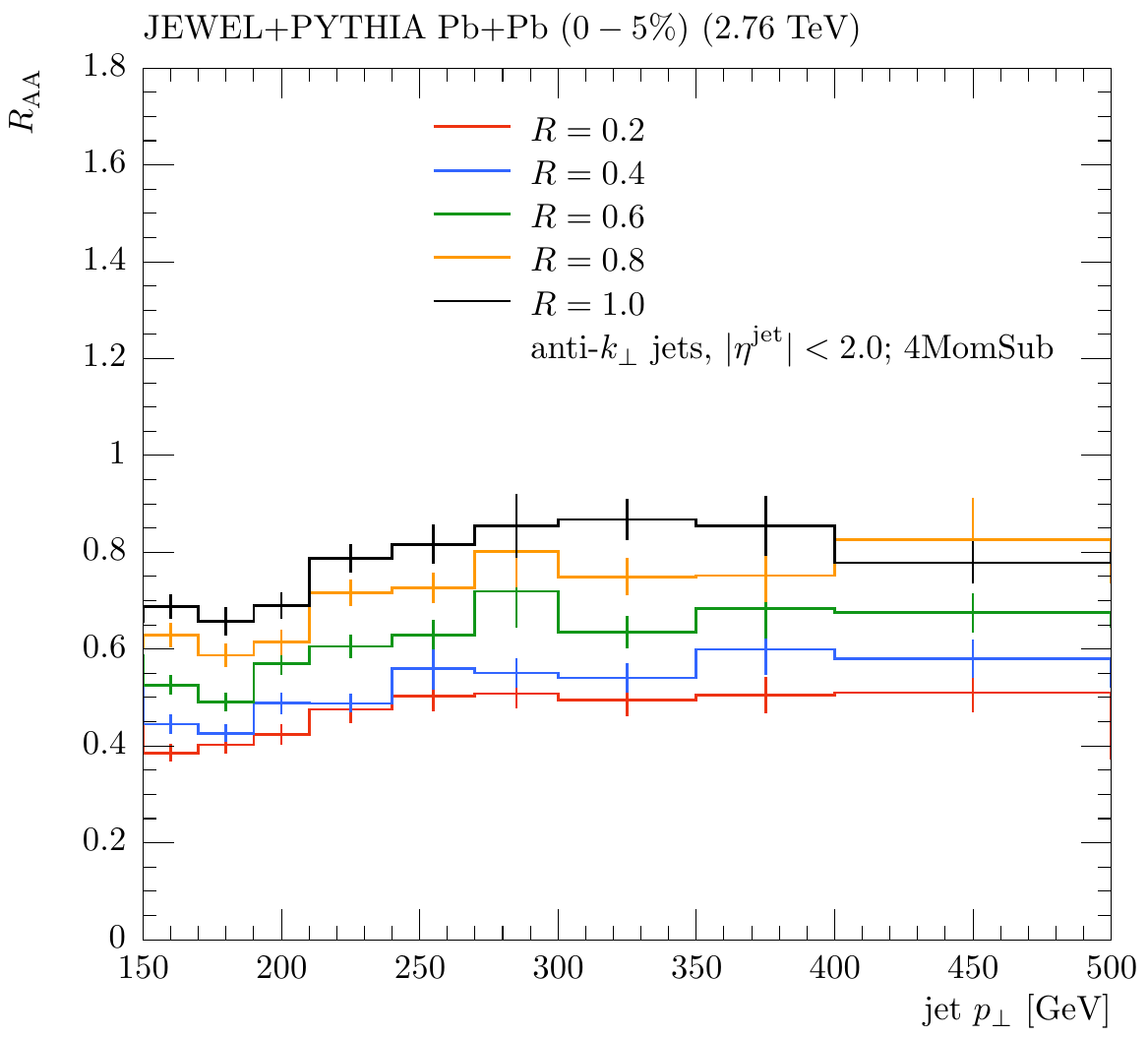} 
		\caption{Inclusive jet nuclear modification factors \raa in Pb+Pb central events in \jwpy~for different jet radii $R$ and including medium response with 4MomSub.}
		\label{fig:RAAdifferentR}
	\end{figure}		

	The jet radius dependence of \raa is shown in Fig.~\ref{fig:RAAdifferentR} with medium response and 4MomSub. The expected increase of \raa with $R$, because with increasing jet radius more and more of the lost energy is recovered, is indeed observed\footnote{This is in contrast to the behaviour observed in~\cite{Casalderrey-Solana:2016jvj}, where \raa decreases with increasing jet radius because wider jets are more easily lost and medium response cannot compensate this loss.}.

	\begin{figure}[h!]
		\centering
		\includegraphics[width=0.47\textwidth]{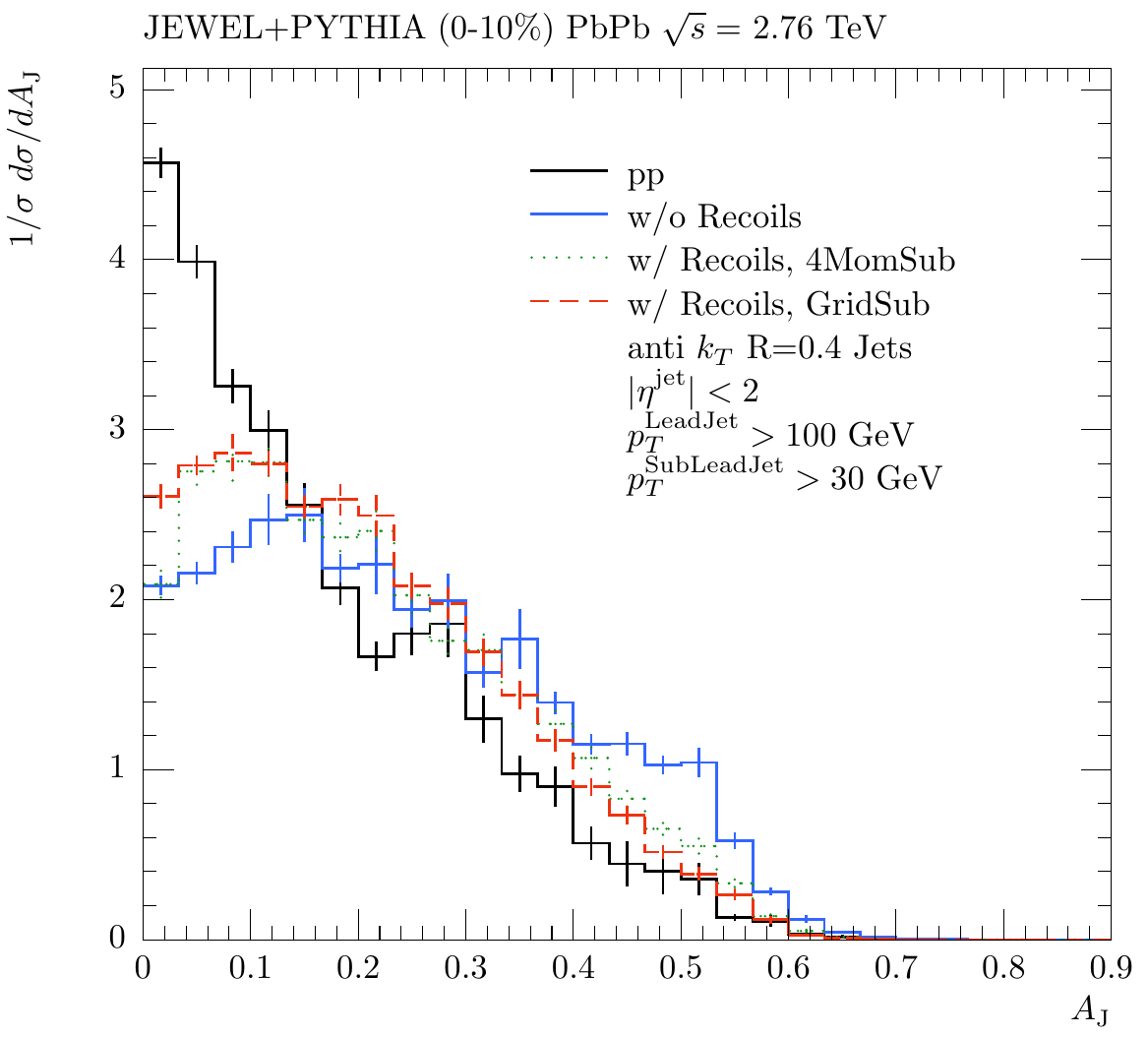} 
		\caption{Di-jet momentum asymmetry $A_J = (p_{T}^\text{LeadJet} - p_{T}^\text{SubLeadJet})/(p_{T}^\text{LeadJet} + p_{T}^\text{SubLeadJet})$ for central Pb+Pb central events in \jwpy. The green line represents \jwpy~without medium response while the blue and red lines show the result including medium response with 4MomSub and GridSub1 respectively.}
		\label{fig:Aj}
	\end{figure}		

	\begin{figure}[h!]
		\centering
		\includegraphics[width=0.47\textwidth]{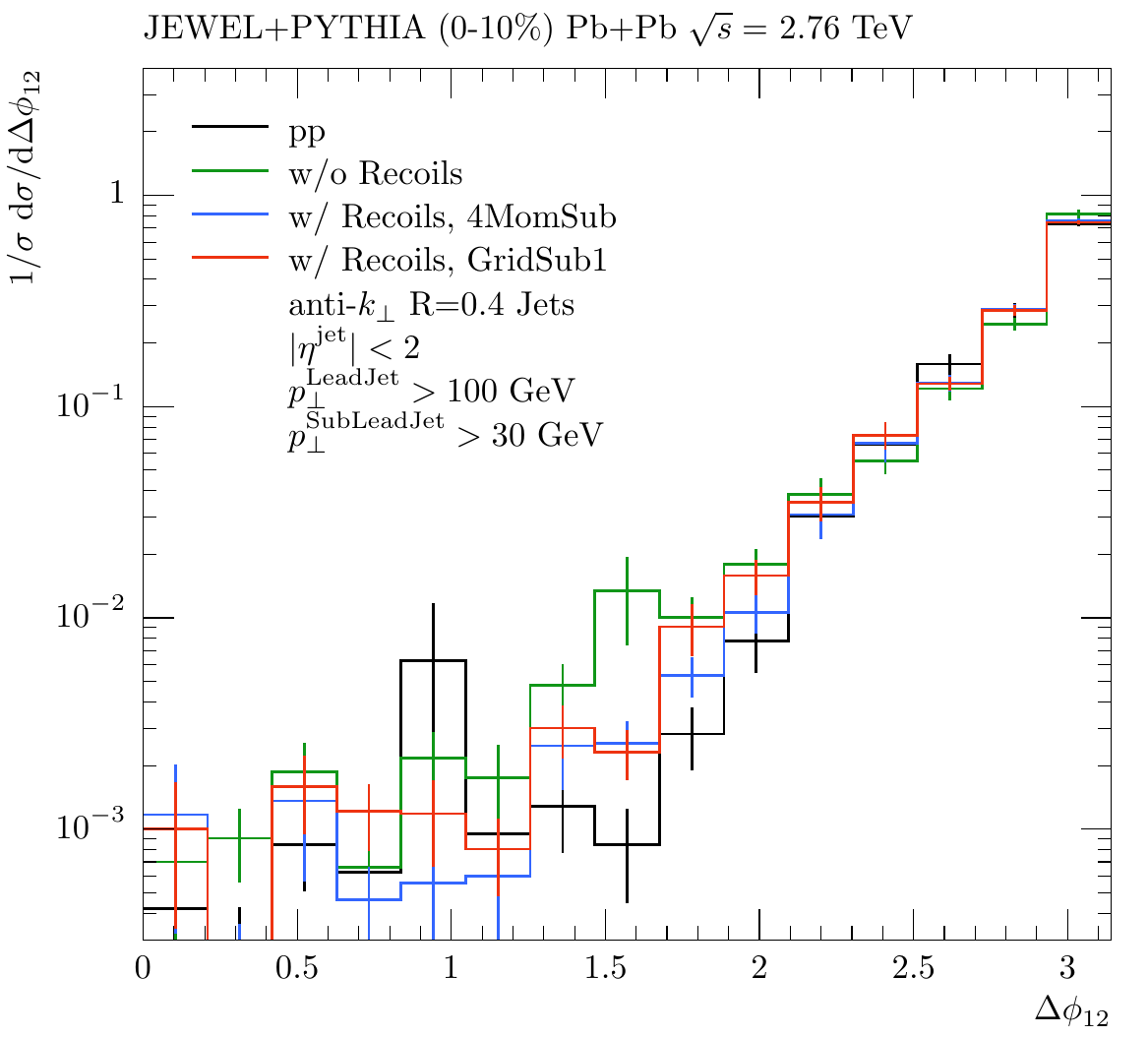} 
		\caption{Di-jet relative azimuthal angle $\Delta \phi_{12}$ for central Pb+Pb central events in \jwpy. The green line represents \jwpy~without medium response while the blue and red lines show the result including medium response with 4MomSub and GridSub1 respectively.}
		\label{fig:Dphi}
	\end{figure}		

	Figs.~\ref{fig:Aj} and ~\ref{fig:Dphi} show the di-jet momentum asymmetry 
	\begin{equation}
		A_J = \frac{p_{T}^\text{LeadJet} - p_{T}^\text{SubLeadJet}}{p_{T}^\text{LeadJet} + p_{T}^\text{SubLeadJet}}
	\end{equation}
	and relative azimuthal angle $\Delta \phi_{12}$, respectively. Here, the leading jet is required to have $p_{T}^\text{LeadJet} > 100$ \gev and the cut on the sub-leading jet is \pt$^\text{SubLeadJet} > 30$\gev \footnote{Analysis cuts are always applied after subtraction.}. The momentum asymmetry $A_J$ is calculated without $\Delta \phi_{1,2}$ cut. The jet axis and thus $\Delta \phi_{1,2}$ are unaffected by medium response, while in the case of $A_J$ it leads to a mild reduction of the medium modification obtained without medium response. 

\section{Application to jet shape observables}
\label{sec::jetshapes}

	In contrast to the observables discussed in the previous section, that aim at characterizing global properties of jet events, jet shape observables are sensitive to the momentum distribution inside the jet. The latter are thus more affected by medium response. The energy in QCD jets is very much concentrated towards the jet axis, while medium response leads to a much broader distribution of relatively soft activity. Also the fluctuations of the two components are different. In this section we discuss a number of jet shape observables and how they are affected by medium response in \jw.

	\subsection{Jet mass}

		\begin{figure}[h!] 
			\centering
			\includegraphics[width=0.47\textwidth]{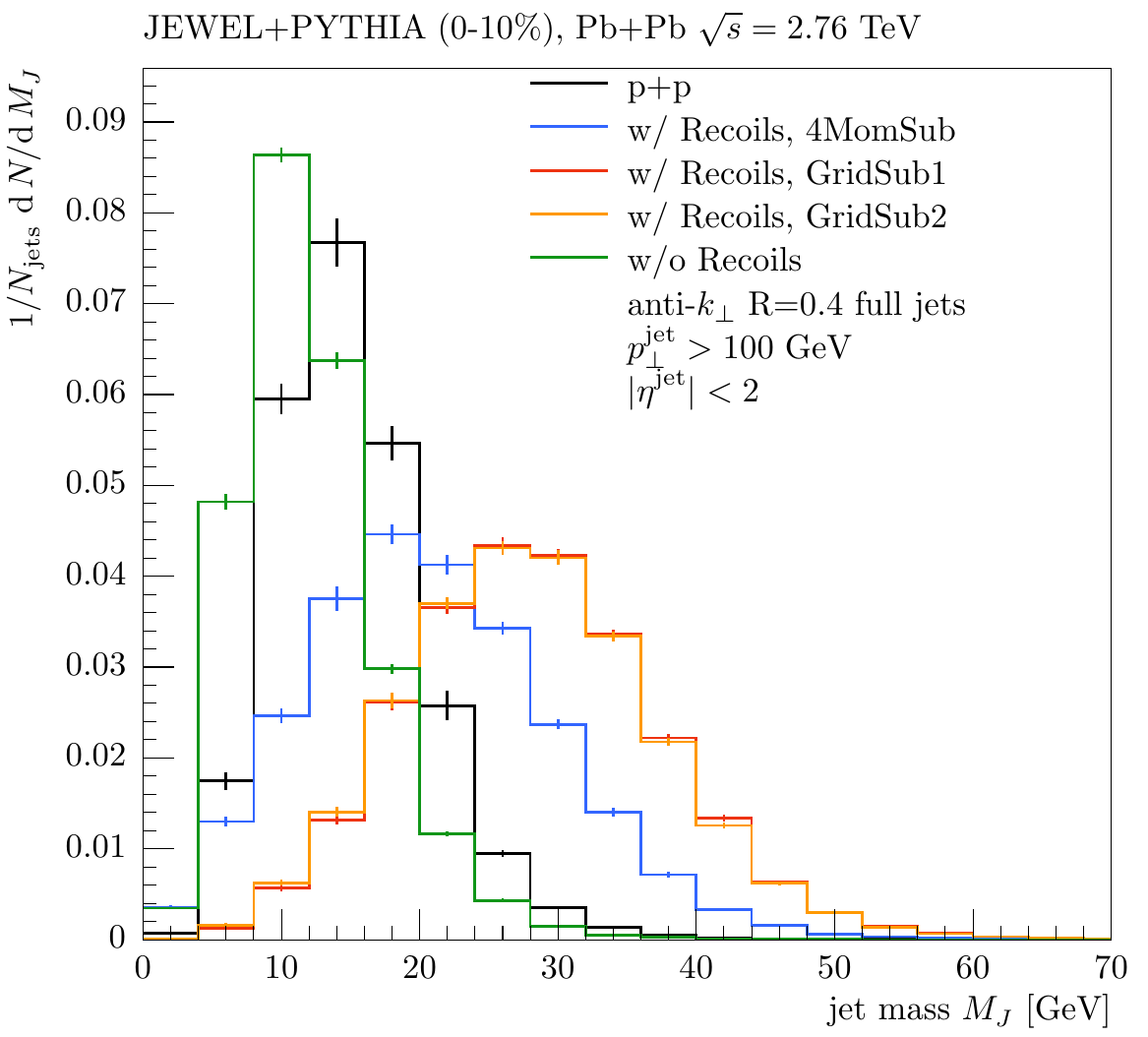} 
			\caption{Jet mass distributions in central Pb+Pb events for \akt full jets with radius parameter $R = 0.4$ and \pt$^\text{jet} > 100$ \gev. The black line represents the mass in corresponding p+p collisions, while the green line is for \jwpy~without medium response and the blue, red and orange lines correspond to \jwpy~including medium response with 4MomSub, GridSub1 and GridSub2 subtraction, respectively.} 
			\label{fig:JetMass}
		\end{figure}

		\begin{figure}[h!] 
			\centering
			\includegraphics[width=0.47\textwidth]{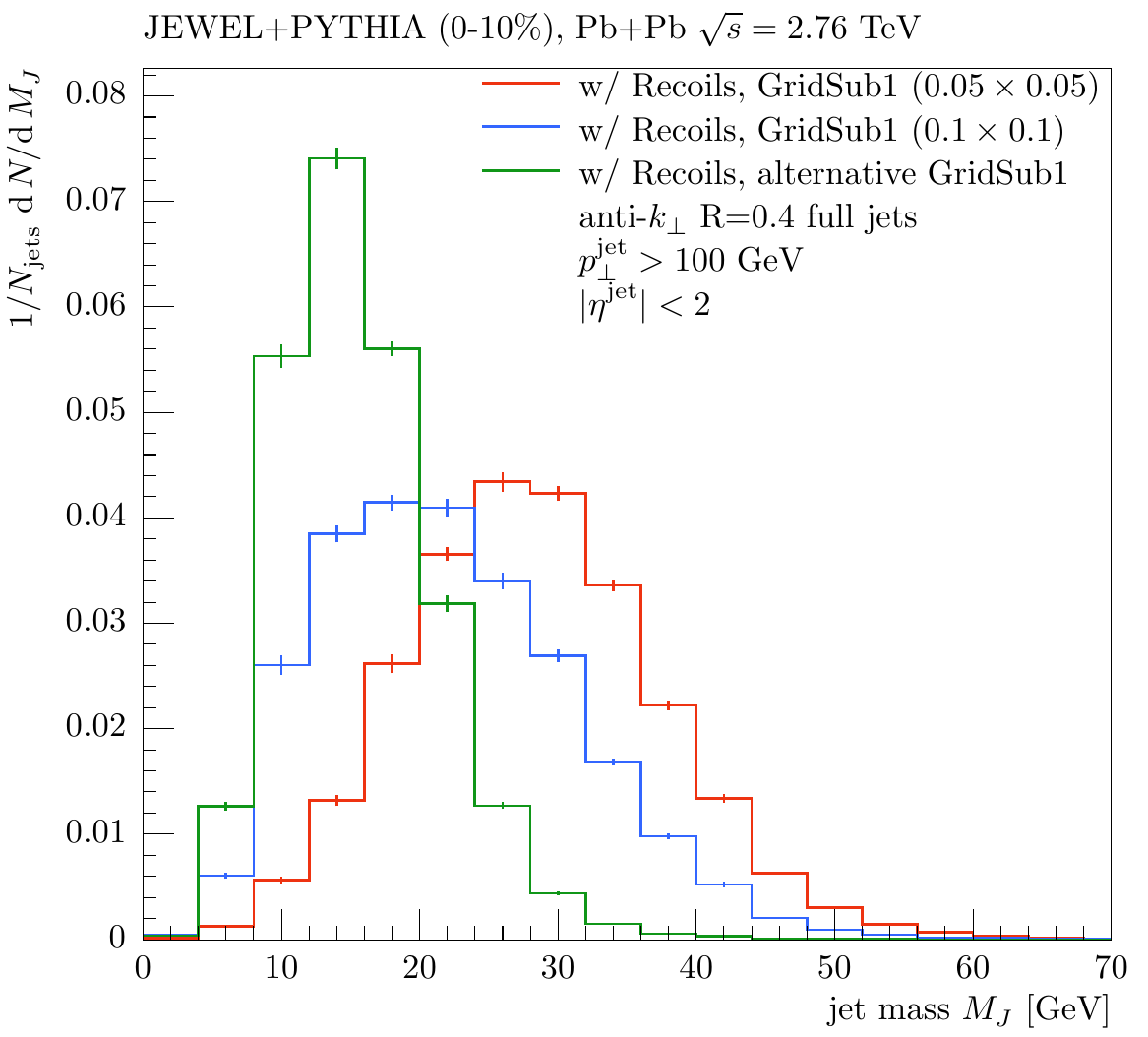} 
			\caption{Jet mass distributions in central Pb+Pb events for \akt full jets with radius parameter $R = 0.4$ and \pt$^\text{jet} > 100$ \gev with medium response and variations of the GridSub1 subtraction. The red histogram is the default (with cell size $0.05 \times 0.05$), in the blue the cell size is increased to $0.1 \times 0.1$, and the green is with default cell size but instead of four-momenta the energies of particles inside the cells are summed and the cell momentum is assumed to be massless.} 
			\label{fig:gridJetMass}
		\end{figure}

		\begin{figure}[h!] 
			\centering
			\includegraphics[width=0.47\textwidth]{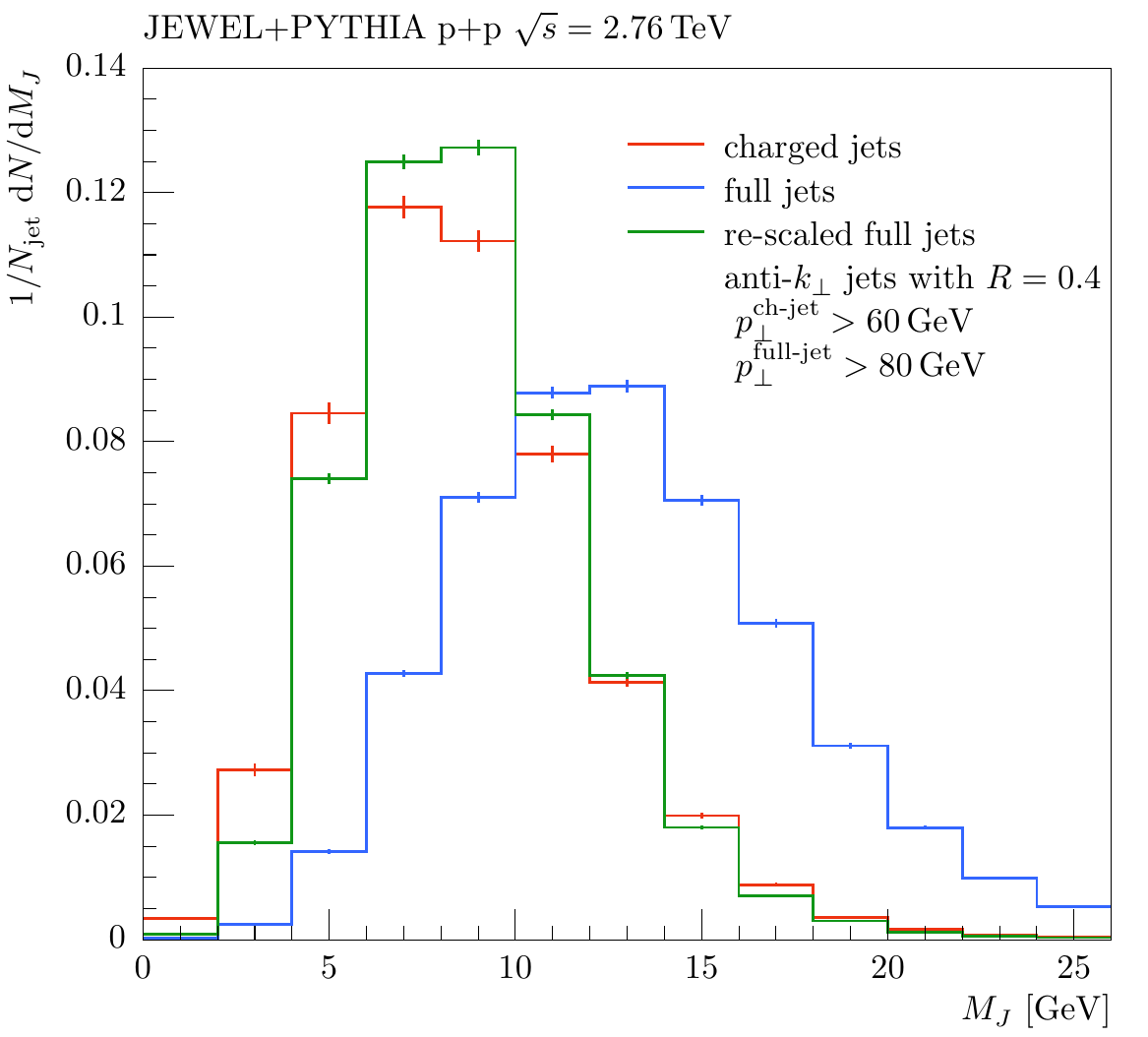} 
			\includegraphics[width=0.47\textwidth]{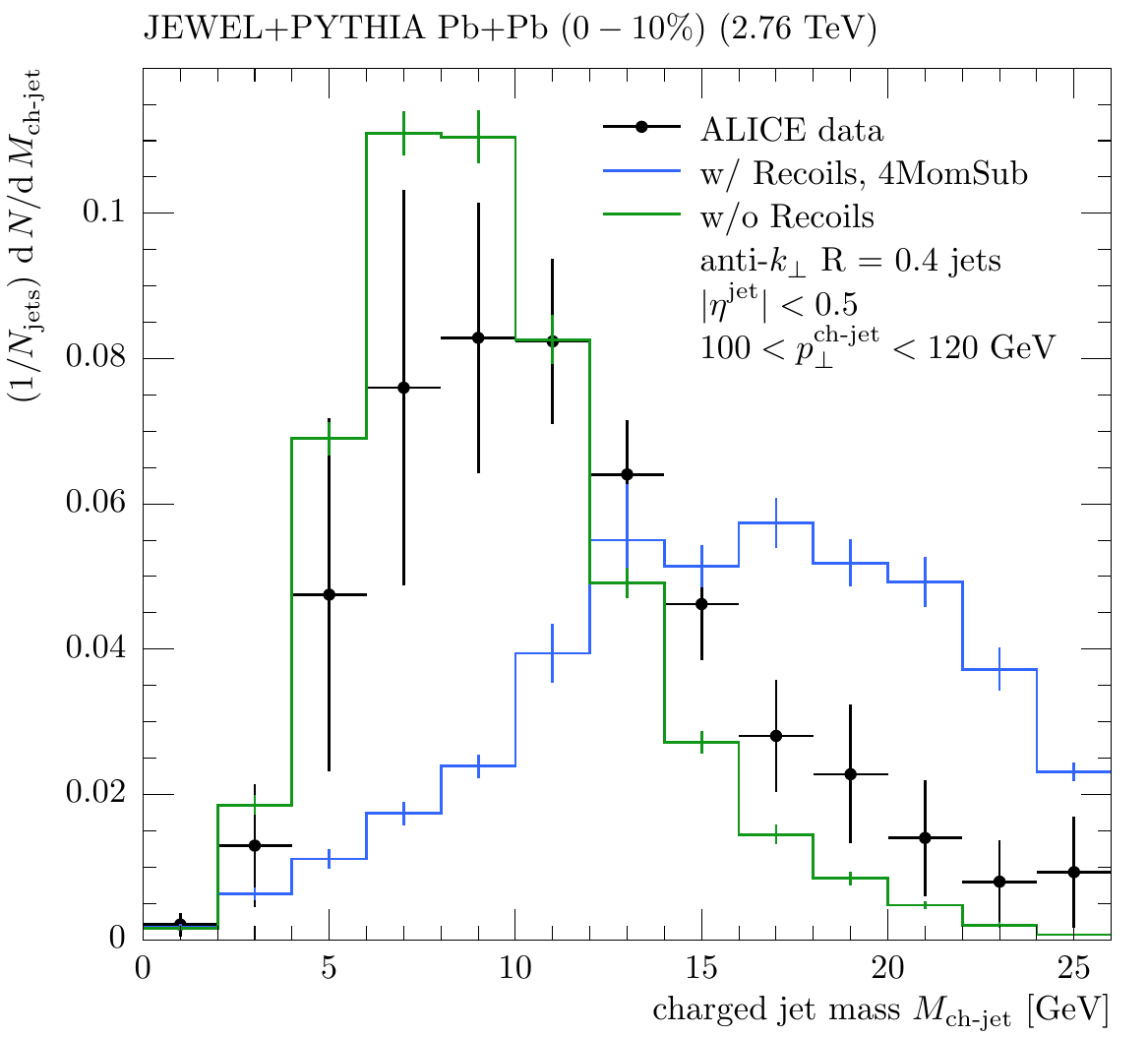} 
			\caption{Left: Jet mass distributions from \jwpy~for p+p. The blue histogram shows the full jet distribution, the red the one for charged jets, and the green histogram is the re-scaled blue histogram. Right: Comparison of the re-scaled full jet mass distribution with recent ALICE data~\cite{Acharya:2017goa} for the charged jet mass.}
			\label{fig:aliceChJetMass}
		\end{figure}

		The reconstructed jet mass is a good probe of medium induced jet modifications and medium response, since it is sensitive to the soft sector. Fig.~\ref{fig:JetMass} shows the \jwpy~results for the jet mass distribution. The Monte Carlo shows a shift towards larger masses when medium response is included, whilst for events generated without recoils, a smaller jet mass is observed for jets belonging to the same kinematic range. The latter is due to the known narrowing of the hard jet core. The partial cancellation between two competing effects -- the narrowing due to energy loss and the broadening due to medium response -- is typical for this kind of observables and also seen in other jet shapes (e.g.\ the jet profile and girth). We observe a large difference between 4MomSub and GridSub subtraction in this observable, but good agreement between the two versions GridSub1 and GridSub2. In fact, the jet mass is very sensitive to the details of the grid subtraction. In Fig.~\ref{fig:gridJetMass} we compare two different cell sizes and two ways of computing the cell momentum. One is the default, which consists of summing the four-momenta of the particles in the cell (and subtracting the thermal momenta), and the other sums the particles' energies and assumes the cell four-momentum to be massless and to point in the direction defined by the cell centre. Both variations lead to large differences in the jet mass distribution (which is not observed in any other observable we studied). We therefore strongly discourage the use of grid subtraction for the jet mass and from here on show results only for 4MomSub subtraction.

		As discussed in section~\ref{sec:limitations}, in order to be able to compare the \jwpy~results to the ALICE data, the charged jet mass has to be estimated from the full jet mass. We do this by re-scaling the full jet mass with a constant factor 2/3 and the jet \pt~with a factor 3/4 (this is needed to match the \pt~cuts in the charged jet sample). The scaling factors are extracted from the \jwpy~p+p sample. The left panel of Fig.~\ref{fig:aliceChJetMass} shows the charged jet, full jet and re-scaled full jet mass distributions in p+p and gives a lower bound on the related systematic uncertainties. We would like to stress once more that this is an ad hoc procedure and that there is no guarantee that it yields meaningful results. The right panel of Fig.~\ref{fig:aliceChJetMass} shows the comparison of the re-scaled full jet mass distribution from \jwpy~to a recent ALICE measurement~\cite{Acharya:2017goa}. The Monte Carlo predicts significantly larger jet masses, but given the uncertainties involved in obtaining the charged jet distribution it is difficult to interpret this comparison with data.

	\subsection{Fragmentation functions}

		\begin{figure}[h!] 
			\centering
			\includegraphics[width=0.47\textwidth]{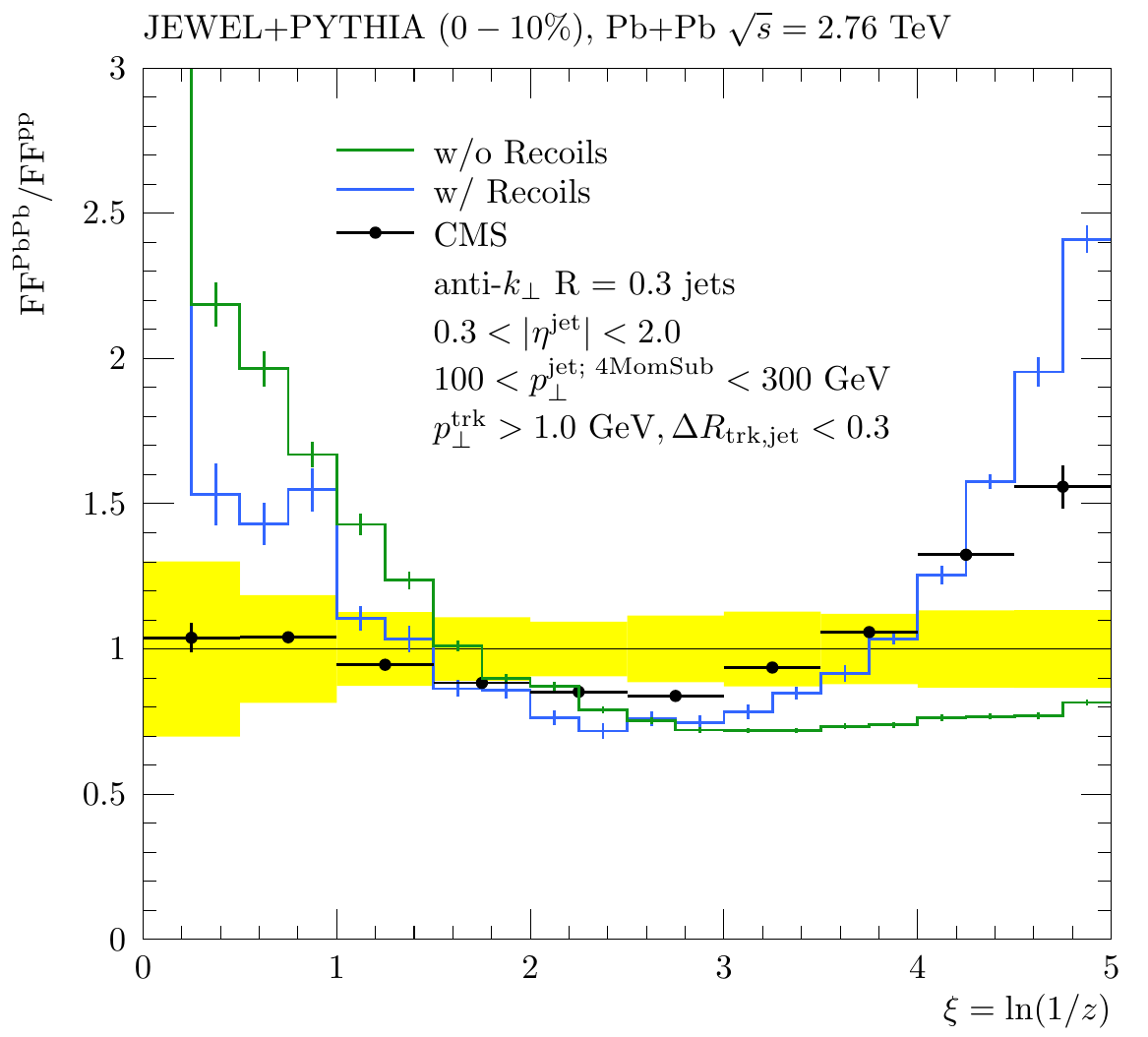} 
			\includegraphics[width=0.47\textwidth]{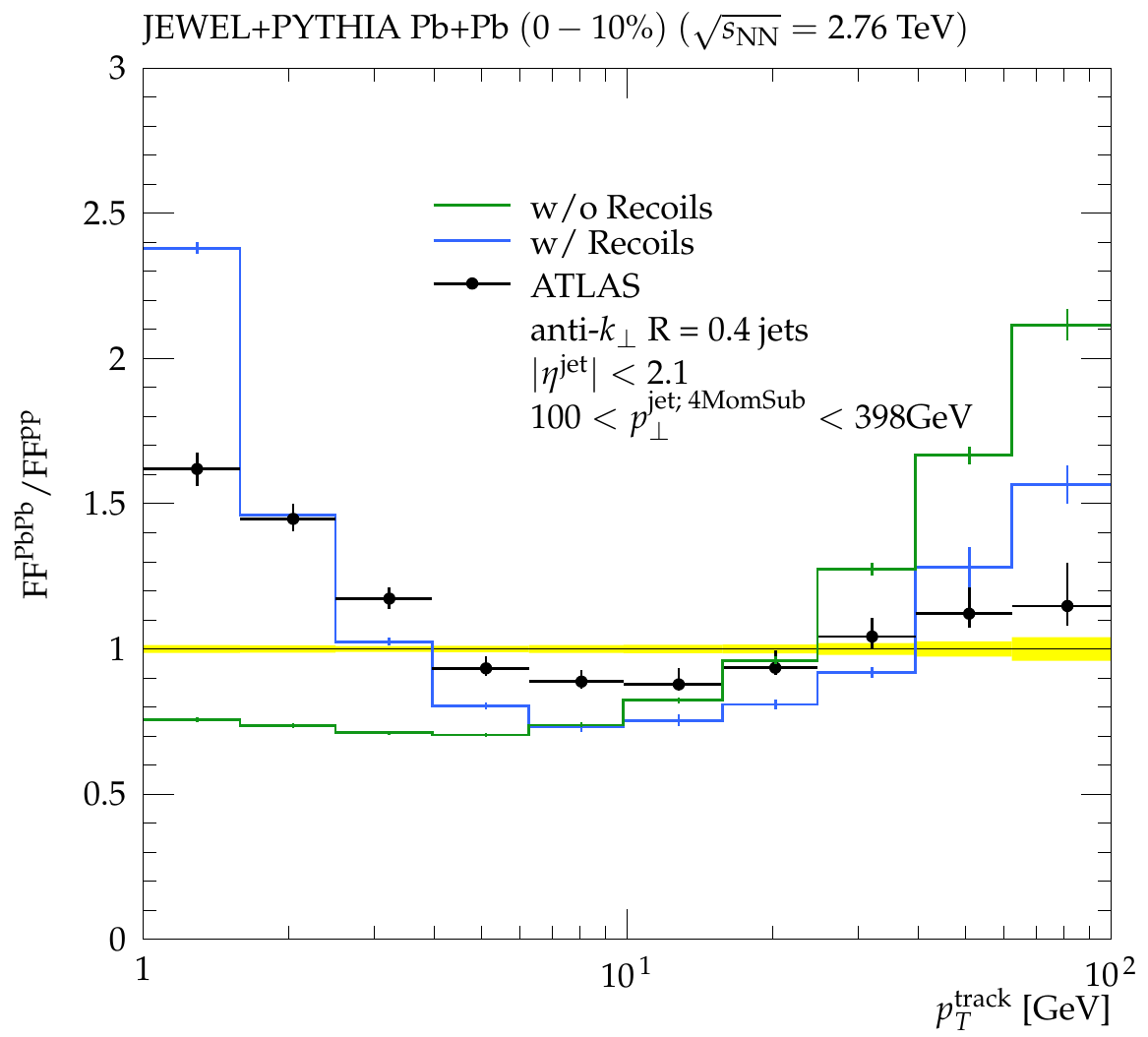} 
			\caption{The ratio of jet fragmentation functions (FF) Pb+Pb to p+p  compared with CMS~\cite{Chatrchyan:2014ava} (left) and ATLAS data~\cite{ATLAS-CONF-2015-055} (right). The data systematic uncertainties are shown in the yellow band around unity. Medium response is included in \jwpy~results shown as blue histograms, but the subtraction (in this case 4MomSub) can only be applied to the jet \pt~and not to the tracks. The corresponding \jwpy~results without medium response are shown as green histograms.}
			\label{fig:jewelFragFunction}
		\end{figure}

		Intra-jet fragmentation function~\cite{Chatrchyan:2014ava,Aad:2014wha,ATLAS-CONF-2015-055} in p+p and Pb+Pb collisions are also an important jet sub-structure observable. However, in \jw~there is no way of doing the subtraction for individual hadrons or, as in this case, tracks. In Fig.~\ref{fig:jewelFragFunction}, which shows the modification of the fragmentation function in Pb+Pb collisions compared to p+p, we therefore in the sample with medium response correct the jet \pt, but all tracks enter the fragmentation function. It is thus expected that \jwpy~overshoots the data in the low $z$ or \pt, corresponding to high $\xi$, region. The sample without medium response in this region shows a suppression as opposed to the enhancement seen in the data and the sample with recoiling partons, confirming the interpretation that the low \pt~(high $\xi$) enhancement seen in the data is due to medium response. The enhancement at high \pt~(low $\xi$) region is caused by the already mentioned narrowing and hardening of the hard jet core, and is more pronounced in \jwpy~than in data. It is stronger without medium response, because the latter does not affect the hard fragments, but slightly increases the jet \pt.

		\subsection{Jet profile}

		\begin{figure}[h!] 
			\centering
			\includegraphics[width=0.47\textwidth]{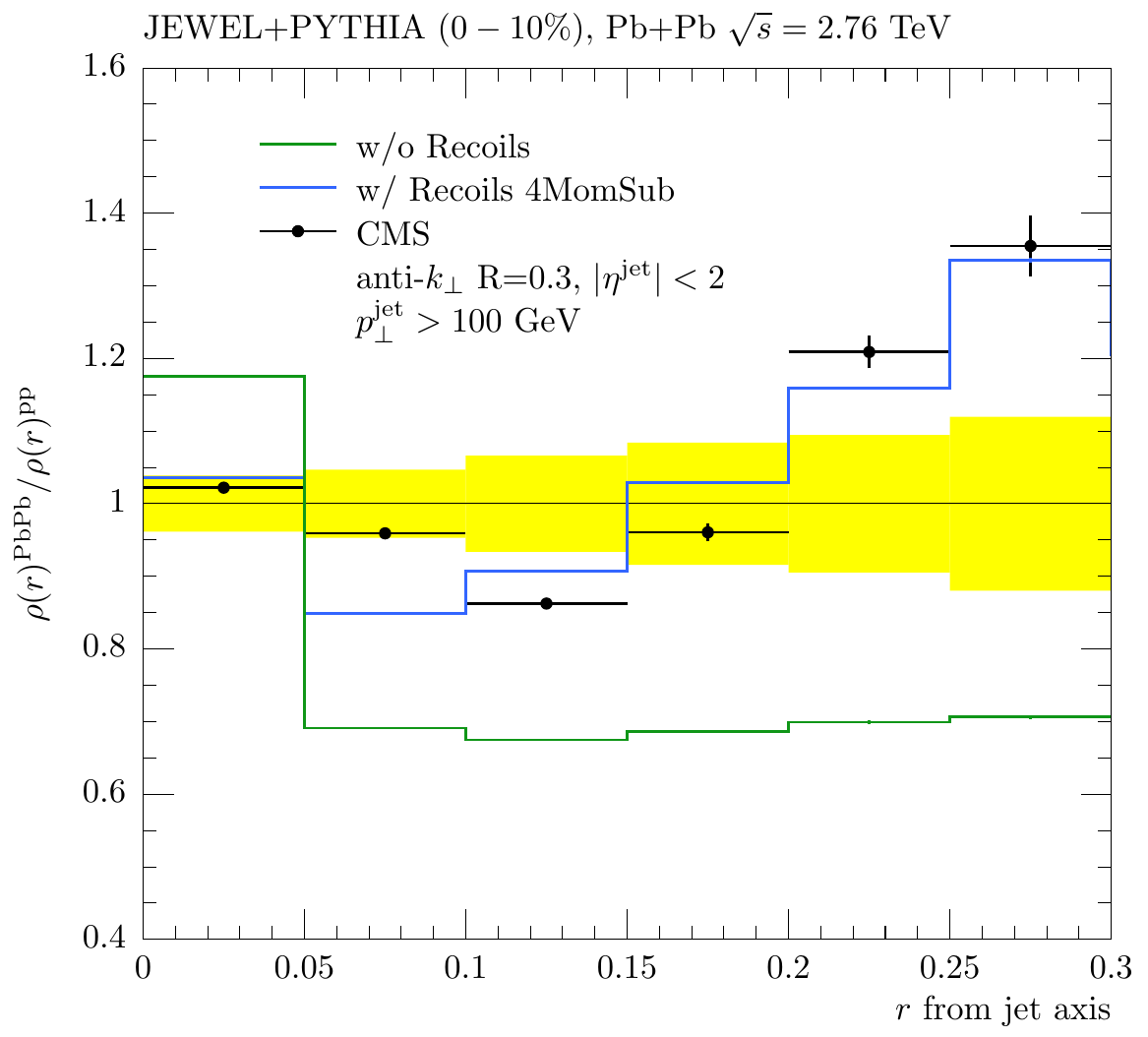} 
			\caption{Ration of the differential jet shape (or jet profile) in Pb+Pb and p+p measured by CMS~\cite{Chatrchyan:2013kwa} (black points) and compared with \jwpy~results with (blue line) and without medium response (green line). The data systematic uncertainties are shown in the yellow band around unity. }
			\label{fig:cmsJetShape}
		\end{figure}

		The differential jet shape or jet profile $\rho(r)$ measures what fraction of the jet \pt~is found at what distance from the jet axis. It is defined as 
		\begin{equation}
			\rho(r) = \frac{1}{p_{T}^\text{jet}} \hspace{-2mm} \sum_{\substack{k \text{\ with\ }\\ \Delta R_{kJ} \in [r, r+\delta r]}} \hspace{-3mm} p_{T}^{(k)} \,,
		\end{equation}
		where the sum runs over all particles in the jet. The CMS measurement~\cite{Chatrchyan:2013kwa} was performed using the full jet \pt, but $\rho(r)$ was built only from tracks. Therefore, as is the case of the fragmentation function, we can do the subtraction for the jet \pt, but not for the charged particles. In this case, however, this is not a problem, since the jet profile built from tracks and the one built form all particles differ only by a constant factor. Assuming this factor to be the same in p+p and Pb+Pb, it will cancel exactly in the ratio of the jet profiles. We can therefore compare \jwpy~results for full jets directly to the CMS data on the jet profile ratio. A more serious problem is that in experimental analysis only tracks with \pt$^\text{trk} > 1$ GeV are included. Since we can only subtract for the inclusive final state, this leads to a small mismatch, that becomes visible only at large $r$ and reaches up to 10\% in the highest $r$ bin.

		Fig.~\ref{fig:cmsJetShape} shows the \jwpy~result compared with CMS data~\cite{Chatrchyan:2013kwa} for the modification of the differential jet shape $\rho^{\text{PbPb}}/\rho^{\text{pp}}$ in Pb+Pb collisions compared to p+p. Including medium response and after performing the subtraction using the 4MomSub method, we are able to reproduce the general trend of the data. \jwpy~with recoiling partons describes the enhancement of the jet shape at large radii mostly due to soft particles (\pt $< 3$ GeV), while without medium response the enhancement is entirely absent. 

	\subsection{Girth}

		\begin{figure}[h!] 
			\centering
			\includegraphics[width=0.47\textwidth]{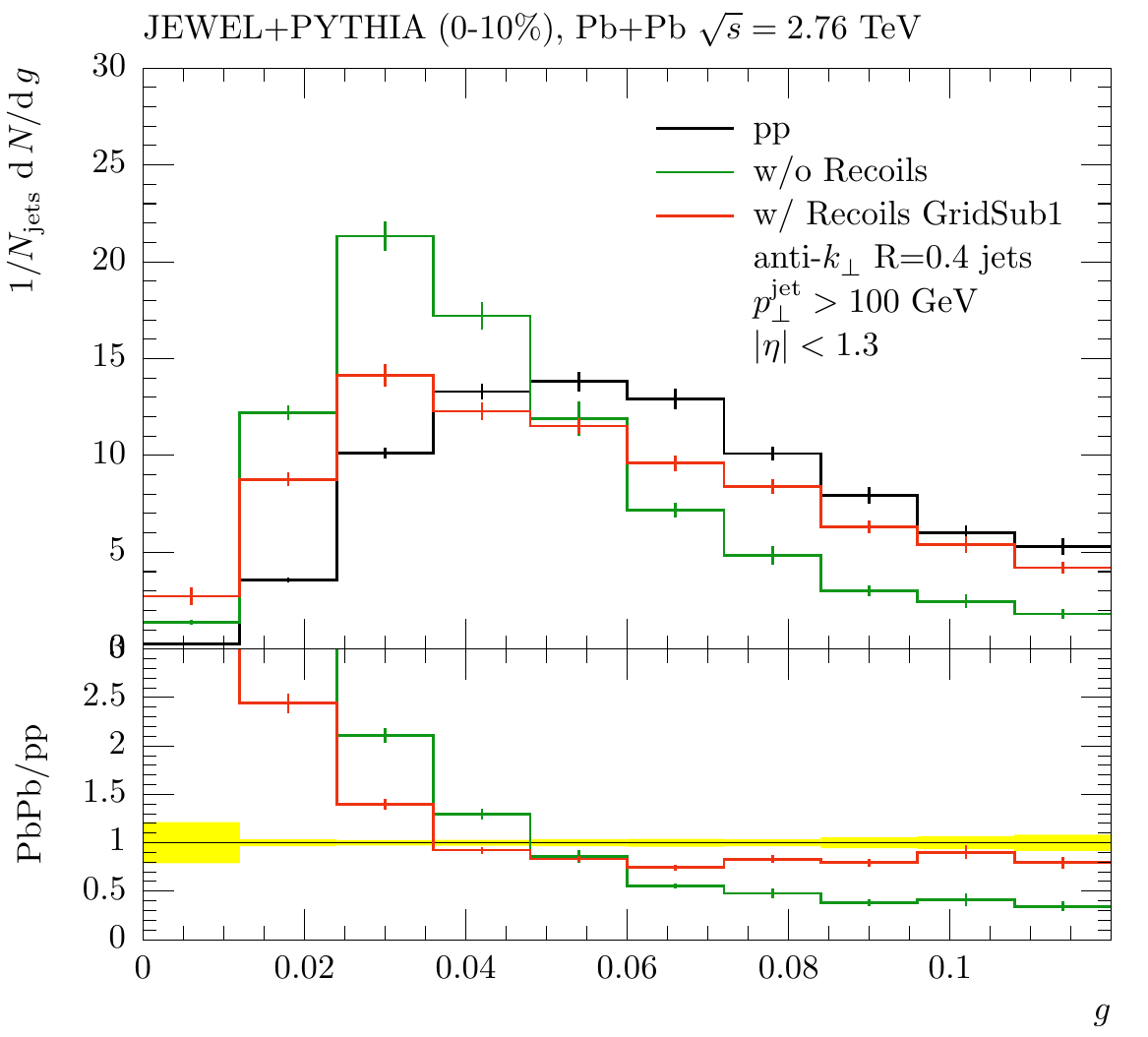} 
			\includegraphics[width=0.47\textwidth]{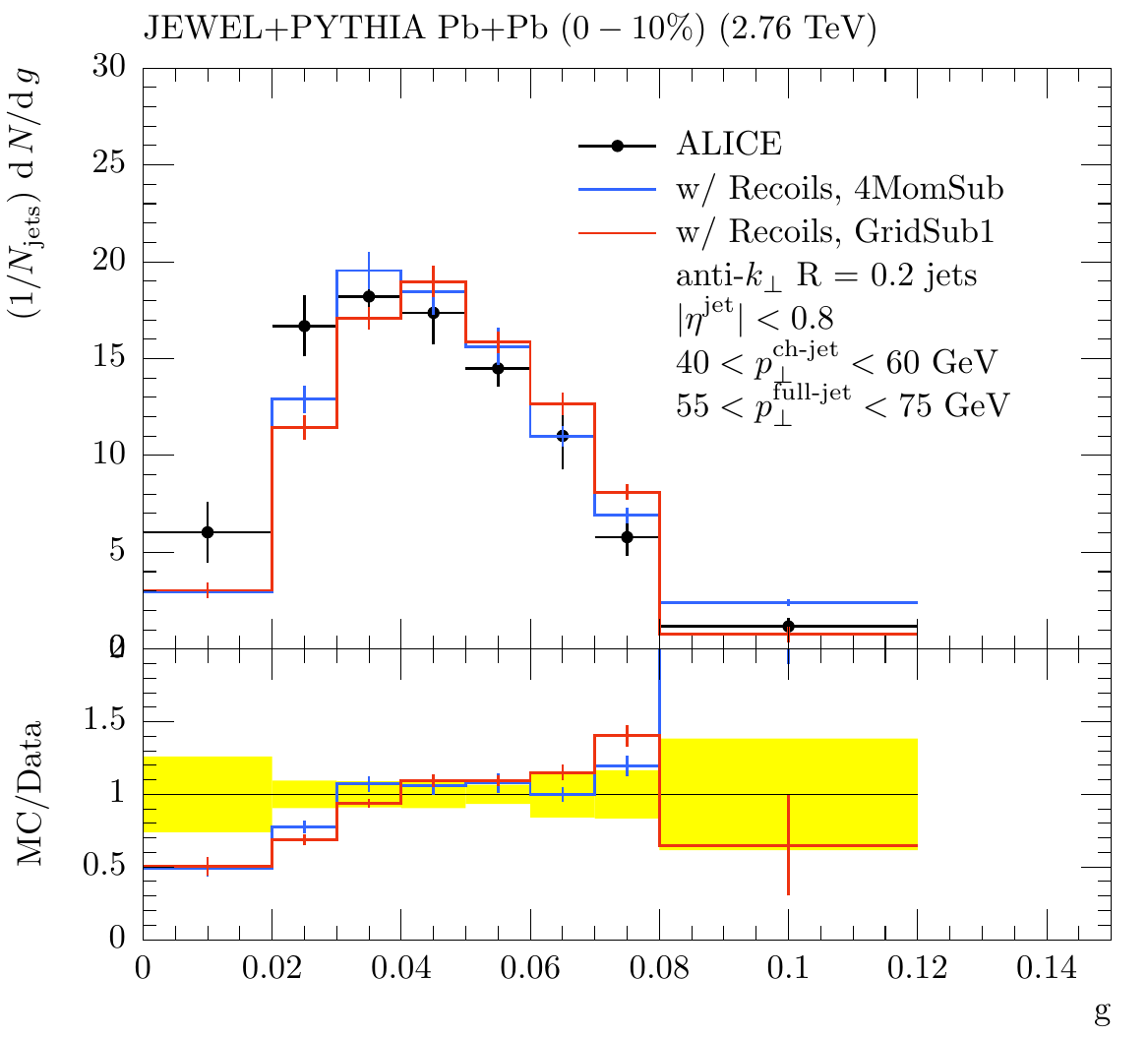} 
			\caption{Left: Distribution of the first radial moment (girth $g$) for $R=0.4$ fully reconstructed jets with \pt$^\text{jet} > 100$ \gev in central Pb+Pb collisions from \jwpy. The black histogram shows the corresponding p+p result, the green Pb+Pb without medium response and the red Pb+Pb including medium response with GridSub1 subtraction. The yellow shaded region around unity on the left panel highlights the statistical uncertainty in the p+p reference. Right: ALICE data~\cite{Cunqueiro:2015dmx} for charged jets ($R = 0.2$ and 40 \gev $<$ \pt$^\text{jet} <$ 60 \gev) compared with \jwpy~for full jets (with adjusted \pt~range). The yellow shaded region around unity represents the data systematic uncertainties. }
			\label{fig:alicegirth}
		\end{figure}

		The first radial moment of the jet profile is called girth~\cite{Giele:1997hd} and is defined as
		\begin{equation}
			g = \frac{1}{p_{T}^\text{jet}} \sum_{k\in J} p_{T}^{(k)} \Delta R_{kJ} \,,
		\end{equation}
		where the numerator sums the distance from the jet axis weighted with \pt$^{(k)}$ of each constituent $k$ of the jet. It characterises the width of the \pt~distribution inside the jet. 

		\jwpy results for girth using GridSub1 subtraction for fully reconstructed jets in central Pb+Pb collisions are shown in the left panel of Fig.~\ref{fig:alicegirth}. We find a shift to smaller values of $g$ due to narrowing of the hard component, which is partly compensated by a broadening of the jet due to medium response. We also compare our results with preliminary ALICE data~\cite{Cunqueiro:2015dmx} for charged jets in the right panel of Fig.~\ref{fig:alicegirth}. Following the same argument as above for the jet profile, the girth of full and charged jets should be the same, provided the \pt~range is adjusted accordingly. We confirmed this in the Monte Carlo for p+p collisions. We therefore in Fig.~\ref{fig:alicegirth} compare \jwpy~results for fully reconstructed jets at a correspondingly higher \pt~with the ALICE data. We find reasonable agreement, but the \jwpy~distribution peaks at slightly higher values than the data.

	\subsection{Groomed shared momentum fraction $z_g$}

		\begin{figure}[h!] 
			\centering
			\includegraphics[width=0.47\textwidth]{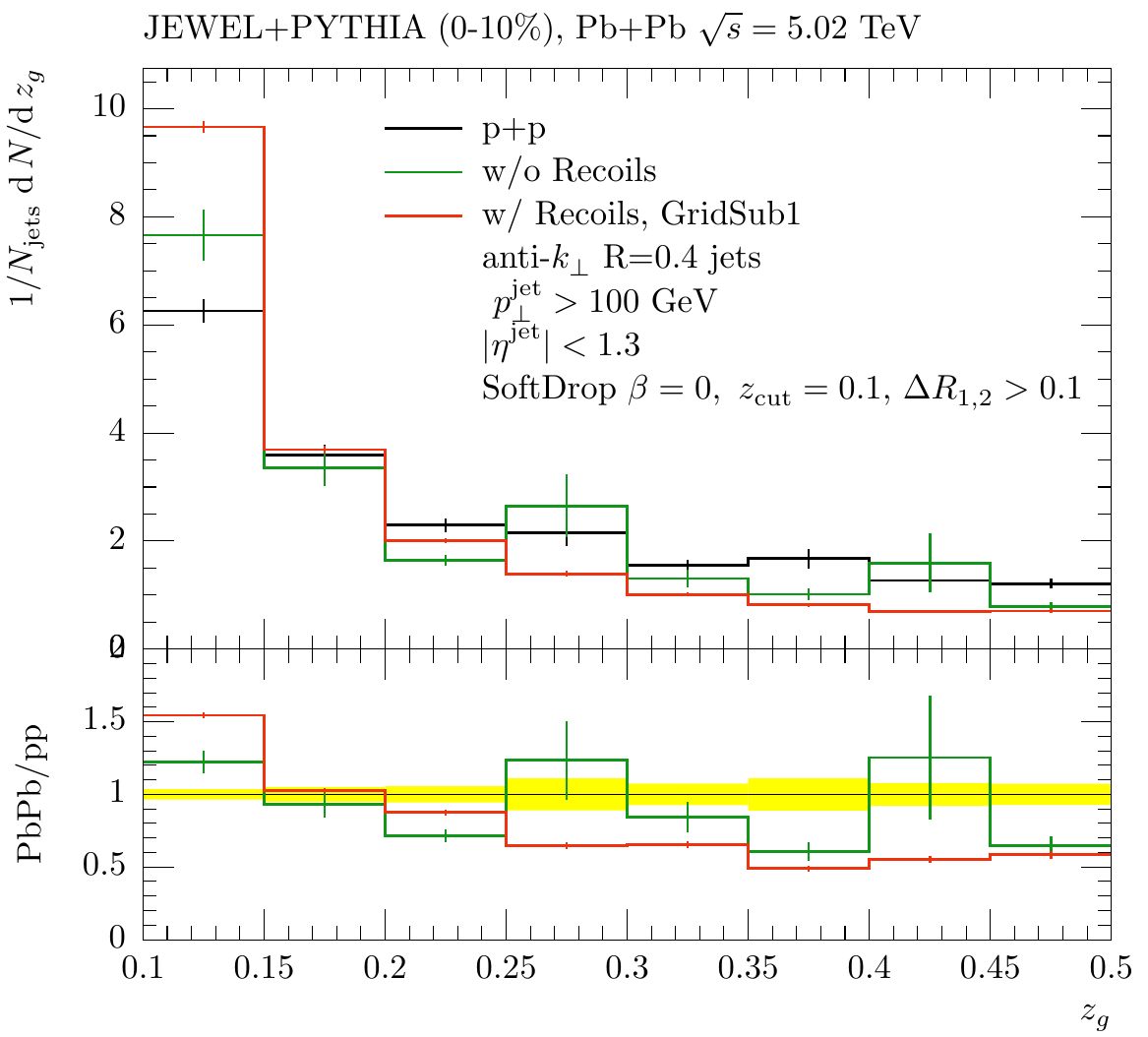} 
			\includegraphics[width=0.47\textwidth]{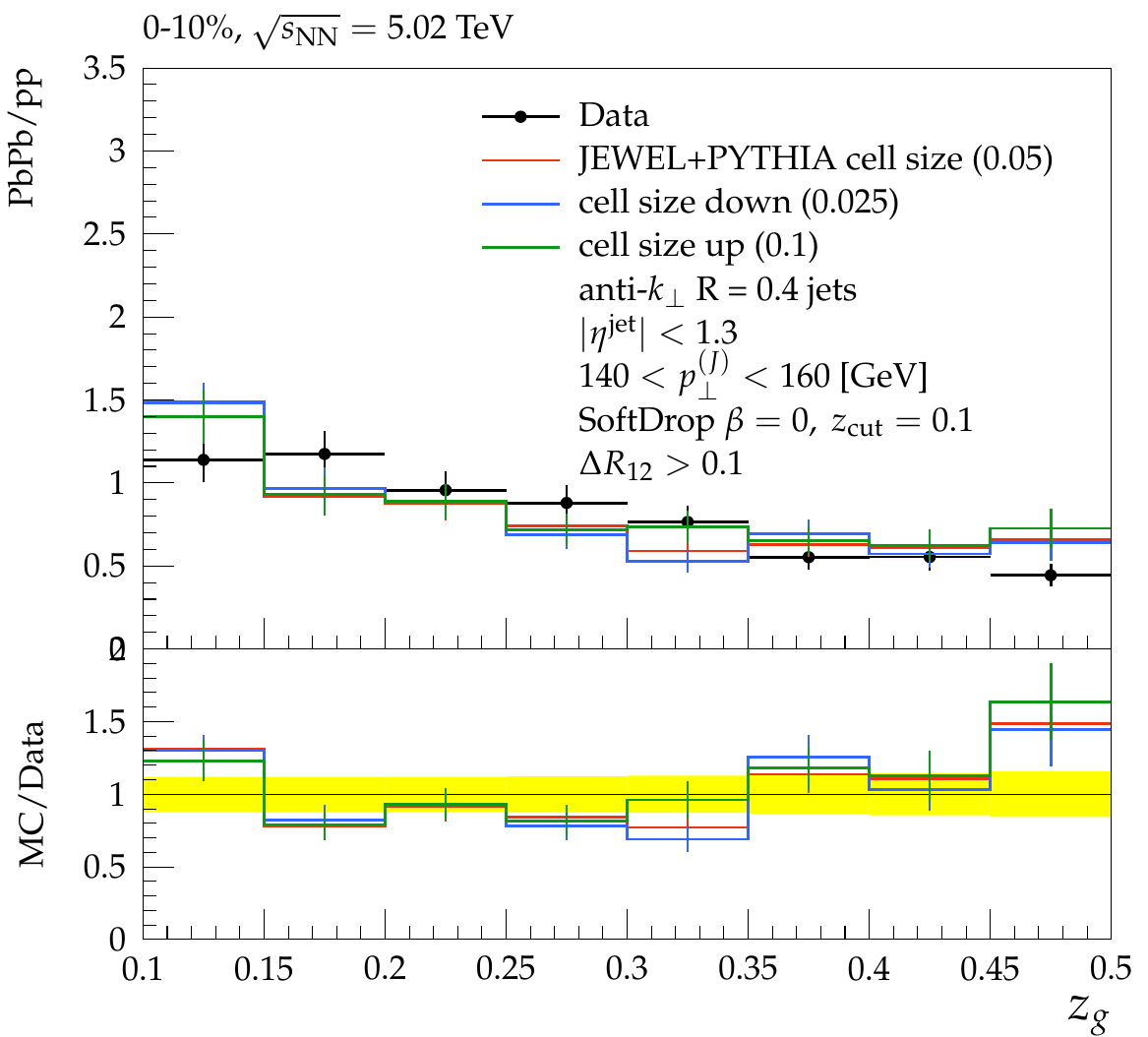} 
			\caption{\jwpy predictions for the groomed shared momentum fraction $z_g$ in central Pb+Pb events and p+p events. Left: $z_g$ distribution in p+p (black), central Pb+Pb collisions without recoiling partons (green) and with medium response and GridSub1 subtraction (red) for jets with \pt$^\text{jet} > 100$ \gev and Soft Drop parameters $z_\text{cut}=0.1$ and $\beta = 0$. Right: Comparison of \jwpy~results with different grid sizes to CMS data~\cite{Sirunyan:2017bsd}. Note that the data is not unfolded, but the resolution is not published so no smearing is applied to the Monte Carlo events. A comparison to properly smeared \jwpy~results can be found in~\cite{Sirunyan:2017bsd}. The yellow shaded region around unity in the left panel highlights the statistical uncertainty in the p+p reference and on the right represents the the data systematic uncertainties.}
			\label{fig:cmsSplitting}
		\end{figure}

		The groomed shared momentum fraction $z_g$ is a measure for the momentum asymmetry in the hardest, i.e.\ largest angle, two-prong structure in the jet. In p+p collisions it is closely related to the Altarelli-Parisi splitting function~\cite{Larkoski:2015lea}. It is defined through the Soft Drop procedure~\cite{Dasgupta:2013ihk,Larkoski:2014wba} detailed below and implemented in \textsc{FastJet}~\cite{Cacciari:2011ma} contrib. First, jets are clustered with the \akt algorithm and re-clustered with Cambridge/Aachen. Then the last clustering step is undone, yielding the largest angle two-prong structure in the jet. If this configuration satisfies the Soft Drop condition
		\begin{equation}
			z_g = \frac{\text{min}(p_{T,1}, p_{T,2})}{p_{T,1}+ p_{T,2}} > z_{\rm{cut}} \left(\frac{\Delta R_{1,2}}{R_{J}}\right)^{\beta}
		\end{equation}
		where $z_{cut}$ and $\beta$ are parameters, it is kept. Otherwise, the softer of the two prongs is discarded and the procedure of un-doing the last clustering step is repeated for the harder prong. In this way soft contaminations are systematically removed from the jet and the hardest two-prong structure is identified. Soft Drop jet grooming thus takes an inclusive jet collection and turns it into a different collection of jets with two-prong structure of a minimum momentum symmetry provided by $z_{\rm{cut}}$.  Varying $z_{\rm{cut}}$ up or down varies the degree of asymmetrical splitting allowed in the  parton's fragmentation, while the $\beta$ controls how collinear the configuration has to be.

		In p+p collisions, this method is has been studied in some detail~\cite{Larkoski:2014wba,Larkoski:2015lea}, but in heavy ion collisions the exact meaning of the grooming procedure is not obvious, due to the presence of the fluctuating underlying heavy ion event and the increased soft sector, that the procedure tries to remove. Recent analytical studies~\cite{Mehtar-Tani:2016aco} have shown that grooming increases the sensitivity to medium induced gluon bremsstrahlung thus experimentally opening up different avenues to directly probe the effect of the medium on a jet by jet basis. In \jw, however, a different story unfolds. 

		As shown in the right panel of Fig.~\ref{fig:cmsSplitting}, there is an increase in asymmetrical splittings in Pb+Pb jets as opposed to p+p jets, which is observed in recent preliminary CMS results and reproduced in \jwpy. The secondary feature observed in this measurement is an apparent reduction of the effect for higher \pt~jets. \jw~reproduces this behavior qualitatively as well, with very high \pt~jets showing very little difference in the momentum fraction of the first splitting~\cite{Sirunyan:2017bsd}. The left panel of Fig.~\ref{fig:cmsSplitting} shows that in the Monte Carlo the modification of the $z_g$ distribution in Pb+Pb collisions is partly due to the narrowing of the jet, as seen in the sample without medium response. The more important contribution, however, comes from adding the recoiling partons\footnote{For a detailed discussion of the origins of the effect in \jw~cf.~\cite{Milhano:2017nzm}.}. In \jwpy~we see no sign of medium induced bremsstrahlung contributing to the effect, as advertised in~\cite{Mehtar-Tani:2016aco}. 

\section{Discussion and conclusions}
\label{sec::conclusions}

	Studies of jet sub-structure modifications in heavy ions probe the intricate interactions between the medium and jets. Due to their sensitivity to medium response, they offer the power to discriminate between several models and shed light on the underlying jet quenching mechanisms as well as the thermalization of the deposited energy and momentum.

	In \jw~it is possible to study medium response in detail by keeping the partons recoiling against interaction with the jet in the event. One has to keep in mind that this is only a limiting case, since these partons do not undergo further interactions in the medium. In order to be able to compare these results to experimental data, the thermal component of the recoiling partons' momenta has to be subtracted. In this paper we introduced two methods for doing this, a four-momentum and a grid based one. With these tools we can for the first time quantitatively study jet shape observables.

	We find that -- at least in \jwpy~-- a number of qualitative feature in the data can only be explained by medium response. These are
	\begin{itemize}
		\item the increase at low $z$ of the ratio of intra-jet fragmentation functions in Pb+Pb compared to p+p,
		\item the increase of the jet profile at large distance from the jet axis in Pb+Pb compared to p+p,
		\item and the enhancement of asymmetric two-prong structures in Pb+Pb compared to p+p as seen in the $z_g$ distribution.
	\end{itemize} 
	This is in line with observations by other authors~\cite{Casalderrey-Solana:2016jvj,Tachibana:2017syd}. In other observables, in particular the jet mass and girth, a non-trivial cancellation between a narrowing of the jet core due to energy loss~\cite{Milhano:2015mng,Rajagopal:2016uip,Casalderrey-Solana:2016jvj} and a broadening due to medium response takes place. Also in the case of girth, including medium response leads to an improvement of the agreement between \jwpy~and ALICE data. 

	For the jet mass we find that the grid based subtraction does not yield reliable results. The 4MomSub subtraction should be more robust, but without grid subtraction we do not have an independent way of cross-checking the results. We therefore recommend not to use GridSub for the jet mass and to take the comparison of \jwpy~results to the ALICE data with a grain of salt.

	Jet shape observables open a new perspective on jet quenching and may also help to address the question of thermalization, and it is important to develop tools capable of quantitatively describing medium response. The present study with \jw~can only be a first step in this direction. As emphasized above, the treatment of recoiling partons is still schematic. The subtraction methods introduced in this paper are solid, but have their limitations, in particular when it comes to the description of charged jets. It is currently also impossible to perform the subtraction for particles (for instance in the fragmentation functions), due to the mix of parton and hadron level in the subtraction. The grid method also introduces systematic uncertainties related to the discretization, that can, however, be quantified (cf. section~\ref{sec::systematics}). Nevertheless, the results for jet shapes obtained with \jwpy~are very promising. In some cases this is the first time that they can be studied quantitatively in a consistent jet quenching model including medium response.

	Upcoming measurements at the LHC will further advance the understanding of jet shapes by utilizing the jet grooming tools, amongst others. This ushers in a new era of sub-structure studies in heavy ion collisions, where correlation between different observables could point the way to the future in decoupling several of the physics features hidden in individual observables.   
		



\clearpage

\chapter{Moving Towards a Quantitative Understanding}
\label{ch_summaryFuture}

\begin{chapquote}{Marie Curie}
``Nothing in life is to be feared, it is only to be understood. \\Now is the time to understand more, so that we may fear less."
\end{chapquote}

\section{Improving current baseline jet measurements}

	Any experimental measurement can be improved by either extending the kinematic reach or better understanding the systematic uncertainties or increasing statistics. Especially jet observables need a lot of statistics since the hard scattering cross section is small compared to the minimum bias. Lets go over the results for each system in consideration and briefly mention how each can be improved.  
	
	\subsection{Jets in pp collisions}

		Measuring the inclusive jet cross section is fundamental to any jet program and we discussed the ability of theory calculations at NNLO and LL to match data for small radii jets. Our result was only looking at the mid rapidity region and thus the next obvious step forward is to look at the Data/Theory comparison across a variety of rapidity windows and radii. Especially studying the forward rapidity starts to go more into the high x region of the PDFs and can expose inconsistencies in the calculations. The overall systematic uncertainty in the jet cross section is probably the smallest its ever been and with the current method of deriving corrections (using MC in a detector simulation), it is improbable that we can reduce it much further. With these comparisons, the theoretical uncertainties will reduce facilitating an improved understanding of the proton and the hard scattering.   

	\subsection{Any nuclear effects in pPb}

		In our inclusive and b-jet studies in pPb collisions in comparison with an extrapolated pp reference, we saw no significant modification due to initial state nuclear effects. This provided us with confidence that \raa~is indeed a final state effect. But pPb has raised more questions than it answered, especially related to the initial state and its composition, accurate centrality determination and effect of multiplicity on jet structure. The parton mass dependence would be interesting upon extending the kinematic range to lower \pt~but at the LHC that is very hard. The overall systematics do reduce when we take into account pp data as the baseline but still it doesn't point to any significant modification. So for pPb collisions, we need to directly push towards jet structure and sub-jet studies as we will discuss in the coming sections. 
	
	\subsection{Jets in the QGP}
		We started this exploration with the goal of extracting QGP properties by using jets as tomographical tools. Have we attained that goal? It is safe to say that so far, that studying jets had proven to be more complicated than assumed. It turns out that we must first understand the baseline in order to use these tools in a heavy ion environment and due to the multi-scale processes involved with jets, there are still several questions that we need to answer when we start to move towards a quantitative reasoning of the QGP interactions.  In the coming sections, we shall briefly summarize what we have learnt from existing measurements and how they can be improved. We will finish the thesis with what I call the new age of jet measurements in heavy ions and how we can refine/upgrade our current measurements in all three systems that we studied in this thesis.  
	
\section{Consensus on jet quenching}

	We have gone through a lot of pages regarding jets and measurements/models so lets try to summarize what we have learnt so far. It is very clear across experiments that jets are indeed quenched in heavy ion collisions when compared with pp collisions. These modifications appear to be final state interactions with the medium as opposed to initial state or cold nuclear matter effects evident by the lack of significant modifications in pPb collisions. These are all broad stroke or rather qualitative features and are quite easily reproduced by a wide spectra of models when taking into account medium induced energy loss via elastic, inelastic and radiative processes. The \raa~turns out to be a sledgehammer approach to the finer details and thus as a natural evolution in jet studies, we also looked at jet structure and even sub-structure measurements. From these jet shape, fragmentation function, mass, moments and even the so-called splitting function results, we start to collect consistent picture of jet structure modification starting from a relatively unmodified core, reduction of medium \pt~particles as we move away from the core and an enhancement of low \pt~particles in the periphery of the jet. As we extend far away from the jet axis, we are even able to recover the lost energy, if we take into account the assumptions involved in the measurement. It is safe to say that there have been many high impact results in the studies of jets in heavy ion collisions in the past decade.

\section{New age of jet measurements}

	Almost all the recent measurements on jet quenching highlight the importance of understanding the correlation between the jet and the medium. These correlations manifest in how jet structure is modified and very recently, how the QGP itself is modified with the presence of the jet. Any standard jet observable is dependent on two fundamental scales in the problem; momentum and angular. This is what I call as the jet phase space and the new age of jet measurements involves a systematic, multifaceted study of this phase space. 

	\subsection{Systematic exploration of the phase space}
	
\begin{figure}[h] 
		   \centering
		   \includegraphics[width=0.9\textwidth]{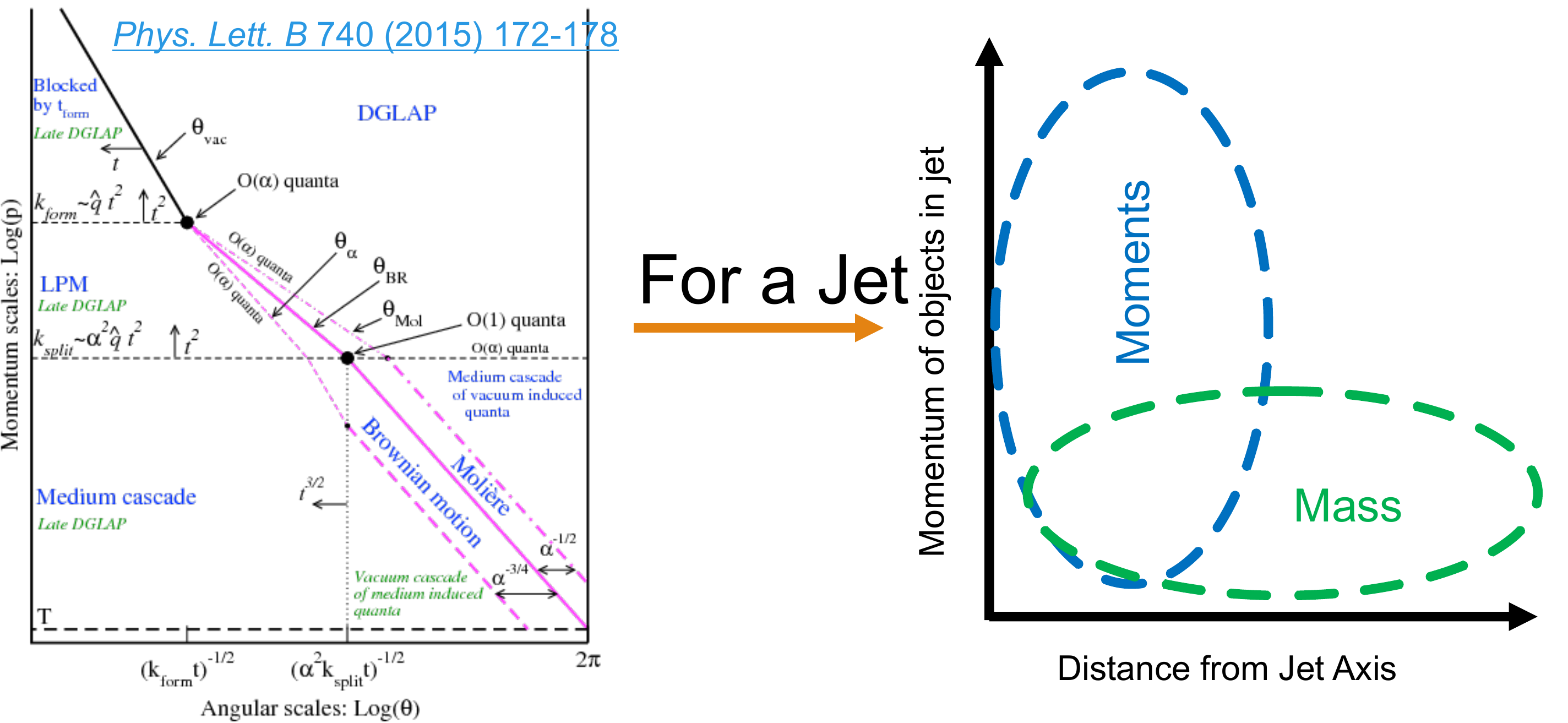} 
		   \caption{Qualitative phase space for a jet and how different observables analyze different parts of the phase space analogous to the overall QGP interactions across momentum and angular scales.}
		   \label{fig:phaseanalysis}
		\end{figure}

		In order to study sub-jet and jet structure observables efficiently in heavy ion collisions, one must first understand their position on the phase space of jet-medium interactions as shown in Fig:~\ref{fig:phaseanalysis}. Observables such as the radial moments and jet mass represent different scales and provide information on different aspects of quenching, such as parton energy loss and the kick away from the jet axis.   
	
		These observables could also be utilized in pPb and pp collisions since they probe QCD properties which could be used to study several important questions such as gluon jet fragmentation, parton shower modifications and hadronization in high multiplicity events amongst others. Another area of recent interest in the pp community is in the study of quark vs gluon identification. If such an identification is possible in PbPb, one can directly isolate the medium interaction properties such as coupling to quarks and gluons and also study jet evolution in the dense medium. A lot of such quark gluon identification is based on understanding the observable and having a machine learning software framework provide classifiers. 
		
		We are in a very exciting time for jet studies in pp and heavy ion collisions wherein both experiment and theory are pushing the frontiers of our current understanding. Within the next decade there are plans for a new modular framework called JETSCAPE, which can incorporate multiple models of the initial state, jets, energy loss all in one framework. This would be very important to take the community forward with the realistic goal of extracting the QGP's inherent properties. In the near future, one also expects a lot from the sPHENIX detector at RHIC, which promises to kinematically overlap with experiments at the LHC, but at the same time, offering an unique opportunity to probe the low \pt~region with high sensitivity to medium-jet interactions. I am personally very excited about all that i've mentioned above and each and every one of these is important for us to elucidate the fundamental properties of the QGP and thus systematically study the early universe within our lifetimes.

\clearpage

\clearpage
\bibliographystyle{ieeetr}
\bibliography{raghav_phd}


\begin{appendices}
\chapter{Collider kinematics}
\label{app_kinematics}

\begin{figure}[htbp] 
   \centering
   \includegraphics[width=0.7\textwidth]{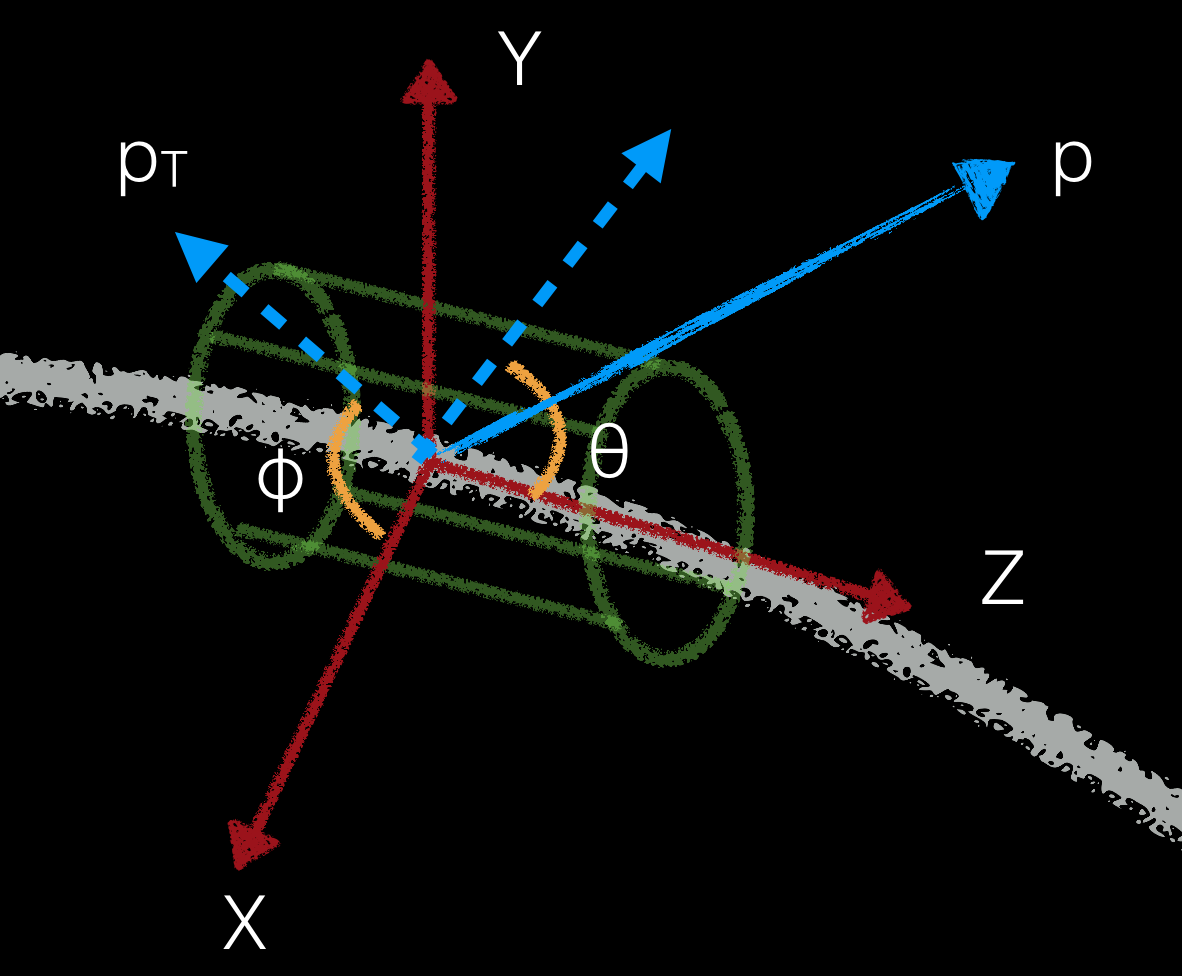} 
   \caption{Cartoon drawing of the collider rest frame and the corresponding axis utilized in high energy physics. A particle produced the collision is shown in the blue arrow.}
   \label{fig:collkine}
\end{figure}

The collider frame of reference is shown in Fig:~\ref{fig:collkine} with the Z axis along the direction of the beam. At the LHC, the convention dictates counter-clockwise beams as the positive Z direction when one looks at the ring from above with the main CERN campus at the 7 o'clock position. The X direction points towards the center of the ring and the Y axis points up towards the skies. The X-Y and Y-Z make up the transverse and longitudinal planes respectively. For a given particle arising from the collision with a given momenta $p$, one can split it into the three components along the directions $p_{x}, p_{y}, p_{z}$. The angle the transverse projection makes with the X axis is $\phi$ and the particle's momenta in the transverse plane is called $p_{T} = \sqrt{p_{x}^{2} + p_{y}^{2}}$ or the transverse momentum. Similarly, the angle with the beam along the longitudinal plane is $\theta$. In particle physics, as opposed to $\theta$, one mostly uses the particle's rapidity $$y = \frac{1}{2}\ln\frac{E+p_{z}}{E-p_{z}}$$ or the pseudorapidity $$\eta = -\ln (\tan (\theta/2))$$ when the particle's mass is small compared to its momenta. 

Since the LHC uses the same dipole magnets to accelerate different particle species such as p or Pb, the individual beams come with different energies. The pPb system at 5.02 TeV was created with a proton beam at 4 TeV and a Pb beam at 1.38 TeV. Thus the mean particle distributions in the lab frame (the frame of the CMS detector) are skewed away from $y=0$. In the CMS convention, for a pPb collision the proton moves to the positive rapidity direction and Pb to the negative rapidity direction. Since the proton is the higher energy beam, the hard scattered distributions (such as high \pt jets) are boosted towards the negative rapidity direction whereas the soft particle or the underlying event distribution is shifted towards to the positive rapidity direction. The boost for the jets can be simply calculated by estimating the rapidity boost as follows 
\begin{equation}
	\beta = (P_{\rm{Pb}} - P_{\rm{p}}) / \sqrt{E_{\rm{Pb}}^{2} + E_{\rm{p}}^{2}}
\end{equation}
and with the CMS convention of not taking into account the mass of the Pb nuclei, one gets a $\beta$ of $-0.434$. The rapidity shift comes from 			
\begin{equation}
	\Delta y = \tanh^{-1}(\beta) = \frac{1}{2} \log\left(\frac{1+\beta}{1-\beta}\right) = -0.465
\end{equation}
where the second equality holds due to $\beta<1$.

\clearpage

\chapter{Jet Algorithms}
\label{app_jetcomposition}

There are essentially two classes of jet algorithms: cone-type and clustering type algorithms. In cone-type algorithms jets are defined by maximizing the amount of energy which can be covered by cones of defined size, whilst in clustering algorithms particles are assigned to jets iteratively according to whether a given energy-angle resolution variable $d_{i,j}$ exceeds a fixed resolution parameter. The iterative cluster algorithms are preferred since they offer infrared and collinear safety. 

\subsubsection{Longitudinally invariant $k_T$}
The Longitudinally invariant $k_T$ jet algorithm comes in inclusive and exclusive variants. the inclusive variant is formulated as follows ($k_T$ represents the parton's transverse momentum/energy. If we choose energy then it is called E-scheme or Energy-scheme recombination):

For each pair of particles i,j calculate the $k_T$ distance. $$ d_{i,j} = min(p_{T,i}^2 , p_{T,j}^2) \Delta R_{i,j}^2 / R^2$$ with $\Delta R_{i,j}^2 = (y_i - y_j)^2 + (\phi_i - \phi_j)^2$, where $p_{t,}$, $y_i$, $\phi_i$ are the transverse momentum, rapidity and azimuth of the particle $i$. $R$ is a jet-radius parameter usually set to 1. For each parton $i$ also calculate the beam distance $d_{i,B} = p_{T,i}^2$. Then we proceed to find the minimum $d_{min}$ of all the $d_{i,j}$, $d_{i,B}$. If $d_{min}$ is a $d_{i,j}$ merge particles $i$ and $j$ into a single particle, summing their four momenta; if it is a $d_{i,B}$ then declare particle $i$ to be a final jet and remove it from the list. This procedure continues till there are no initial state particles remaining and all have been converted to beam particles. 

The exclusive variant of the longitudinally invariant $k_T$ algorithm is similar except that when $d_{i,B}$ is the smallest value, that particle is considered to become part of the beam jet (i.e. discarded) and clustering is stopped when all $d_{i,j}$ and $d_{i,B}$ are above some $d_{cut}$.

\subsubsection {Cambridge/Aachen}

The $pp$ Cambridge/Aachen (C/A) jet algorithm is provided only in an inclusive version whose formulation is identical to that of the $k_T$ jet algorithm, except the distance measure which are:$$ d_{i,j} = \Delta R_{i,j}^2 /R^2$$ with $d_{i,B} = 1$. 
This clusters jets purely based on their distance towards each other with disregard to the transverse momentum of the respective particles. 

\subsubsection {anti $k_T$}

\begin{figure}[h]
	   \centering
	   \includegraphics[width=0.5\textwidth]{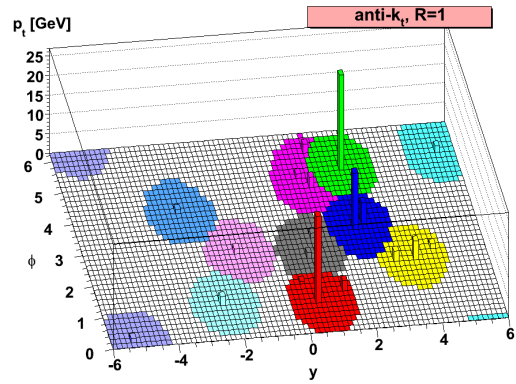} 
	   \caption{Jets clustered with the anti-k$_{T}$ algorithm for a given event with the fastjet algorithm.}
	   \label{fig:antiktfastjetpicture}
	\end{figure}

The most popular algorithm in high energy physics recently is the anti-k$_{t}$ and its defined exactly like the standard $k_T$ algorithm, except for the distance measures which are now given by 
$$ d_{i,j} = min(1/p_{T,i}^2 , 1/p_{T,j}^2)  \Delta R_{i,j}^2 / R^2$$ with $d_{i,B} = 1/p_{T,i}^2$. While it is a sequential recombination algorithm like $k_T$ and Cambridge/Aachen, the anti-$k_T$ algorithm behaves in some sense like a perfect cone algorithm, in that its hard jets tend to be circular on the $y-\phi$ plane as shown in Fig:~\ref{fig:antiktfastjetpicture}. 

All references and further details concerning the different algorithms and their implementation can be found in the fastjet package manual~\cite{Cacciari:2011ma}. 
\clearpage

\chapter{Early QGP Signatures}
\label{app_qgpsignatures}
\begin{figure}[h!]
	   \centering
	   \includegraphics[width=0.7\textwidth]{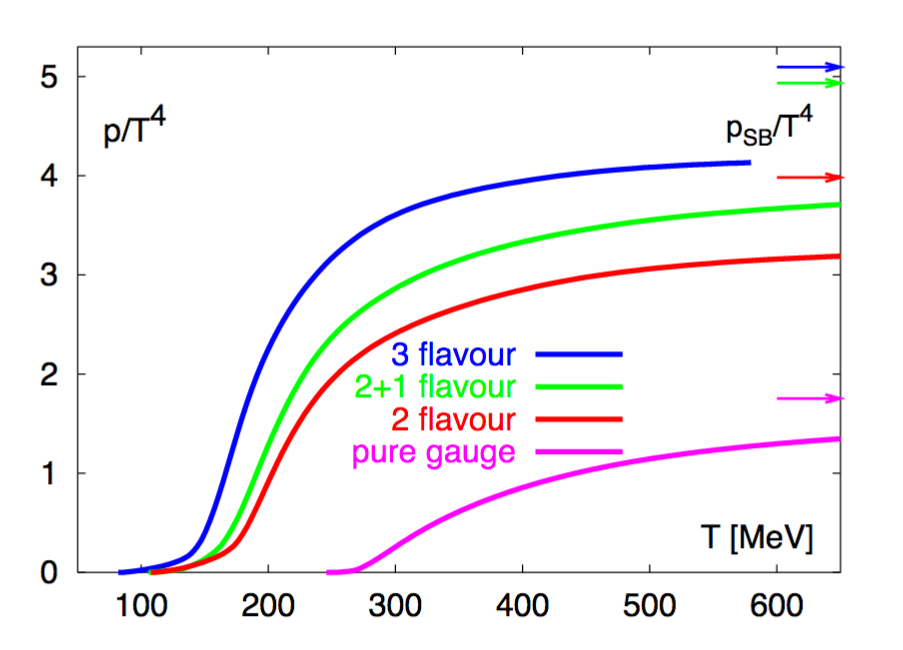}
	   \caption{Lattice QCD calculations showing the pressure vs temperature for a variety of systems as mentioned in the text. The respective arrows on the right axis show the steffan boltzman estimation which is well above the estimate from the calculations~\cite{Karsch:2003jg}.}
	   \label{fig:latticePrevsTemp}
	\end{figure}

Lattice QCD has to tools for \emph{ab-initio} calculation of non perturbative physics at the energy scales relevant to high energy physics~\cite{Karsch:2003jg}. 
	Lattice calculations as shown in Fig:~\ref{fig:latticePrevsTemp} show the pressure vs temperature curve for a variety of systems composed of 3 light flavor quarks, two light flavor plus one heavy flavor, two light flavors and a pure gauge field as shown in blue, green, red and purple respectively. The Steffan-Boltzman limit, as shown in the arrows on the right axis, is well above the lattice calculations, marking the level of non-ideal fluid behavior. The curves show a flattening after crossing the critical temperature threshold i.e. around 180-200 MeV for the three or two flavor calculations. Experiments at RHIC are working on measuring fluctuations in particle densities and production rates to help pinpoint the phase transition and its relevant temperature scale. 
	
\begin{figure}[h]
	   \centering
	   \includegraphics[width=0.7\textwidth]{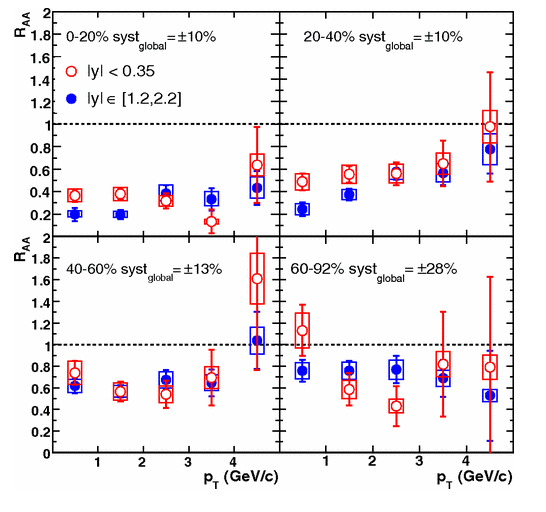}
	   \caption{Melting of J/Psi yield in Au-Au collisions at RHIC measured by the PHENIX collaboration~\cite{PhysRevLett.98.232301}. The open red circles and filled blue circles represent mid rapidity and forward rapidity while the panels show head-on in top left to peripheral collisions in the bottom right.}
	   \label{fig:phenixJPSIsuppresion}
	\end{figure}

 	We trust binary scaling, as introduced in chapter 2, since \raa~for electroweak boson yields have been measured and shown to be consistent with unity~\cite{Florent:2014qja}. Mesonic and baryonic states are characterized by their binding energy which is different for different quark flavors such as the J/$\psi$ and $\Upsilon$ particles. By measuring the reduced yield of such states in heavy ion collisions compared to p-p, one can affirm the presence of the QGP medium indirectly. That is why the melting of J/$\psi$ and $\Upsilon$ yields are colloquially called a QGP thermometer, since by measuring the yield of sequentially heavier states, one can infer (via model fitting) the temperature at various stages of the QGP development. The PHENIX collaboration at RHIC has measured this result~\cite{PhysRevLett.98.232301} as shown in Fig:~\ref{fig:phenixJPSIsuppresion} where the \raa~is plotted as a function of the transverse momenta (\pt) of the identified J/$\psi$ at mid rapidity (for a discussion of the kinematic variables see Appendix \ref{app_kinematics}) in red open circles and forward/backward in blue filled circles. The different panels represent central or head-on collisions from the top left to peripheral collisions in the bottom right. There is a stark reduction in the particle yield in the central collisions with an increasing \raa~as we go from central to peripheral collisions.  

	Along with a suppression in J/$\psi$, another famous signature of the QGP is strangeness enhancement where the yield of strange mesons/baryons are increased compared to light mesons, for example, pions. This particular enhancement was shown in a variety of early experiments followed by theoretical calculations~\cite{Bass:1998vz, Ahle:1998gv, Soff:1999et}.   

	One of the major discoveries of the QGP was its remarkable perfect fluid like properties. QGP seemed to flow with a small viscosity and any initial state anisotropy seemed to directly translate to the final state particles. The flow coefficients ($v_{n}$) of the QGP can be extracted by fourier transforming the final state identified particle's azimuthal distribution; 
	
	\begin{equation}
		\frac{dN}{d\phi} \approx 1 + \sum_{n} 2 v_{n} \cos n (\phi - \psi_{\rm{EP}}) 
	\end{equation}  
	where $\phi$ is the azimuthal angle and $\psi_{\rm{EP}}$ is the angle of the particle in the event plane. 
	
\begin{figure}[h]
	   \centering
	   \includegraphics[width=0.7\textwidth]{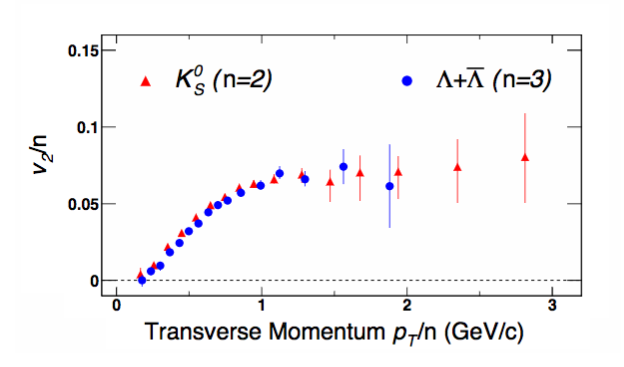}
	   \caption{Constituent quark scaling for identified particle $v_{2}$ as measured by the STAR collaboration~\cite{PhysRevLett.92.052302}. The blue circles and the red triangles show the constituent quark scaled $v_{2}$ as a function of the transverse momentum for $\Lambda+\bar{\Lambda}$ and $K^{0}_{s}$.}
	   \label{fig:starv2ncqscaling}
	\end{figure}

	The STAR collaboration measured the elliptical flow coefficient ($v_{2}$) for identified particles~\cite{PhysRevLett.92.052302}, such as $K^{0}_{s}$ and $\Lambda$ as a function of their transverse momentum. A remarkable property of these distributions was that if these were scaled by the number of constituent quarks, then the distributions match up as shown in Fig:~\ref{fig:starv2ncqscaling}. This scaling suggests that these particles were created in an environment after the collision with quark and gluon degrees of freedom strongly indicating the presence of the QGP. 

\clearpage


\chapter{Status of Jet Structure Observables}
\label{app_jetstructure}
	There are several jet structure measurements published from all three major experiments at the LHC and we will highlight a few here. With regards to jet structure, observables are split into two broad categories depending on their analysis of the momentum scale or the angular scale in the jet constituents. 
	
\begin{figure}[h!] 
	   \centering
	   \includegraphics[width=0.9\textwidth]{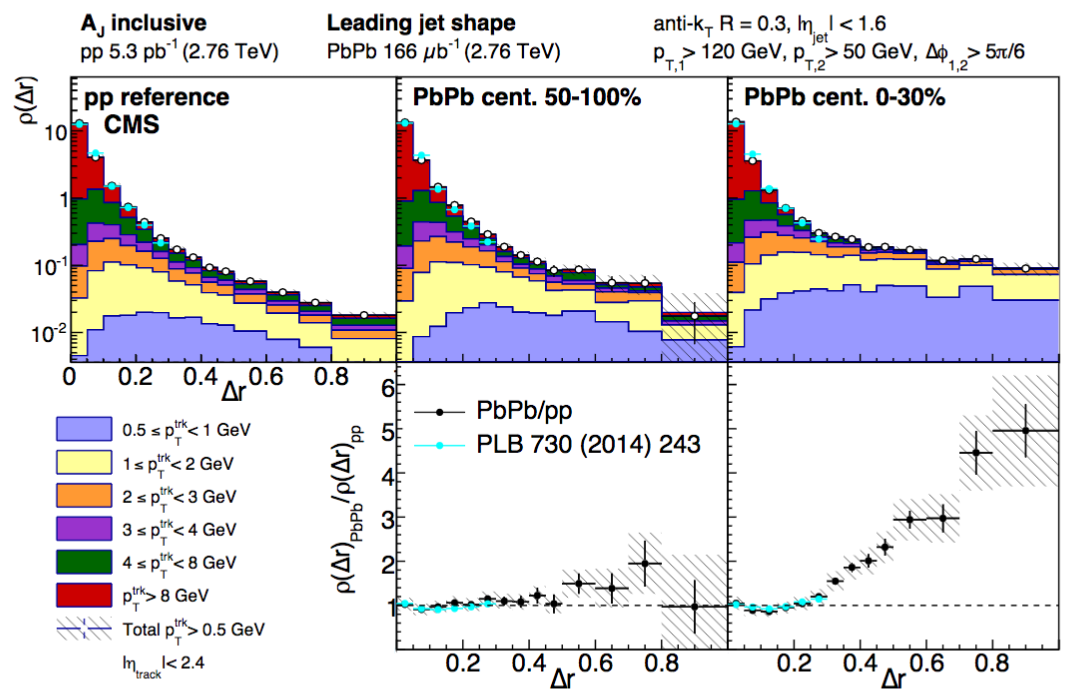} 
	   \caption{CMS jet shape measurement~\cite{Khachatryan:2016tfj} of the \pt~ weighted track density as a function of the distance from the jet axis $\Delta r$ for pp (left), peripheral (middle) and central PbPb events (right). The bottom panels show the ratio PbPb/pp.}
	   \label{fig:cmsjetshape}
	\end{figure}

	The jet shape is defined as the track density as a function of angular distance from the jet axis, weighted by the $p^{track}_{T}/p^{jet}_{T}$ within a given $\Delta r$ region starting from the jet axis and outwards. This density is called $\rho$ and in the top panels of Fig:~\ref{fig:cmsjetshape}, we see the distribution for data in pp (left), peripheral PbPb (middle) and central PbPb (right). The distributions are split into different track momenta bins as showed in the different colors from low \pt~ in the purple and high \pt~ tracks in the red. The ratio of PbPb to pp is shown in the bottom two panels for peripheral and central events. The black points are the latest CMS~\cite{Khachatryan:2016tfj} data and the light blue points are from an earlier result. We see a very large increase of particle density at the large $\Delta r$ region away from the jet axis dominated by the low \pt~ tracks. 
		
\begin{figure}[h!] 
	   \centering
	   \includegraphics[width=0.8\textwidth]{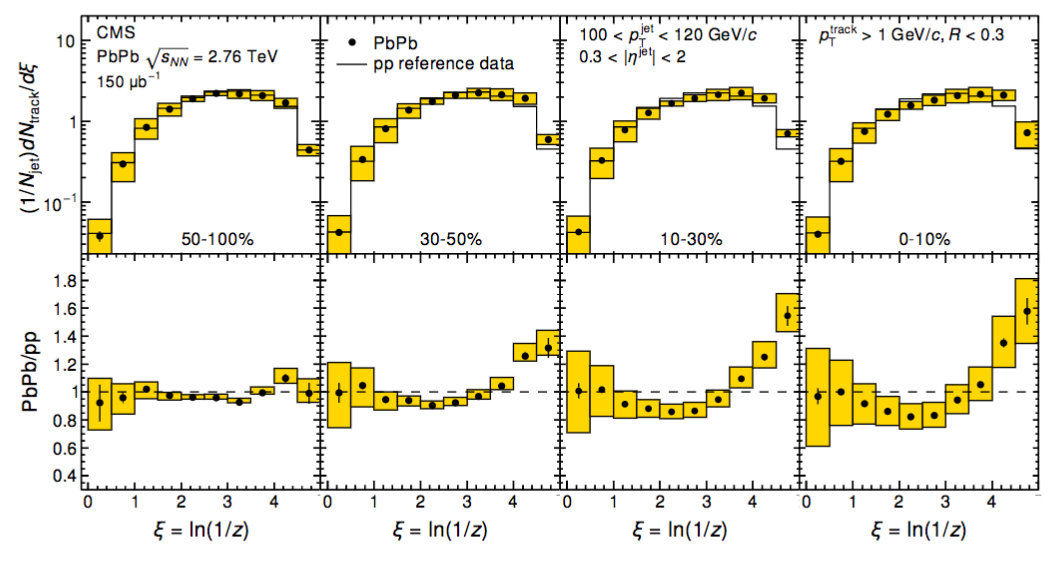} 
	   \includegraphics[width=0.7\textwidth]{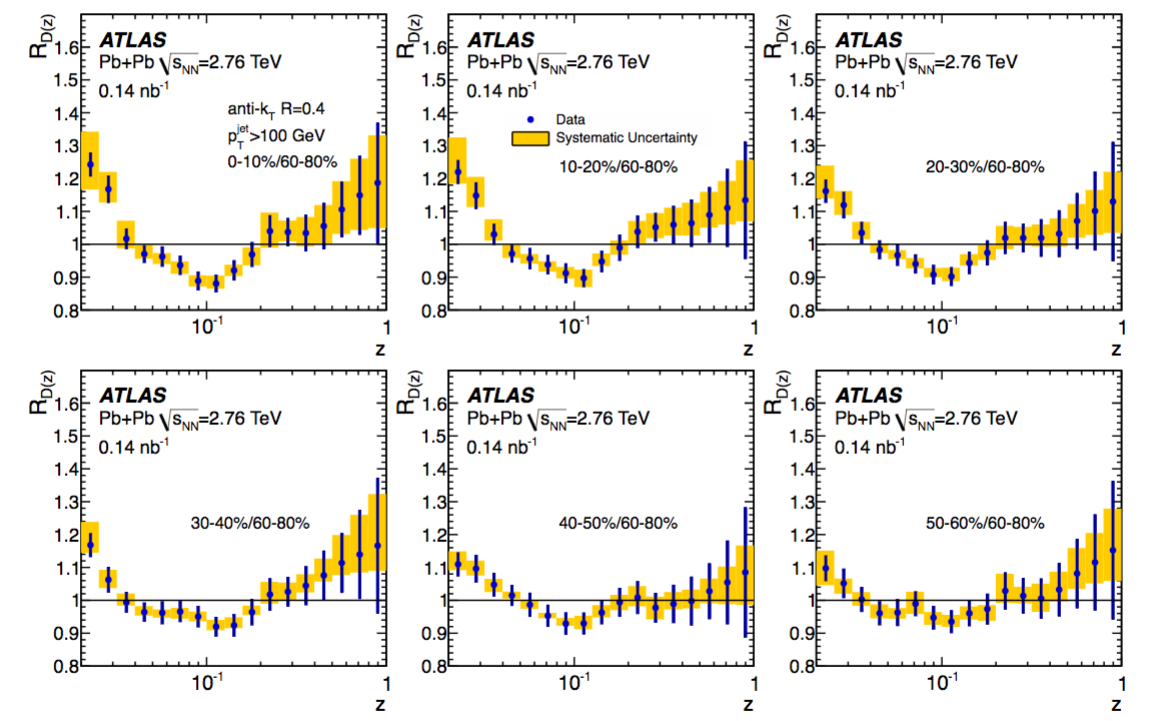} 
	   \caption{Fragmentation function measurements from CMS~\cite{Chatrchyan:2014ava} and ATLAS~\cite{Aad:2014wha}. The yellow shaded region is the data systematic uncertainty. Please refer text for details regarding the measurement.}
	   \label{fig:ffcmsatlas}
	\end{figure}

	On the other hand, the fragmentation function plots the distribution of the track's momenta, either as $\xi = \ln(1/z)$ preferred by CMS~\cite{Chatrchyan:2014ava} or $z = p^{track, \parallel}_{T}/p^{jet}_{T}$, preferred by ATLAS~\cite{Aad:2014wha} within a $\Delta r$ cutoff and a low \pt~ cutoff based on fake rejection and track reconstruction efficiency. This is the same fragmentation function that we introduced in an earlier chapter and is a distribution, normalized by the number of jets, of finding a track with a given fractional momenta $z$. The CMS result is shown in Fig:~\ref{fig:ffcmsatlas}, where the top panels show the PbPb and pp distributions and the bottom panel the ratio PbPb/pp, with central bins on the right and peripheral on the left. The ATLAS result is plotted a central/peripheral for six different centrality bins with central being the top left and peripheral in the bottom right. The large $\xi$, or low track momenta and low $z$, we can see an increase for central collisions which is consistent from the previous jet shape analysis and at low $\xi$, or large track momenta and high $z$, the data is consistent with unity with possibly an updated behavior that is not disallowed.  

	Another jet structure observable measured by ALICE~\cite{Cunqueiro:2015dmx} is the first radial moment of the jet. The radial moments are a class of jet structure observables defined as follows 
	\begin{equation}
		g^{\beta} = \sum_{k \in J} \frac{p^{k}_{T} \left(\Delta R\right)^{\beta}}{p^{J}_{T}}.
	\end{equation} 
	For $\beta = 1$, its the first moment and by tuning this parameter, one can increase or decrease the sensitivity to the angular scale in a jet. The ALICE measurement for PbPb charged jets is shown in Fig:~\ref{fig:alicegirth} and its compared to a \py tune. The PbPb data, in the kinematic region measured, appears to be slightly shifted to the left or smaller values of $g$ which corresponds to a narrowing of the core of the jet.   

	\begin{figure}[h!] 
	   \centering
	   \includegraphics[width=0.7\textwidth]{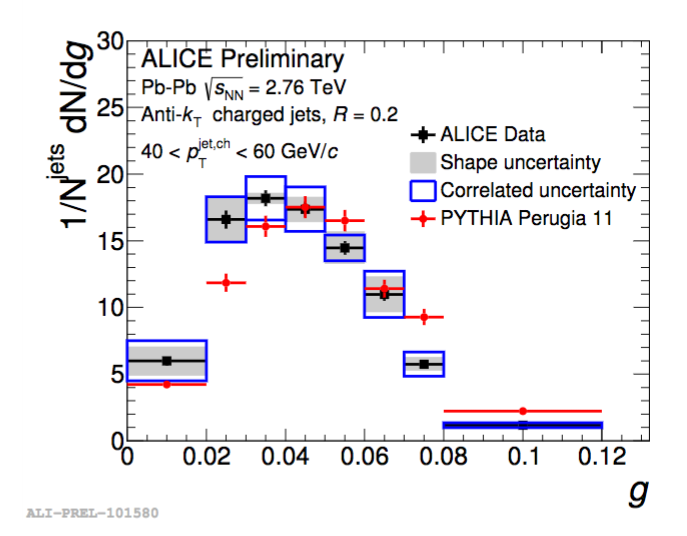} 
	   \caption{Measurement of the jet girth or first radial moment from ALICE~\cite{Cunqueiro:2015dmx}. The central PbPb data is for R=0.2 charged jets and its compared with a \py tune.}
	   \label{fig:alicegirth}
	\end{figure}
		
		Two of the recent measurements that have generated a lot of interest and discussion in the community are the splitting functions and the jet mass. Both these results probe the structural composition of a jet and its modification, whether it be the core or periphery, as a result of interaction with the medium. 
		
		\begin{figure}[h!] 
		   \centering
		   \includegraphics[width=0.9\textwidth]{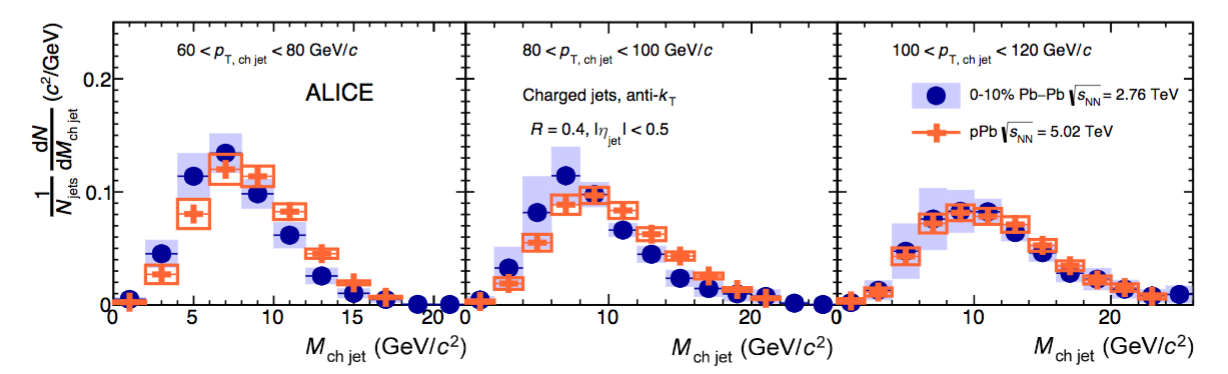} 
		   \caption{Charged jet mass distributions for pPb and central PbPb collisions at ALICE in three different jet \pt~ bins, increasing from left to right.}
		   \label{fig:alicejetmass}
		\end{figure}
	
		Lets begin with the charged jet mass result from ALICE~\cite{Acharya:2017goa} in bins of jet \pt~ for PbPb and pPb. These are 2-D unfolded measurements, in mass and \pt~ for R=0.4 jets only composed of charged particles. The three panels shown in Fig:~\ref{fig:alicejetmass} correspond to different jet \pt~ bins and we see the distributions normalized to number of jets for pPb in yellow diamonds and central PbPb in blue circles. Even though, they are at different center of mass energies, the distributions seem to agree with each other with the sudakov peak\footnote{The jet mass is directly proportional to the splitting fraction times the angular scale squared. Thus the most common or peak of a mass distribution is dominated by the sudakov factor.} nicely placed around a tenth of the average jet \pt~ in the momentum range studied. It is not shown in the figure here, but they also have the distributions from a \py pp tune in their paper and it is also comparable to the heavy ion results. The physics interpretation from the result is somewhat difficult to extract, due to the nature of the mass being sensitive to multiple effects, but qualitatively the core of a jet appears to be unmodified. We also see this behavior in the CMS jet shape result where for very small $\Delta r$, the ratio of PbPb/pp appears to be around one.    

		\begin{figure}[h!] 
		   \centering
		   \includegraphics[width=0.8\textwidth]{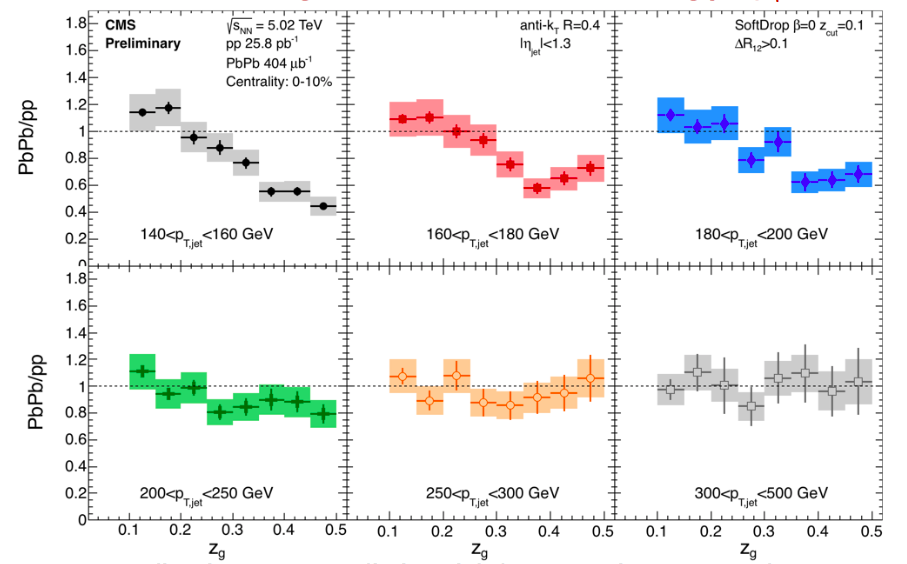} 
		   \caption{Ratio of the subjet groomed shared momentum fraction $z_{g}$ in central PbPb over pp collisions at 5 TeV from the CMS collaboration. The different panels correspond to jet \pt~ ranges from 140-160 \gev~ in the top left to 300-500 \gev~ in the bottom right. Figure taken from here~\cite{Sirunyan:2017jic}.}
		   \label{fig:cmssplitting}
		\end{figure}

		The art of grooming jets is well know and utilized in high energy physics but it has recently found its use in heavy ions to remove soft contributions to a jet's periphery. We take a standard \akt clustered jet and re-cluster its constituents with the C/A algorithm highlighting the angular structure of the jet. Once this is done, we walk backwards along the cluster tree and for each leg, we keep it if it passes the condition 
		\begin{equation}
			z_{g} = \frac{\rm{min}(p^{1}_{T}, p^{2}_{T})}{p^{1}_{T} + p^{2}_{T}} > z_{0} \frac{\Delta R^{\beta}}{R_{0}}.  
		\end{equation} 
		where $z_{0}$ is the energy threshold and $\beta$ is the angular exponent weighting the distance between the two legs $p^{1}_{T}, p^{2}_{T}$. For a special criteria of grooming, called softdrop~\cite{Larkoski:2014wba, Larkoski:2015lea}, the threshold is set at $z_{g} = 10\%$ and the angular exponent is set to $\beta = 0$If the leg fails this condition, then we remove that leg and move on to the next step in the de-clustering. The goal is to potentially hit the first splitting in the jet with this approach and theoretically with the use of grooming, the resulting $z_{g}$ distribution in vacuum, can reproduce the AP splitting. 
		The CMS result~\cite{Sirunyan:2017jic} for the subjet groomed shared momentum fraction is shown in Fig:~\ref{fig:cmssplitting} as a ratio of PbPb to pp collisions in different \pt~ bins. For low \pt~ jets, the splitting seems to favor an enhanced number of asymmetric splitting ($z_{g} < 0.2$) in PbPb collisions followed by suppression of democratic splitting ($z_{g} > 0.3$). Another trend we see in the result, is that the suppression is strongly dependent on the jet \pt~ and disappears when we consider very high \pt~ jets above 300 \gev~. The core of the jet appears to be actually enhanced in PbPb collisions based on this measurement consistent with the jet mass and jet shape, but we need to take this with a grain of salt. This measurement is not unfolded and there is a distinct selection bias related to the smallest $\Delta r$ cutoff allowed due to CMS geometry, leading to an enhanced selection of di-core jets.

\clearpage

\chapter{Abbreviations}
\label{app_abb}

{\setlength{\parindent}{0cm}
CERN : European Center for Research in Nuclear Physics

LHC : Large Hadron Collider

CMS : Compact Muon Solenoid 

ATLAS : A Toroidal LHC ApparatuS 

ALICE : A Large Ion Collider Experiment 

pp : Proton - Proton 

pPb : proton - Lead

PbPb : Lead - Lead

BNL : Brookhaven National Lab 

RHIC : Relativistic Heavy Ion Collider 

STAR : Solenoidal Tracker At RHIC

PHENIX : Pioneering High Energy Nuclear Interaction eXperiment

BES : Beam Energy Scan 

AuAu : Gold - Gold 

d-Au : deutron - Gold 

DESY : Deutsches Elektronen-Synchrotron 

QED : Quantum Electro Dynamics 

QCD : Quantum Chromo Dynamics 

pQCD : perturbative QCD 

QGP : Quark Gluon Plasma

SM : Standard Model 

PDF : Parton Distribution Function 

LO : Leading Order

NLO : Next to Leading Order 

NNLO : Next to Next to Leading Order (and so on and so forth)

LL : Leading Log

NLL : Next to Leading Log 

NP : Non Perturbative

nPDF : nuclear PDF 

MC : Monte Carlo 

JEWEL : Jet Evolution With Energy Loss 

ECAL : Electromagnetic Calorimeter 

HCAL : Hadronic Calorimeter 

HF : Hadronic Forward calorimeter 

ZDC : Zero Degree Calorimeter 

BSC : Beam Scintillator Counters

PU : Pile-Up 

L1 : Level 1 trigger

HLT : High Level Trigger 

Calo : Calorimeter towers 

PF : Particle Flow 

Gen : Generator 

Reco : Reconstruction

JES : Jet Energy Scale

JER : Jet Energy Resolution/Response  

JEC : Jet Energy Correction 

ak : anti-k$_{t}$ algorithm 

PS : Parton Shower 

HAD : Hadronization 

MPI : Multi-Parton Interaction 

SCET : Soft Collinear Effective Theory 

}

\clearpage

\chapter{Publications}
\label{app_pub}

\subsubsection{Journals}
\begin{itemize}
\item \href{https://link.springer.com/article/10.1007/JHEP07(2017)141}{Medium response in JEWEL and its impact on jet shape observables in heavy ion collisions}\\Raghav Kunnawalkam Elayavalli, Korinna Christine Zapp\\JHEP 1707 (2017) 141 \href{http://arxiv.org/abs/1707.01539}{arXiv:1707.01539}
\item \href{https://journals.aps.org/prc/abstract/10.1103/PhysRevC.96.015202}{Measurement of inclusive jet cross-sections in pp and PbPb collisions at $\sqrt{s_{\rm{NN}}}$=2.76 TeV} \\CMS Collaboration\\Phys. Rev. C 96, 015202 \href{http://arxiv.org/abs/1609.05383}{arXiv:1609.05383}
\item \href{http://epjc.epj.org/articles/epjc/abs/2016/12/10052_2016_Article_4534/10052_2016_Article_4534.html}{Simulating V+jet processes in heavy ion collisions with JEWEL}\\Raghav Kunnawalkam Elayavalli, Korinna Christine Zapp\\Eur. Phys. J. C. (2016) 76: 695 \href{http://arxiv.org/abs/1608.03099}{arXiv:1608.03099} 
\item \href{http://link.springer.com/article/10.1140\%2Fepjc\%2Fs10052-016-4205-7}{Measurement of inclusive jet production and nuclear modifications in pPb collisions at $\sqrt{s_{\rm{NN}}}$=5.02 TeV}\\CMS Collaboration\\Eur. Phys. J. C (2016) 76: 372 \href{http://arxiv.org/abs/1601.02001}{arXiv:1601.02001}
\item \href{http://www.sciencedirect.com/science/article/pii/S0370269316000149}{Transverse momentum spectra of inclusive b jets in pPb collisions at $\sqrt{s_{\rm{NN}}}$=5.02 TeV}\\CMS Collaboration\\Phys. Lett. B754 (2016) 59 \href{http://arxiv.org/abs/1510.03373}{arXiv:1510.03373}
\end{itemize}

\subsubsection{Proceedings}
\begin{itemize}
\item \href{}{Medium Recoils and background subtraction in JEWEL} (HP'16)\\Raghav Kunnawalkam Elayavalli, Korinna Christine Zapp\\Accepted by Nuclear and Particle Physics Proceedings \#\# (2017) \#\#\# \href{https://arxiv.org/abs/1612.05116}{arXiv:1612.05116}
\item \href{http://iopscience.iop.org/article/10.1088/1742-6596/832/1/012004/meta;jsessionid=FDFAD94D230B4DC1A7BF5F304ACA0E0C.c3.iopscience.cld.iop.org}{Jet structure modifications in heavy-ion collisions with JEWEL} (HQ'16)\\Raghav Kunnawalkam Elayavalli\\ J. Phys. Conf. Ser. 832 (2017) no.1, 012004  \href{http://arxiv.org/abs/1610.09364}{arXiv:1610.09364}
\item \href{http://iopscience.iop.org/article/10.1088/1742-6596/612/1/012008/meta;jsessionid=8DF75609E70B49952FC7A1FECAB740EE.c4.iopscience.cld.iop.org}{Jet Measurements in Heavy-ion Collisions with CMS} (HQ'14)\\Raghav Kunnawalkam Elayavalli (on behalf of the CMS collaboration)\\J. Phys. Conf. Ser. 612 (2015) no.1, 012008  \href{http://arxiv.org/abs/1607.03281}{arXiv:1607.03281}
\end{itemize}

\clearpage

\end{appendices}

\pagestyle{empty}

\end{document}